\pdfoutput=1

\documentclass[12pt,a4paper]{article}

\usepackage{amsmath}
\usepackage{breqn}
\usepackage{ifthen} 
\newboolean{pdflatex}
\setboolean{pdflatex}{true} 

\newboolean{articletitles}
\setboolean{articletitles}{true} 

\newboolean{uprightparticles}
\setboolean{uprightparticles}{false} 

\newboolean{inbibliography}
\setboolean{inbibliography}{false} 

\usepackage{microtype}
\usepackage{lineno}  
\usepackage{xspace} 
\usepackage{caption} 

\usepackage{graphicx}  
\usepackage{color}
\usepackage{colortbl}
\graphicspath{{./figs/}} 

\usepackage{amsmath} 
\usepackage{amssymb}
\usepackage{amsfonts}
\usepackage{upgreek} 

\newcommand*\patchAmsMathEnvironmentForLineno[1]{%
\expandafter\let\csname old#1\expandafter\endcsname\csname #1\endcsname
\expandafter\let\csname oldend#1\expandafter\endcsname\csname
end#1\endcsname
 \renewenvironment{#1}%
   {\linenomath\csname old#1\endcsname}%
   {\csname oldend#1\endcsname\endlinenomath}%
}
\newcommand*\patchBothAmsMathEnvironmentsForLineno[1]{%
  \patchAmsMathEnvironmentForLineno{#1}%
  \patchAmsMathEnvironmentForLineno{#1*}%
}
\AtBeginDocument{%
\patchBothAmsMathEnvironmentsForLineno{equation}%
\patchBothAmsMathEnvironmentsForLineno{align}%
\patchBothAmsMathEnvironmentsForLineno{flalign}%
\patchBothAmsMathEnvironmentsForLineno{alignat}%
\patchBothAmsMathEnvironmentsForLineno{gather}%
\patchBothAmsMathEnvironmentsForLineno{multline}%
\patchBothAmsMathEnvironmentsForLineno{eqnarray}%
}

\usepackage{hyperref}    
\usepackage[all]{hypcap} 

\def\MagUp {\mbox{\em Mag\kern -0.05em Up}\xspace}

\ifthenelse{\boolean{uprightparticles}}%
{

 \def\Pmu         {\ensuremath{\upmu}\xspace}

 \def\Ppi         {\ensuremath{\uppi}\xspace}

 \def\PDelta      {\ensuremath{\Delta}\xspace}                 
 \def\PXi      {\ensuremath{\Xi}\xspace}                 
 \def\PLambda      {\ensuremath{\Lambda}\xspace}                 
 \def\PSigma      {\ensuremath{\Sigma}\xspace}                 
 \def\POmega      {\ensuremath{\Omega}\xspace}                 
 \def\PUpsilon      {\ensuremath{\Upsilon}\xspace}                 
 
 \def\PB      {\ensuremath{\mathrm{B}}\xspace}                 
                  
 \def\PD      {\ensuremath{\mathrm{D}}\xspace}

 \def\PK      {\ensuremath{\mathrm{K}}\xspace}

 \def\Pi      {\ensuremath{\mathrm{i}}\xspace}

}
{

 \def\Pmu         {\ensuremath{\mu}\xspace}

 \def\Ppi         {\ensuremath{\pi}\xspace}

 \mathchardef\PDelta="7101
 \mathchardef\PXi="7104
 \mathchardef\PLambda="7103
 \mathchardef\PSigma="7106
 \mathchardef\POmega="710A
 \mathchardef\PUpsilon="7107
                  
 \def\PB      {\ensuremath{B}\xspace}                 
                  
 \def\PD      {\ensuremath{D}\xspace}

 \def\PK      {\ensuremath{K}\xspace}

 \def\Pi      {\ensuremath{i}\xspace}

}

\makeatletter
\ifcase \@ptsize \relax
  \newcommand{\miniscule}{\@setfontsize\miniscule{4}{5}}
\or
  \newcommand{\miniscule}{\@setfontsize\miniscule{5}{6}}
\or
  \newcommand{\miniscule}{\@setfontsize\miniscule{5}{6}}
\fi
\makeatother

\DeclareRobustCommand{\optbar}[1]{\shortstack{{\miniscule (\rule[.5ex]{1.25em}{.18mm})}
  \\ [-.7ex] $#1$}}

\def\mup        {{\ensuremath{\Pmu^+}}\xspace}
\def\mun        {{\ensuremath{\Pmu^-}}\xspace} 

\def\pion   {{\ensuremath{\Ppi}}\xspace}

\def\pip    {{\ensuremath{\pion^+}}\xspace}
\def\pipOne    {{\ensuremath{\pion^+_1}}\xspace}
\def\pipTwo    {{\ensuremath{\pion^+_2}}\xspace}
\def\pim    {{\ensuremath{\pion^-}}\xspace}

\def\kaon    {{\ensuremath{\PK}}\xspace}
  \def\Kbar    {{\kern 0.2em\overline{\kern -0.2em \PK}{}}\xspace}

\def\KorKbar    {\kern 0.18em\optbar{\kern -0.18em K}{}\xspace}

\def\Kp      {{\ensuremath{\kaon^+}}\xspace}
\def\Km      {{\ensuremath{\kaon^-}}\xspace}

\def\Kstarz  {{\ensuremath{\kaon^{*0}}}\xspace}
\def\Kstarzb {{\ensuremath{\Kbar{}^{*0}}}\xspace}

  \def\Dbar    {{\kern 0.2em\overline{\kern -0.2em \PD}{}}\xspace}
\def\D       {{\ensuremath{\PD}}\xspace}

\def\DorDbar    {\kern 0.18em\optbar{\kern -0.18em D}{}\xspace}
\def\Dz      {{\ensuremath{\D^0}}\xspace}

\def\B       {{\ensuremath{\PB}}\xspace}
\def\Bbar    {{\ensuremath{\kern 0.18em\overline{\kern -0.18em \PB}{}}}\xspace}

\def\BorBbar    {\kern 0.18em\optbar{\kern -0.18em B}{}\xspace}
\def\Bz      {{\ensuremath{\B^0}}\xspace}
\def\Bzb     {{\ensuremath{\Bbar{}^0}}\xspace}

  \def\Y#1S{\ensuremath{\PUpsilon{(#1S)}}\xspace}

\def\Lbar        {{\ensuremath{\kern 0.1em\overline{\kern -0.1em\PLambda}}}\xspace}
\def\LorLbar    {\kern 0.18em\optbar{\kern -0.18em \PLambda}{}\xspace}

\newcommand{\decay}[2]{\ensuremath{#1\!\to #2}\xspace}         

\def\to                 {\ensuremath{\rightarrow}\xspace}

\def\norm    {\decay{\Dz}{\Km\pip\pip\pim}}
\def\kstmumu    {\decay{\Bz}{\Kstarz(\rightarrow\Kp\pim)\mup\mun}}
\def\Kspipi    {\decay{\Dz}{K^0_S\pip\pim}}

\def\AT#1     {\ensuremath{A_{\mathrm{T}}^{#1}}\xspace}           

\def\C#1      {\ensuremath{\mathcal{C}_{#1}}\xspace}                       
\def\Cp#1     {\ensuremath{\mathcal{C}_{#1}^{'}}\xspace}                    
\def\Ceff#1   {\ensuremath{\mathcal{C}_{#1}^{\mathrm{(eff)}}}\xspace}        
\def\Cpeff#1  {\ensuremath{\mathcal{C}_{#1}^{'\mathrm{(eff)}}}\xspace}       
\def\Ope#1    {\ensuremath{\mathcal{O}_{#1}}\xspace}                       
\def\Opep#1   {\ensuremath{\mathcal{O}_{#1}^{'}}\xspace}                    

\newcommand{\tev}{\ifthenelse{\boolean{inbibliography}}{\ensuremath{~T\kern -0.05em eV}\xspace}{\ensuremath{\mathrm{\,Te\kern -0.1em V}}}\xspace}
\newcommand{\gev}{\ensuremath{\mathrm{\,Ge\kern -0.1em V}}\xspace}
\newcommand{\mev}{\ensuremath{\mathrm{\,Me\kern -0.1em V}}\xspace}
\newcommand{\kev}{\ensuremath{\mathrm{\,ke\kern -0.1em V}}\xspace}
\newcommand{\ev}{\ensuremath{\mathrm{\,e\kern -0.1em V}}\xspace}
\newcommand{\gevc}{\ensuremath{{\mathrm{\,Ge\kern -0.1em V\!/}c}}\xspace}
\newcommand{\mevc}{\ensuremath{{\mathrm{\,Me\kern -0.1em V\!/}c}}\xspace}
\newcommand{\gevcc}{\ensuremath{{\mathrm{\,Ge\kern -0.1em V\!/}c^2}}\xspace}
\newcommand{\gevgevcccc}{\ensuremath{{\mathrm{\,Ge\kern -0.1em V^2\!/}c^4}}\xspace}
\newcommand{\mevcc}{\ensuremath{{\mathrm{\,Me\kern -0.1em V\!/}c^2}}\xspace}

\def\mum  {\ensuremath{{\,\upmu\rm m}}\xspace}

\def\gsim{{~\raise.15em\hbox{$>$}\kern-.85em
          \lower.35em\hbox{$\sim$}~}\xspace}
\def\lsim{{~\raise.15em\hbox{$<$}\kern-.85em
          \lower.35em\hbox{$\sim$}~}\xspace}

\def\pt         {\mbox{$p_{\rm T}$}\xspace}

\def\tell1  {TELL1\xspace}
\def\ukl1   {UKL1\xspace}

\usepackage{cite} 
\usepackage{mciteplus}

\usepackage{longtable} 

\usepackage{mciteplus}
\usepackage{float}
\usepackage[toc,page]{appendix}

\begin{document}

\renewcommand{\thefootnote}{\fnsymbol{footnote}}
\setcounter{footnote}{1}


\begin{titlepage}
\pagenumbering{roman}

\vspace*{0.cm}
\noindent
\begin{tabular*}{\linewidth}{lc@{\extracolsep{\fill}}r@{\extracolsep{0pt}}}
\vspace*{-2.cm}\mbox{\!\!\!\includegraphics[width=.3\textwidth]{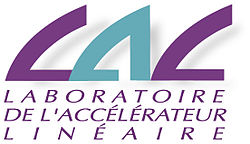}} & &%
\\
 & & LAL 16-009 \\  
 & & \today \\ 
 & & \\
\end{tabular*}

\vspace*{1.5cm}

{\bf\boldmath\huge
\begin{center}
On the potential of multivariate techniques for the determination 
of multidimensional efficiencies. 
\end{center}
}

\vspace*{0.4cm}

\begin{center}
B. Viaud$^{a}$

{\footnotesize \it$^{a}$ LAL, Univ. Paris-Sud, CNRS/IN2P3, Universit\'e Paris-Saclay, 91400 Orsay, France}

\end{center}

\vspace{0.4cm}

\begin{abstract}
  \noindent

Differential measurements of particle collisions or decays can provide
stringent constraints on physics beyond the Standard Model of particle
physics. In particular, the distributions of the kinematical and angular
variables that characterise heavy meson multibody decays are non trivial and 
can sign the underlying interaction physics. 
In the era of high luminosity opened by the advent of the Large Hadron Collider and of Flavor Factories, 
differential measurements are less and less dominated by statistical precision and require a precise determination of
efficiencies that depend simultaneously on several variables and do not factorise in these variables.
This document is a reflection on the potential of multivariate techniques for the determination 
of such multidimensional efficiencies. 
We carried out two case studies that show that multilayer perceptron neural networks can determine and correct
for the distortions introduced by reconstruction and selection criteria in the multidimensional phase space
of the decays \kstmumu and \norm, at the price of a minimal analysis effort. We conclude that this method
can already be used for measurements which statistical precision does not yet reach the percent level and that with 
more sophisticated machine learning methods, the aforementioned potential is very promising. 
\end{abstract}

\vspace*{2.0cm}

\end{titlepage}


\newpage
\setcounter{page}{2}
\mbox{~}
%
%

\cleardoublepage


\renewcommand{\thefootnote}{\arabic{footnote}}
\setcounter{footnote}{0}



\pagestyle{plain} 
\setcounter{page}{1}
\pagenumbering{arabic}

\section{Introduction}
\label{sec:Introduction}

With the advent of the LHC and, a few years before, of Flavor 
Factories, High Energy Physics (HEP) entered an era of high statistical precision.
More and more differential measurements of particle collisions 
or decays are now possible. For instance, stringent 
constraints can be imposed on the models predicting the dynamics of the decay \kstmumu by
measuring the distribution of this decay in the four-dimensional 
space defined by $q^2$, cos$\theta_l$, cos$\theta_K$ and $\phi$, a
set of independent variables that provide a full description of the 
decay dynamics and that are defined in Sect.~\ref{sec:TechSummClass}. Deviations from the 
Standard Model (SM) predictions in specific regions of the phase space can be 
detected this way and sign the action of a physics beyond the SM.
Loop-mediated rare decays like \kstmumu are particularly sensitive to this New Physics.  
Many models in which particles do not couple to weak interaction via only their 
left chiral component predict angular distributions that differ from the SM ones. 
More detail can be found, for instance, in Ref.~\cite{LHCb_KstMuMu_conf}.

In analyses of the kind introduced above, one has to account for the distortion of the phase space
caused by reconstruction and selection criteria. In the example above, this is the distortion of the  
($q^2$, cos$\theta_l$, cos$\theta_K$, $\phi$) distribution. The most straightforward method would be to use a 
sample of simulated \mbox{\kstmumu} phase-space decays.
A four-dimensional binning could be defined, and the efficiency in each bin determined
by the ratio between the yield of reconstructed and selected events and the yield of generated events.  
If this determination is made in terms of all the kinematic variables describing the 
decay and if the granularity of the binning is fine enough, the result does not
depend on the distributions assumed by the simulation for these variables. This assumption often 
relies on decay models poorly predicted by theory. However, this method would necessit to generate
a huge sample. Even with only 10 bins per dimension, 10000 four-dimensional bins have to be defined.
It takes typically millions of generated events to determine the efficiency bins with less than a 5\% uncertainty. 
Instead, sophisticated methods are available to account for efficiencies that depend simultaneously on several 
variables and that do not factorise in these variables (see Sect.~\ref{sec:TechSummClass}).

We propose in this document to explore the potential of another approach, suggested by 
rapid progresses  in machine learning and multivariate analysis observed in the last decade.
Techniques such as Neural Networks (NN)~\cite{Haykin:2009} or Boosted Decision Trees (BDT)~\cite{Roe:2004na}, 
among others,  
can now detect with high sensitivity differences between two samples of events  characterised by a 
large number of variables. They are routinely used nowdays in particle physics measurements
to tell signal from background events. Said otherwise, these techniques are performant at comparing 
$n$-dimensional distributions to detect even subtle differences between signal and background samples 
(that traditional ``by eye'' studies would miss) and discrimate between event types via the NN or BDT score, 
a single variable incorporating all the information
found in $n$ dimensions. They should naturally be sensitive also to distortions introduced in the phase space of
particle collisions or decays by reconstruction and selection criteria. 

We do not take for granted that multivariate techniques reach the same level of precision as
methods such as the principal moment analysis described in~\cite{LHCb_KstMuMu_conf}, nor that such a result can be
obtained without a certain expertise in MVA techniques or without a time-consumming optimisation of 
the parameters that rule the technique's behavior and performance. However, the rythm at which MVA 
techniques have been progressing recently, their growing availability to basic users in the form of user-friendly packages, and 
the increasing typical expertise of high energy physicists suggest MVAs might soon become 
very valuable tools, and easy to use, for multidimensional efficiency determination. It is therefore
interesting to start exploring the potential of this approach. 

This document only starts this exploration, with the ambition to answer the following question:
what can be achieved if multivariate techniques are used in a very simple manner, i.e.
without devoting more than a few hours to their optimisation --- therefore using
generic settings, like the default ones found in packages like~\cite{TMVA}? In the case
of analyses which do not require the same precision as the analysis of \mbox{\kstmumu} introduced above, 
is this approach enough ? Also, our goal is not to provide quantitative results, but an illustration of 
the potential of this approach.

In this document, we will first describe briefly a typical technique used to treat 
multidimensional efficiencies, and describe the approach we propose (Sect.~\ref{sec:TechSumm}).
Then, we will apply it to a first test case, involving the decay \norm  (Sect.~\ref{sec:AppliedToDK3pi}).
This decay can be used to improve our knowledge of $D$-meson mixing. For that purpose, one needs
to determine the selection efficiency across the space defined by five independent variables in terms
of which the decay dynamics can be expressed.  Another test case will involve \mbox{\kstmumu} (Sect.~\ref{sec:AppliedToKstMuMu}). 
The results observed in these two cases will be summarized and discussed in the conclusion (Sect.~\ref{sec:Conclusion}).

\section{Techniques for multidimensional efficiencies}
\label{sec:TechSumm}

\subsection{A classical technique}
\label{sec:TechSummClass}

\begin{figure}[tb]
  \begin{center}
    \hspace*{-1.3cm}
    \includegraphics[width=0.6\linewidth]{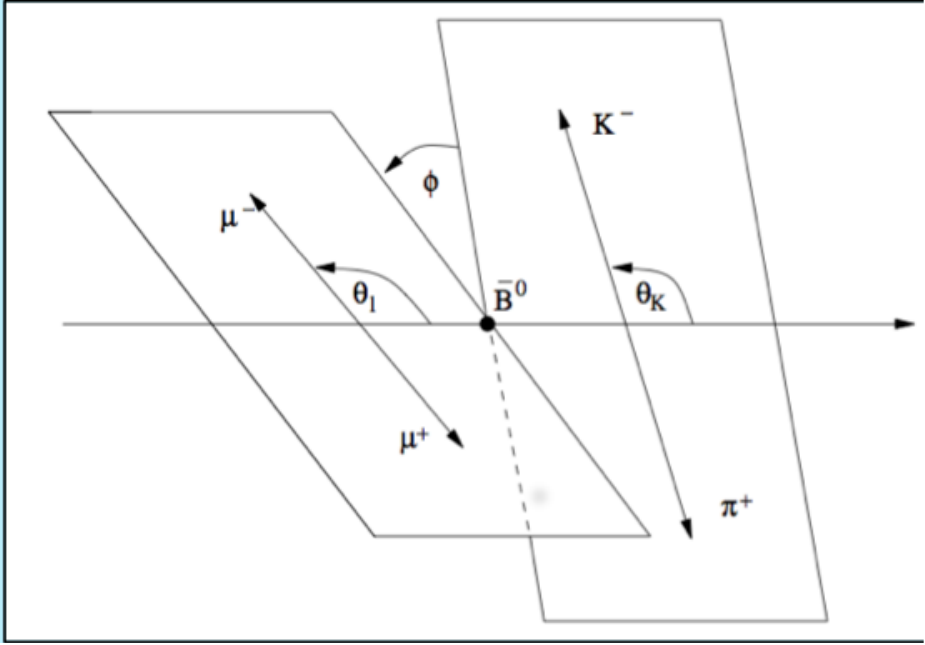}
    \vspace*{-0.3cm}
  \end{center}
  \caption{Graphical representation of cos$\theta_l$, cos$\theta_K$ and $\phi$. Their precise definitions are given in the text.} 
  \label{fig:bkstarmumu_angles}
\end{figure}

Methods of a certain complexity can be used to account for efficiencies that depend simultaneously on several 
variables and {\it that do not factorise in these variables}. The study of the decay \kstmumu provides a typical example~\cite{LHCb_KstMuMu_conf}.
The dymamics of this decay is fully described by four independent variables, $q^2$, cos$\theta_l$, cos$\theta_K$ and $\phi$. They are 
defined as the invariant mass of the dimuon system squared, the cosine of the angle between the \mup (\mun) and the direction 
opposite the \Bz (\Bzb) in the rest frame of the dimuon system, the cosine of the angle between the
direction of the $K^+$ ($K^{-}$) and the \Bz (\Bzb)  in the rest frame of the \Kstarz 
(\Kstarzb) system, and the angle between the plane defined by the \mup and \mun and the plane
defined by the kaon and pion in the \Bz (\Bzb) rest frame, respectively. The definition of the three
angles above is illustrated on Fig.~\ref{fig:bkstarmumu_angles}.
In~\cite{LHCb_KstMuMu_conf}, 
a long sum of products of Legendre polynomials in the concerned variables is used :
\begin{equation}
\label{eq:principalmoment}
\epsilon\left(cos\theta_l,cos\theta_K,\phi,q^{2}\right) = \sum_{klmn}c_{klmn}P_{k}(cos\theta_l)P_{l}(cos\theta_K)P_{m}(\phi)P_{n}(q^{2}),
\end{equation}
where the terms $P_{i}(x)$ stand for Legendre polynomials of order $i$. The coefficients $c_{klmn}$ are 
evaluated by performing a principal moment analysis of simulated \mbox{\kstmumu} phase-space decays:
\begin{dmath}
c_{klmn}=\frac{1}{N^{'}}\sum_{i}^{N}\omega_i\left[\left(\frac{2k+1}{2}\right)\left(\frac{2l+1}{2}\right)\left(\frac{2m+1}{2}\right)\left(\frac{2n+1}{2}\right)
\times P_{k}(cos\theta_l)P_{l}(cos\theta_K)P_{m}(\phi)P_{n}(q^{2})\right],
\end{dmath}
where $N$ is the total number of events and $\omega_{i}$ are  weights imposed for instance to correct 
for known discrepancies between data and simulation. Their sum provides the total normalisation $N^{'}$. 
For $cos\theta_l$, $cos\theta_K$, the angle $\phi$ and $q^{2}$, polynomials up
to order 5, 6, 6 and 7 are used, respectively (see Ref.~\cite{LHCb_KstMuMu_conf}).  
Designing the method, understanding the properties of the  Legendre 
polynomials and,  more generally,  of the sum in Eq.~\ref{eq:principalmoment}, implementing the software to compare 
this parametrisation with the efficiencies observed for simulated decays so as to determine the highest orders to include until 
a proper description of the efficiency is obtained and to determine the $c_{klmn}$ coefficients, and interpretating the 
results are time demanding tasks. In the end, hundreds of $c_{klmn}$ are necessary. This might be worse in cases where 
more than four variables have to be dealt with, and it's not obvious the method will always work accurately.

\subsection{A new approach}
\label{sec:TechSummNew}

The approach we propose is based on the idea that if a BDT or a NN is very powerful 
at detecting differences between a signal and a background sample based on a given set 
of $n$ variables, not by considering them individually, but by also exploiting
their correlations, i.e. by comparing $n$-dimensional distributions, it should also be powerful at finding 
the differences between two samples differing only due to 
reconstruction and selection biases. In this case, instead of training the multivariate 
discriminator by comparing a $signal$ and a $background$ sample, one would compare samples of the 
same decay: an $original$ sample made of generator level decays with a $distorted$ sample made of decays that satisfied 
reconstruction and selection criteria. Instead of comparing discriminating variables, one would focus on 
the phase space of the decay, or in other words 
the $n$-dimensional distribution of events in the space defined by a set of independent variables that
can describe fully the decay dynamics. In the example of the decay \kstmumu, this is 
($q^2$, cos$\theta_l$, cos$\theta_K$, $\phi$). 

In a four-stage approach, we first generate an original MC sample and a distorted one. The latter is generated in
the same way as the former, save that reconstruction and selection cuts are applied. 
This stage is mandatory in most if not all physics analyses in HEP. The approaches like the 
ones in Sect.~\ref{sec:TechSummClass} need that too. When high precision is necessary
one generates samples containing up to a few million events. In most HEP collaborations,  
generating larger samples is challenging due to limited CPU and data storage capabilities.  
In the test cases presented in Sect.~\ref{sec:AppliedToDK3pi} and~\ref{sec:AppliedToKstMuMu}, 
we use the \texttt{ROOT} package~\cite{Brun:1997pa}, and more specifically the \texttt{TGenPhaseSpace} class to generate these samples. 

The second step is to train a multivariate analysis by comparing the samples generated
above. In the case studies we carried out, we used the \texttt{TMVA}~\cite{TMVA} package to train Multilayers Perceptron NNs. 
The only variables we provided the NNs with are the phase space variables.  

The third stage is to generate additional original and distorted samples, independent of those
used above for training, and to compute for each event the NN score.
It summarises into one single variable all the difference detected between the original and 
distorted phase spaces. The reconstruction and the selection efficiency can then be parameterised in terms of only this single 
variable, which makes the task of accounting for this multidimensional 
efficiency far more practical. This is the fourth stage, where one computes the ratio of the NN score distribution obtained 
in the distorted sample to the distribution in the original sample. Fitting this ratio provides the parameterization, and therefore
a per-candidate efficiency as a function of the score, that can be used, e.g., to weight the distorted sample so as to reproduce the phase 
space (and the various characteristic distributions) of the original sample.

\section{\mbox{Application to the study of {\bf\boldmath \norm}}}
\label{sec:AppliedToDK3pi}

Multibody \Dz decays provide examples where multidimensional efficiencies must be determined. 
These decays can be used, for instance, to improve our understanding 
of \Dz mixing~\cite{PDG2014}. One of the most studied decay is \Kspipi. Its phase space can be described 
in terms of two invariant masses squared: $m^{2}(K^{0}_{S}\pip)$ and $m^{2}(K^{0}_{S}\pim)$. 
The distribution of the  \Kspipi decays in the Dalitz plane~\cite{Dalitz:1953cp} defined by $m^{2}(K^{0}_{S}\pip)$ and $m^{2}(K^{0}_{S}\pim)$
varies in time under the action of the mixing. By measuring this variation, one can access directly the mixing
parameters $x=\frac{M_1-M_2}{\Gamma}$ and $y=\frac{\Gamma_1-\Gamma_2}{2\Gamma}$~\cite{Asner:2005sz,Abe:2007rd,delAmoSanchez:2010xz}, 
where $M_{1}$($M_{2}$),  
$\Gamma_1$($\Gamma_2$) and $\Gamma$ are the masses of the two neutral $D$ physical states, the two
corresponding widths, and the average \Dz width, respectively. In simpler approaches, only a combination 
of $x$ and $y$ with a phase due to the strong interaction can be measured. Such Dalitz analyses also open the 
possibility to measure CP violation in the mixing. They can be extended to four-body decays, where the 
phenomenology is richer and can therefore provide more information. In this case, a five-dimensional Dalitz analysis 
must be carried out. Our case study in this document is the \norm decay, which dynamics can be described by the 
following set of masses:
\begin{equation*}
\begin{array}{c}
m_{12}=m(\pipOne\pim) \\
m_{23}=m(\pipTwo\pim) \\
m_{34}=m(\pipTwo\Km) \\
m_{123}=m(\pipOne\pim\pipTwo) \\
m_{234}=m(\pim\pipTwo\Km). \\
\end{array}
\end{equation*}

\subsection{Generation of the distorted and original samples}

We generated an original sample and a distorted sample of \norm decays.
Both contain $\sim$ 230000 decays. A sample of reconstructed and selected 
events of this size is a significant computing effort if a full simulation 
(physics, detector response and reconstruction) has to be performed. However, this is still in the 
scope of what a collaboration like LHCb, the present world leading experiment in Flavor Physics~\cite{LHCb-DP-2014-002}, 
is ready to produce even for measurements of intermediate importance.

To produce samples that can be compared with the samples used by LHCb in~\cite{LHCb_D2K3pi_ANA}, 
an important element is to generate 
decays that have the same kinematics as in LHCb's laboratory frame. This is not possible
with the \texttt{TGenPhaseSpace} class alone, which knows nothing of the physics of the 
proton-proton collision where $D^0$ mesons are produced. Using this class one 
can generate the four-momentum of each daughter particles given the four-momentum of the decaying particle, 
assuming a flat phase space ( i.e. assuming that all the combinations of daughter four-momenta that respect 
energy and momentum conservation are equally allowed). To reproduce the kinematics of \Dz mesons 
produced in proton-proton collisions at a center-of-mass energy of 7 or 8 TeV, we combined 
several samples, each generated assuming a different value of the \Dz transverse momentum and rapidity. 
The relative contribution of each sample to the final one was decided according the \Dz production cross sections
measured in Ref.~\cite{Aaij:2013mga} as a function of these quantities. This is how the original
sample was produced. 

To produce the distorted sample, we first generated decays as described above, 
and filtered them according to a selection which aims to be as close as possible to 
the selection used in Ref.~\cite{LHCb_D2K3pi_ANA}.
At LHCb, the selection of a given $M$ meson decay typically involves:
\begin{itemize}
\item The momentum ($p$) and transverse momentum ($p_{\mathrm{T}}$) of the mother and daughter particles.
\item The minimum distance of a track to a proton-proton primary vertex (PV), called the impact parameter (IP).
\item The difference in $\chi^{2}$ of the closest PV reconstructed with and without the particle under
consideration ($\chi^{2}_{\mathrm{IP}}$). This particle can either be $M$ or one of its decay products. 
\item The $\chi^{2}$ evaluating the quality of the $M$ decay vertex ($\chi^{2}_{\mathrm{Vtx}}$).
\item The invariant mass of the $M$ candidate $m(M)$.
\item The angle between the $M$ momentum and the line joining the PV and the decay vertex of $M$ (DIRA), which 
has to be consistent with  zero to ensure the $M$ candidate originates from the PV.
\item The significance of the distance between the $M$ decay vertex and the PV ($\chi^{2}_{\mathrm{FD}}$).
\item The reconstructed lifetime $\tau_M$.
\item Variables describing the isolation of the final state tracks. 
\item Statistical estimators of the nature of the decay products, which combine the information from LHCb's RICH, 
calorimeters and muon system (\texttt{PID($\pi$)}, \texttt{ProbNNmu}).
\end{itemize}
In the present study, only the generation phase is simulated. Therefore, only the ``true'', generator-level, $p$ and $p_{\mathrm{T}}$ 
of the \Dz and decay products are available in our samples, and the selection we apply to 
produce a realistic distorted sample relies only on these variables. 
We checked that the distortion we introduced in the phase space is consistent with that observed in the analysis presently 
carried out by LHCb~\cite{LHCb_D2K3pi_ANA}, even though we can use only the true information, 
rather than the reconstructed one, and no information on vertexing nor on decay topology.
To achieve this, some cuts on momenta and transverse momenta are tightened with respect to the selection 
in~\cite{LHCb_D2K3pi_ANA}.
We require all the decay products to be in the acceptance of the LHCb detector (i.e. the angle between their 
momentum and the nominal beam line should lie in the range 0.01-0.4 rad), and the candidates must  
satisfy the following criteria~:
\begin{itemize}
\item $p_{\mathrm{T}}(D^0) > 3 $~GeV$/c$.
\item $p_{\mathrm{T}} > 0.5 $~GeV$/c$,   $p > 3 $~GeV$/c$ for all the decay products.
\item max($p(K^-),p(\pi^+),p(\pi^+),p(\pi^-)$)$ > 10$~GeV$/c$.
\item max($p_{\mathrm{T}}(K^-),p_{\mathrm{T}}(\pi^+),p_{\mathrm{T}}(\pi^+),p_{\mathrm{T}}(\pi^-)$)$ > 1.7$~GeV$/c$.
\item max($p_{\mathrm{T}}(K^-),p_{\mathrm{T}}(\pi^+),p_{\mathrm{T}}(\pi^+),p_{\mathrm{T}}(\pi^-)$)$ > 3.7$~GeV$/c$ with the hardware trigger reconstruction.
\end{itemize}
By construction, the generator-level $p$ and $p_{\mathrm{T}}$'s available in our samples do not account for the finite resolution 
of the reconstruction.
Since in LHCb the momentum resolution of the offline reconstruction is very small (from 0.5\% at low momentum to 1\% at 200 GeV$/c$), 
we consider that we can safely neglect it. One expection is made for the first stage of LHCb's trigger system. 
It's a hardware trigger that searches for high $p_{T}$ objects, based on a partial event reconstruction carried out by the Front End 
electronics~\cite{LHCb-DP-2012-004}.
Several kinds of objects are searched for, involving several trigger $lines$: 
candidates are looked for in the muons system, while in the electromagnetic calorimeter clusters
due to photons or electrons are sought. In the case of a decay like \norm, the hardware trigger 
looks for high $p_{T}$ clusters ($>$ 3.7 GeV$/c$) in the 
Hadron Calorimeter, which resolution is $\sigma_{E}/E=80\%/\sqrt{E}\oplus 10\%$ with $E$ expressed in GeV. We apply the trigger cut
to particles which $p_{T}$ has been smeared in order to reproduce this resolution.
This trigger cut is applied to only a third of the events since LHCb events 
in which a \norm decay is produced are often triggered on independently, due to the decay products of
the second charm hadron produced in the event (proton-proton collisions actually produce $c-\bar{c}$ $pairs$).  

\begin{figure}[tb]
    \hspace*{-3cm}
    \includegraphics[width=1.4\linewidth]{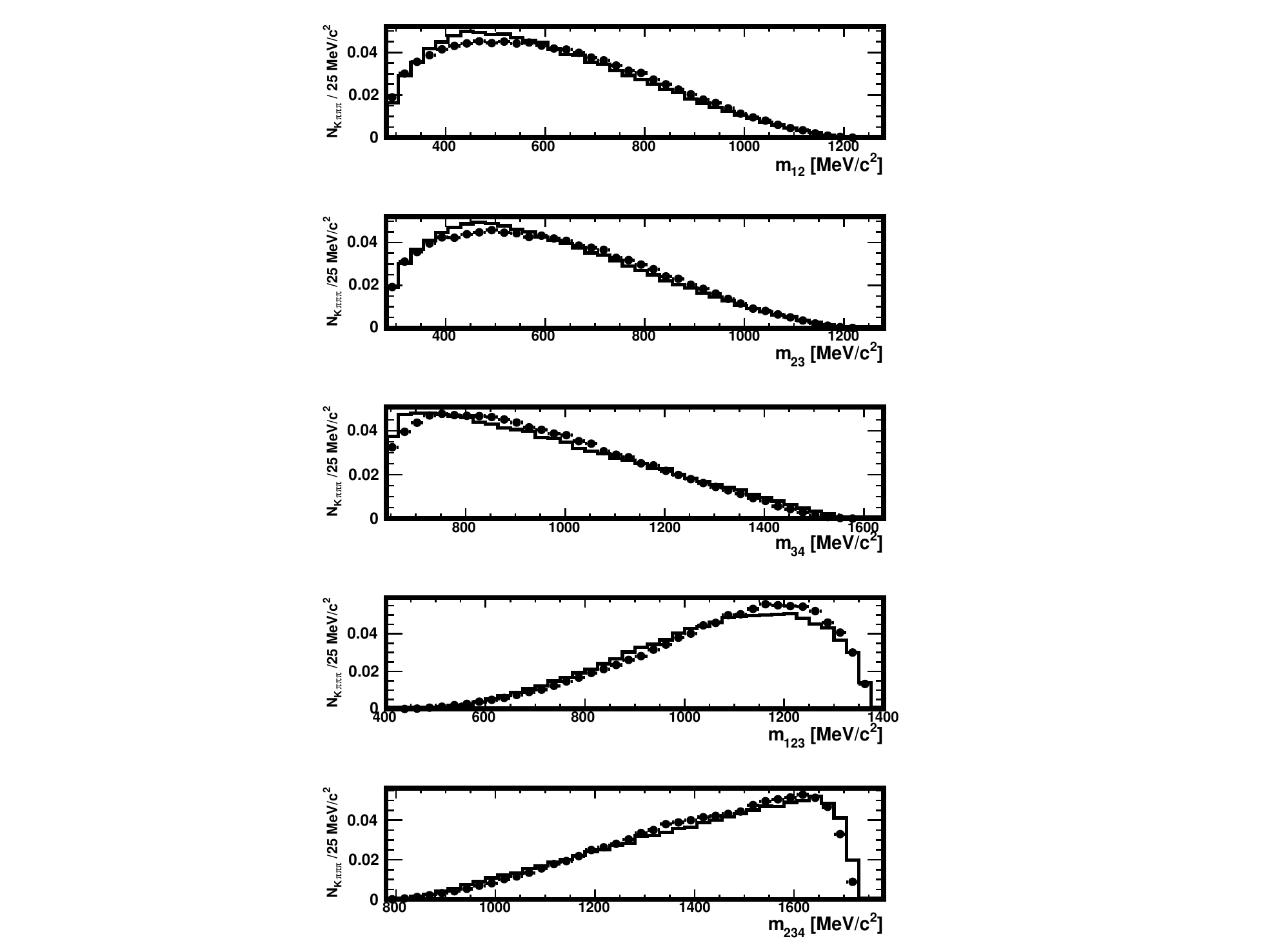}
    \vspace*{-0.5cm}
  \caption{Distributions of $m_{12}$, $m_{23}$, $m_{34}$, $m_{123}$ and $m_{234}$ in the original sample (histogram) and 
in the distorted one (full circles). The normalisation is arbitrary.}  
  \label{fig:Distr_whole_NoCorr}
\end{figure}

\begin{figure}[tb]
    \hspace*{-3cm}
    \includegraphics[width=1.4\linewidth]{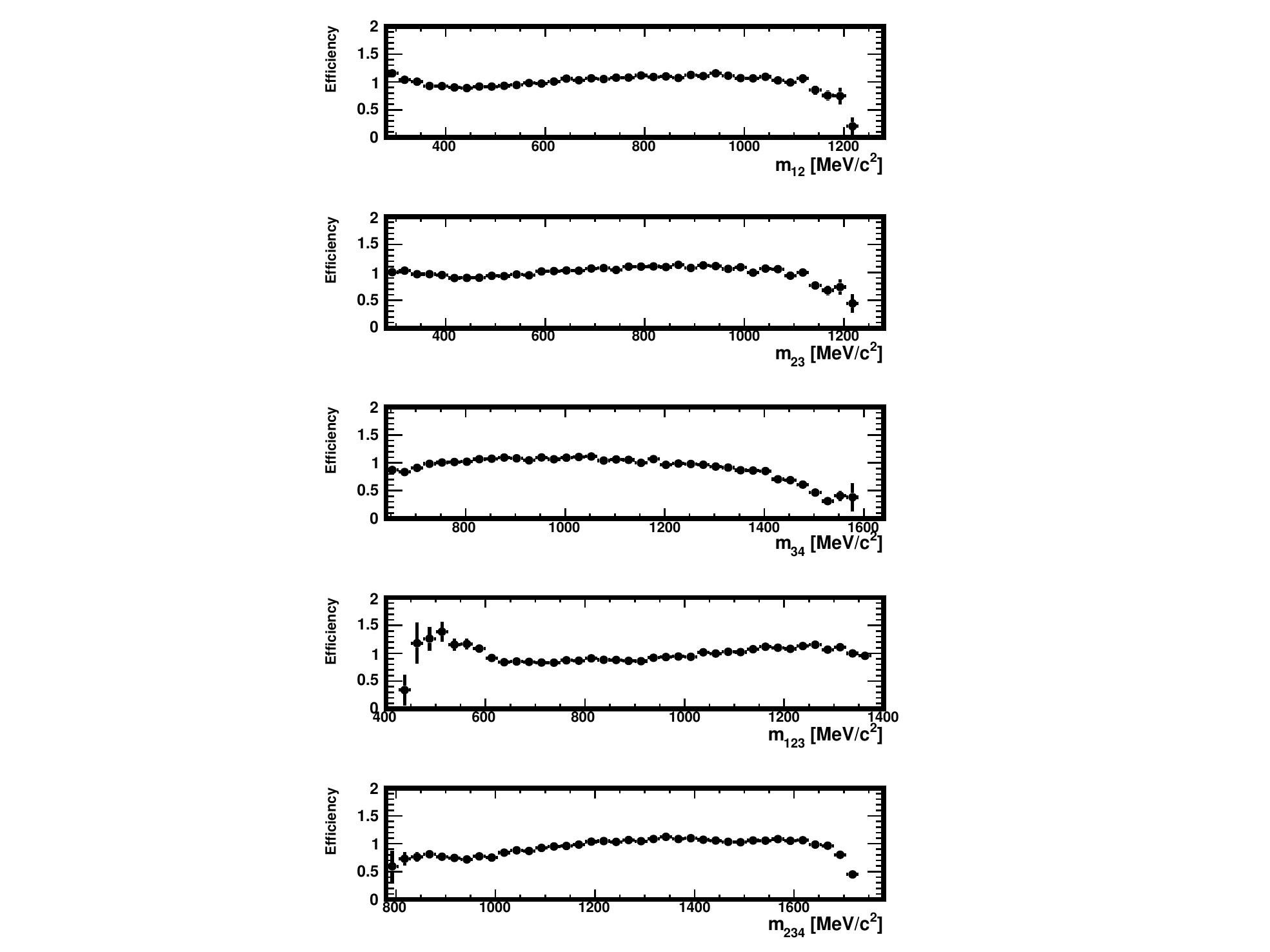}
    \vspace*{-0.5cm}
  \caption{Efficiency in $m_{12}$, $m_{23}$, $m_{34}$, $m_{123}$ and $m_{234}$ observed in the data generated for this study. 
Shown are the ratios of the distributions found in the distorted and original samples. The normalisation of these distributions was arbitrary.} 
  \label{fig:Ratio_whole_NoCorr}
\end{figure}

On Fig.~\ref{fig:Distr_whole_NoCorr} we compare the distributions of $m_{12},~m_{23},~m_{34},~m_{123}$ and $m_{234}$ 
in the distorted and original samples. The selection efficiencies as a function of each of these variables 
(i.e., the ratio between the distributions superimposed on Fig.~\ref{fig:Distr_whole_NoCorr}) are also shown on Fig.~\ref{fig:Ratio_whole_NoCorr}. 
This illustrates the effect of the selection on the phase space. 

\clearpage 

\subsection{Neural network training}

We chose to use the \texttt{MLPBNN} NN provided by the \texttt{TMVA} package. 
This Multilayer Perceptron (MLP) NN is trained using the BFGS method instead of a 
simple back-propagation method. The definition of a MLP, and a description of the latter method
can be found in Ref.~\cite{TMVA}. Also, a Bayesian regulation technique is 
employed. It is described in Ref.~\cite{Zhong:2011xm}.
We used the default configuration found in the example macro downloaded with the \texttt{TMVA} package. 
To the attention of the expert reader, we specify that in this case the MLP involves only one hidden layer, 
which comprises $N+10$ neurons, where $N$ is the number of input variables ($m_{12},~m_{23},~m_{34},~m_{123}$ and $m_{234}$ in our case). 
The neuron activation function is $tanh$. Input variables are  linearly scaled to lie with $\mathrm{\left[-1;1\right]}$, 
and 600 training cycles are performed. An overtraining test is run every 5 cycles. 
All the other settings can be found in Table~19 of Ref.~\cite{TMVA}. 

\begin{figure}[tb]
  \begin{center}
    \hspace*{-2.3cm}
    \includegraphics[width=1.3\linewidth]{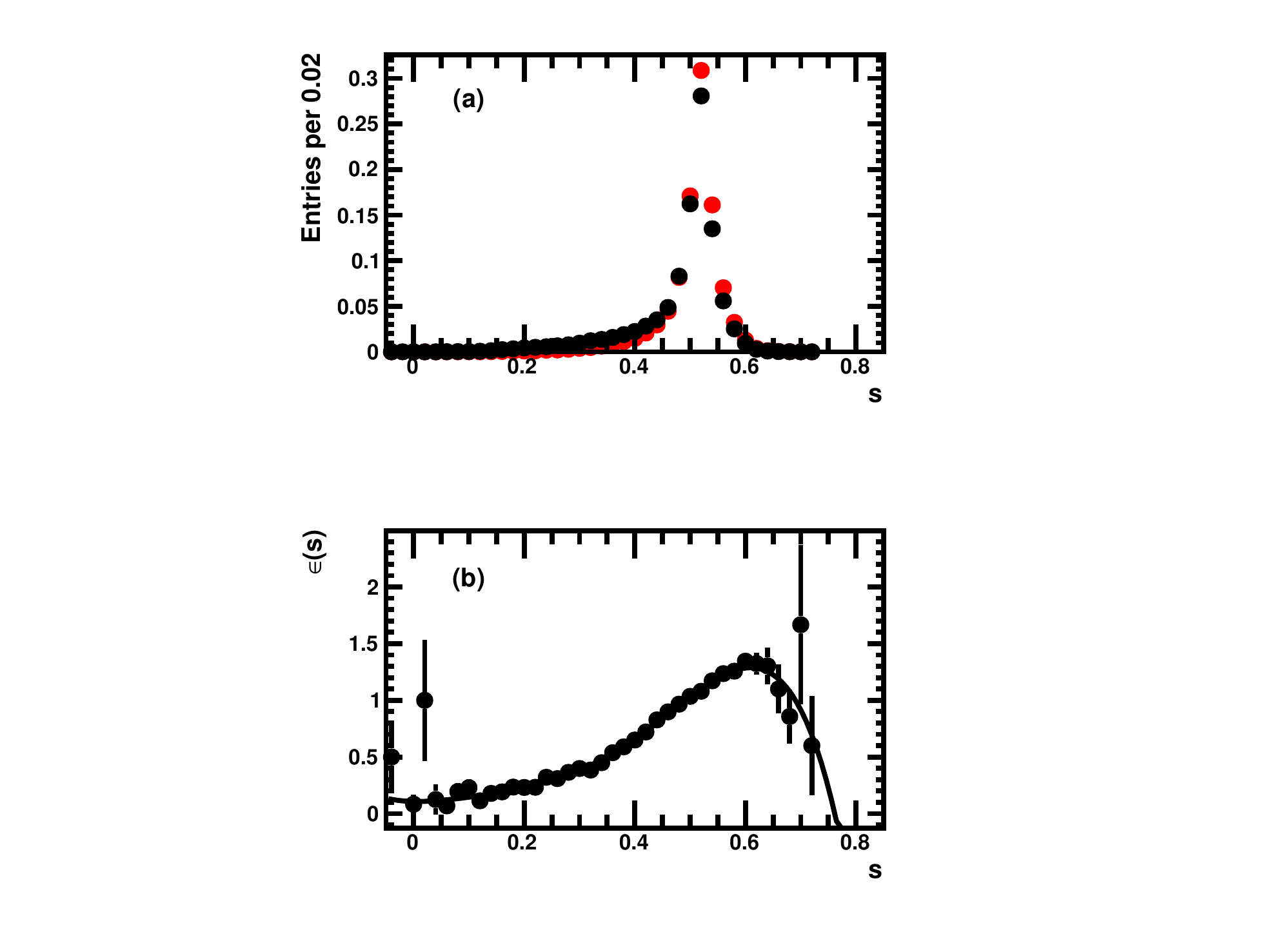}
    \vspace*{-0.5cm}
  \end{center}
  \caption{Distributions and distribution ratio showing (a) the NN score $s$  in the distorted (red) and original (black)  \norm samples, and (b) 
the selection efficiency as a function of $s$ (with an arbitrary normalisation). The fit leading to $\epsilon(s)$ is superimposed to 
the measured efficiencies.} 
  \label{fig:Score_K3pi_Sel1}
\end{figure}

It took two hours to train this MLP to distinguish between decays from the distorted or original samples (both contain 230000 events).
We used a machine equipped with a Intel i7-640M 2.8 GHz 2-Core processor. 
Two additional distorted and original samples, produced in the same way as those used for training, and of
identical size, have been produced to test the resulting NN score. The distribution of this score in both 
samples is shown on Fig.~\ref{fig:Score_K3pi_Sel1}, which also displays the ratio these distributions,
to which we fitted a parameterised efficiency, $\epsilon\left(s\right)$, where $s$ is the NN score.
We used for that a fifth order polynomial. The $\epsilon\left(s\right)$ function should match 
$\epsilon\left(m_{12},m_{23},m_{34},m_{123},m_{234}\right)$, the multidimensional efficiency we aim at.

\subsection{Results}
\label{subsec:ResultsK3pi}

\begin{figure}[tb]
    \hspace*{-3cm}
    \includegraphics[width=1.4\linewidth]{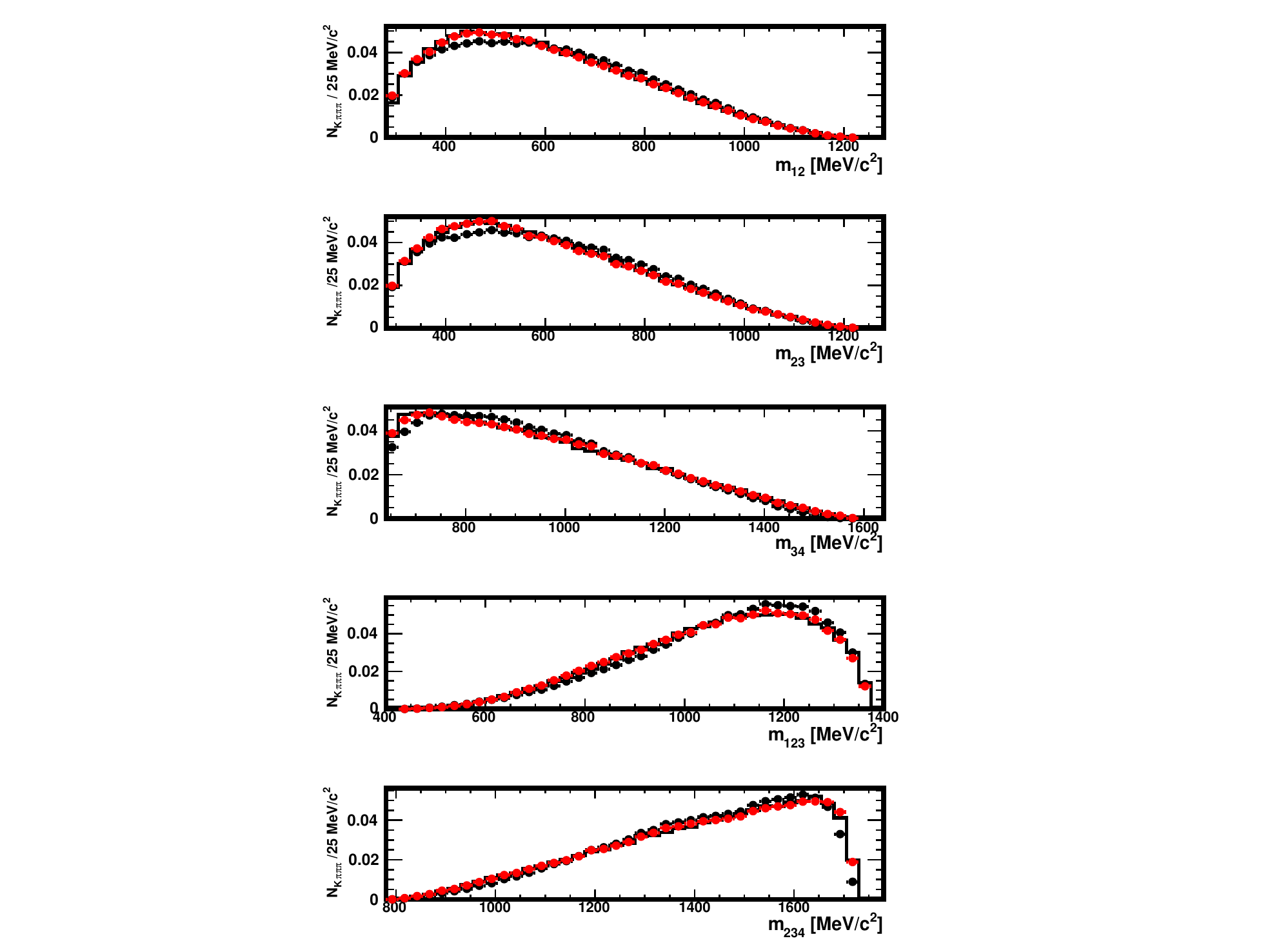}
    \vspace*{-0.5cm}
  \caption{Distributions of $m_{12}$, $m_{23}$, $m_{34}$, $m_{123}$ and $m_{234}$ in the original sample (histogram),
in the distorted one (full black circles) and in the distorted sample where the decays have been re-weighted using the $\omega_i$ weights (red), 
as explained in the text. The absolute normalisation is arbitrary when the correction is not applied and natural when it is applied (red).}  
  \label{fig:Distr_whole}
\end{figure}

\begin{figure}[tb]
    \hspace*{-3cm}
    \includegraphics[width=1.4\linewidth]{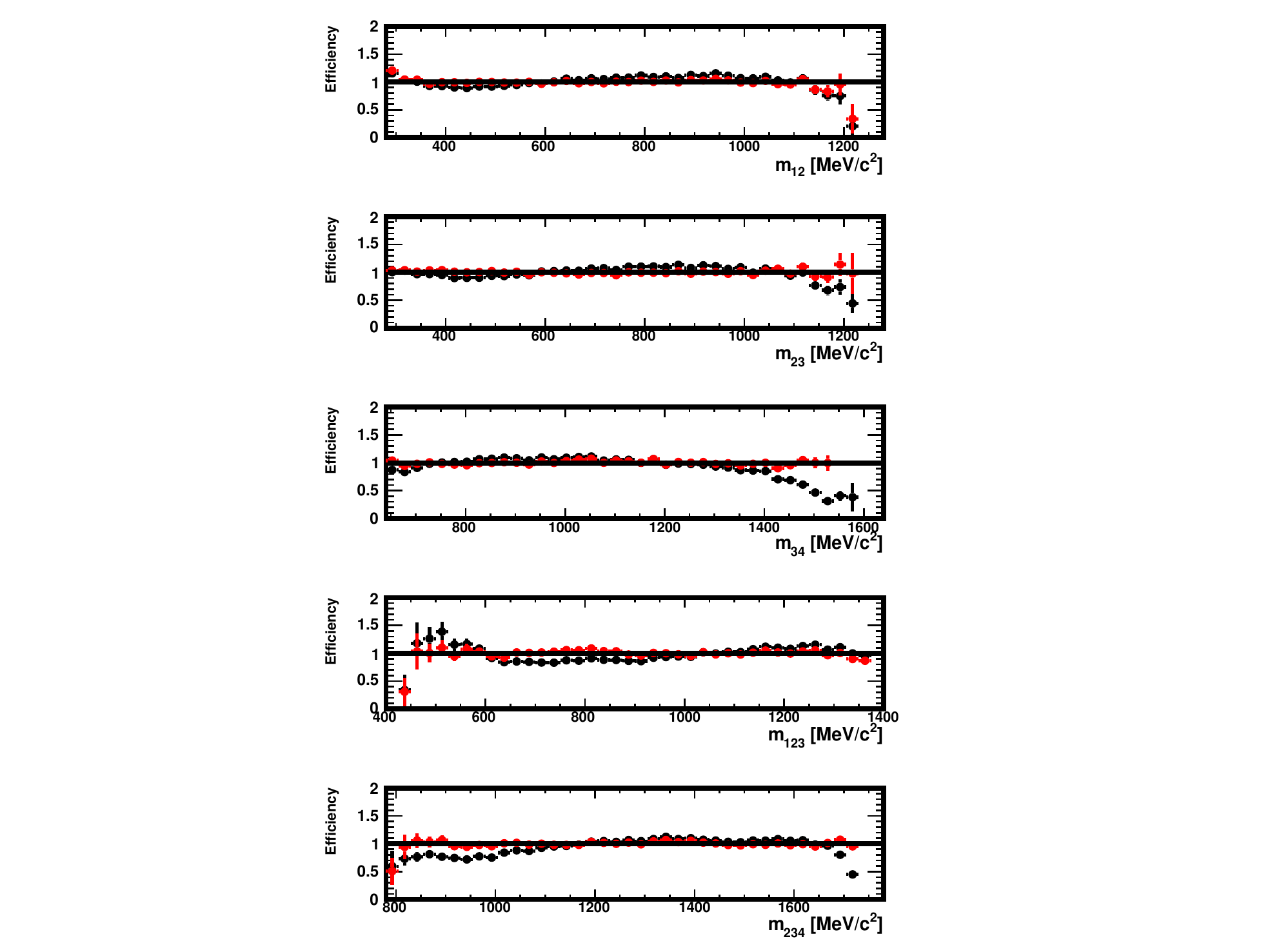}
    \vspace*{-0.5cm}
  \caption{Efficiency in $m_{12}$, $m_{23}$, $m_{34}$, $m_{123}$ and $m_{234}$ in the data generated for this study. 
Shown are the ratios of the distributions found in the distorted and 
original samples, with no correction (black) and for decays re-weighted using $\omega_i$ weights (red)
as explained in the text. The absolute normalisation is arbitrary when the correction is not applied and natural when it is applied (red).}  
  \label{fig:Ratio_whole}
\end{figure}

To test the efficiency obtained above, we weighted each \norm decay $i$ in the distorted test sample with 
$\omega_i = 1/\epsilon\left(s_i\right)$. The result can be found in Fig.~\ref{fig:Distr_whole}. 
It is the same as Fig.~\ref{fig:Distr_whole_NoCorr} with distributions from the re-weighted distorted sample superimposed. 
One can see that these distributions match
closely those observed in the original sample,  before the distortions due to the selection were imposed. 
The ratios on Fig.~\ref{fig:Ratio_whole} are consistent with 1 all over the $m_{12},~m_{23},~m_{34},~m_{123}$ and $m_{234}$ spectra, 
unlike the un-weighted ratios on Fig.~\ref{fig:Ratio_whole_NoCorr}, which show clear distortions. 
Consequently, the MLP we trained produces a single variable $s$ to fully encompass the 5-dimensional information 
on the distortion of the  ($m_{12},m_{23},m_{34},m_{123},m_{234}$) space.  
The $\epsilon\left(s\right)$ efficiency accounts simultaneously for the 5 individual 
efficiencies as a function of $m_{12},~m_{23},~m_{34},~m_{123}$ and $m_{234}$. The yield in each bin of the corrected distributions in 
Fig.~\ref{fig:Distr_whole} is calculated as the sum of the $\omega_i$ weights over 
all the events that lie in this bin. In principle, this corrected yield can be correct even if $\epsilon\left(s\right)$
is not an accurate evaluation of $\epsilon\left(m_{12},m_{23},m_{34},m_{123},m_{234}\right)$ everywhere in the 
$\left(m_{12},m_{23},m_{34},m_{123},m_{234}\right)$ space. A compensation is possible in the sum. Also, the evaluation
could be wrong in regions containing too little statistics for the effect to appear significantly
 on Figs.~\ref{fig:Distr_whole} and~\ref{fig:Ratio_whole}. To constrain this possibility, we repeated the 
test of Figs.~\ref{fig:Distr_whole} and~\ref{fig:Ratio_whole} in various restricted regions of the $\left(m_{12},m_{23},m_{34},m_{123},m_{234}\right)$
space. The results are on Figs.~\ref{fig:Distr_m12_1} to~\ref{fig:Ratio_m234_4} of Appendix~\ref{App:AppA}. 
The corrected distributions once more match the original ones. Note that 
we still use the efficiency from Fig.~\ref{fig:Score_K3pi_Sel1} for these corrections. The polynomial's parameters were not 
re-evaluated by performing in each region specific fits or new trainings.
We conclude that $\epsilon\left(s\right)$ = $\epsilon\left(m_{12},m_{23},m_{34},m_{123},m_{234}\right)$, 
in the limit of the precision with which it can be verified with our data.

\begin{figure}[tb]
    \hspace*{-3cm}
    \includegraphics[width=1.4\linewidth]{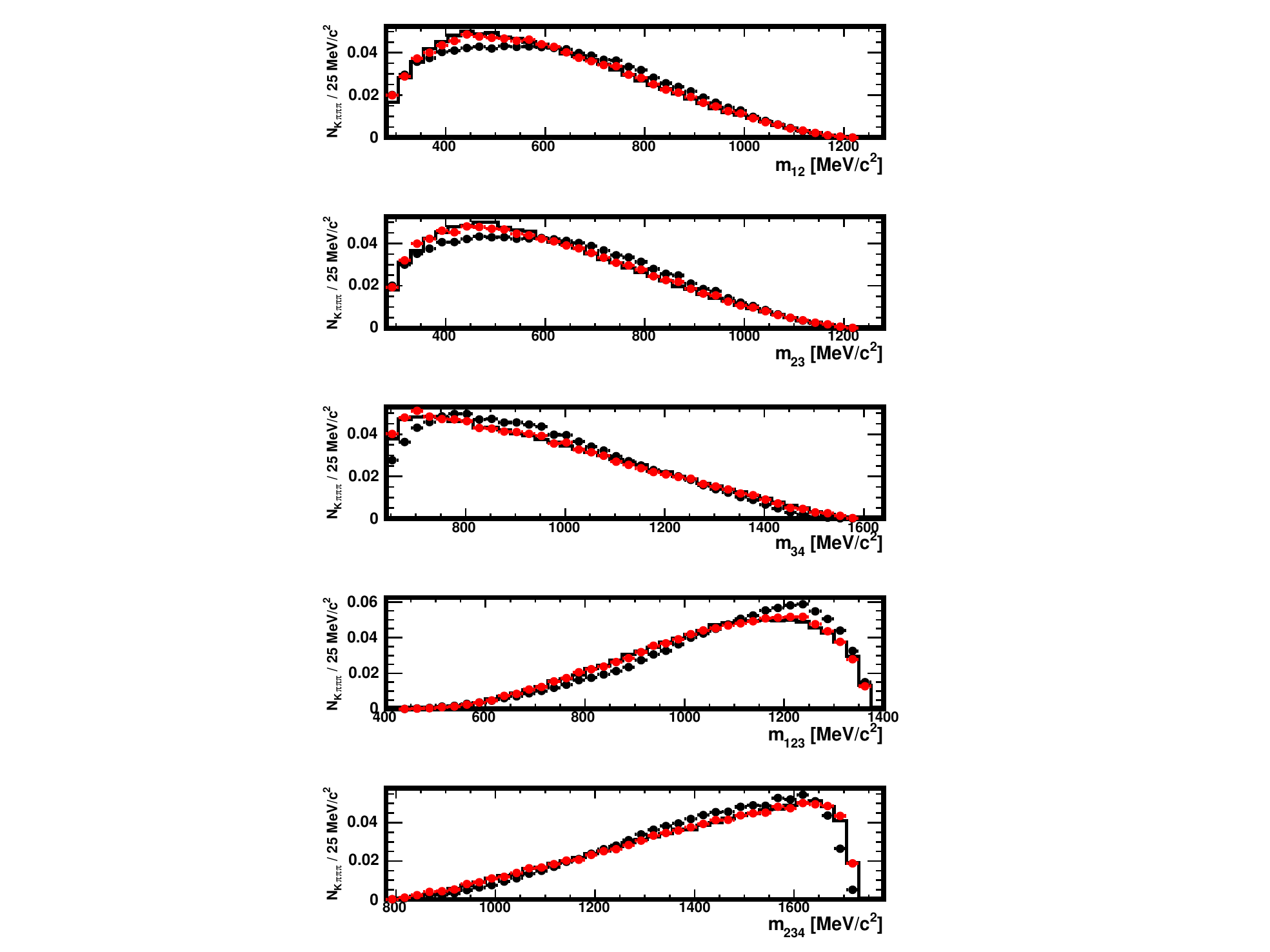}
    \vspace*{-0.5cm}
  \caption{Distributions of $m_{12}$, $m_{23}$, $m_{34}$, $m_{123}$ and $m_{234}$ in the original sample (histogram),
in the distorted one obtained with a tighter selection (full black circles) and in the same sample where the decays have 
been re-weighted using the $\omega_i$ weights (red), 
as explained in the text. The absolute normalisation is arbitrary when the correction is not applied and natural when it is applied (red).}  
  \label{fig:Distr_whole_Sel2}
\end{figure}

\begin{figure}[tb]
    \hspace*{-3cm}
    \includegraphics[width=1.4\linewidth]{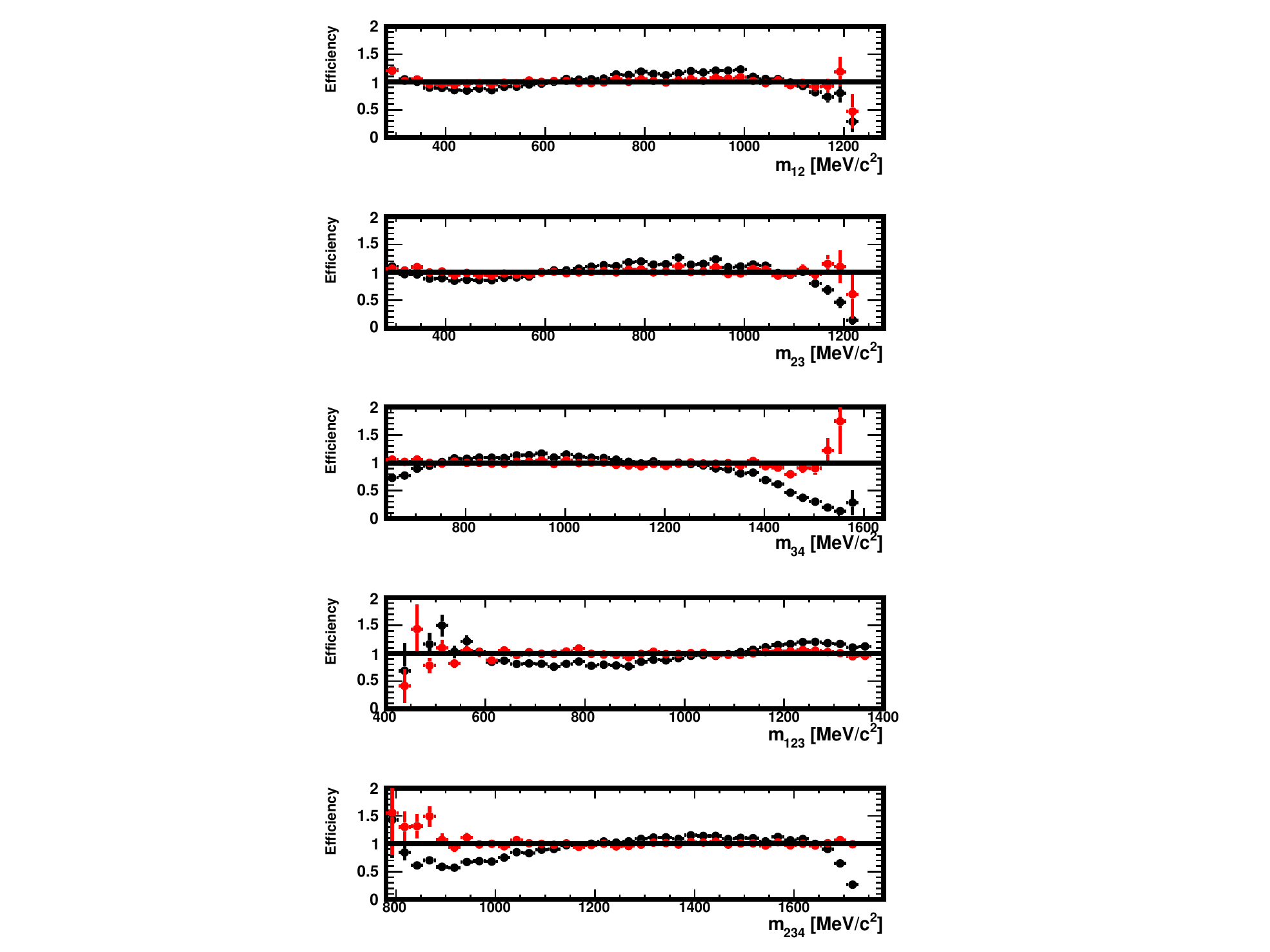}
    \vspace*{-0.5cm}
  \caption{Efficiency in $m_{12}$, $m_{23}$, $m_{34}$, $m_{123}$ and $m_{234}$ in the data generated for this study, 
with a tighter selection for the distorted sample. Shown are the ratios of the distributions found in the distorted and 
original samples, with no correction (black) and for decays re-weighted using $\omega_i$ weights (red)
as explained in the text. The absolute normalisation is arbitrary when the correction is not applied and natural  
when it is applied (red).}  
  \label{fig:Ratio_whole_Sel2}
\end{figure}

We reach the same conclusion when applying the same technique to correct an alternative distorted sample, 
obtained with tighter cuts and therefore showing stronger distortions in the phase space. 
We tightened one cut on the \norm decay products: $p_{\mathrm{T}} > 0.6 $~GeV$/c$.
Also, the hardware trigger selection is applied to all events instead of only a third. 
The results we obtained can be judged with the help of Figs.~\ref{fig:Distr_whole_Sel2} and~\ref{fig:Ratio_whole_Sel2}.
What we obtained in particular regions of the phase space is shown on Figs.~\ref{fig:Distr_m12_1_Sel2} to~\ref{fig:Ratio_m234_4_Sel2} in
Appendix~\ref{App:AppA}.

\section{\mbox{Application to the study of {\bf\boldmath \kstmumu}}}
\label{sec:AppliedToKstMuMu}

We performed a second case study to explore the potential of MVAs to treat multidimensional efficiencies.
It is based on the decay \kstmumu.  A similar procedure to that described in Sect.~\ref{sec:AppliedToDK3pi} 
has been carried out for that purpose.  

\subsection{Sample generation }

We generated a distorted and an original sample, both containing 600\,000 decays. 
This is less than in the sample used by the LHCb collaboration for the result in Ref.~\cite{LHCb_KstMuMu_conf}.
Because the study of this mode is one of LHCb's priorities, it was actually acceptable 
produce as much as 1.406M reconstructed and selected events.  

The \texttt{ROOT} \texttt{TGenPhaseSpace} class is used once more to generate decays. 
The kinematics of the $B^{0}$ meson is derived from the differential production
cross-section  as a function of transverse momentum and rapidity, measured in proton-proton 
collisions at a center-of-mass energy of 7 TeV~\cite{Aaij:2013noa}. 

\begin{figure}[tb]
  \begin{center}
    \hspace*{-1.3cm}
    \includegraphics[width=1.1\linewidth]{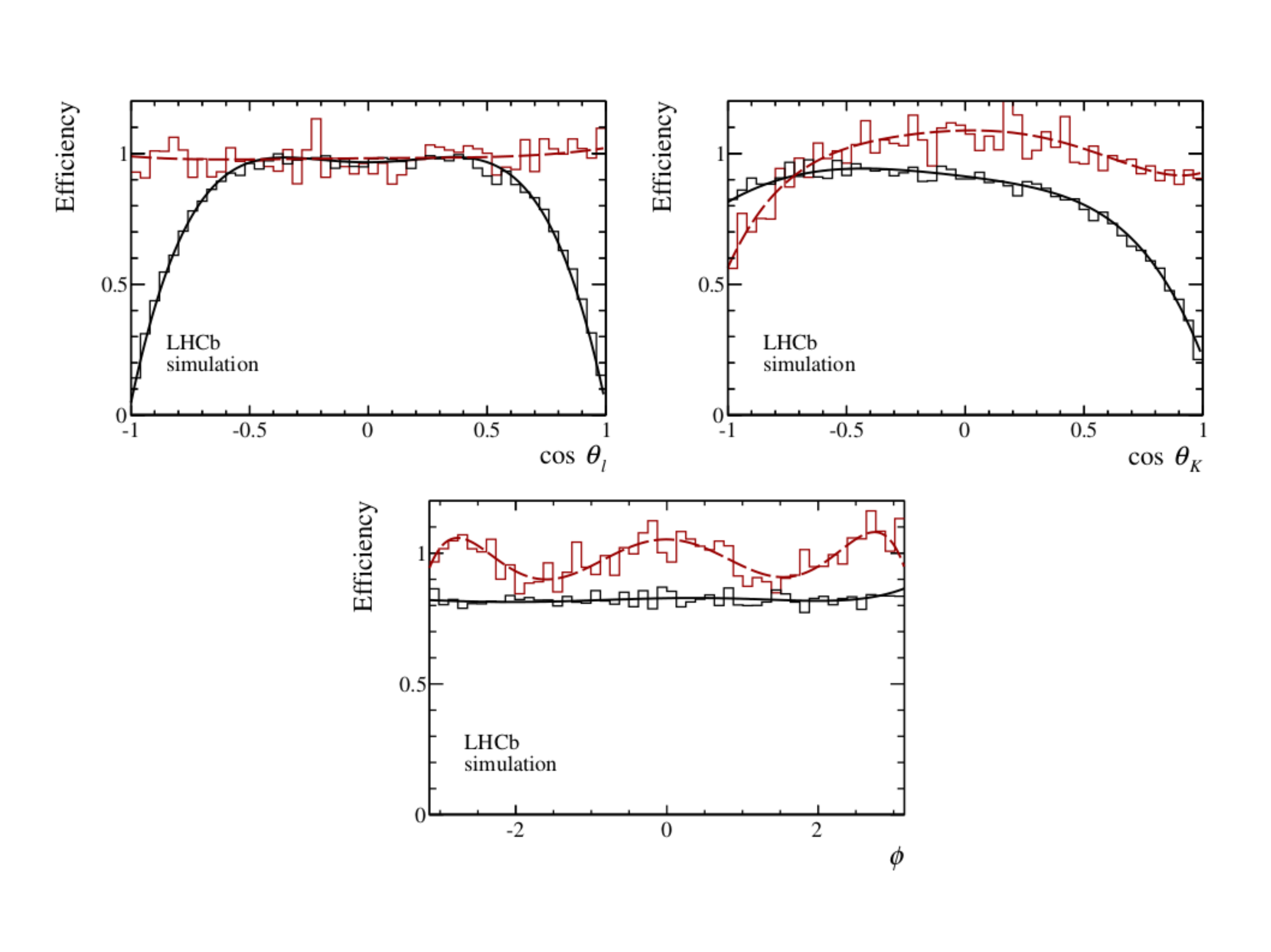}
    \vspace*{-0.3cm}
  \end{center}
  \caption{Angular efficiency in cos$\theta_l$, cos$\theta_K$ and $\phi$, as determined from a principal moment
analysis of simulated three-body \kstmumu phase-space decays (black solid and red long-dashed lines), and compared to simulated 
data (histograms). The efficiency is shown for the regions: $0.1 < q^2 < 0.98$ GeV$^{2}/c^4$ (black) and 
$18.0 < q^2 < 19.0$ GeV$^{2}/c^4$ (red). The absolute normalisation is arbitrary. This figure is reproduced from Ref.~\cite{LHCb_KstMuMu_conf}.} 
  \label{fig:Fig-KstMuMu-1}
\end{figure}

\begin{figure}[tb]
  \begin{center}
    \hspace*{-2.3cm}
    \includegraphics[width=1.3\linewidth]{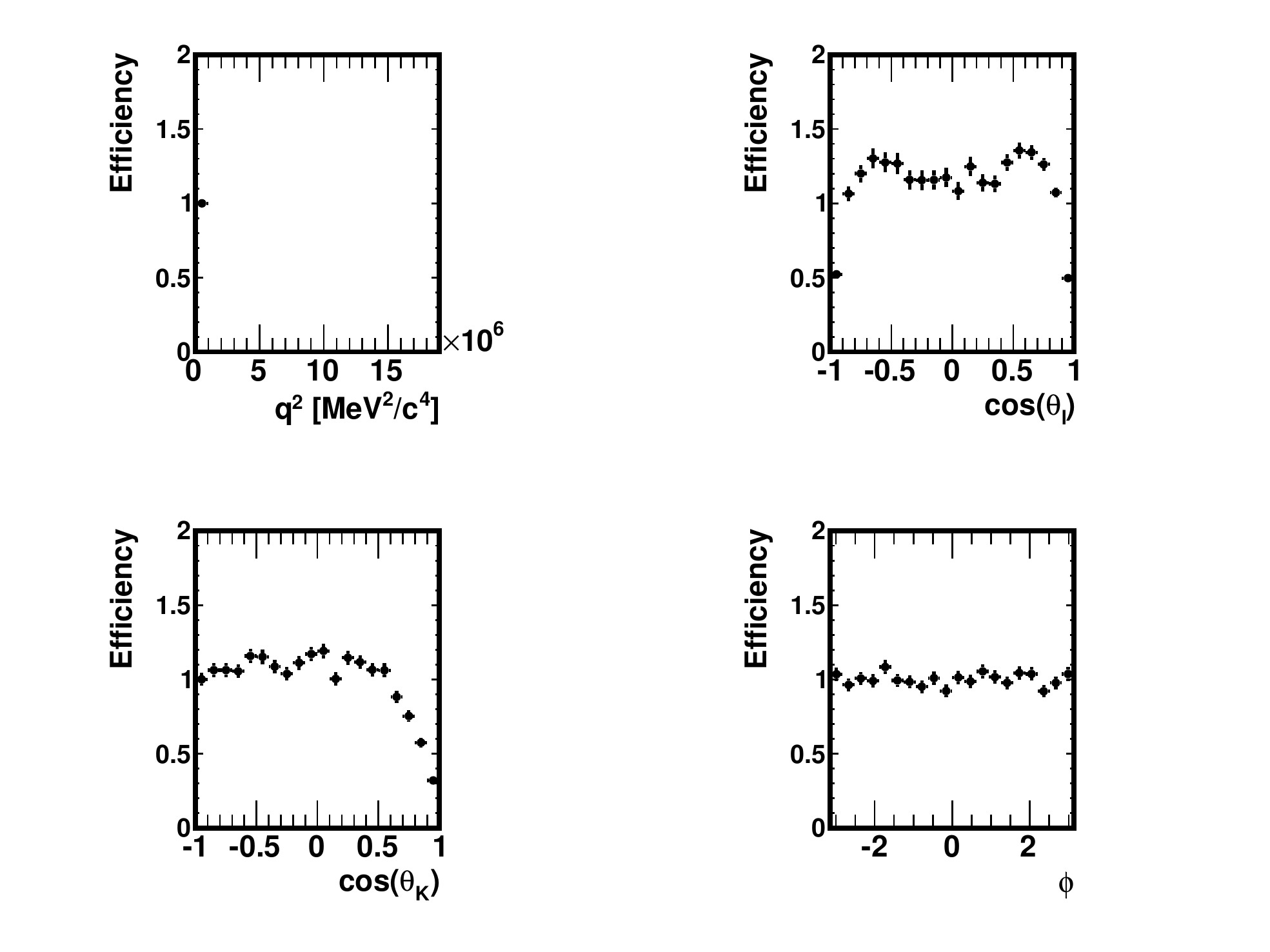}
    \vspace*{-0.5cm}
  \end{center}
  \caption{Efficiency in $q^{2}$, cos$\theta_l$, cos$\theta_K$ and $\phi$ in the data generated for the present study, 
in the region $0.1 < q^2 < 0.98$ GeV$^{2}/c^4$. The absolute normalisation is arbitrary.} 
  \label{fig:Ratio_q2_1_NoCorr}
\end{figure}

\begin{figure}[tb]
  \begin{center}
    \hspace*{-2.3cm}
    \includegraphics[width=1.3\linewidth]{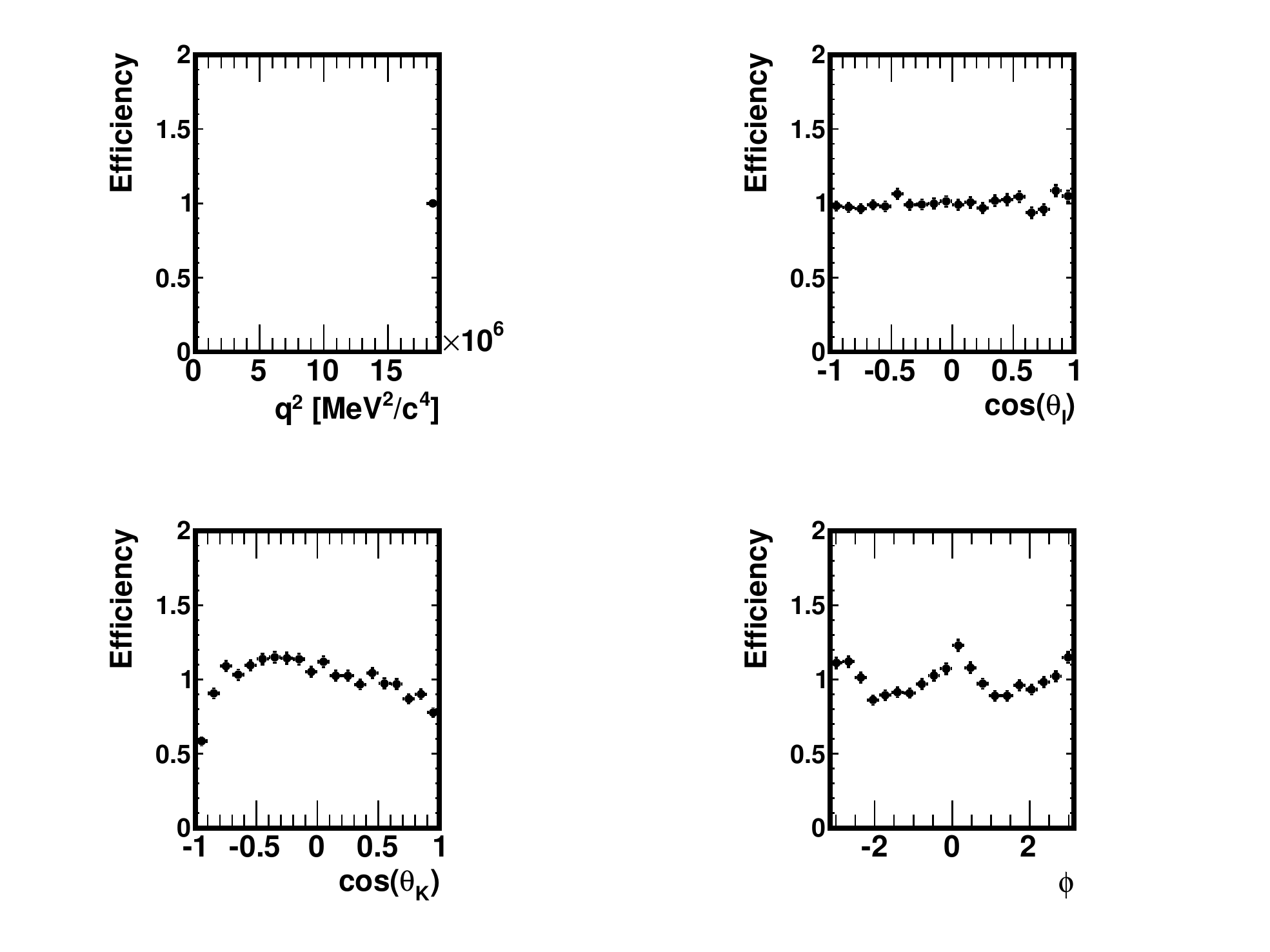}
    \vspace*{-0.5cm}
  \end{center}
  \caption{Efficiency in $q^{2}$, cos$\theta_l$, cos$\theta_K$ and $\phi$ in the data generated for the present study, 
in the region $18.0 < q^2 < 19.0$ GeV$^{2}/c^4$. The absolute normalisation is arbitrary.} 
  \label{fig:Ratio_q2_19_NoCorr}
\end{figure}

To produce the distorted sample, we applied a selection as similar 
as possible to that in~\cite{LHCb_KstMuMu_conf}. The following criteria are used:
\begin{itemize}
\item All the decay products must be in the acceptance of the LHCb detector.
\item One of the muons should satisfy  $p_{\mathrm{T}}(\mu) > 1.8 $~GeV$/c$ in order to reproduce the cut used by the 
muon-specific line which dominates the hardware trigger in the case of decays such as \kstmumu.
\item max($p(K^+),p(\pi^-),p(\mu^+),p(\mu^-)$)$ > 10$~GeV$/c$. 
\item max($p_{\mathrm{T}}(K^+),p_{\mathrm{T}}(\pi^-),p_{\mathrm{T}}(\mu^+),p_{\mathrm{T}}(\mu^-)$)$ > 0.2$~GeV$/c$.
\item $p_{\mathrm{T}}(B^0) > 4 $~GeV$/c$.
\item $p(B^0) > 40 $~GeV$/c$.
\item max($IP(K^+),IP(\pi^-),IP(\mu^+),IP(\mu^-)$)$ > 1$~mm.
\item For all decay products, $\chi^{2}_{\mathrm{IP}}>9$.
\end{itemize}
Only momenta and transverse momenta are directly accessible in the data generated with the 
\texttt{TGenPhaseSpace} class. The cuts on these variables, even tighter than those 
in~\cite{LHCb_KstMuMu_conf}, could not reproduce the distortion of the phase space variables
actually observed in this analysis (Fig.~\ref{fig:Fig-KstMuMu-1}). This was true in particular of the 
acceptance as a function of $\phi$, which was flatter.
In the hope to reproduce a comparable distortion, we included cuts that affect angular variables 
more directly. This is the reason of the presence of cuts on $IP$ and $\chi^{2}_{\mathrm{IP}}$ in
the above list. The $IP$ was determined based on the angle between the \kstmumu decay products and 
the momentum of the $B^{0}$, assuming the latter travelled a typical 8 mm before its decay. 
We approximated $\chi^{2}_{\mathrm{IP}}$ by the squared ratio between $IP$ and its uncertainty. 
The latter was computed based on the transverse momentum of the particle under consideration and on 
the typical uncertainty observed in LHCb. We used the uncertainty 
advertised by LHCb in recent publications (see for instance~\cite{LHCb_KstMuMu_conf}), 
$\sigma_{IP}=(15+29/\pt)\mum$ with $\pt$ in GeV$/c$. These cuts did not yet suffice to reproduce the amplitude of the distortion in the
$\phi$ distribution. We obtained this by discarding decays which do not satisfy $\left|m(K\pi)-892\right|>0.09\times\left(q^2/(19.10^{6})\mathrm{sin^2}(\phi)\times 100\right)$ where $m(K\pi)$ and $q$ are expressed in MeV$/c^{2}$. With this set of criteria, 
the distortion of the ($q^2$, cos$\theta_l$, cos$\theta_K$, $\phi$) space, although not idenditical, is similar to that 
observed in Ref.~\cite{LHCb_KstMuMu_conf}. This can be seen in on Figs.~\ref{fig:Ratio_q2_1_NoCorr} and~\ref{fig:Ratio_q2_19_NoCorr}, which show the
distortion in our generated data, in extreme regions of $q^{2}$: $0.1 < q^2 < 0.98$ GeV$^{2}/c^4$ and 
$18.0 < q^2 < 19.0$ GeV$^{2}/c^4$. They can be compared with Fig.~\ref{fig:Fig-KstMuMu-1} to notice that the evolution of the distortion 
between these two regions is reproduced. We checked that the same conclusion holds in all 
$q^{2}$ regions by comparing Figs.~\ref{fig:Ratio_q2_1} to~\ref{fig:Ratio_q2_19} (see Appendix~\ref{App:AppB}) with the equivalent figures in~\cite{LHCb_KstMuMu_ANA}. 

\subsection{Neural network training}

We trained the \texttt{MLP} NN provided by the \texttt{TMVA} package. 
It is the same NN as that described in Sect.~\ref{sec:AppliedToDK3pi}, modulo 
two differences: the classical back-propagation technique described in~\cite{TMVA}
is used instead of the BFGS method, and we use no Bayesian regulation technique. 

The distortion of the ($q^2$, cos$\theta_l$, cos$\theta_K$, $\phi$) space is more challenging
to correct as that of the ($m_{12},~m_{23},~m_{34},~m_{123}$, $m_{234}$) space in the previous section.
Indeed, as can be seen on Figs.~\ref{fig:Fig-KstMuMu-1}, \ref{fig:Ratio_q2_1_NoCorr}  and~\ref{fig:Ratio_q2_19_NoCorr}, 
the distortion the cos$\theta_l$ and $\phi$ distributions is
symmetric in these variables. As a consequence, the discriminative power of these variables is low:
the distorted and original samples cannot be distinguished 
via a strong ``preference'' of one of them for the higher or the lower end of the cos$\theta_l$ or $\phi$ 
distributions. In other words, the efficiencies with which distorted and original events would be selected
by a cut on cos$\theta_l$ or $\phi$ would not differ nor vary much as a function of the value of this cut.
Another difficulty complicates the determination of the multidimensional efficiency 
$\epsilon\left(q^2, cos\theta_l, cos\theta_K, \phi\right)$ with the approach presented in this 
document. It stems from the fast variation as a function of $q^{2}$ of the distortion in the 3 other variables 
and the particular pattern it follows: 
at low $q^{2}$, the distortion is strong in cos$\theta_l$ and 
very light in $\phi$, while the opposite is observed at high $q^{2}$ (this pattern 
can be seen again on  Figs.~\ref{fig:Fig-KstMuMu-1}, \ref{fig:Ratio_q2_1_NoCorr}  and~\ref{fig:Ratio_q2_19_NoCorr}). 
There is therefore a more complicated 
structure to be understood by the NN. Also, the fact that in some regions in $q^{2}$ only 
two variables are available to discrimate the distorted sample against the original one is a difficulty in
itself. Moreover, it is not trivial for the NN to adapt to this varying behavior since it is trained using the whole 
distorted and original samples. It is not instructed to treat differently the various $q^{2}$ regions. 
This difficulty affects the determination of $\epsilon\left(q^2, cos\theta_l, cos\theta_K, \phi\right)$ in 
the whole sample and becomes more acute when it comes to determine the multidimensional efficiency in restricted regions 
of $q^2$.

To overcome these difficulties, the input variables mapped to the NN's first layer are 
not directly $q^2$, cos$\theta_l$, cos$\theta_K$, and $\phi$. We replace cos$\theta_l$ and $\phi$ by two transformed,  
non-symmetric variables: $e^{-\theta_l^2/4}$ and $e^{-\mathrm{sin}^2(\phi)/4}$. Also, unlike in Sect.~\ref{sec:AppliedToDK3pi},
we had to optimise the NN settings instead of using blindly the ones provided by default by \texttt{TMVA}. 
We devoted a limited effort (a day of work) to this. In practice~:
\begin{itemize}
\item We further transformed the input variables so as to make their distributions gaussian, which helps
the decorrelation algorithms used by the NN~\cite{TMVA}.
\item We used two hidden layers instead of only one. The number of neurons constituting the first and second layers is 14 and 6, 
respectively. It has to be compared with 4, the number of input variables. 
\item The number of training cycle was raised to 7000. 
\item Overtraining tests were run every 5 cycles. Each time, the convergence
is also tested. If 10 consecutive tests fail to observe an improvement of the error function, 
the training is considered optimal and stopped. 
\end{itemize}
All the other settings can be found in Table~19 of Ref.~\cite{TMVA}. 
With these settings, it took $\sim$12 hours to train the NN on the same machine as in Sect.~\ref{sec:AppliedToDK3pi}.

\subsection{Results}
\label{subsec:resultsB2KstMuMu}

\begin{figure}[tb]
  \begin{center}
    \hspace*{-1.5cm}
    \includegraphics[width=1.2\linewidth]{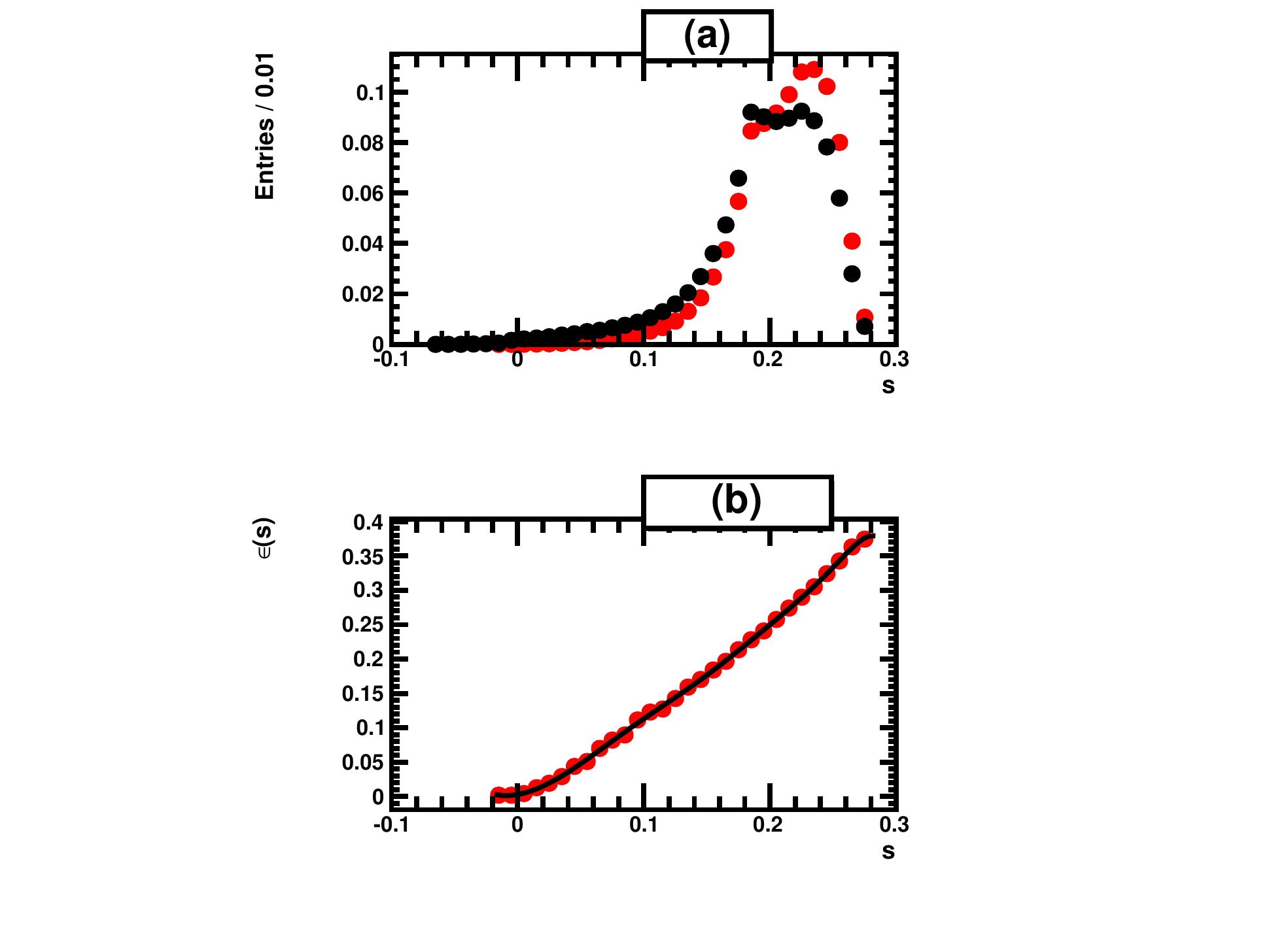}
    \vspace*{-0.5cm}
  \end{center}
  \caption{Distributions and distribution ratio showing (a) the NN score $s$  in the distorted (red) and original (black)  \kstmumu samples, and (b) 
the selection efficiency as a function of $s$ (with an arbitrary normalisation). The fit providing to $\epsilon\left(s\right)$ is superimposed to 
the measured efficiencies.} 
  \label{fig:Score_KstMuMu}
\end{figure}

\begin{figure}[tb]
  \begin{center}
    \hspace*{-2.3cm}
    \includegraphics[width=1.3\linewidth]{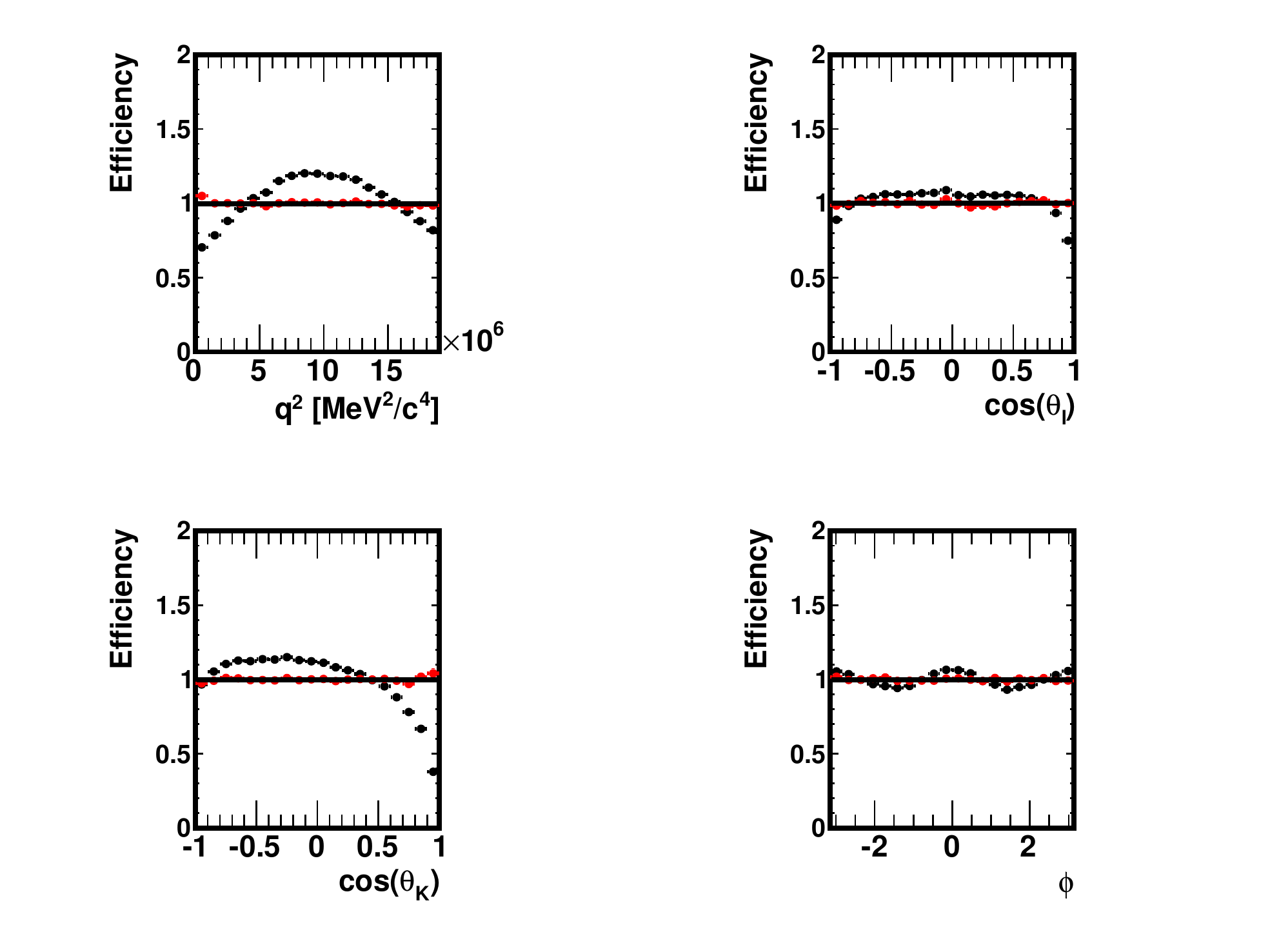}
    \vspace*{-0.5cm}
  \end{center}
  \caption{Efficiency in $q^{2}$, cos$\theta_l$, cos$\theta_K$ and $\phi$ in the data generated for the present study. 
Shown are the ratios of the distributions found in the distorted and original samples, with no correction (black) and for 
decays re-weighted using $\omega_i$ weights (red) as explained in the text. 
The absolute normalisation is arbitrary when the correction is not applied, natural when it is (red).} 
  \label{fig:Ratio_whole_KstMuMu}
\end{figure}

\begin{figure}[tb]
  \begin{center}
    \hspace*{-2.3cm}
    \includegraphics[width=1.3\linewidth]{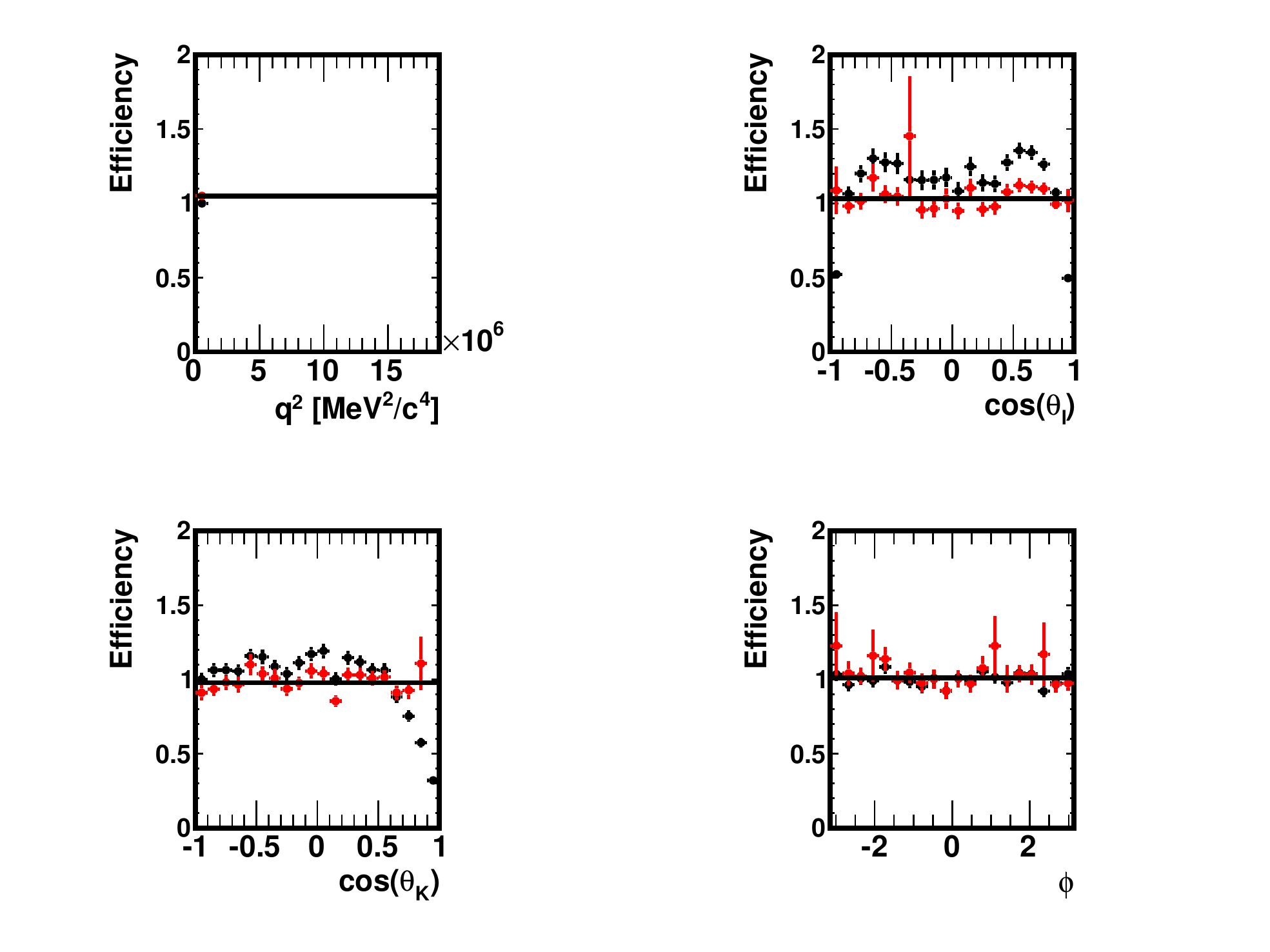}
    \vspace*{-0.5cm}
  \end{center}
  \caption{Efficiency in $q^{2}$, cos$\theta_l$, cos$\theta_K$ and $\phi$ in the data generated for the present study, 
in the region $0.1 < q^2 < 0.98$ GeV$^{2}/c^4$. Shown are the ratios of the distributions found in the distorted and 
original samples, with no correction (black) and for decays re-weighted using $\omega_i$ weights (red)
as explained in the text. The absolute normalisation is arbitrary when the correction is not applied, natural 
when it is (red).} 
  \label{fig:Ratio_q2_1_dem}
\end{figure}

\begin{figure}[tb]
  \begin{center}
    \hspace*{-2.3cm}
    \includegraphics[width=1.3\linewidth]{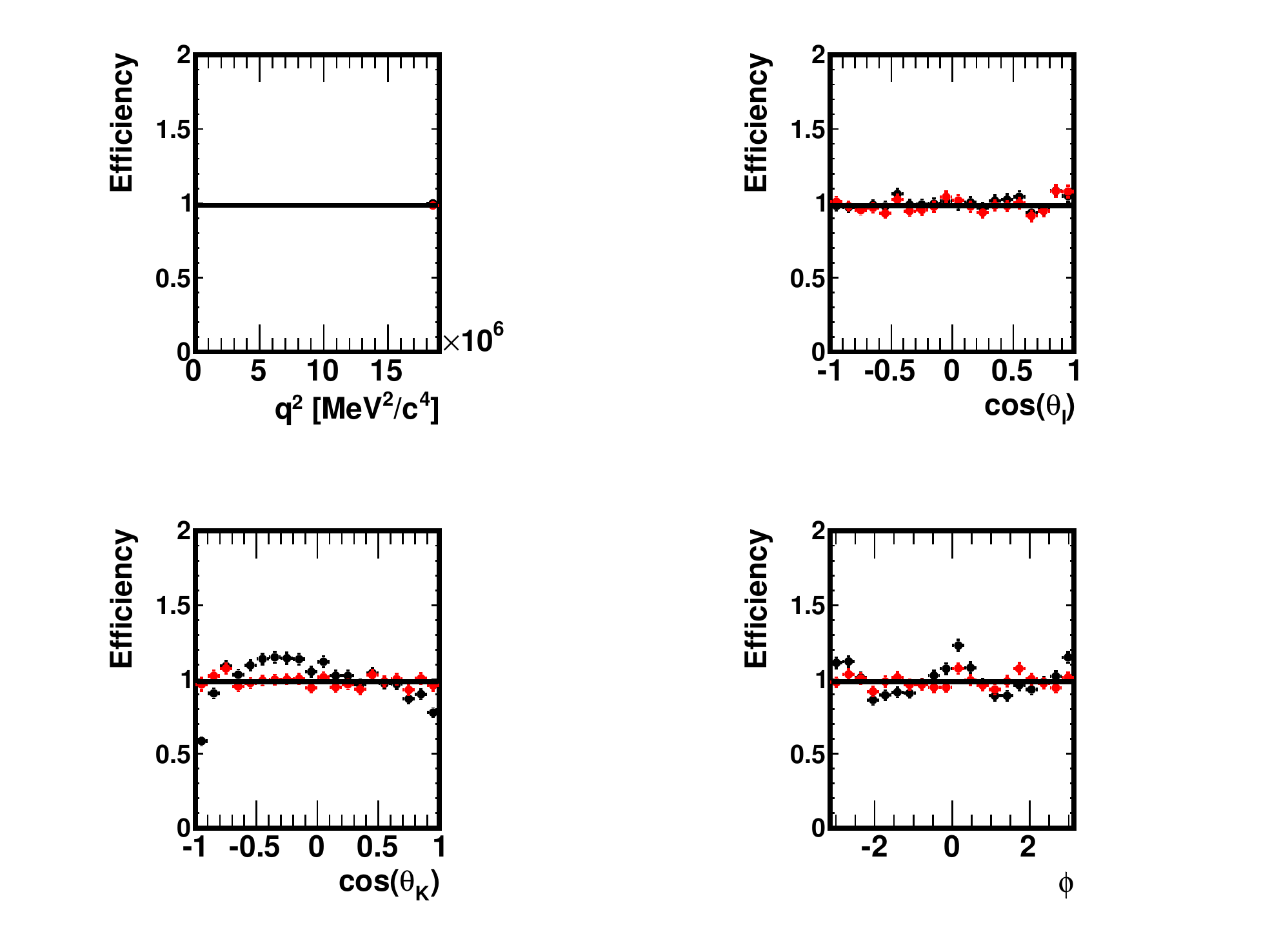}
    \vspace*{-0.5cm}
  \end{center}
  \caption{Efficiency in $q^{2}$, cos$\theta_l$, cos$\theta_K$ and $\phi$ in the data generated for the present study, 
in the region $18.0 < q^2 < 19.0$ GeV$^{2}/c^4$. Shown are the ratios of the distributions found in the distorted and 
original samples, with no correction (black) and for decays re-weighted using $\omega_i$ weights (red)
as explained in the text. The absolute normalisation is arbitrary when the correction is not applied, natural 
when it is (red).} 
  \label{fig:Ratio_q2_19_dem}
\end{figure}

We show on Fig.~\ref{fig:Score_KstMuMu} the distribution of the NN score $s$ obtained in  original and distorted test samples, 
generated independently from the training samples. This figure also shows $\epsilon\left(s\right)$, the parameterised 
efficiency fitted to the ratio of the original and distorted distributions.
We tested this efficiency in the same way as in Sect.~\ref{subsec:ResultsK3pi}.
Fig.~\ref{fig:Ratio_whole_KstMuMu} shows the efficiency in $q^2$, cos$\theta_l$, cos$\theta_K$ and $\phi$. Also shown are the efficiencies 
obtained with the corrected distorted sample, in which each decay $i$ is weigted by $\omega_i = 1/\epsilon\left(s_i\right)$.
The corrections works precisely: in no bin does the latter ratio differ from 1 by more than a few percents. This difference
is never statiscally significant. The latter statement also holds in 
specific regions of the $q^{2}$ distribution, as can be seen on Fig.~\ref{fig:Ratio_q2_1_dem}, which 
shows this efficiency in the region $0.1 < q^2 < 0.98$ GeV$^{2}/c^4$ and on Fig.~\ref{fig:Ratio_q2_19_dem}
which focusses on the region $18.0 < q^2 < 19.0$ GeV$^{2}/c^4$. This can be compared to what was obtained by the 
analysis reported in~\cite{LHCb_KstMuMu_conf} (Fig.~\ref{fig:Fig-KstMuMu-1}), 
with the principal moment analysis briefly described in Sect.~\ref{sec:TechSummClass}. The correction we obtained is less 
precise statistically --- which is natural with more than twice less statistics --- but is of comparable accuracy. This is 
a promising result since it was obtained with very limited efforts and MVA-related competences. The quality 
of the correction was confirmed in 19 $q^{2}$ regions, as can be seen on Figs.~\ref{fig:Ratio_q2_1} to~\ref{fig:Ratio_q2_19} in 
Appendix~\ref{App:AppB}. We remind that
in each $q^{2}$ region the weights are still calculated from the $\epsilon\left(s\right)$ function fitted to the efficiency
shown on Fig.~\ref{fig:Score_KstMuMu}. No specific training and no 
re-evaluation of the polynomial is performed in individual regions. 

\clearpage 

As in Sect.~\ref{sec:AppliedToDK3pi}, we conclude the $\epsilon\left(s\right)$
obtained with the approach proposed in this document can be used to evaluate the effiency 
at a given point of the decay phase space.

\section{Conclusion}
\label{sec:Conclusion}

We proposed a novel approach to the determination of multidimensional efficiencies 
and explored its potential with two realistic examples, typical of the need
of modern Heavy Flavor physics measurements. We used Neural Networks to characterize
the differences introduced in a 4 or 5-dimensional phase space by typical 
reconstruction and selecion criteria. These tools were trained and tested using 
simulated samples of similar size to that of the samples used by the LHCb collaboration
for published (or soom published) measurements of the decay modes \norm and \kstmumu.
In both test cases studied in this document, the NN score allows to correct
the selected samples in order to reproduce the phase space distributions
observed in samples that have not undergone any selection. This is an evidence
the approach developped here allows to evaluate the efficiency at any point of 
a multidimensional phase space. Compared to elaborate techniques like the 
principal moment method used in~\cite{LHCb_KstMuMu_conf}, this new appraoch seems
less precise statistically although as accurate. It would probably suffice for many 
measurements that do not require the same level of precision as the angular
analysis of \kstmumu reported in~\cite{LHCb_KstMuMu_conf}. In such cases, it would
represent a considerable gain of working time since satisfactory results 
can be achieved with  minimum  skills and knowledge of MVA techniques, using
packages already routinely used within the HEP community, with no need of any
elaborate optimisation nor of more than a few days of work and 
a few hours of CPU consumption. This is the conclusion of this study. It also
suggests that with more expertise and cutting-edge MVA techniques, a precise
treatment of multidimensional efficiencies is possible and could be applied
to measurements of primary importance. 

Other applications of MVAs to HEP, besides signal vs. background discrimination, can 
be considered in the future. Selection efficiencies often rely on simulations that 
match real data only imperfectly and require systematic Data/MC comparisons to correct 
the simulation.  When the correction must be applied to many variables, using a MVA to 
compare data with MC and derive ``automatically'' a unique number to reweight the
simulation would be a valuable tool. 

In a given analysis, if one knows the efficiency at each point of the 
phase space, the signal events found in data can be corrected to obtained the 
distributions of interest before any selection bias, without having to 
use an imperfect simulation. This is not possible in cases where a rare decay is searched 
for. There, very few signal events are found, if any, and there is nothing to re-weight. The technique proposed
in this document could be used to guide the selection design in order to obtain
a flat $\epsilon(s)$. For that purpose, one could re-train the MVA developped 
for the signal selection with a weight applied to signal training events, derived from the $\epsilon\left(s\right)$
observed when the original selection is applied.
If this goal is achieved, even a simulation that does not reproduce correctly
the real phase space of the decay can be used to determine the efficiency. Upper limits
are often normalised to the branching fraction of a well known
non-suppressed decay which decay products are identical to the signal's. It differs 
from the signal only due to a different phase space. Ensuring for both modes a flat 
efficiency across the phase space would make their efficiency ratio closer to one and 
more robust against systematic uncertainties.

\newpage

\bibliographystyle{LHCb}

\bibliography{main}

\newpage
\begin{appendices}
\section{ Additional figures related to the \norm test case}
\label{App:AppA}

\begin{figure}[tb]
    \hspace*{-3cm}
       \includegraphics[width=1.4\linewidth]{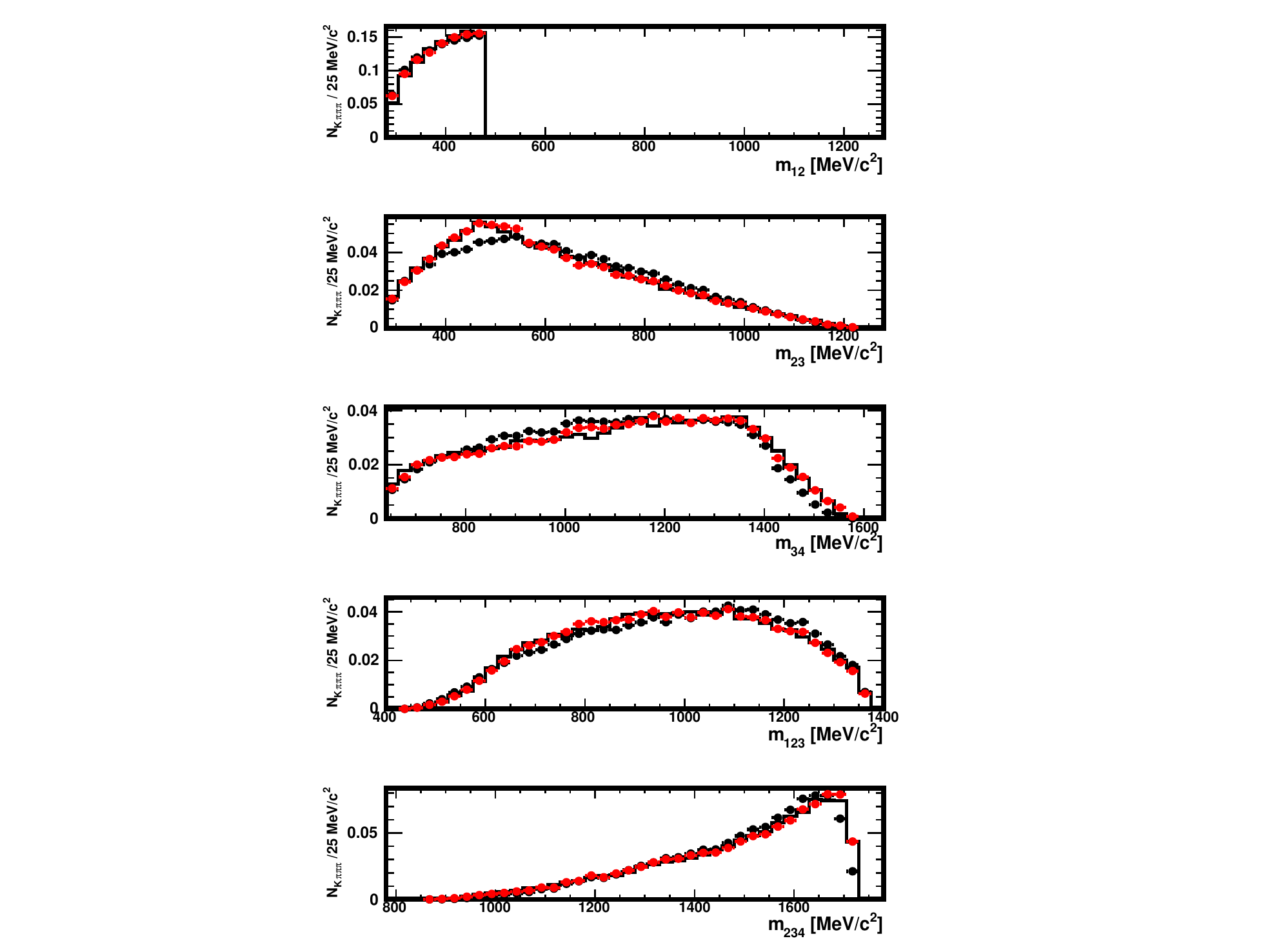}
    \vspace*{-0.5cm}
  \caption{Distributions of $m_{12}$, $m_{23}$, $m_{34}$, $m_{123}$ and $m_{234}$ in the original sample (histogram),
in the distorted one (full black circles) and in the distorted sample where the decays have been re-weighted using the $\omega_i$ weights (red), 
as explained in Sect.~\ref{subsec:ResultsK3pi}. The data used here are restricted to the region $ 0 < m_{12} < 480$ MeV$/c^2$.
The absolute normalisation is arbitrary when the correction is not applied and natural when it is applied (red).} 
  \label{fig:Distr_m12_1}
\end{figure}

\clearpage

\begin{figure}[tb]
    \hspace*{-3cm}
       \includegraphics[width=1.4\linewidth]{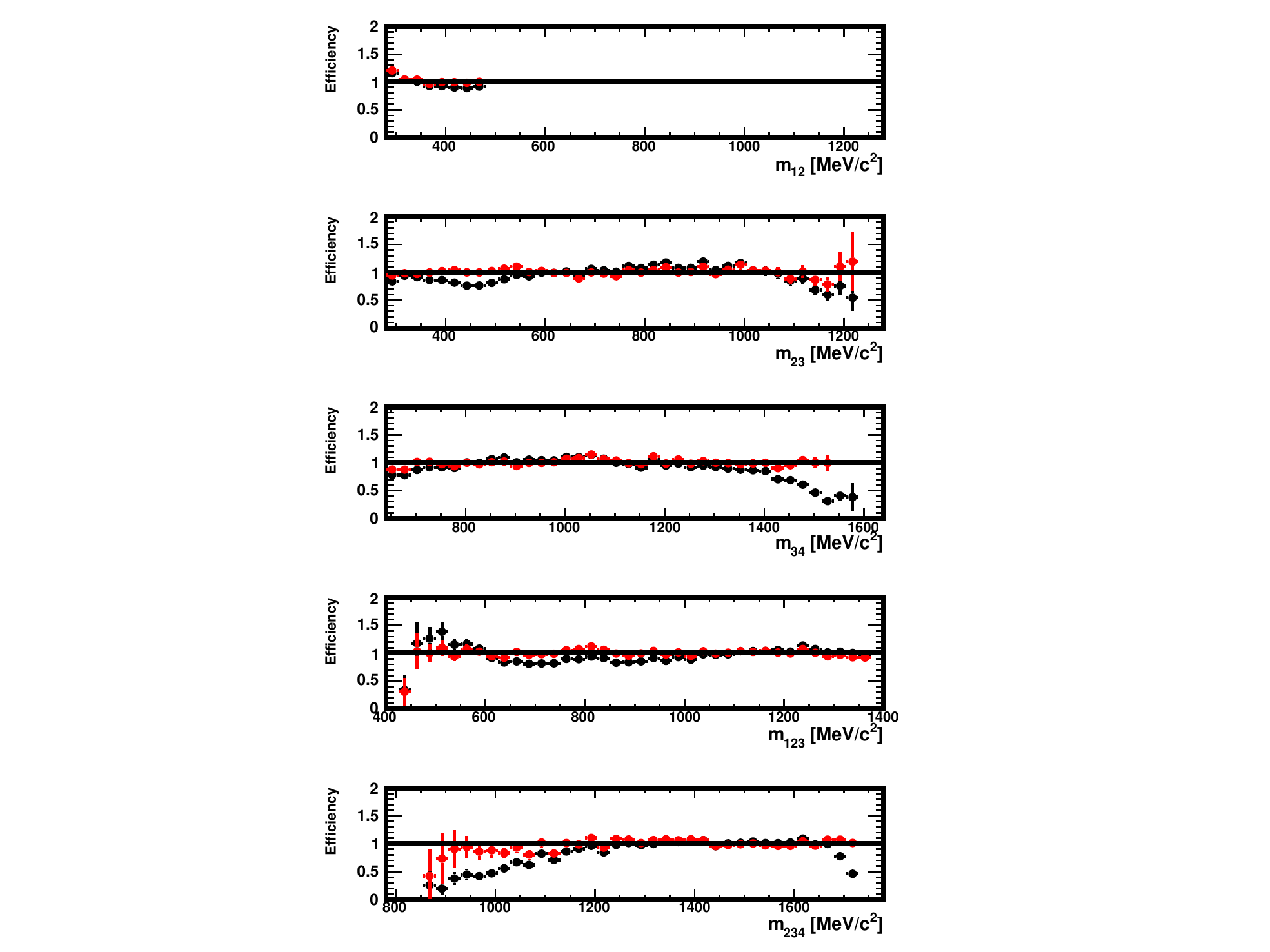}
    \vspace*{-0.5cm}
  \caption{Efficiency in $m_{12}$, $m_{23}$, $m_{34}$, $m_{123}$ and $m_{234}$ in the data generated for this study, 
in the region $0 < m_{12} < 480$ MeV$/c^2$. Shown are the ratios of the distributions found in the distorted and 
original samples, with no correction (black) and for decays re-weighted using $\omega_i$ weights (red)
as explained in Sect.~\ref{subsec:ResultsK3pi}. The absolute normalisation is arbitrary when the correction is not applied and natural when it is applied (red).} 
  \label{fig:Ratio_m12_1}
\end{figure}

\clearpage

\begin{figure}[tb]
    \hspace*{-3cm}
       \includegraphics[width=1.4\linewidth]{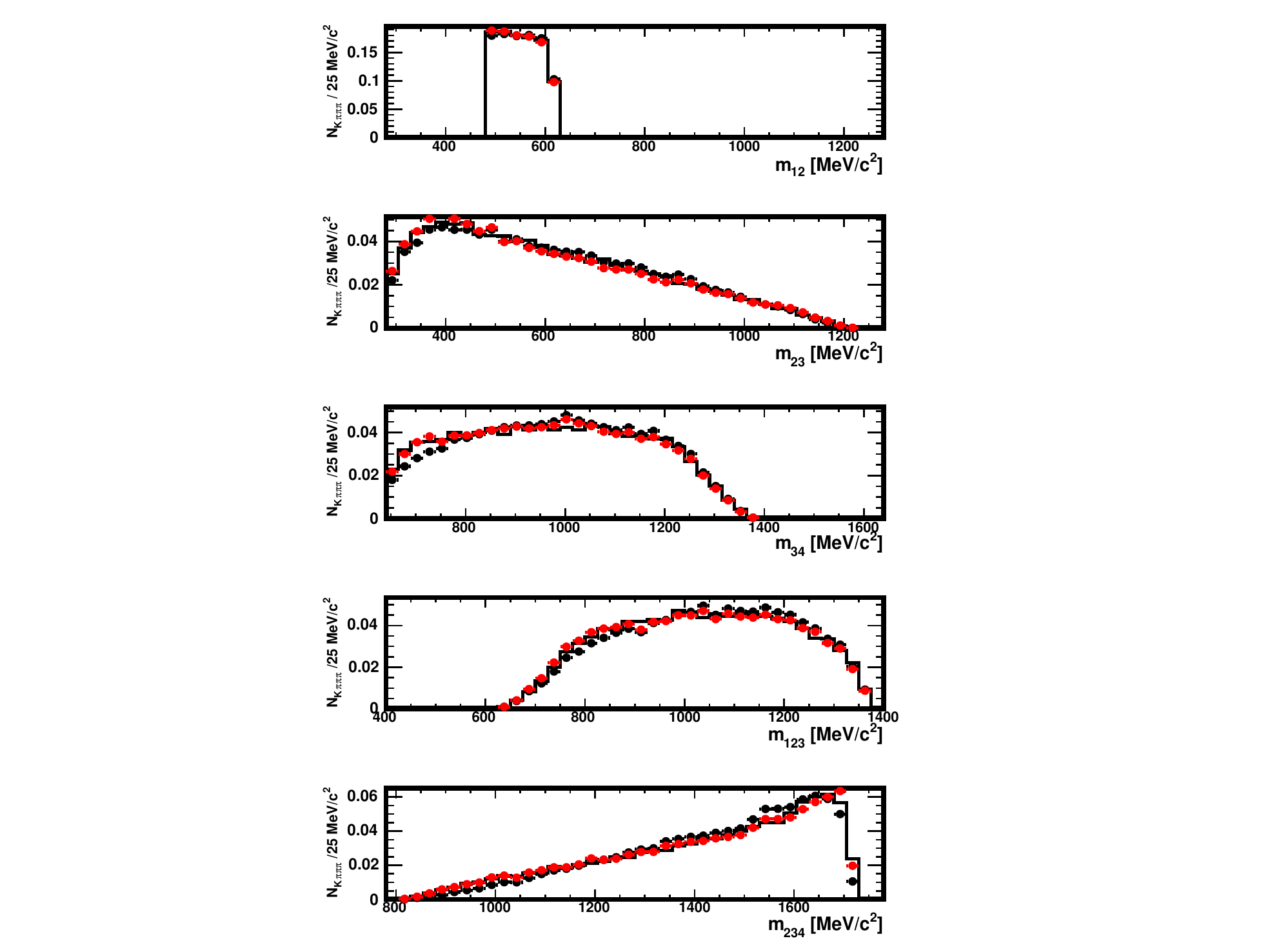}
    \vspace*{-0.5cm}
  \caption{Distributions of $m_{12}$, $m_{23}$, $m_{34}$, $m_{123}$ and $m_{234}$ in the original sample (histogram),
in the distorted one (full black circles) and in the distorted sample where the decays have been re-weighted using the $\omega_i$ weights (red), 
as explained in Sect.~\ref{subsec:ResultsK3pi}. The data used here are restricted to the region $480 < m_{12} < 620$ MeV$/c^2$.
The absolute normalisation is arbitrary when the correction is not applied and natural when it is applied (red).} 
  \label{fig:Distr_m12_2}
\end{figure}

\clearpage

\begin{figure}[tb]
    \hspace*{-3cm}
       \includegraphics[width=1.4\linewidth]{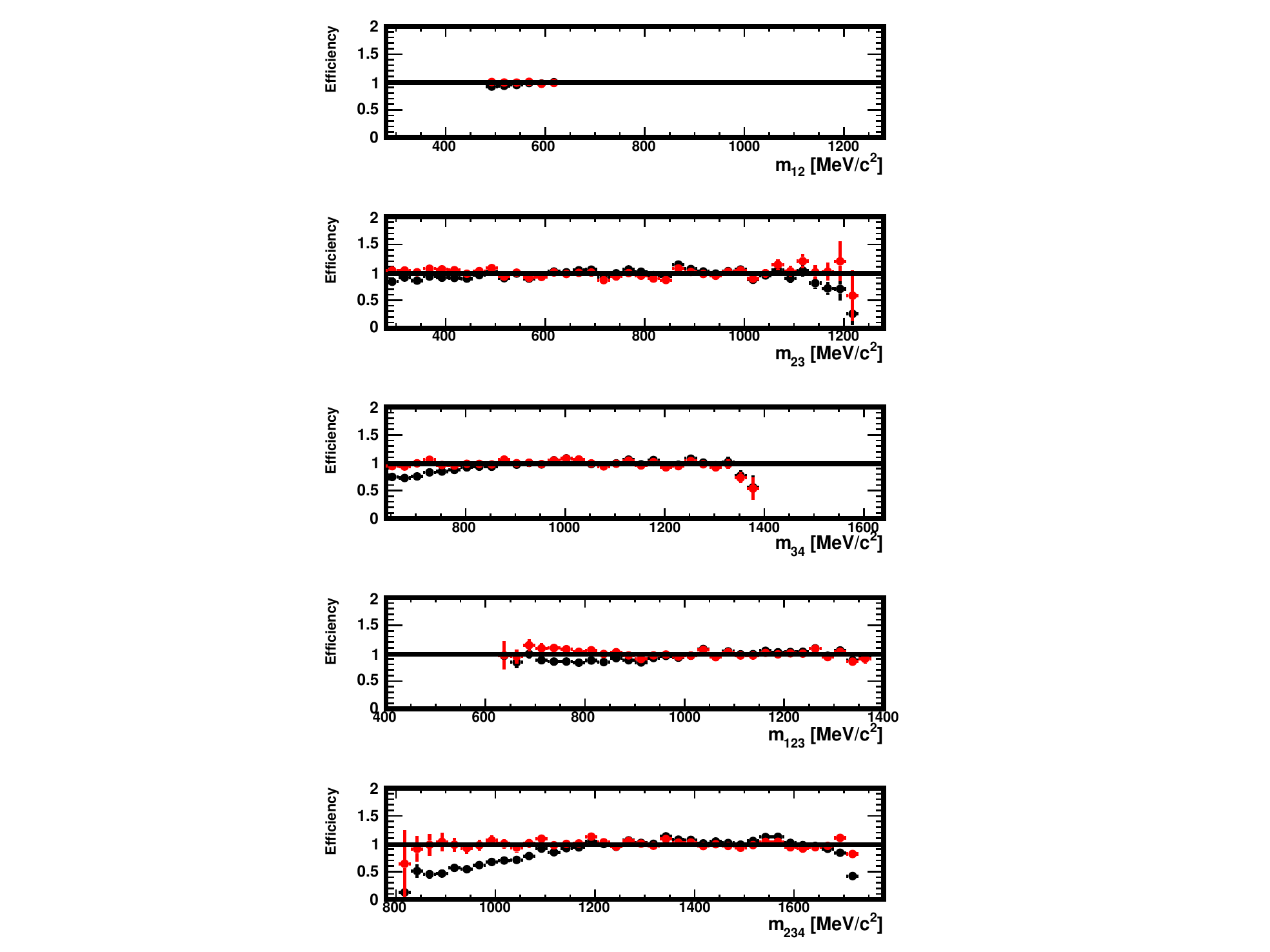}
    \vspace*{-0.5cm}
  \caption{Efficiency in $m_{12}$, $m_{23}$, $m_{34}$, $m_{123}$ and $m_{234}$ in the data generated for this study, 
in the region $ <480 m_{12} < 620$ MeV$/c^2$. Shown are the ratios of the distributions found in the distorted and 
original samples, with no correction (black) and for decays re-weighted using $\omega_i$ weights (red)
as explained in Sect.~\ref{subsec:ResultsK3pi}. The absolute normalisation is arbitrary when the correction is not applied and natural when it is applied (red).} 
  \label{fig:Ratio_m12_2}
\end{figure}

\clearpage

\begin{figure}[tb]
    \hspace*{-3cm}
       \includegraphics[width=1.4\linewidth]{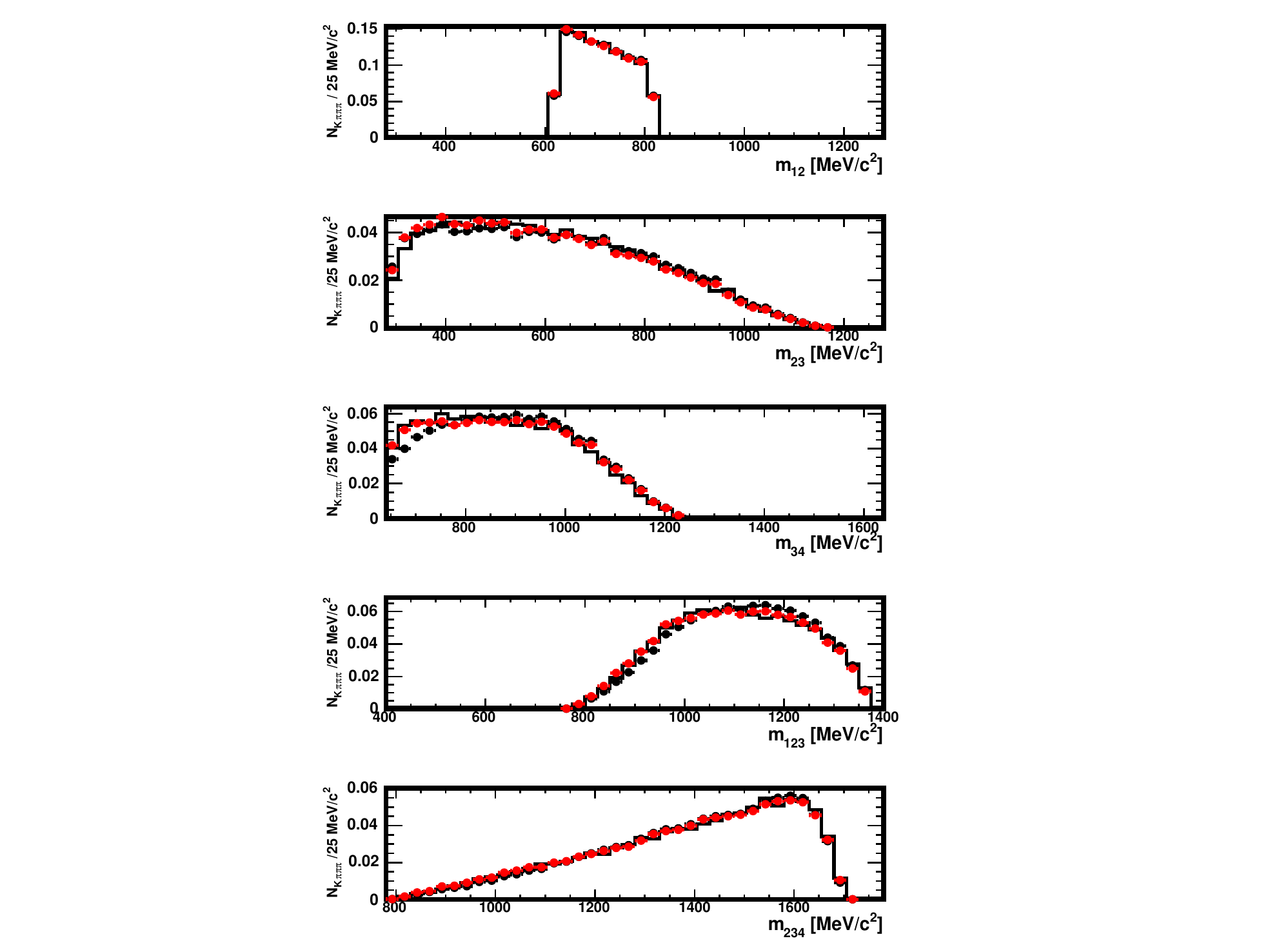}
    \vspace*{-0.5cm}
  \caption{Distributions of $m_{12}$, $m_{23}$, $m_{34}$, $m_{123}$ and $m_{234}$ in the original sample (histogram),
in the distorted one (full black circles) and in the distorted sample where the decays have been re-weighted using the $\omega_i$ weights (red), 
as explained in Sect.~\ref{subsec:ResultsK3pi}. The data used here are restricted to the region $620 < m_{12} < 820$ MeV$/c^2$.
The absolute normalisation is arbitrary when the correction is not applied and natural when it is applied (red).} 
  \label{fig:Distr_m12_3}
\end{figure}

\clearpage

\begin{figure}[tb]
    \hspace*{-3cm}
       \includegraphics[width=1.4\linewidth]{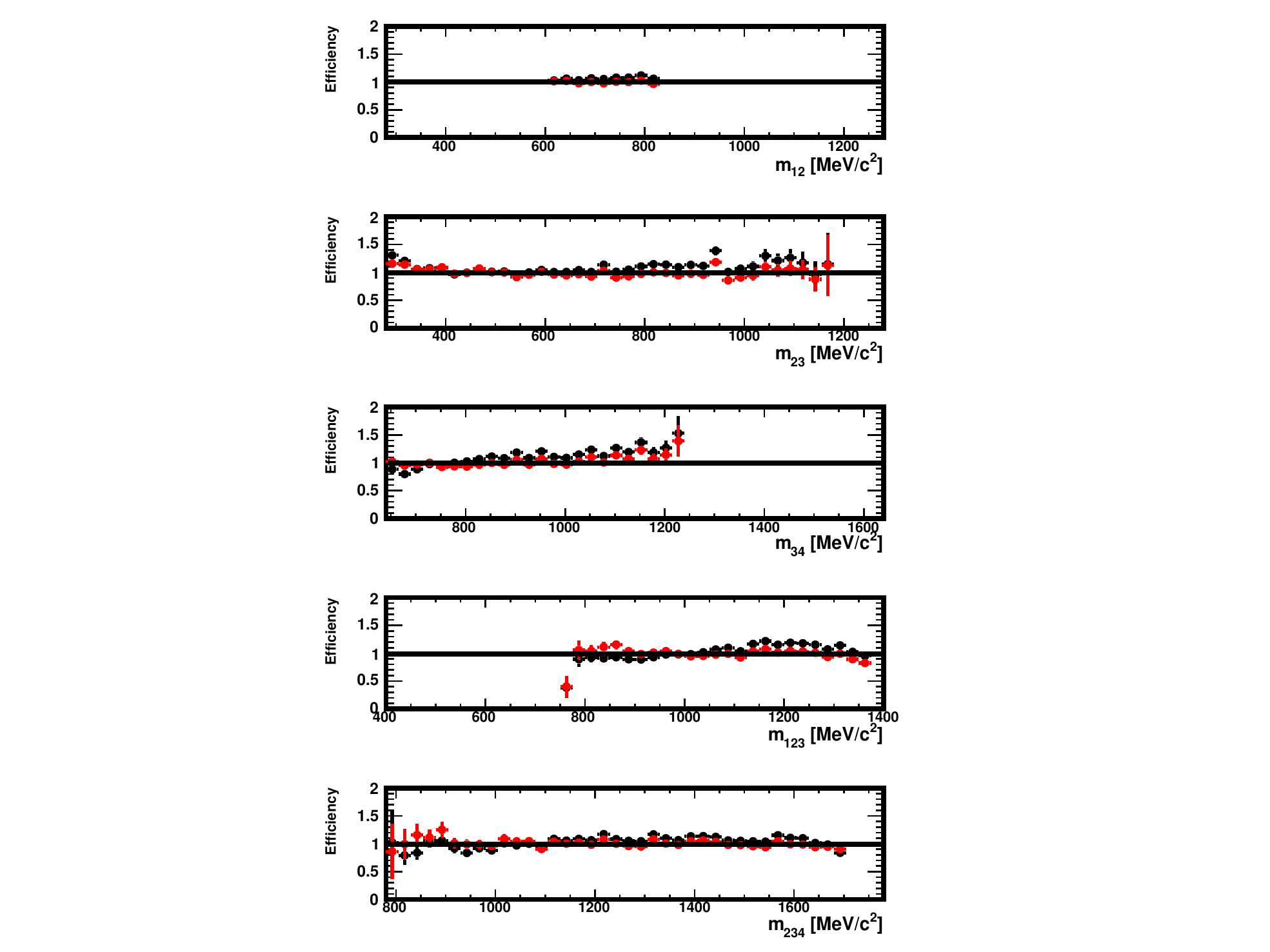}
    \vspace*{-0.5cm}
  \caption{Efficiency in $m_{12}$, $m_{23}$, $m_{34}$, $m_{123}$ and $m_{234}$ in the data generated for this study, 
in the region $620 < m_{12} < 820$ MeV$/c^2$. Shown are the ratios of the distributions found in the distorted and 
original samples, with no correction (black) and for decays re-weighted using $\omega_i$ weights (red)
as explained in Sect.~\ref{subsec:ResultsK3pi}. The absolute normalisation is arbitrary when the correction is not applied and natural when it is applied (red).} 
  \label{fig:Ratio_m12_3}
\end{figure}

\clearpage

\begin{figure}[tb]
    \hspace*{-3cm}
       \includegraphics[width=1.4\linewidth]{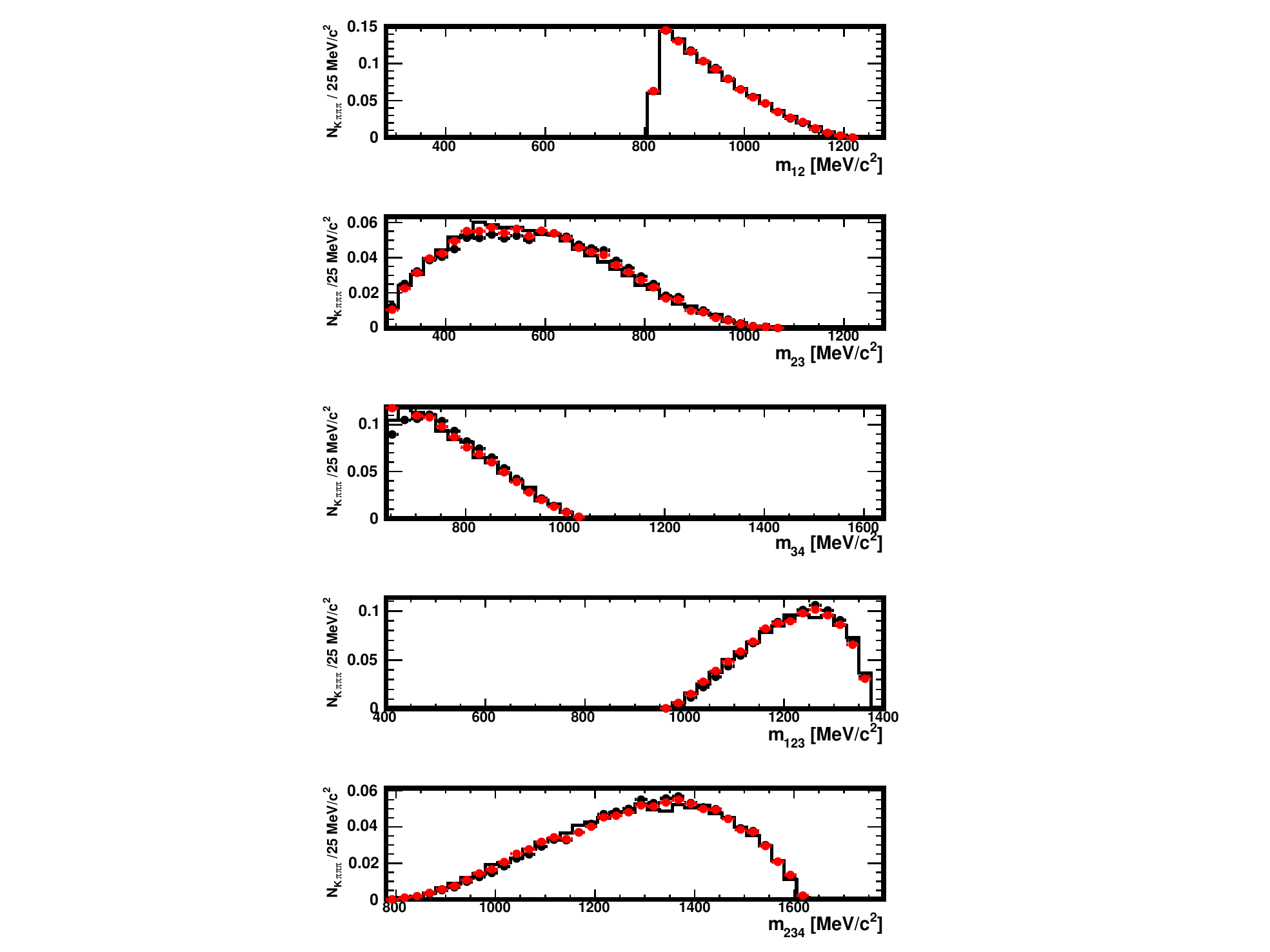}
    \vspace*{-0.5cm}
  \caption{Distributions of $m_{12}$, $m_{23}$, $m_{34}$, $m_{123}$ and $m_{234}$ in the original sample (histogram),
in the distorted one (full black circles) and in the distorted sample where the decays have been re-weighted using the $\omega_i$ weights (red), 
as explained in Sect.~\ref{subsec:ResultsK3pi}. The data used here are restricted to the region $820 < m_{12}$ MeV$/c^2$.
The absolute normalisation is arbitrary when the correction is not applied and natural when it is applied (red).} 
  \label{fig:Distr_m12_4}
\end{figure}

\clearpage

\begin{figure}[tb]
    \hspace*{-3cm}
       \includegraphics[width=1.4\linewidth]{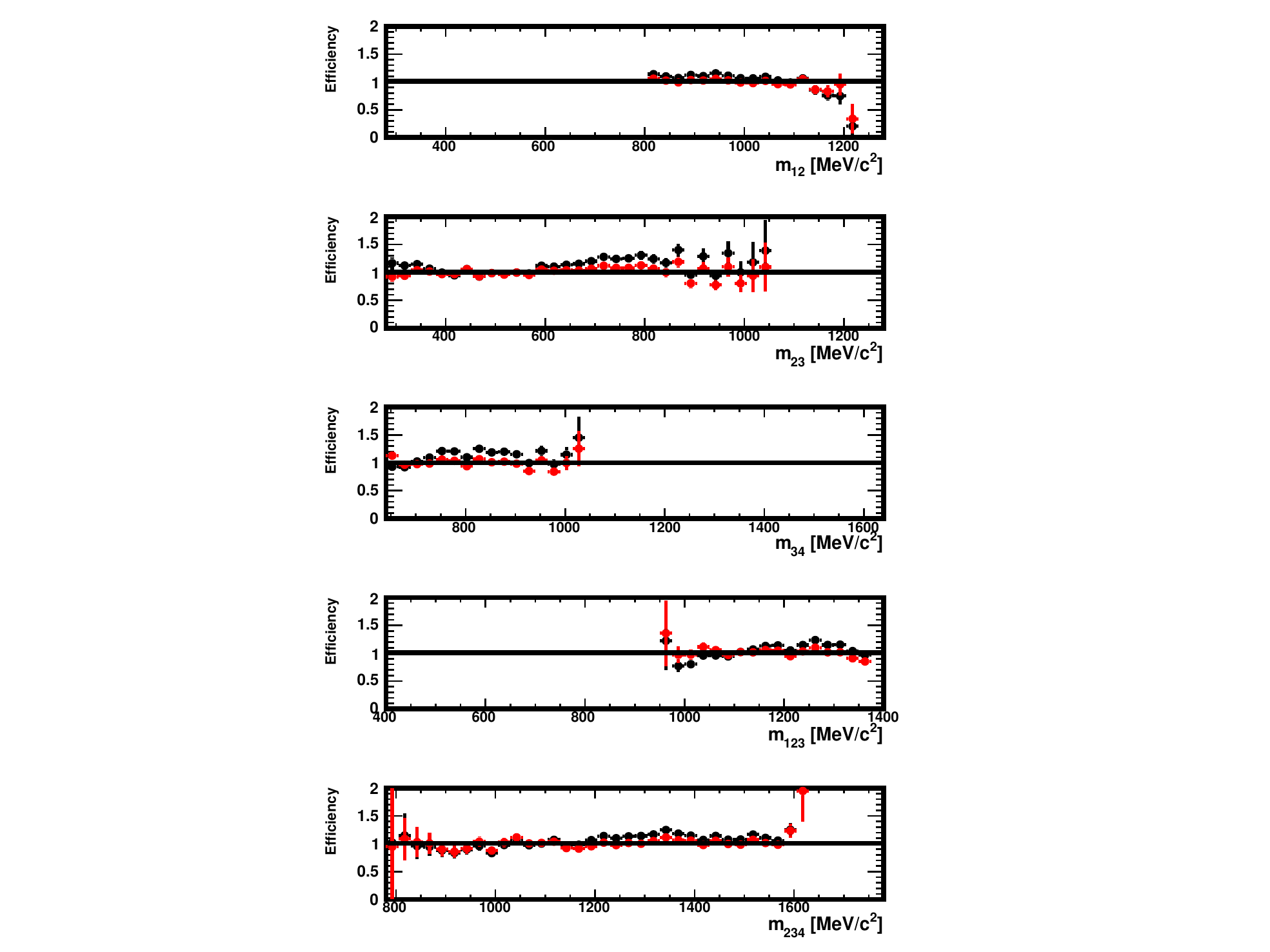}
    \vspace*{-0.5cm}
  \caption{Efficiency in $m_{12}$, $m_{23}$, $m_{34}$, $m_{123}$ and $m_{234}$ in the data generated for this study, 
in the region $820 < m_{12}$ MeV$/c^2$. Shown are the ratios of the distributions found in the distorted and 
original samples, with no correction (black) and for decays re-weighted using $\omega_i$ weights (red)
as explained in Sect.~\ref{subsec:ResultsK3pi}. The absolute normalisation is arbitrary when the correction is not applied and natural when it is applied (red).} 
  \label{fig:Ratio_m12_4}
\end{figure}

\clearpage

\begin{figure}[tb]
    \hspace*{-3cm}
       \includegraphics[width=1.4\linewidth]{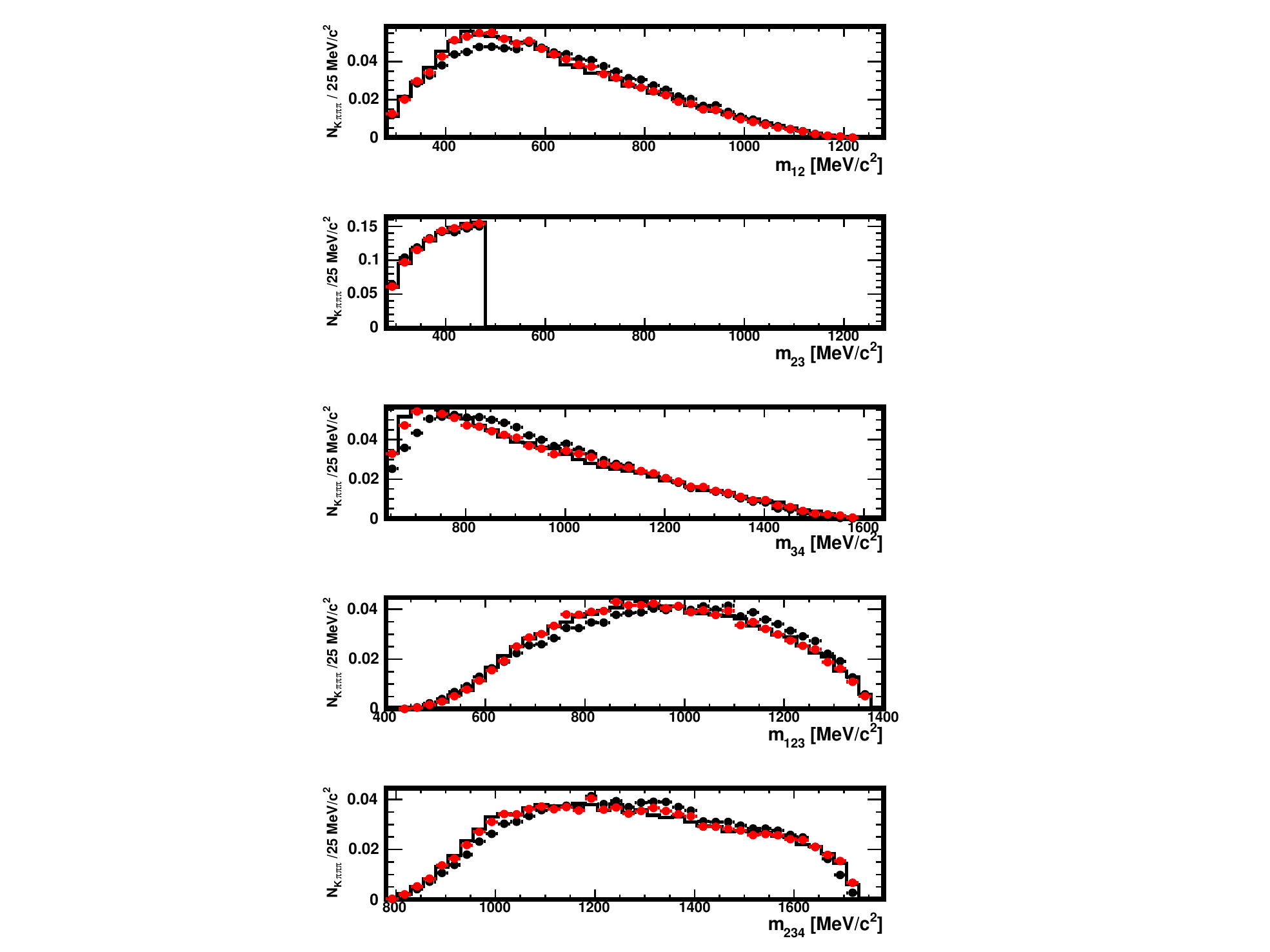}
    \vspace*{-0.5cm}
  \caption{Distributions of $m_{12}$, $m_{23}$, $m_{34}$, $m_{123}$ and $m_{234}$ in the original sample (histogram),
in the distorted one (full black circles) and in the distorted sample where the decays have been re-weighted using the $\omega_i$ weights (red), 
as explained in Sect.~\ref{subsec:ResultsK3pi}. The data used here are restricted to the region $ 0 < m_{23} < 480$ MeV$/c^2$.
The absolute normalisation is arbitrary when the correction is not applied and natural when it is applied (red).} 
  \label{fig:Distr_m23_1}
\end{figure}

\clearpage

\begin{figure}[tb]
    \hspace*{-3cm}
       \includegraphics[width=1.4\linewidth]{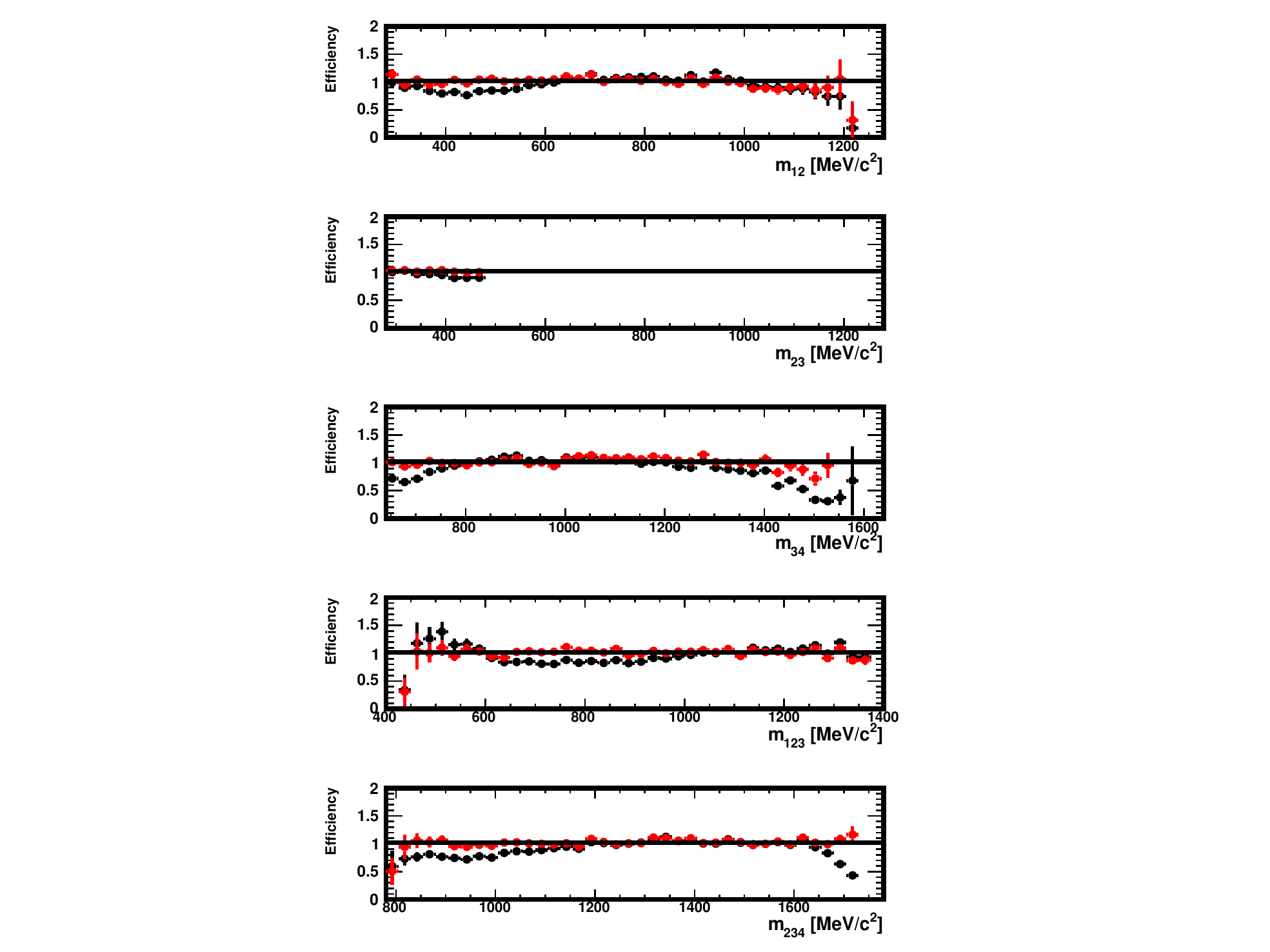}
    \vspace*{-0.5cm}
  \caption{Efficiency in $m_{12}$, $m_{23}$, $m_{34}$, $m_{123}$ and $m_{234}$ in the data generated for this study, 
in the region $0 < m_{23} < 480$ MeV$/c^2$. Shown are the ratios of the distributions found in the distorted and 
original samples, with no correction (black) and for decays re-weighted using $\omega_i$ weights (red)
as explained in Sect.~\ref{subsec:ResultsK3pi}. The absolute normalisation is arbitrary when the correction is not applied and natural when it is applied (red).} 
  \label{fig:Ratio_m23_1}
\end{figure}

\clearpage

\begin{figure}[tb]
    \hspace*{-3cm}
       \includegraphics[width=1.4\linewidth]{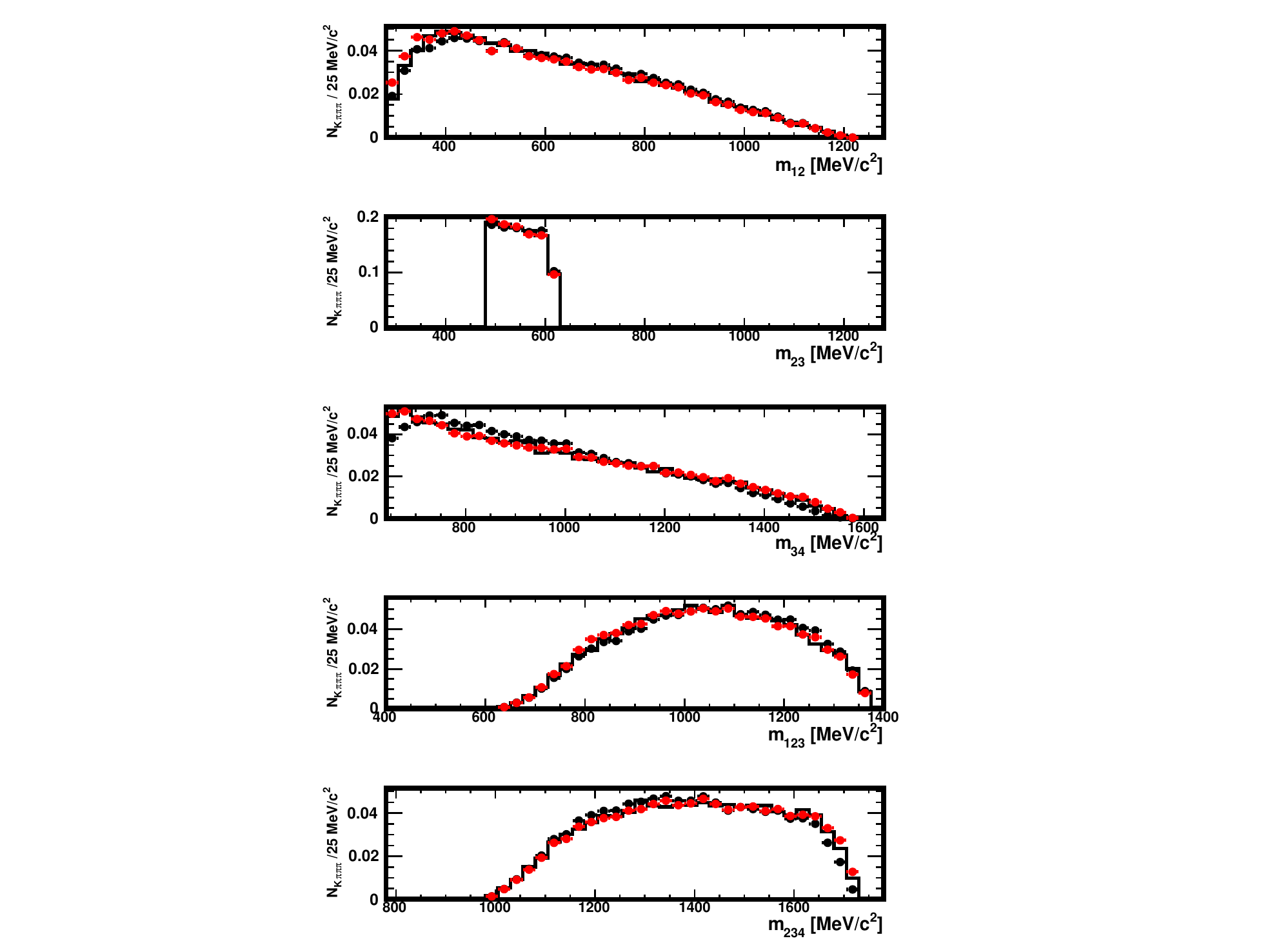}
    \vspace*{-0.5cm}
  \caption{Distributions of $m_{12}$, $m_{23}$, $m_{34}$, $m_{123}$ and $m_{234}$ in the original sample (histogram),
in the distorted one (full black circles) and in the distorted sample where the decays have been re-weighted using the $\omega_i$ weights (red), 
as explained in Sect.~\ref{subsec:ResultsK3pi}. The data used here are restricted to the region $480 < m_{23} < 620$ MeV$/c^2$.
The absolute normalisation is arbitrary when the correction is not applied and natural when it is applied (red).} 
  \label{fig:Distr_m23_2}
\end{figure}

\clearpage

\begin{figure}[tb]
    \hspace*{-3cm}
       \includegraphics[width=1.4\linewidth]{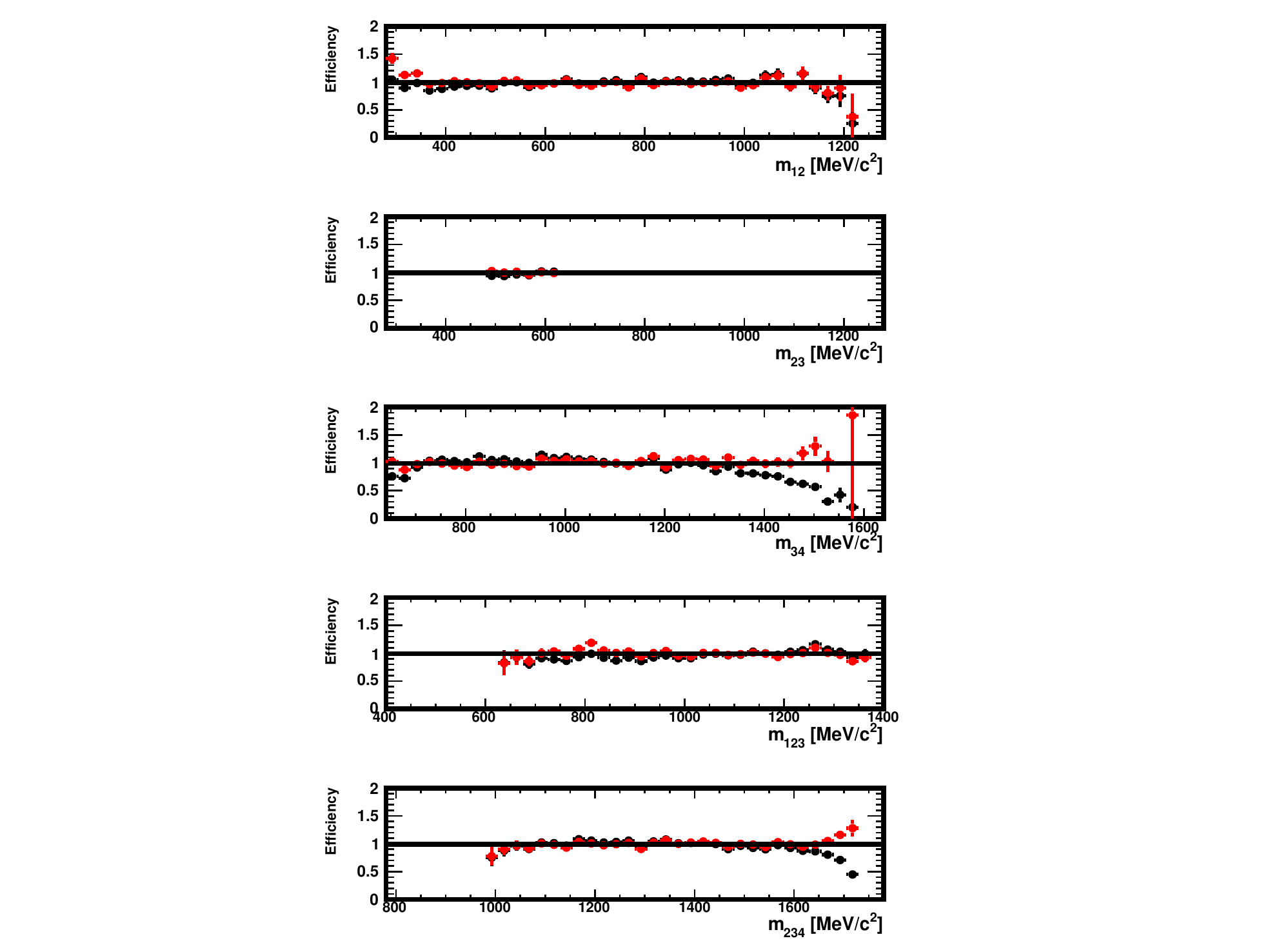}
    \vspace*{-0.5cm}
  \caption{Efficiency in $m_{12}$, $m_{23}$, $m_{34}$, $m_{123}$ and $m_{234}$ in the data generated for this study, 
in the region $ <480 m_{23} < 620$ MeV$/c^2$. Shown are the ratios of the distributions found in the distorted and 
original samples, with no correction (black) and for decays re-weighted using $\omega_i$ weights (red)
as explained in Sect.~\ref{subsec:ResultsK3pi}. The absolute normalisation is arbitrary when the correction is not applied and natural when it is applied (red).} 
  \label{fig:Ratio_m23_2}
\end{figure}

\clearpage

\begin{figure}[tb]
    \hspace*{-3cm}
       \includegraphics[width=1.4\linewidth]{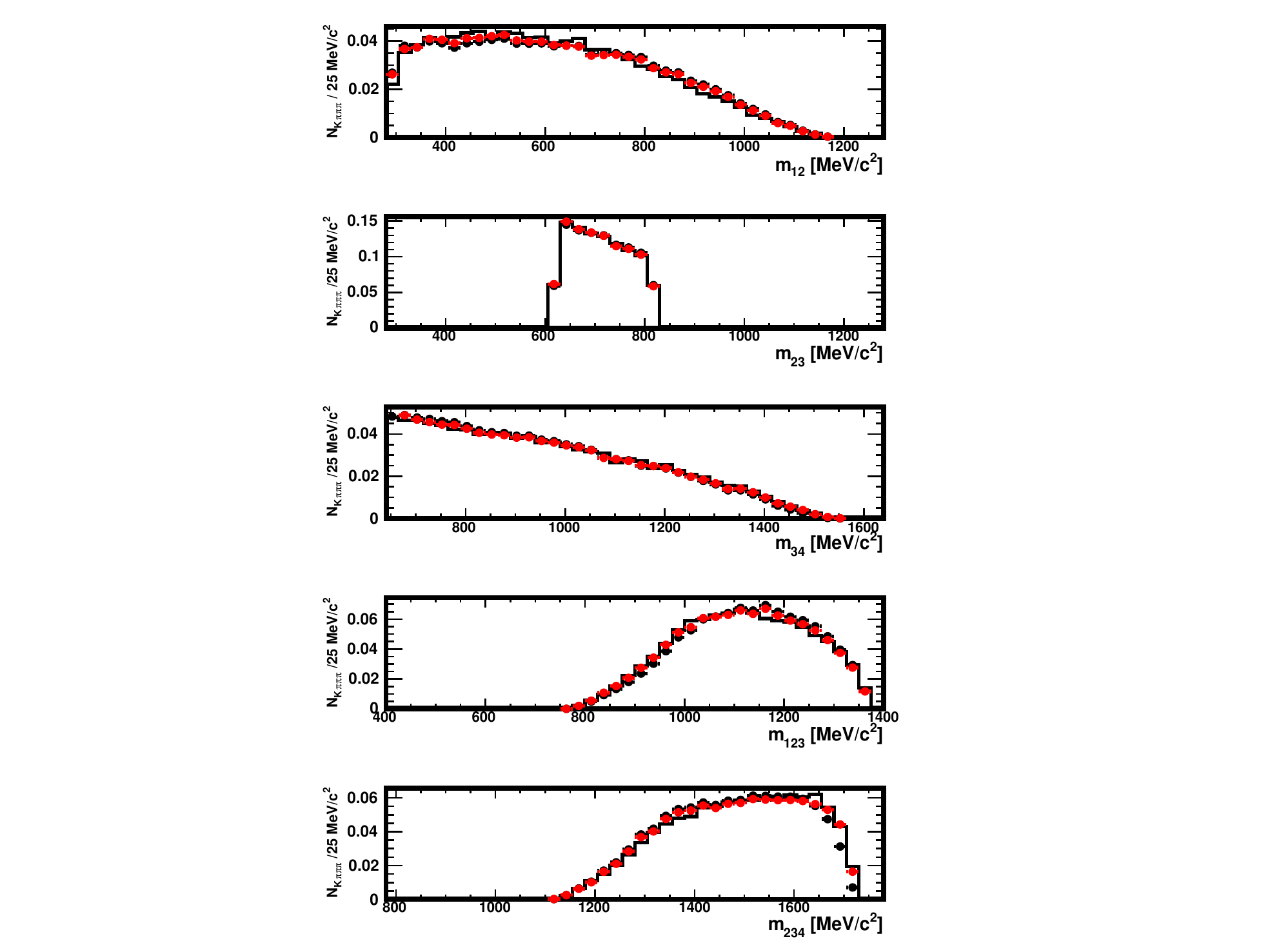}
    \vspace*{-0.5cm}
  \caption{Distributions of $m_{12}$, $m_{23}$, $m_{34}$, $m_{123}$ and $m_{234}$ in the original sample (histogram),
in the distorted one (full black circles) and in the distorted sample where the decays have been re-weighted using the $\omega_i$ weights (red), 
as explained in Sect.~\ref{subsec:ResultsK3pi}. The data used here are restricted to the region $620 < m_{23} < 820$ MeV$/c^2$.
The absolute normalisation is arbitrary when the correction is not applied and natural when it is applied (red).} 
  \label{fig:Distr_m23_3}
\end{figure}

\clearpage

\begin{figure}[tb]
    \hspace*{-3cm}
       \includegraphics[width=1.4\linewidth]{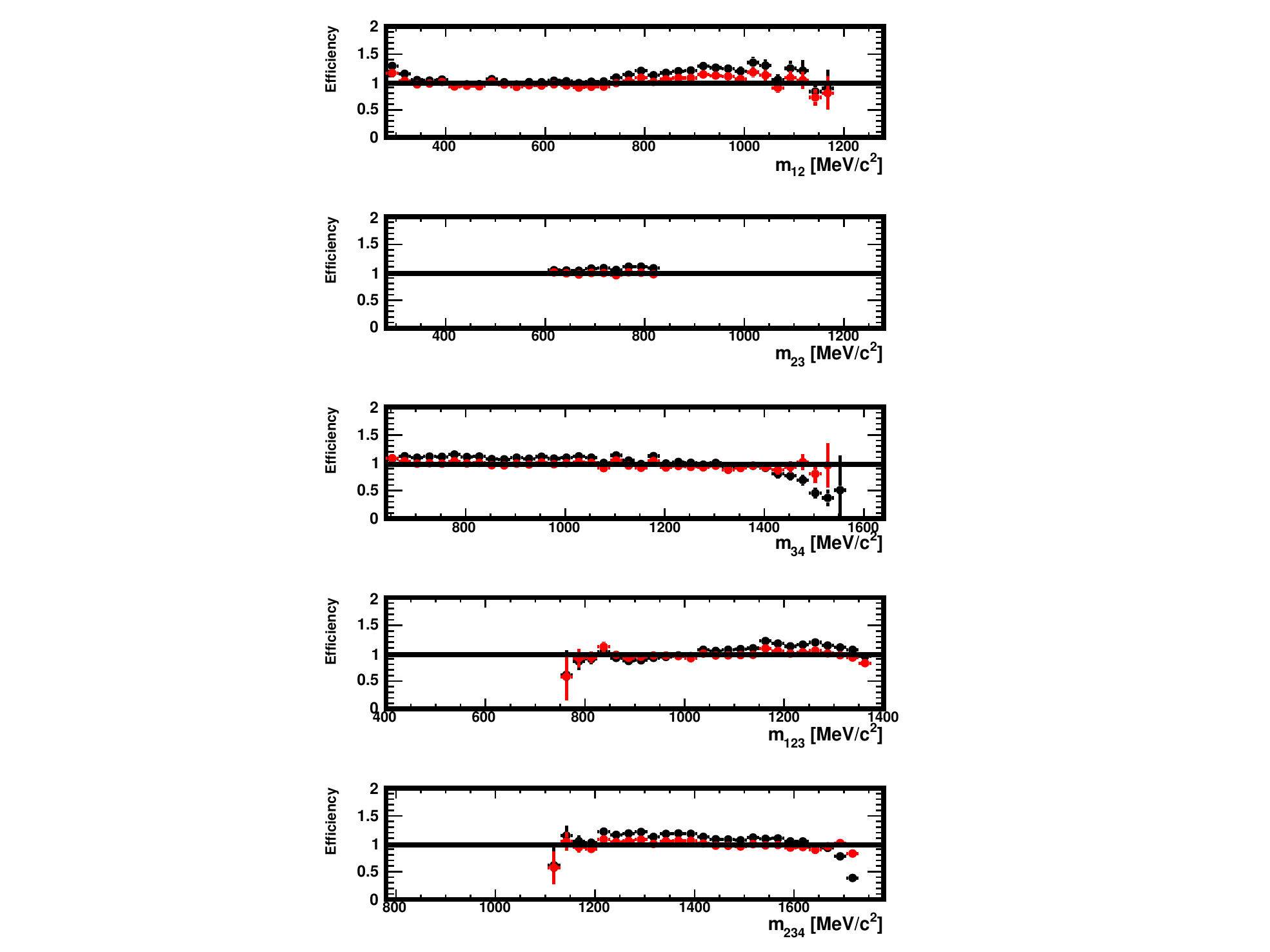}
    \vspace*{-0.5cm}
  \caption{Efficiency in $m_{12}$, $m_{23}$, $m_{34}$, $m_{123}$ and $m_{234}$ in the data generated for this study, 
in the region $620 < m_{23} < 820$ MeV$/c^2$. Shown are the ratios of the distributions found in the distorted and 
original samples, with no correction (black) and for decays re-weighted using $\omega_i$ weights (red)
as explained in Sect.~\ref{subsec:ResultsK3pi}. The absolute normalisation is arbitrary when the correction is not applied and natural when it is applied (red).} 
  \label{fig:Ratio_m23_3}
\end{figure}

\clearpage

\begin{figure}[tb]
    \hspace*{-3cm}
       \includegraphics[width=1.4\linewidth]{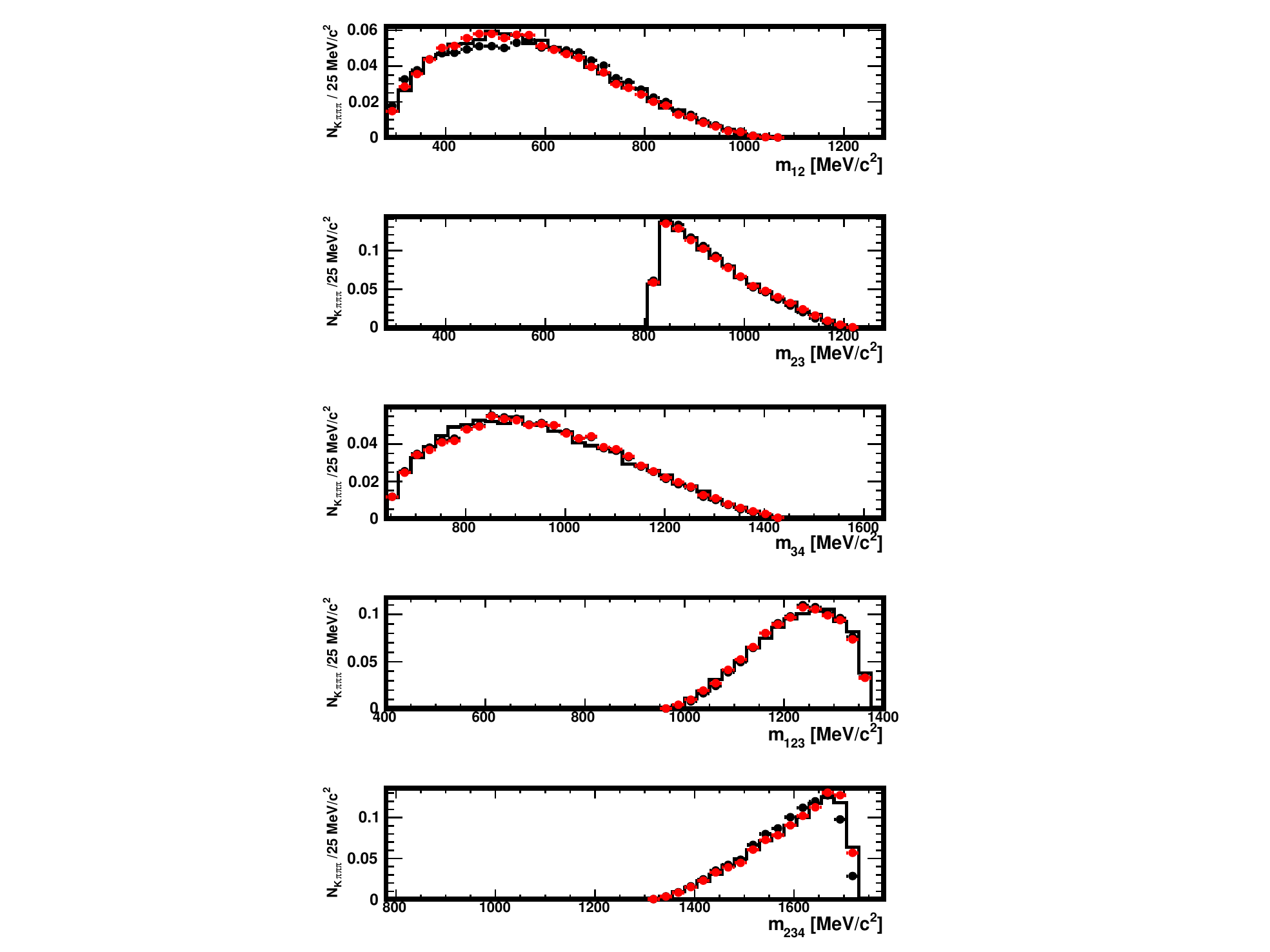}
    \vspace*{-0.5cm}
  \caption{Distributions of $m_{12}$, $m_{23}$, $m_{34}$, $m_{123}$ and $m_{234}$ in the original sample (histogram),
in the distorted one (full black circles) and in the distorted sample where the decays have been re-weighted using the $\omega_i$ weights (red), 
as explained in Sect.~\ref{subsec:ResultsK3pi}. The data used here are restricted to the region $820 < m_{23}$ MeV$/c^2$.
The absolute normalisation is arbitrary when the correction is not applied and natural when it is applied (red).} 
  \label{fig:Distr_m23_4}
\end{figure}

\clearpage

\begin{figure}[tb]
    \hspace*{-3cm}
       \includegraphics[width=1.4\linewidth]{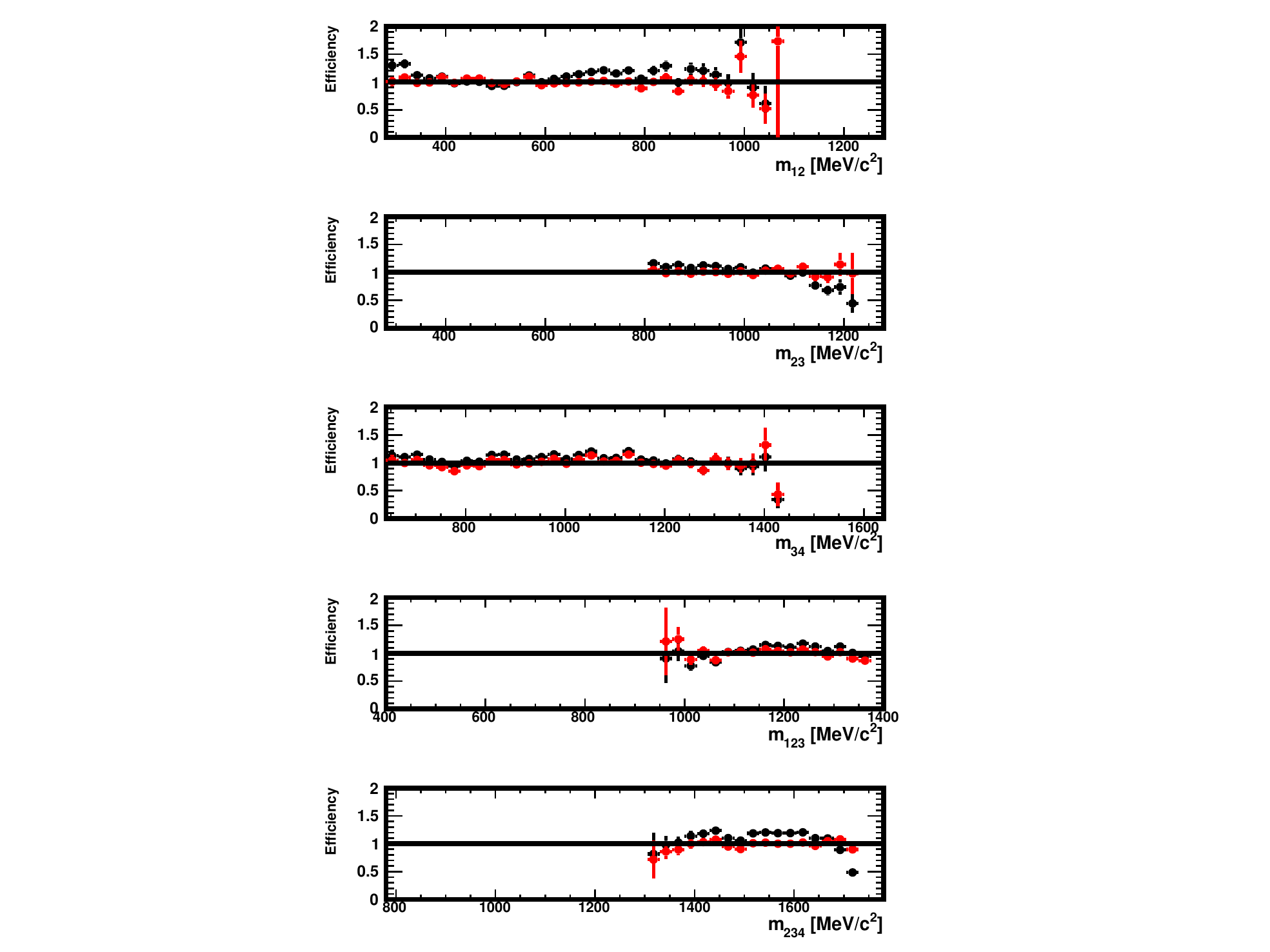}
    \vspace*{-0.5cm}
  \caption{Efficiency in $m_{12}$, $m_{23}$, $m_{34}$, $m_{123}$ and $m_{234}$ in the data generated for this study, 
in the region $820 < m_{23}$ MeV$/c^2$. Shown are the ratios of the distributions found in the distorted and 
original samples, with no correction (black) and for decays re-weighted using $\omega_i$ weights (red)
as explained in Sect.~\ref{subsec:ResultsK3pi}. The absolute normalisation is arbitrary when the correction is not applied and natural when it is applied (red).} 
  \label{fig:Ratio_m23_4}
\end{figure}

\clearpage

\begin{figure}[tb]
    \hspace*{-3cm}
       \includegraphics[width=1.4\linewidth]{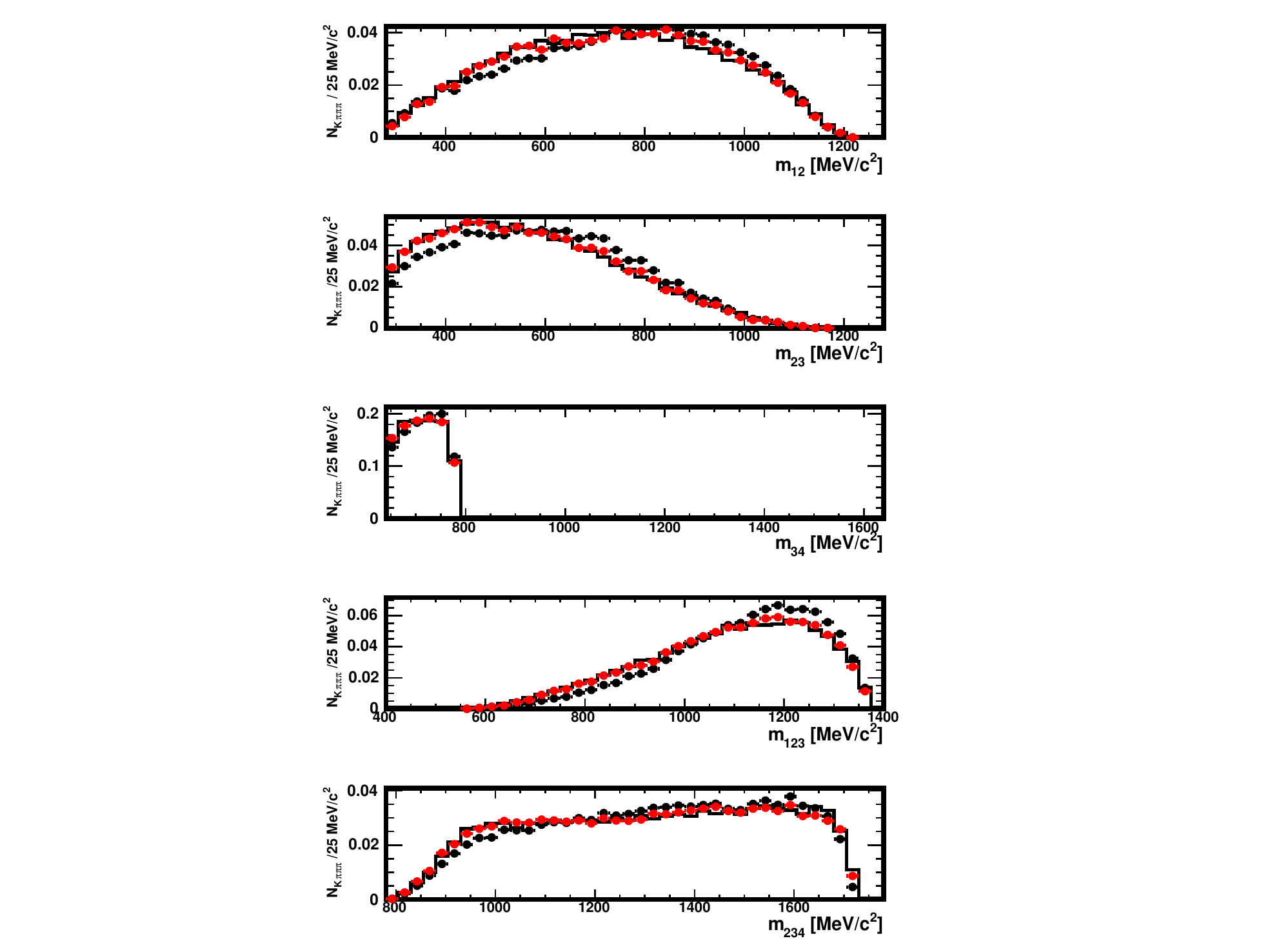}
    \vspace*{-0.5cm}
  \caption{Distributions of $m_{12}$, $m_{23}$, $m_{34}$, $m_{123}$ and $m_{234}$ in the original sample (histogram),
in the distorted one (full black circles) and in the distorted sample where the decays have been re-weighted using the $\omega_i$ weights (red), 
as explained in Sect.~\ref{subsec:ResultsK3pi}. The data used here are restricted to the region $ 0 < m_{34} < 780$ MeV$/c^2$.
The absolute normalisation is arbitrary when the correction is not applied and natural when it is applied (red).} 
  \label{fig:Distr_m34_1}
\end{figure}

\clearpage

\begin{figure}[tb]
    \hspace*{-3cm}
       \includegraphics[width=1.4\linewidth]{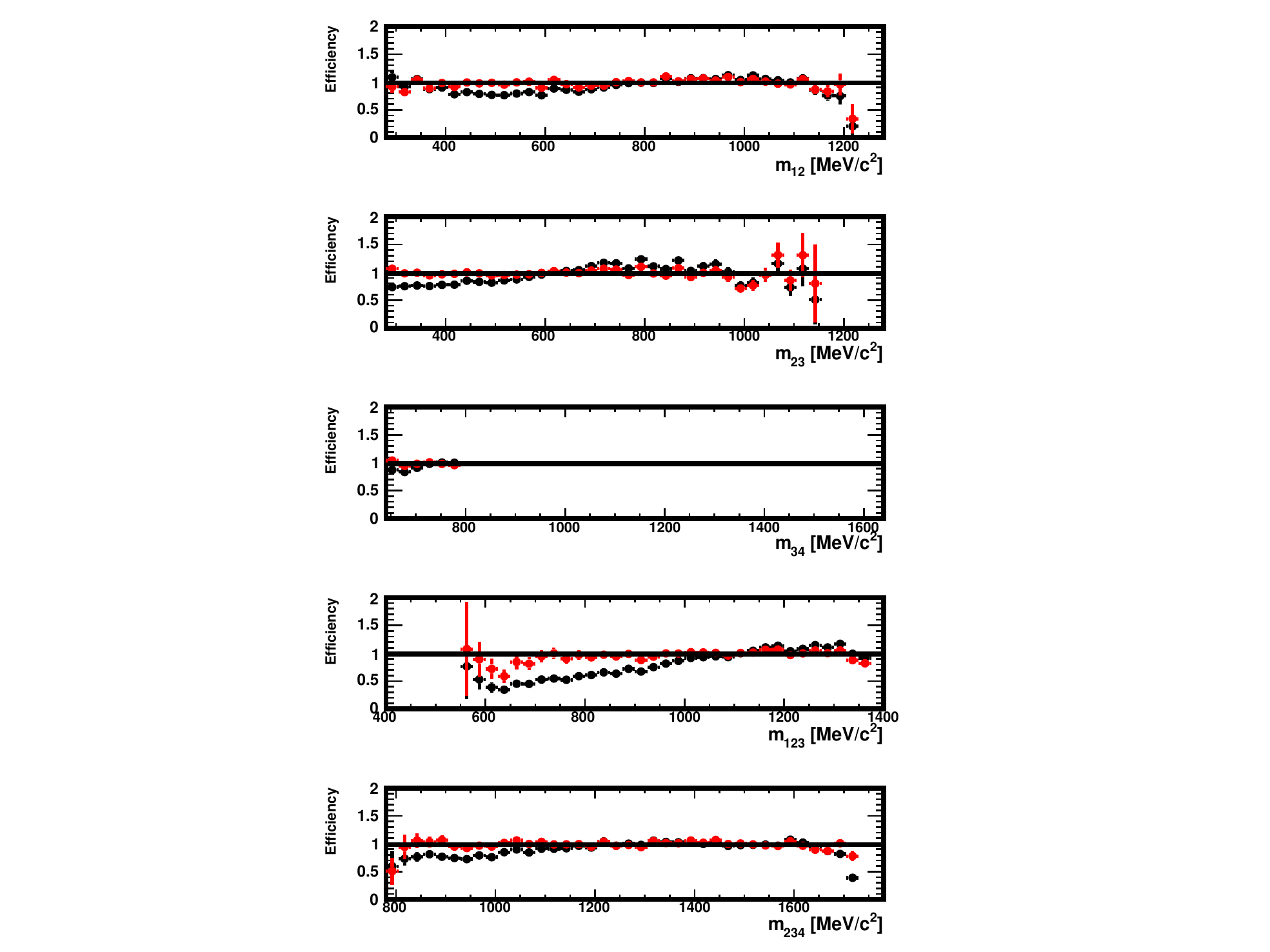}
    \vspace*{-0.5cm}
  \caption{Efficiency in $m_{12}$, $m_{23}$, $m_{34}$, $m_{123}$ and $m_{234}$ in the data generated for this study, 
in the region $0 < m_{34} < 780$ MeV$/c^2$. Shown are the ratios of the distributions found in the distorted and 
original samples, with no correction (black) and for decays re-weighted using $\omega_i$ weights (red)
as explained in Sect.~\ref{subsec:ResultsK3pi}. The absolute normalisation is arbitrary when the correction is not applied and natural when it is applied (red).} 
  \label{fig:Ratio_m34_1}
\end{figure}

\clearpage

\begin{figure}[tb]
    \hspace*{-3cm}
       \includegraphics[width=1.4\linewidth]{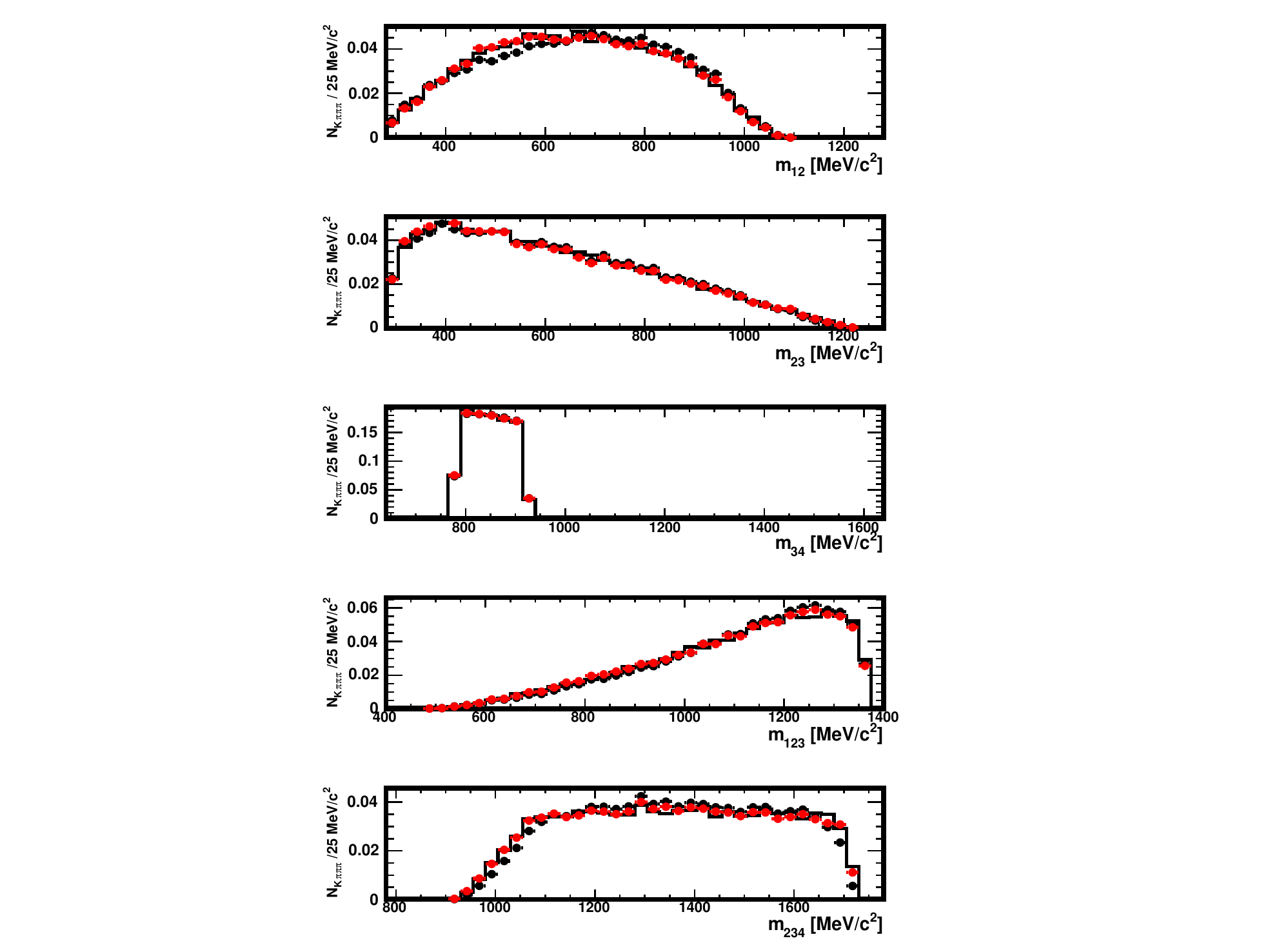}
    \vspace*{-0.5cm}
  \caption{Distributions of $m_{12}$, $m_{23}$, $m_{34}$, $m_{123}$ and $m_{234}$ in the original sample (histogram),
in the distorted one (full black circles) and in the distorted sample where the decays have been re-weighted using the $\omega_i$ weights (red), 
as explained in Sect.~\ref{subsec:ResultsK3pi}. The data used here are restricted to the region $780 < m_{34} < 920$ MeV$/c^2$.
The absolute normalisation is arbitrary when the correction is not applied and natural when it is applied (red).} 
  \label{fig:Distr_m34_2}
\end{figure}

\clearpage

\begin{figure}[tb]
    \hspace*{-3cm}
       \includegraphics[width=1.4\linewidth]{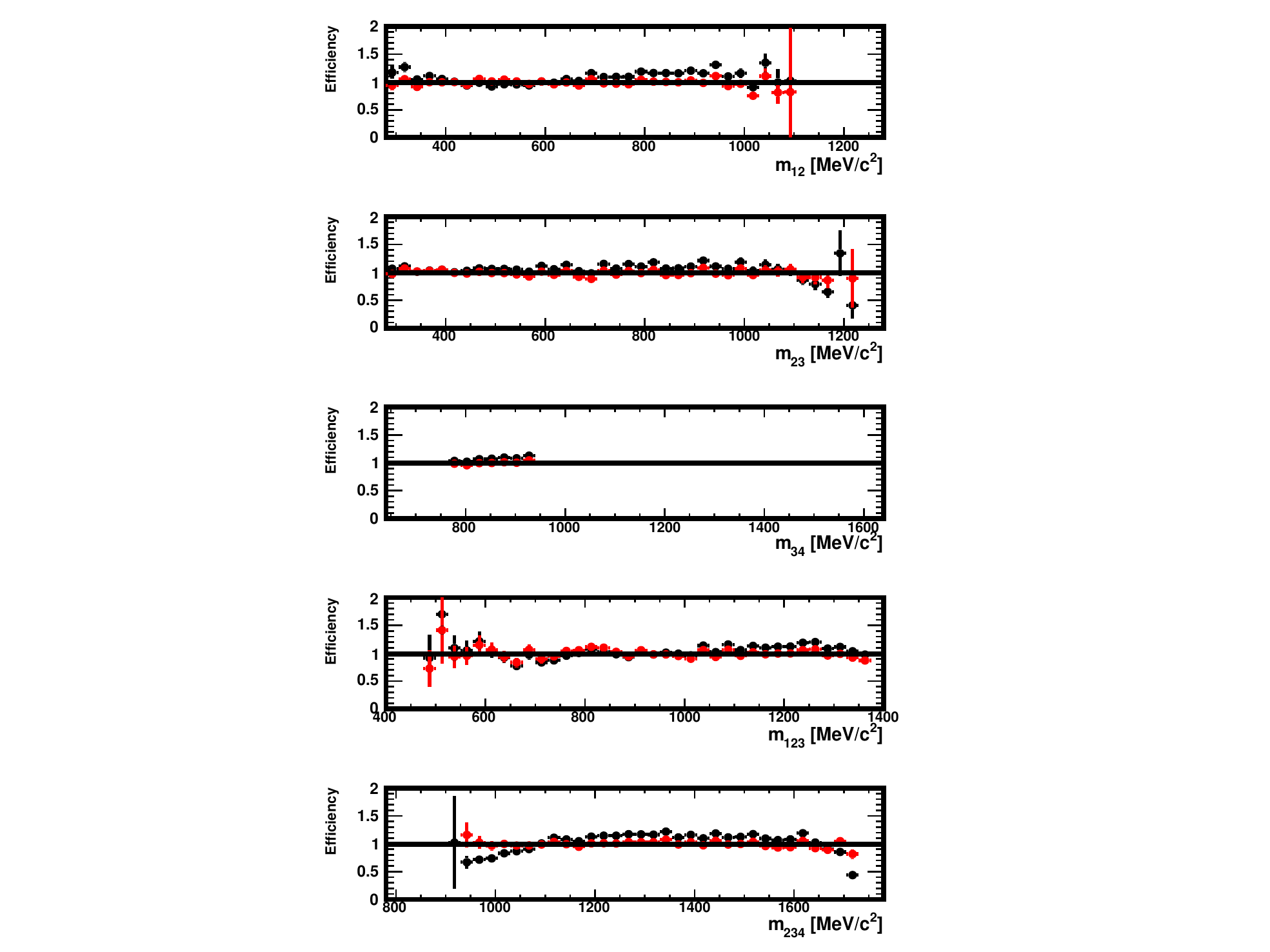}
    \vspace*{-0.5cm}
  \caption{Efficiency in $m_{12}$, $m_{23}$, $m_{34}$, $m_{123}$ and $m_{234}$ in the data generated for this study, 
in the region $ <780 m_{34} < 920$ MeV$/c^2$. Shown are the ratios of the distributions found in the distorted and 
original samples, with no correction (black) and for decays re-weighted using $\omega_i$ weights (red)
as explained in Sect.~\ref{subsec:ResultsK3pi}. The absolute normalisation is arbitrary when the correction is not applied and natural when it is applied (red).} 
  \label{fig:Ratio_m34_2}
\end{figure}

\clearpage

\begin{figure}[tb]
    \hspace*{-3cm}
       \includegraphics[width=1.4\linewidth]{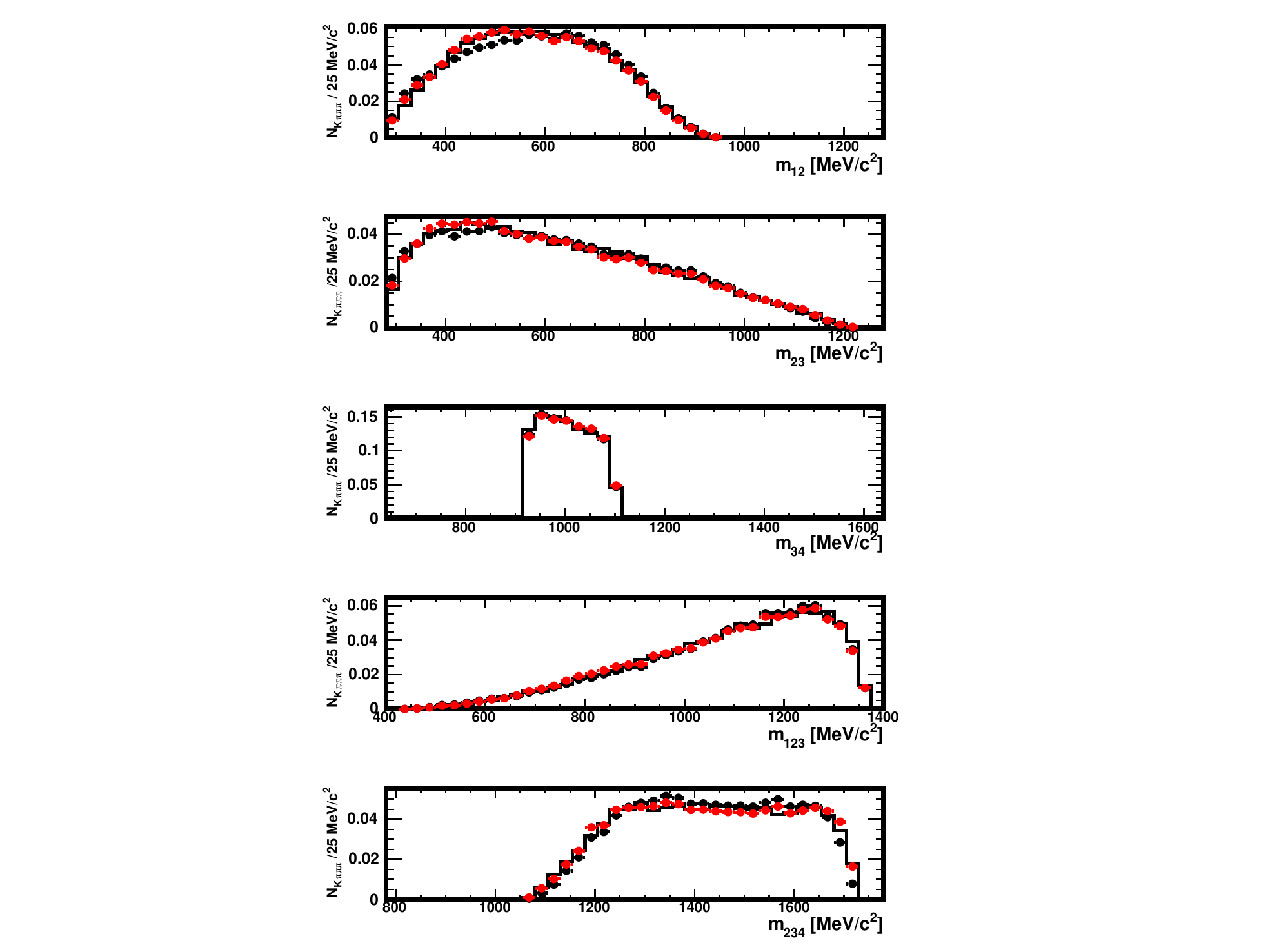}
    \vspace*{-0.5cm}
  \caption{Distributions of $m_{12}$, $m_{23}$, $m_{34}$, $m_{123}$ and $m_{234}$ in the original sample (histogram),
in the distorted one (full black circles) and in the distorted sample where the decays have been re-weighted using the $\omega_i$ weights (red), 
as explained in Sect.~\ref{subsec:ResultsK3pi}. The data used here are restricted to the region $920 < m_{34} < 1100$ MeV$/c^2$.
The absolute normalisation is arbitrary when the correction is not applied and natural when it is applied (red).} 
  \label{fig:Distr_m34_3}
\end{figure}

\clearpage

\begin{figure}[tb]
    \hspace*{-3cm}
       \includegraphics[width=1.4\linewidth]{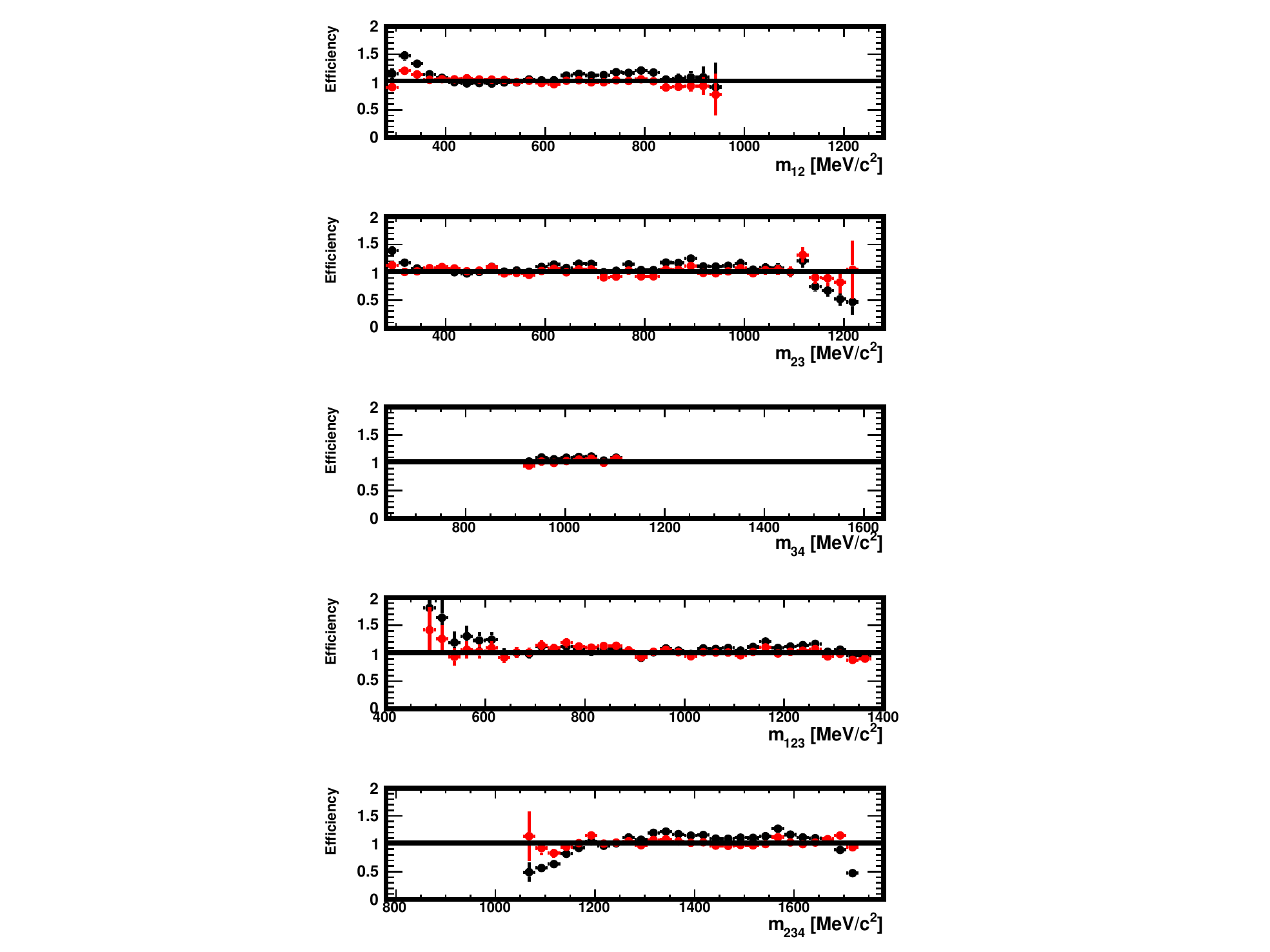}
    \vspace*{-0.5cm}
  \caption{Efficiency in $m_{12}$, $m_{23}$, $m_{34}$, $m_{123}$ and $m_{234}$ in the data generated for this study, 
in the region $920 < m_{34} < 1100$ MeV$/c^2$. Shown are the ratios of the distributions found in the distorted and 
original samples, with no correction (black) and for decays re-weighted using $\omega_i$ weights (red)
as explained in Sect.~\ref{subsec:ResultsK3pi}. The absolute normalisation is arbitrary when the correction is not applied and natural when it is applied (red).} 
  \label{fig:Ratio_m34_3}
\end{figure}

\clearpage

\begin{figure}[tb]
    \hspace*{-3cm}
       \includegraphics[width=1.4\linewidth]{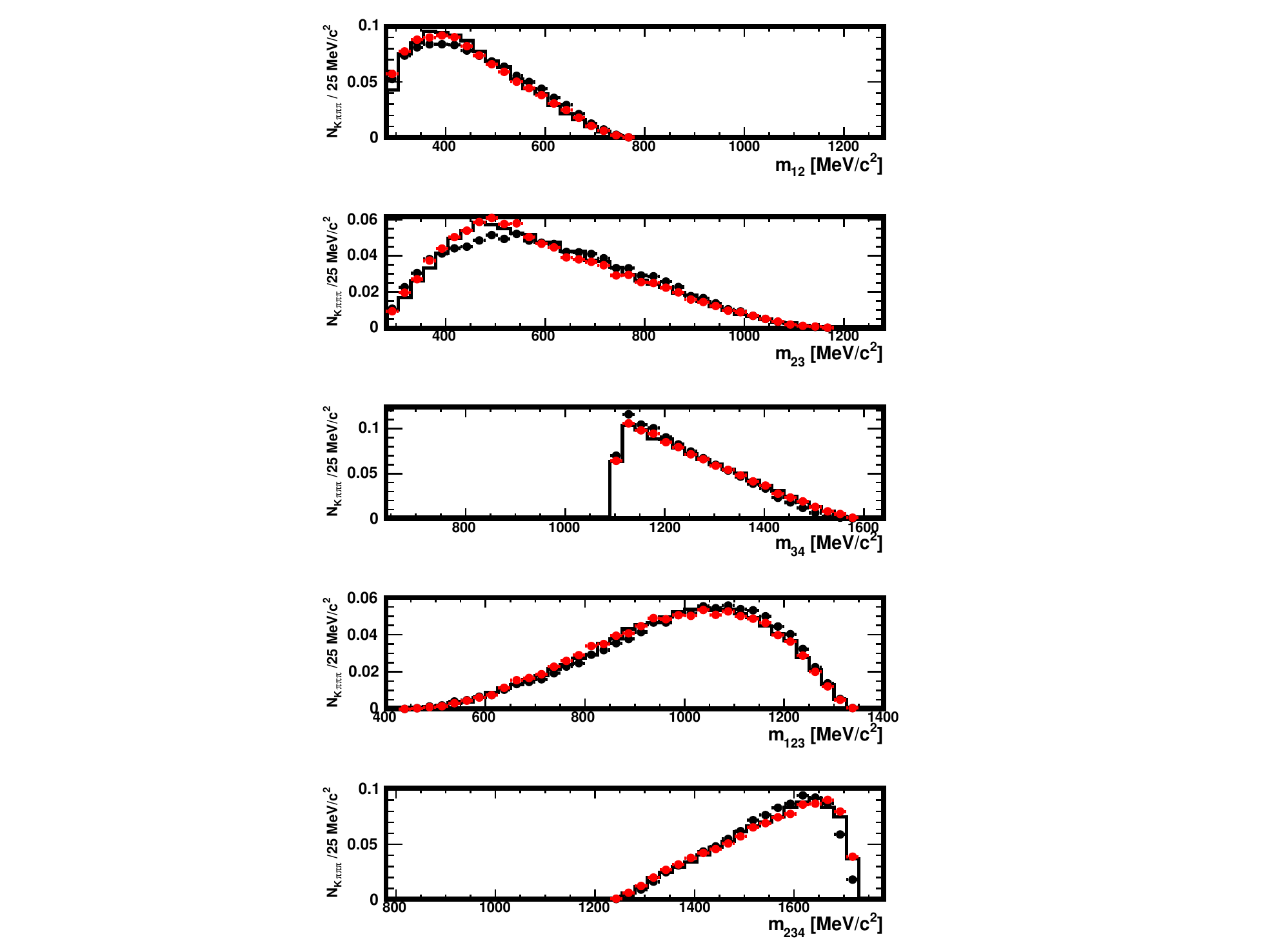}
    \vspace*{-0.5cm}
  \caption{Distributions of $m_{12}$, $m_{23}$, $m_{34}$, $m_{123}$ and $m_{234}$ in the original sample (histogram),
in the distorted one (full black circles) and in the distorted sample where the decays have been re-weighted using the $\omega_i$ weights (red), 
as explained in Sect.~\ref{subsec:ResultsK3pi}. The data used here are restricted to the region $1100 < m_{34}$ MeV$/c^2$.
The absolute normalisation is arbitrary when the correction is not applied and natural when it is applied (red).} 
  \label{fig:Distr_m34_4}
\end{figure}

\clearpage

\begin{figure}[tb]
    \hspace*{-3cm}
       \includegraphics[width=1.4\linewidth]{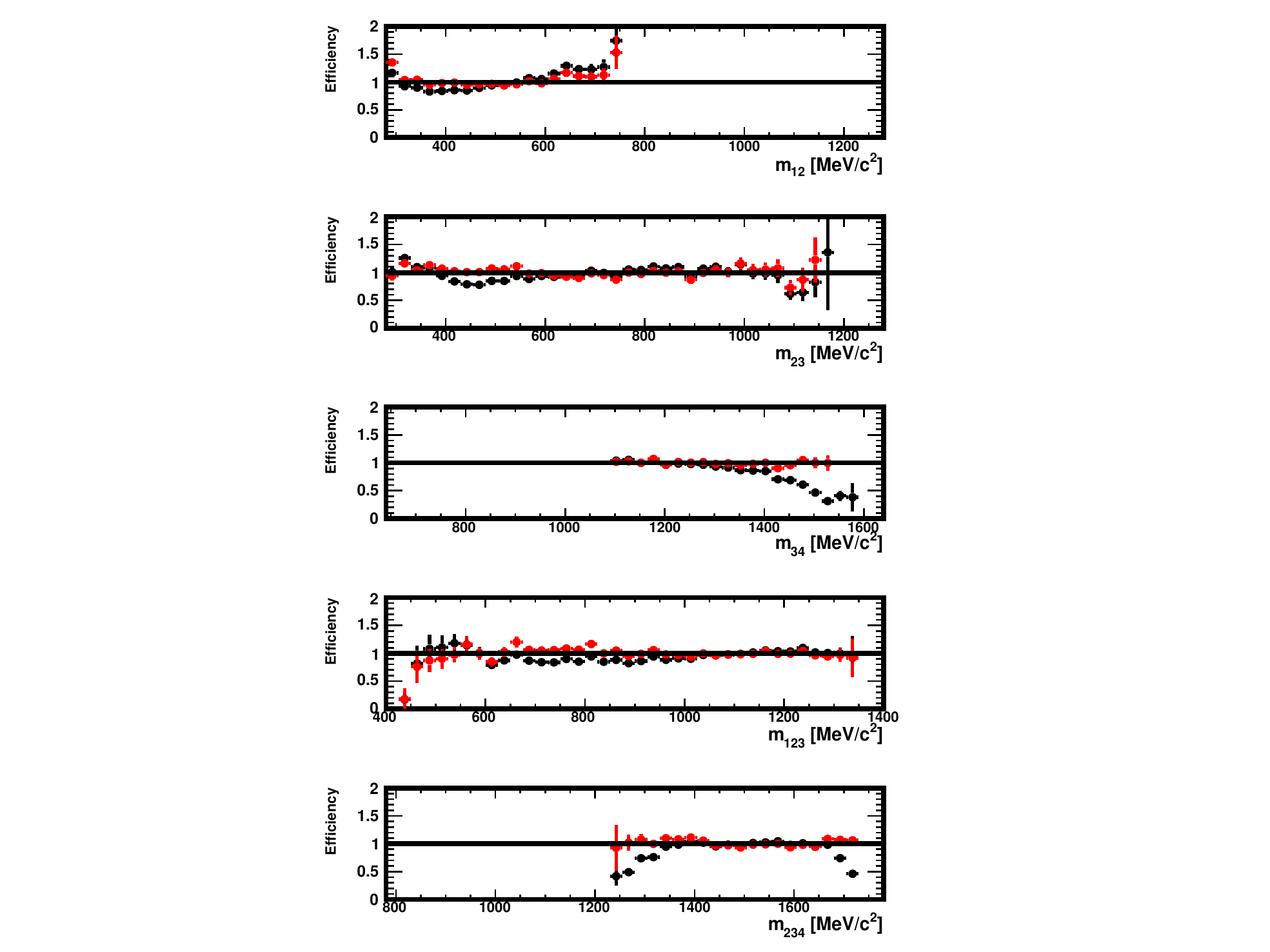}
    \vspace*{-0.5cm}
  \caption{Efficiency in $m_{12}$, $m_{23}$, $m_{34}$, $m_{123}$ and $m_{234}$ in the data generated for this study, 
in the region $1100 < m_{34}$ MeV$/c^2$. Shown are the ratios of the distributions found in the distorted and 
original samples, with no correction (black) and for decays re-weighted using $\omega_i$ weights (red)
as explained in Sect.~\ref{subsec:ResultsK3pi}. The absolute normalisation is arbitrary when the correction is not applied and natural when it is applied (red).} 
  \label{fig:Ratio_m34_4}
\end{figure}

\clearpage

\begin{figure}[tb]
    \hspace*{-3cm}
       \includegraphics[width=1.4\linewidth]{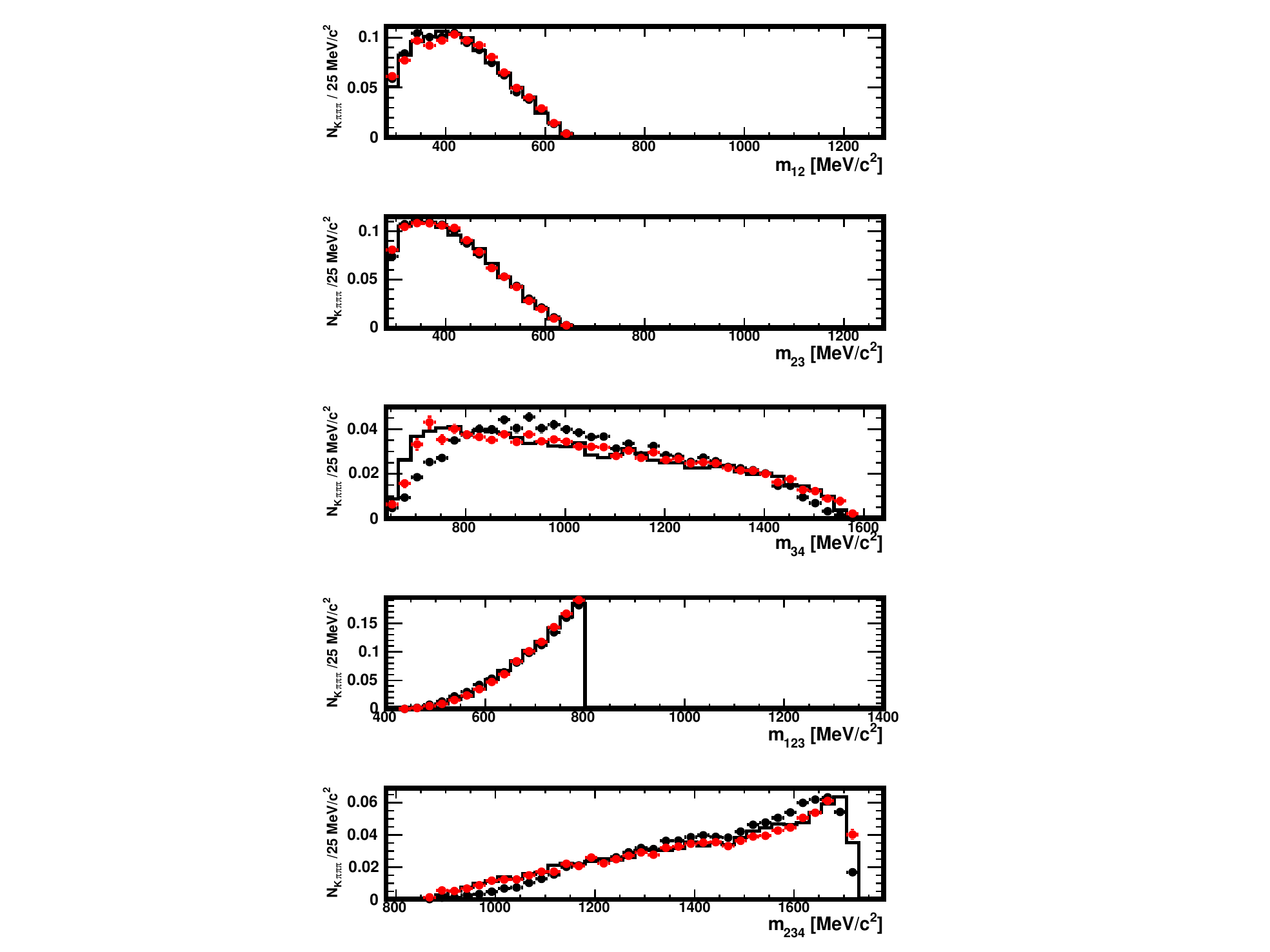}
    \vspace*{-0.5cm}
  \caption{Distributions of $m_{12}$, $m_{23}$, $m_{34}$, $m_{123}$ and $m_{234}$ in the original sample (histogram),
in the distorted one (full black circles) and in the distorted sample where the decays have been re-weighted using the $\omega_i$ weights (red), 
as explained in Sect.~\ref{subsec:ResultsK3pi}. The data used here are restricted to the region $ 0 < m_{123} < 800$ MeV$/c^2$.
The absolute normalisation is arbitrary when the correction is not applied and natural when it is applied (red).} 
  \label{fig:Distr_m123_1}
\end{figure}

\clearpage

\begin{figure}[tb]
    \hspace*{-3cm}
       \includegraphics[width=1.4\linewidth]{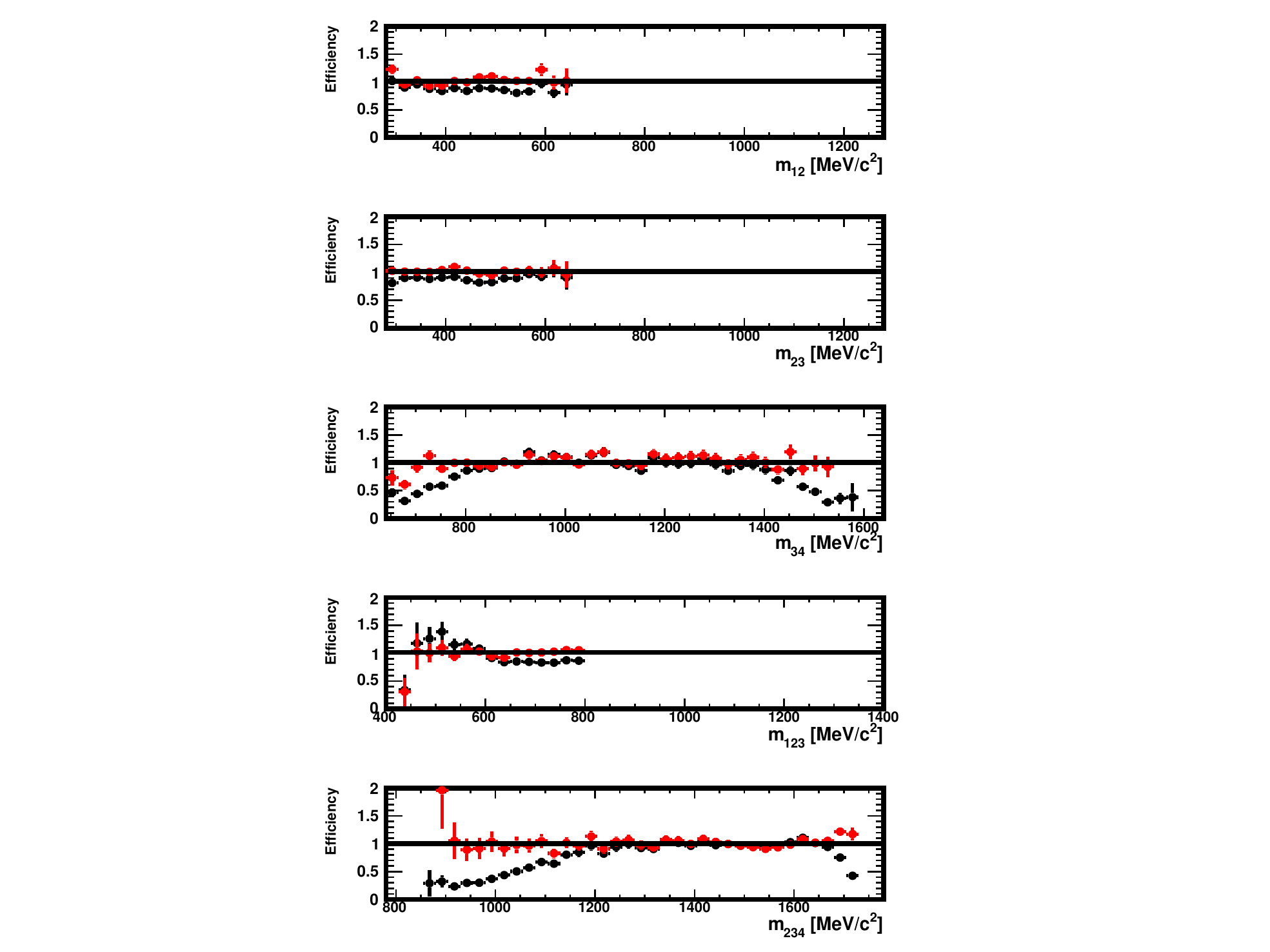}
    \vspace*{-0.5cm}
  \caption{Efficiency in $m_{12}$, $m_{23}$, $m_{34}$, $m_{123}$ and $m_{234}$ in the data generated for this study, 
in the region $0 < m_{123} < 800$ MeV$/c^2$. Shown are the ratios of the distributions found in the distorted and 
original samples, with no correction (black) and for decays re-weighted using $\omega_i$ weights (red)
as explained in Sect.~\ref{subsec:ResultsK3pi}. The absolute normalisation is arbitrary when the correction is not applied and natural when it is applied (red).} 
  \label{fig:Ratio_m123_1}
\end{figure}

\clearpage

\begin{figure}[tb]
    \hspace*{-3cm}
       \includegraphics[width=1.4\linewidth]{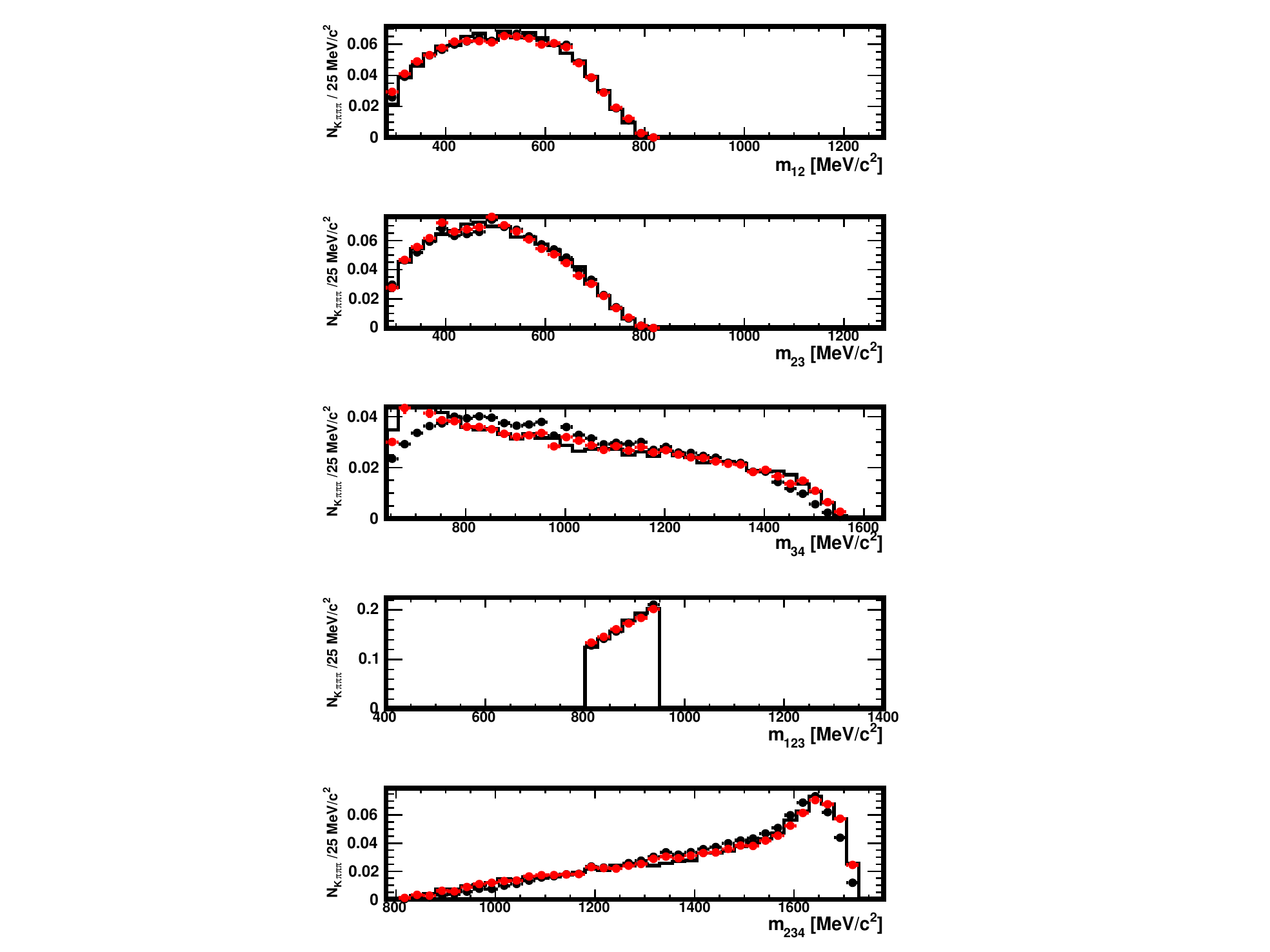}
    \vspace*{-0.5cm}
  \caption{Distributions of $m_{12}$, $m_{23}$, $m_{123}$, $m_{123}$ and $m_{234}$ in the original sample (histogram),
in the distorted one (full black circles) and in the distorted sample where the decays have been re-weighted using the $\omega_i$ weights (red), 
as explained in Sect.~\ref{subsec:ResultsK3pi}. The data used here are restricted to the region $800 < m_{123} < 950$ MeV$/c^2$.
The absolute normalisation is arbitrary when the correction is not applied and natural when it is applied (red).} 
  \label{fig:Distr_m123_2}
\end{figure}

\clearpage

\begin{figure}[tb]
    \hspace*{-3cm}
       \includegraphics[width=1.4\linewidth]{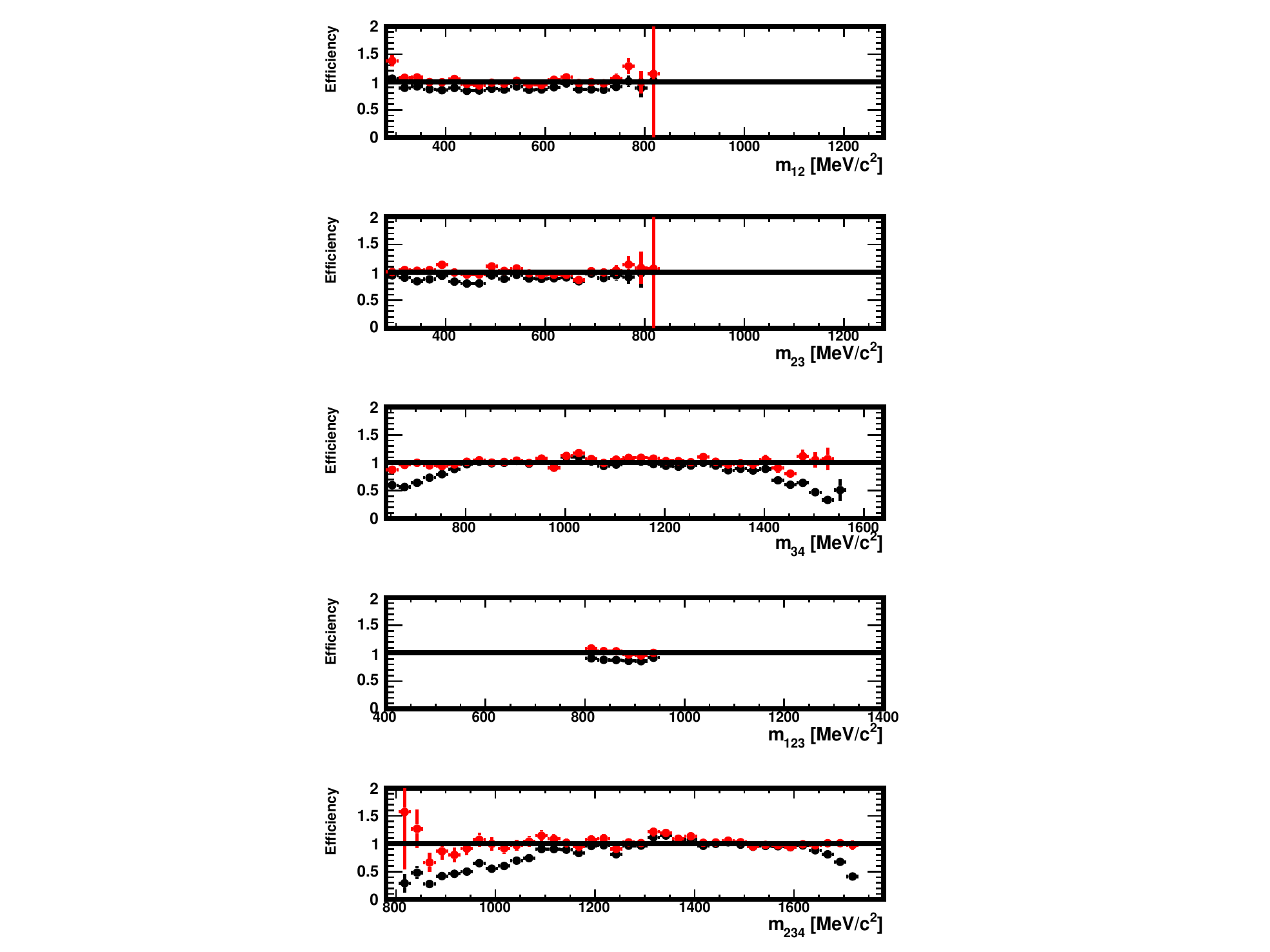}
    \vspace*{-0.5cm}
  \caption{Efficiency in $m_{12}$, $m_{23}$, $m_{34}$, $m_{123}$ and $m_{234}$ in the data generated for this study, 
in the region $ <800 m_{123} < 950$ MeV$/c^2$. Shown are the ratios of the distributions found in the distorted and 
original samples, with no correction (black) and for decays re-weighted using $\omega_i$ weights (red)
as explained in Sect.~\ref{subsec:ResultsK3pi}. The absolute normalisation is arbitrary when the correction is not applied and natural when it is applied (red).} 
  \label{fig:Ratio_m123_2}
\end{figure}

\clearpage

\begin{figure}[tb]
    \hspace*{-3cm}
       \includegraphics[width=1.4\linewidth]{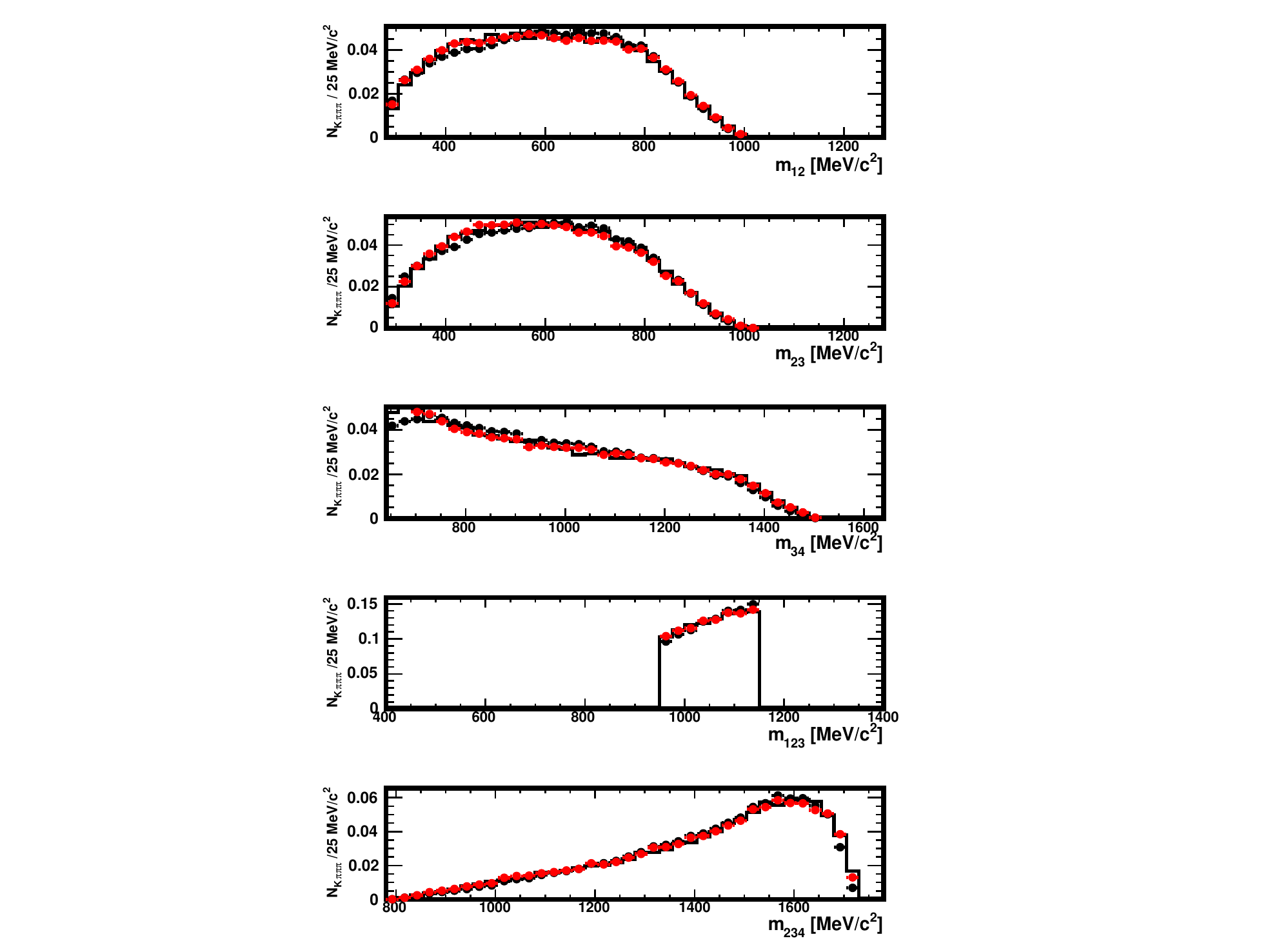}
    \vspace*{-0.5cm}
  \caption{Distributions of $m_{12}$, $m_{23}$, $m_{34}$, $m_{123}$ and $m_{234}$ in the original sample (histogram),
in the distorted one (full black circles) and in the distorted sample where the decays have been re-weighted using the $\omega_i$ weights (red), 
as explained in Sect.~\ref{subsec:ResultsK3pi}. The data used here are restricted to the region $950 < m_{123} < 1150$ MeV$/c^2$.
The absolute normalisation is arbitrary when the correction is not applied and natural when it is applied (red).} 
  \label{fig:Distr_m123_3}
\end{figure}

\clearpage

\begin{figure}[tb]
    \hspace*{-3cm}
       \includegraphics[width=1.4\linewidth]{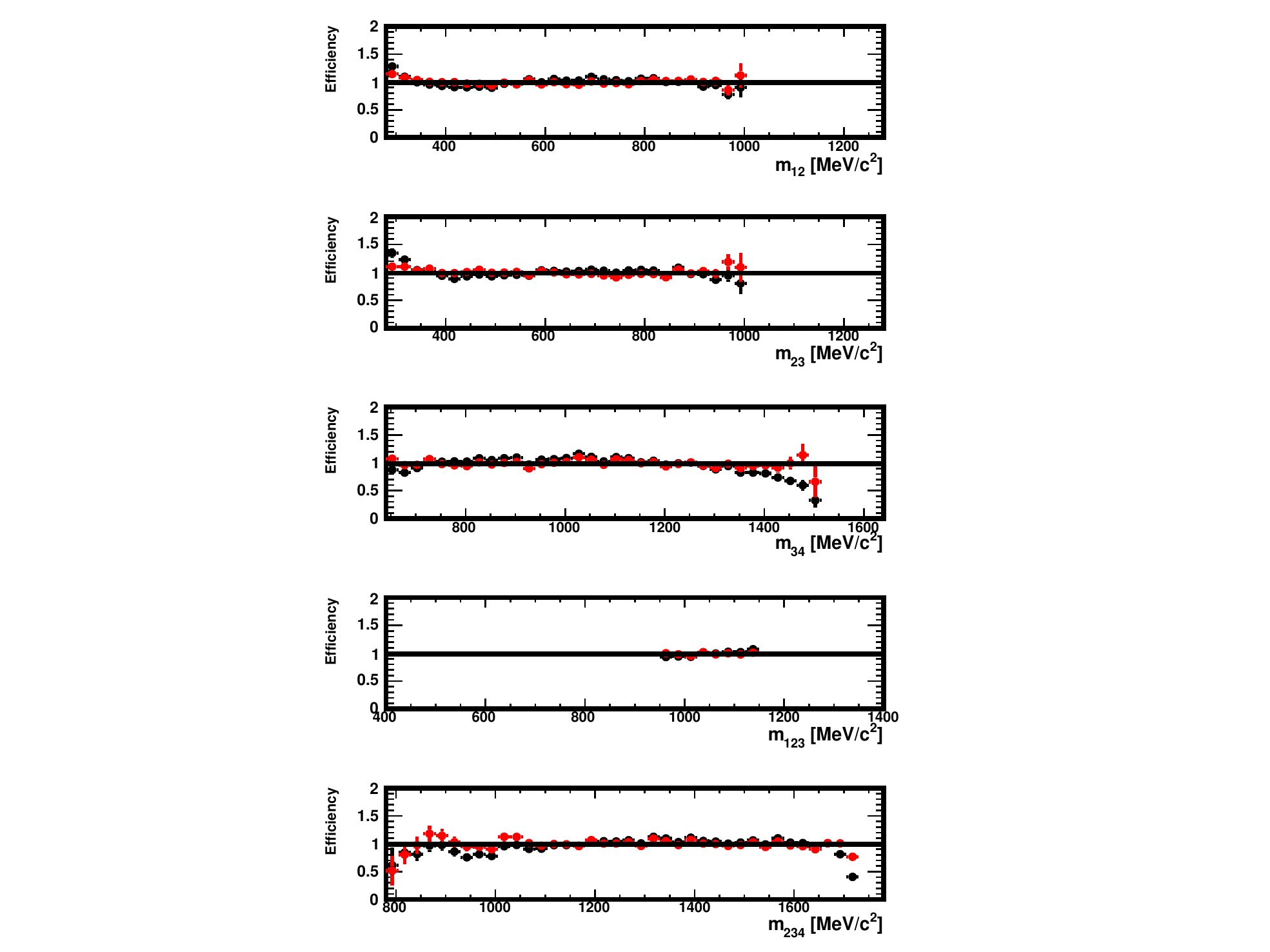}
    \vspace*{-0.5cm}
  \caption{Efficiency in $m_{12}$, $m_{23}$, $m_{34}$, $m_{123}$ and $m_{234}$ in the data generated for this study, 
in the region $950 < m_{123} < 1150$ MeV$/c^2$. Shown are the ratios of the distributions found in the distorted and 
original samples, with no correction (black) and for decays re-weighted using $\omega_i$ weights (red)
as explained in Sect.~\ref{subsec:ResultsK3pi}. The absolute normalisation is arbitrary when the correction is not applied and natural when it is applied (red).} 
  \label{fig:Ratio_m123_3}
\end{figure}

\clearpage

\begin{figure}[tb]
    \hspace*{-3cm}
       \includegraphics[width=1.4\linewidth]{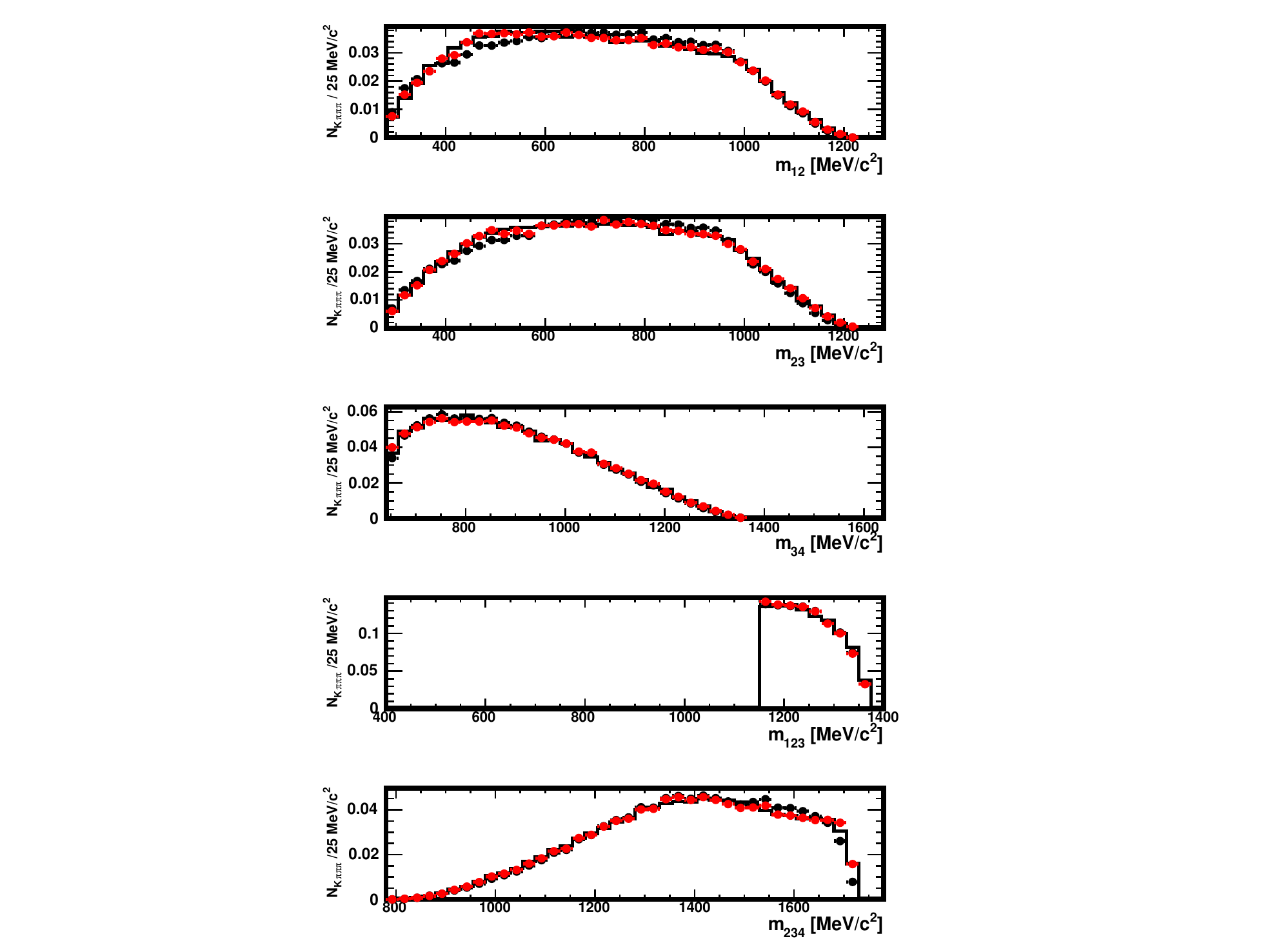}
    \vspace*{-0.5cm}
  \caption{Distributions of $m_{12}$, $m_{23}$, $m_{34}$, $m_{123}$ and $m_{234}$ in the original sample (histogram),
in the distorted one (full black circles) and in the distorted sample where the decays have been re-weighted using the $\omega_i$ weights (red), 
as explained in Sect.~\ref{subsec:ResultsK3pi}. The data used here are restricted to the region $1150 < m_{123}$ MeV$/c^2$.
The absolute normalisation is arbitrary when the correction is not applied and natural when it is applied (red).} 
  \label{fig:Distr_m123_4}
\end{figure}

\clearpage

\begin{figure}[tb]
    \hspace*{-3cm}
       \includegraphics[width=1.4\linewidth]{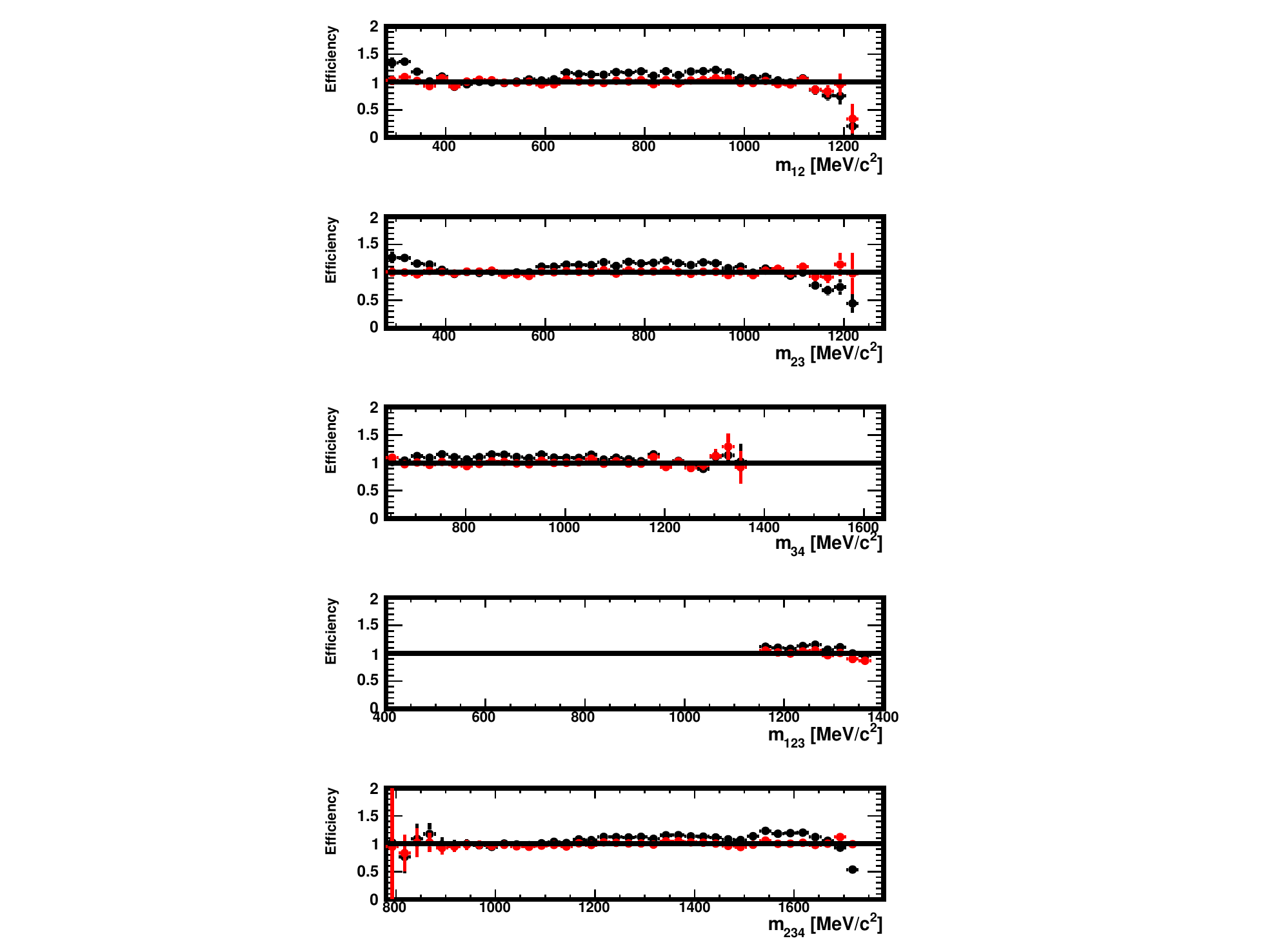}
    \vspace*{-0.5cm}
  \caption{Efficiency in $m_{12}$, $m_{23}$, $m_{34}$, $m_{123}$ and $m_{234}$ in the data generated for this study, 
in the region $1150 < m_{123}$ MeV$/c^2$. Shown are the ratios of the distributions found in the distorted and 
original samples, with no correction (black) and for decays re-weighted using $\omega_i$ weights (red)
as explained in Sect.~\ref{subsec:ResultsK3pi}. The absolute normalisation is arbitrary when the correction is not applied and natural when it is applied (red).} 
  \label{fig:Ratio_m123_4}
\end{figure}

\clearpage

\begin{figure}[tb]
    \hspace*{-3cm}
       \includegraphics[width=1.4\linewidth]{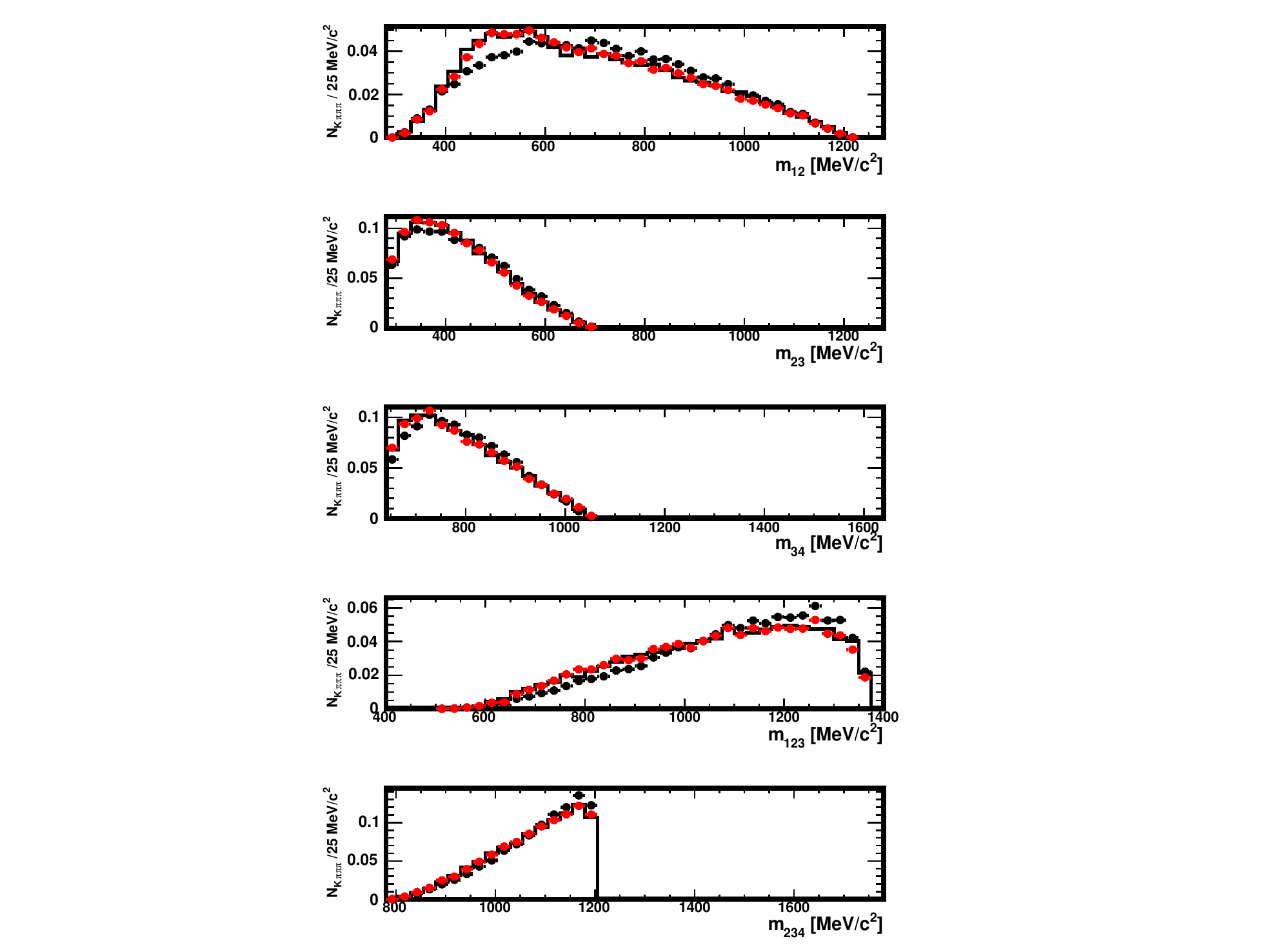}
    \vspace*{-0.5cm}
  \caption{Distributions of $m_{12}$, $m_{23}$, $m_{34}$, $m_{123}$ and $m_{234}$ in the original sample (histogram),
in the distorted one (full black circles) and in the distorted sample where the decays have been re-weighted using the $\omega_i$ weights (red), 
as explained in Sect.~\ref{subsec:ResultsK3pi}. The data used here are restricted to the region $ 0 < m_{234} < 800$ MeV$/c^2$.
The absolute normalisation is arbitrary when the correction is not applied and natural when it is applied (red).} 
  \label{fig:Distr_m234_1}
\end{figure}

\clearpage

\begin{figure}[tb]
    \hspace*{-3cm}
       \includegraphics[width=1.4\linewidth]{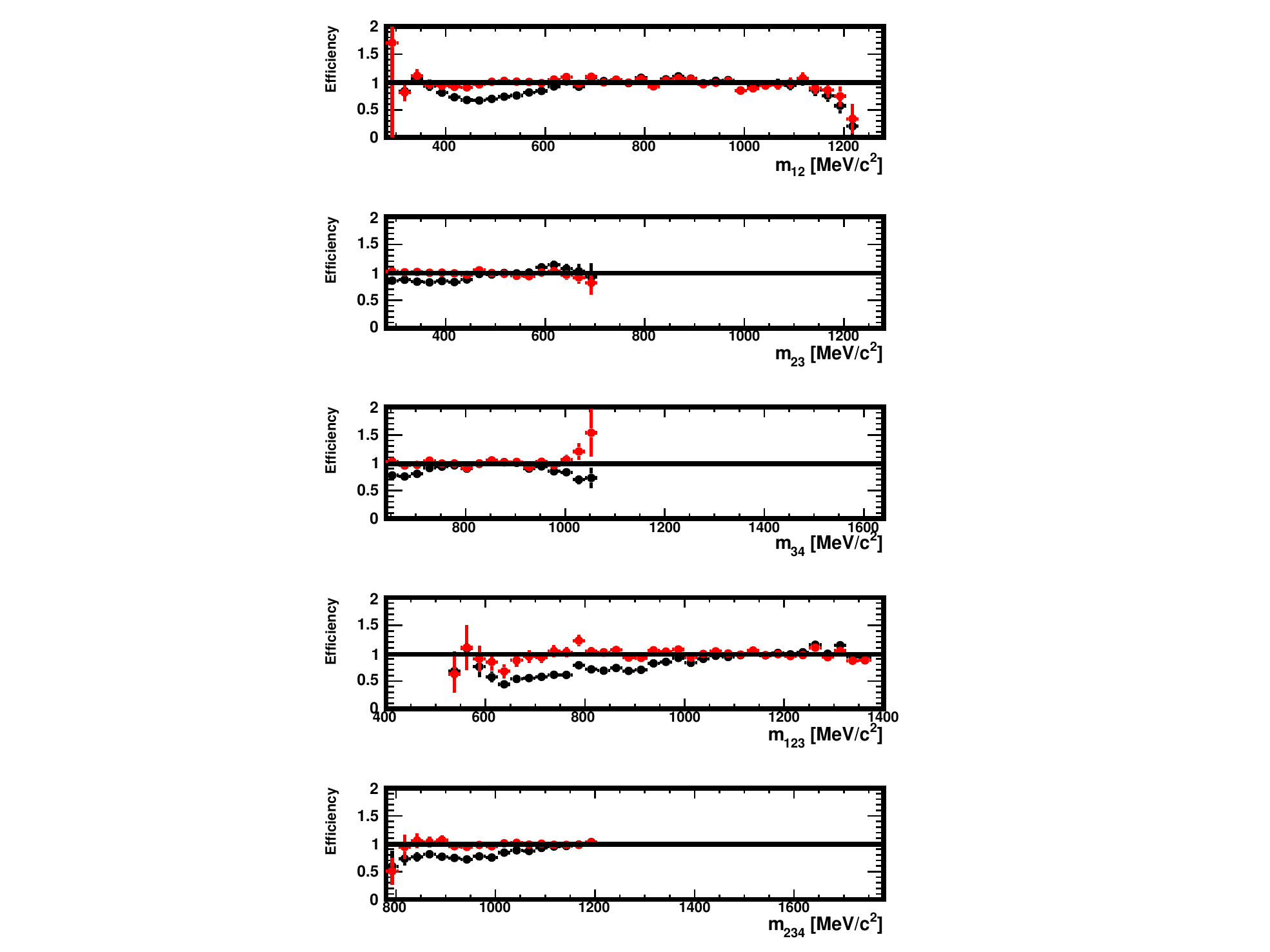}
    \vspace*{-0.5cm}
  \caption{Efficiency in $m_{12}$, $m_{23}$, $m_{34}$, $m_{123}$ and $m_{234}$ in the data generated for this study, 
in the region $0 < m_{234} < 800$ MeV$/c^2$. Shown are the ratios of the distributions found in the distorted and 
original samples, with no correction (black) and for decays re-weighted using $\omega_i$ weights (red)
as explained in Sect.~\ref{subsec:ResultsK3pi}. The absolute normalisation is arbitrary when the correction is not applied and natural when it is applied (red).} 
  \label{fig:Ratio_m234_1}
\end{figure}

\clearpage

\begin{figure}[tb]
    \hspace*{-3cm}
       \includegraphics[width=1.4\linewidth]{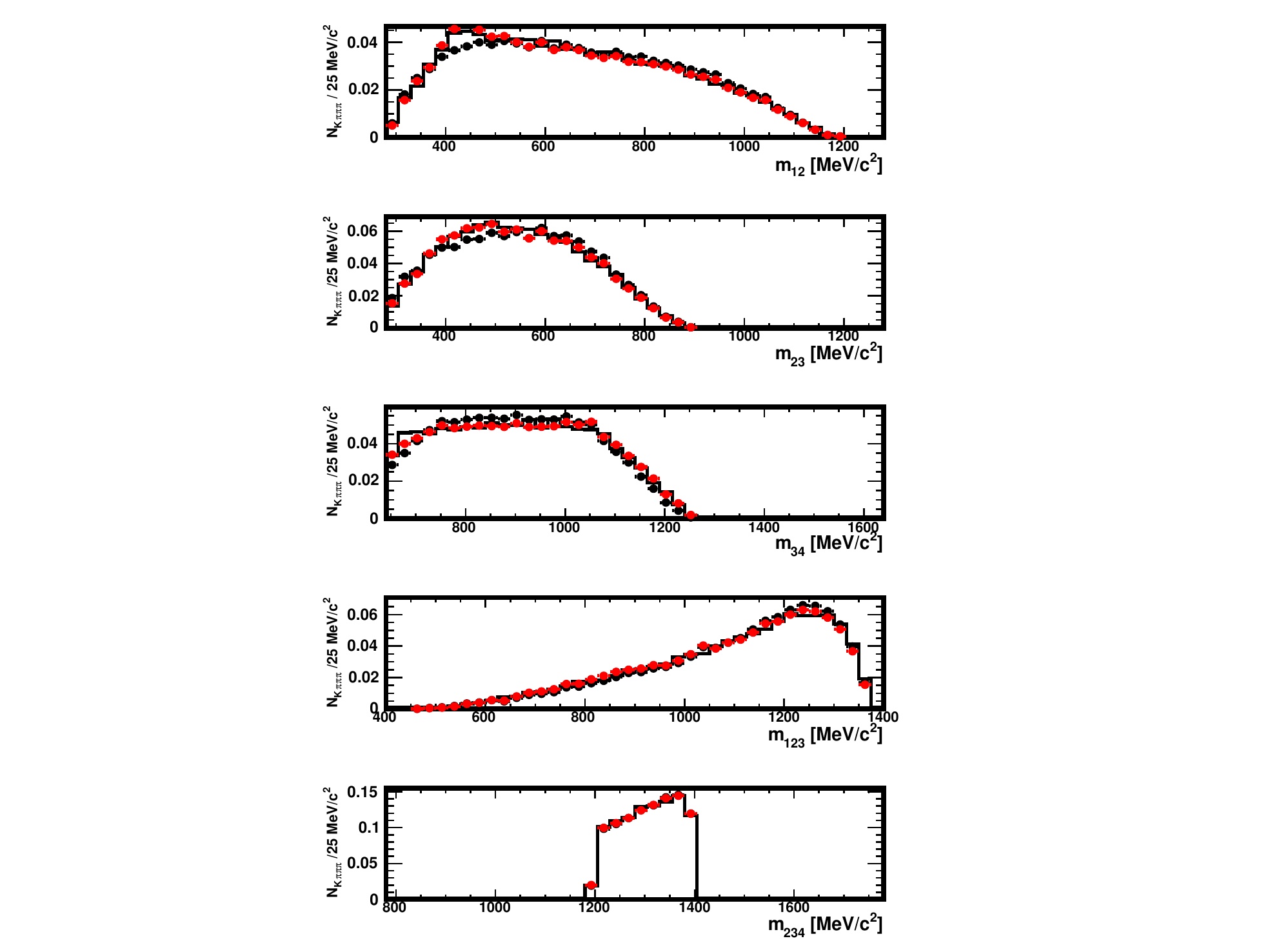}
    \vspace*{-0.5cm}
  \caption{Distributions of $m_{12}$, $m_{23}$, $m_{123}$, $m_{123}$ and $m_{234}$ in the original sample (histogram),
in the distorted one (full black circles) and in the distorted sample where the decays have been re-weighted using the $\omega_i$ weights (red), 
as explained in Sect.~\ref{subsec:ResultsK3pi}. The data used here are restricted to the region $800 < m_{234} < 950$ MeV$/c^2$.
The absolute normalisation is arbitrary when the correction is not applied and natural when it is applied (red).} 
  \label{fig:Distr_m234_2}
\end{figure}

\clearpage

\begin{figure}[tb]
    \hspace*{-3cm}
       \includegraphics[width=1.4\linewidth]{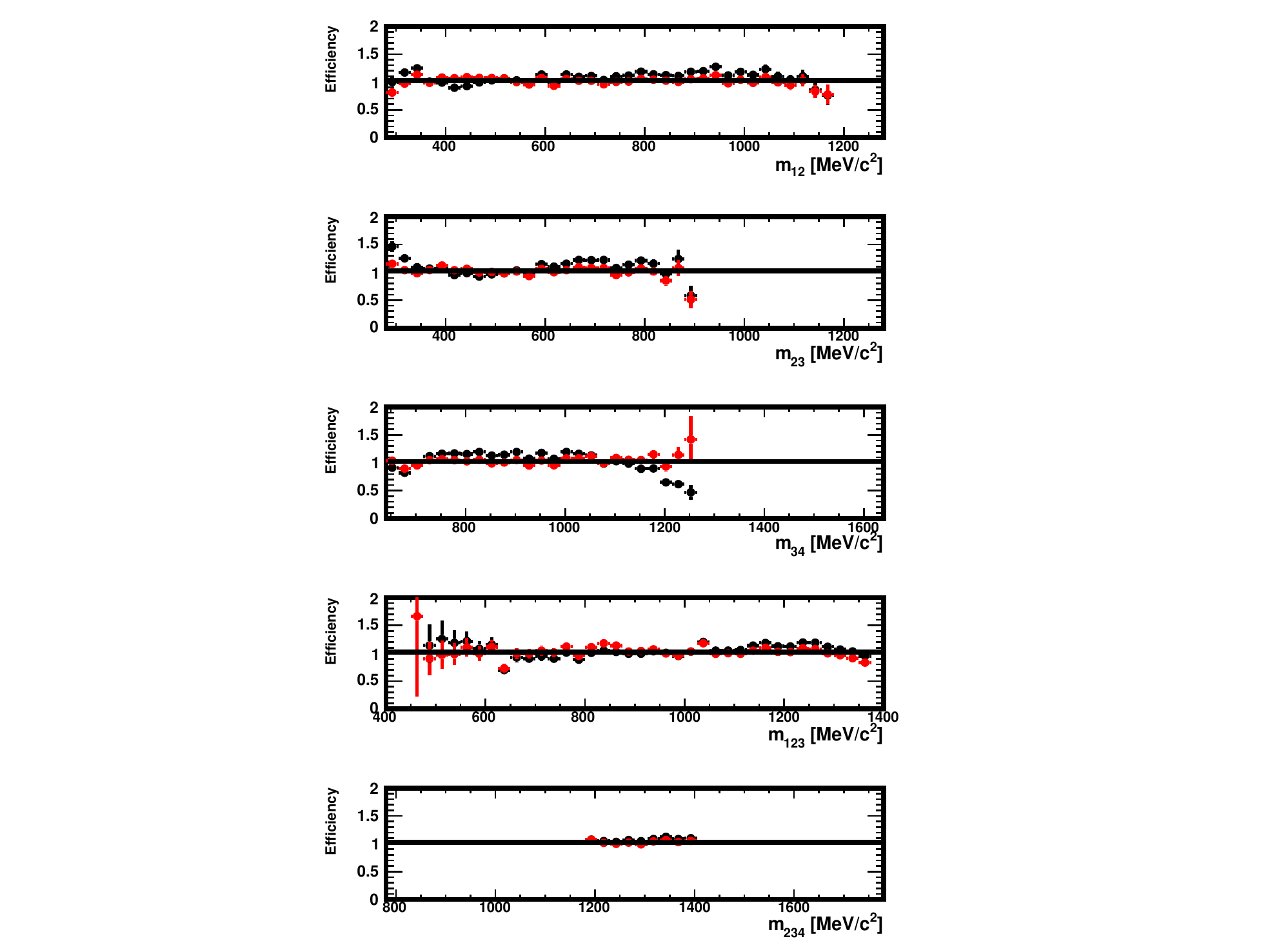}
    \vspace*{-0.5cm}
  \caption{Efficiency in $m_{12}$, $m_{23}$, $m_{34}$, $m_{123}$ and $m_{234}$ in the data generated for this study, 
in the region $ <800 m_{234} < 950$ MeV$/c^2$. Shown are the ratios of the distributions found in the distorted and 
original samples, with no correction (black) and for decays re-weighted using $\omega_i$ weights (red)
as explained in Sect.~\ref{subsec:ResultsK3pi}. The absolute normalisation is arbitrary when the correction is not applied and natural when it is applied (red).} 
  \label{fig:Ratio_m234_2}
\end{figure}

\clearpage

\begin{figure}[tb]
    \hspace*{-3cm}
       \includegraphics[width=1.4\linewidth]{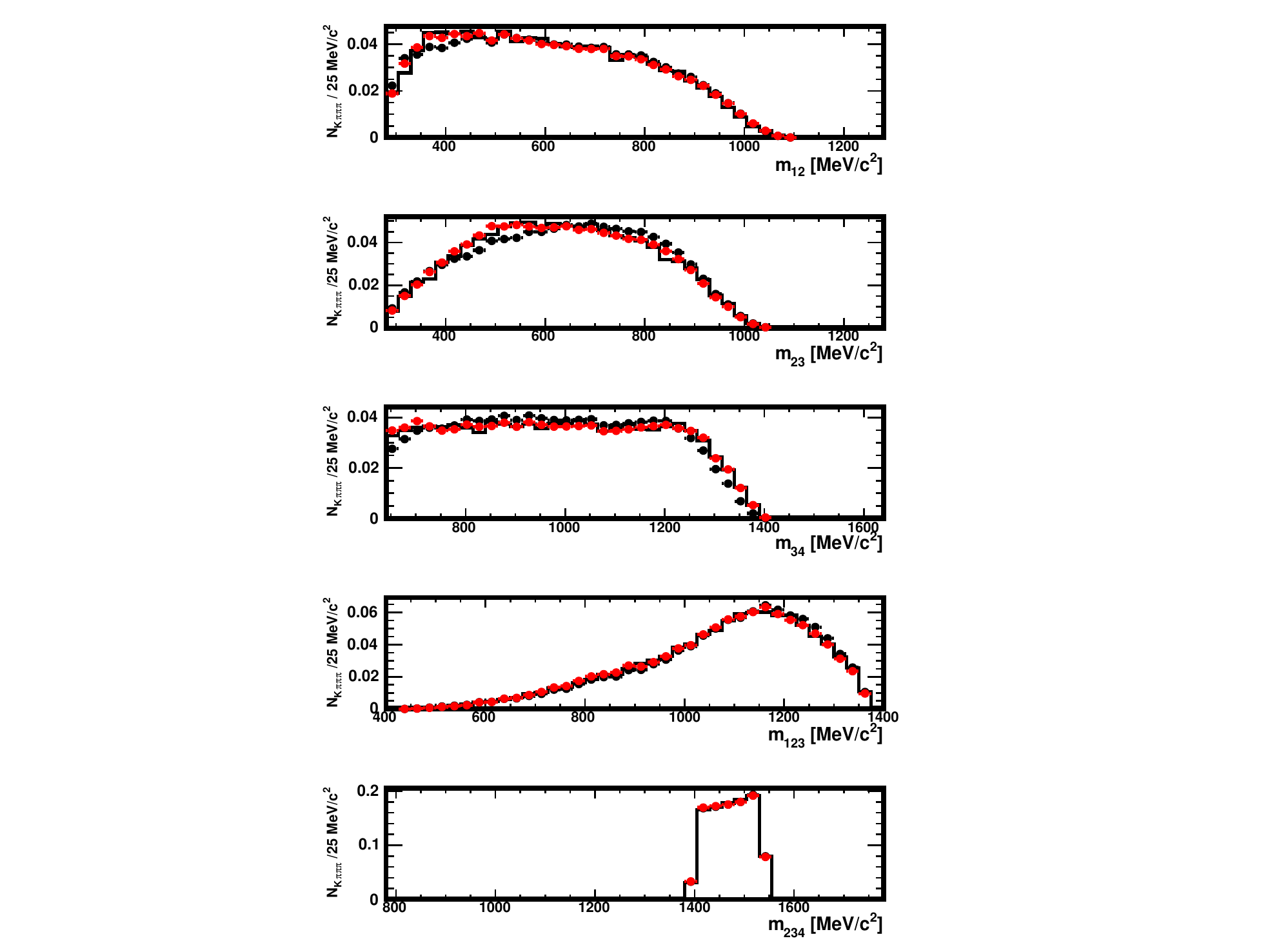}
    \vspace*{-0.5cm}
  \caption{Distributions of $m_{12}$, $m_{23}$, $m_{34}$, $m_{123}$ and $m_{234}$ in the original sample (histogram),
in the distorted one (full black circles) and in the distorted sample where the decays have been re-weighted using the $\omega_i$ weights (red), 
as explained in Sect.~\ref{subsec:ResultsK3pi}. The data used here are restricted to the region $950 < m_{234} < 1150$ MeV$/c^2$.
The absolute normalisation is arbitrary when the correction is not applied and natural when it is applied (red).} 
  \label{fig:Distr_m234_3}
\end{figure}

\clearpage

\begin{figure}[tb]
    \hspace*{-3cm}
       \includegraphics[width=1.4\linewidth]{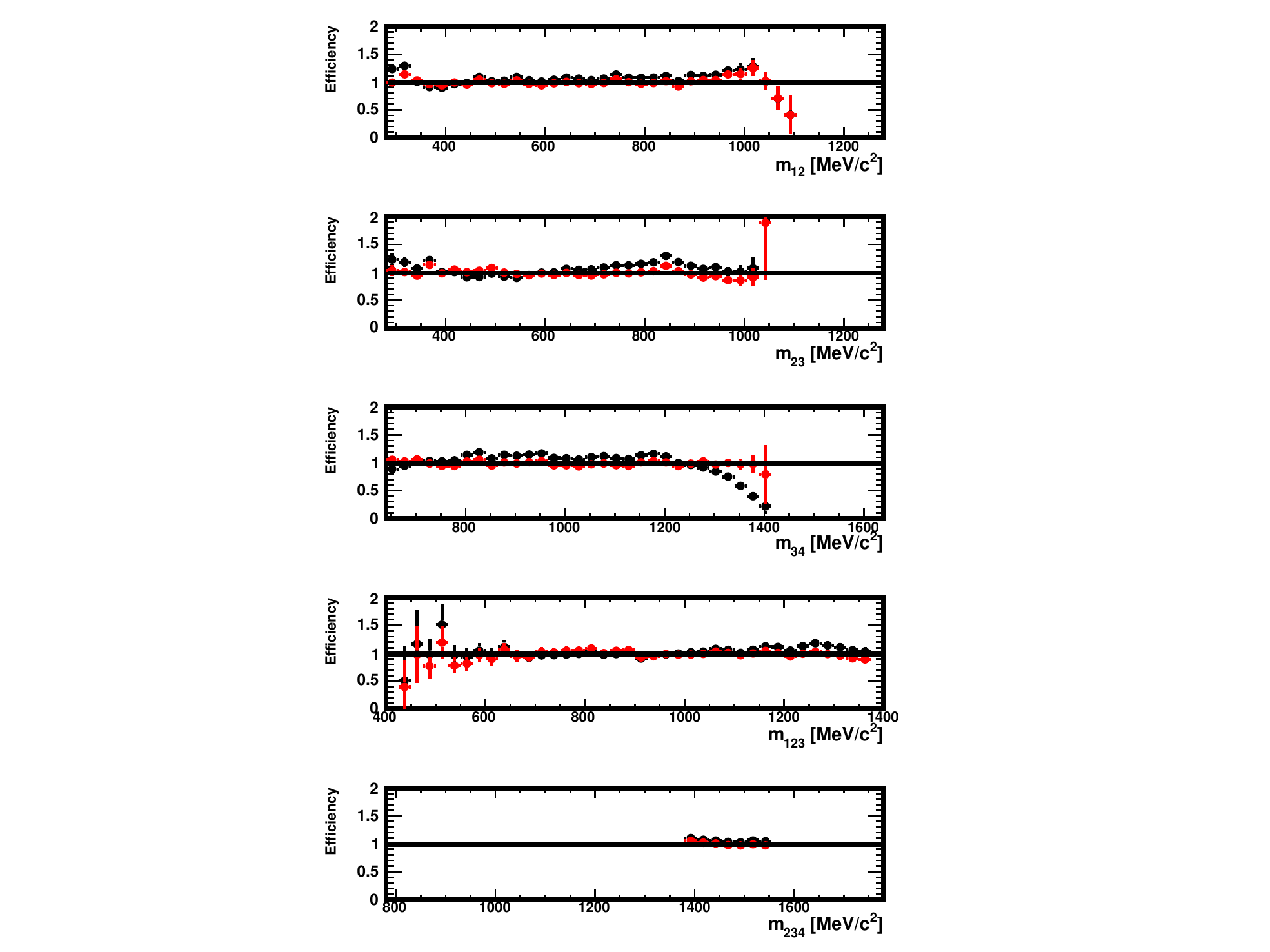}
    \vspace*{-0.5cm}
  \caption{Efficiency in $m_{12}$, $m_{23}$, $m_{34}$, $m_{123}$ and $m_{234}$ in the data generated for this study, 
in the region $950 < m_{234} < 1150$ MeV$/c^2$. Shown are the ratios of the distributions found in the distorted and 
original samples, with no correction (black) and for decays re-weighted using $\omega_i$ weights (red)
as explained in Sect.~\ref{subsec:ResultsK3pi}. The absolute normalisation is arbitrary when the correction is not applied and natural when it is applied (red).} 
  \label{fig:Ratio_m234_3}
\end{figure}

\clearpage

\begin{figure}[tb]
    \hspace*{-3cm}
       \includegraphics[width=1.4\linewidth]{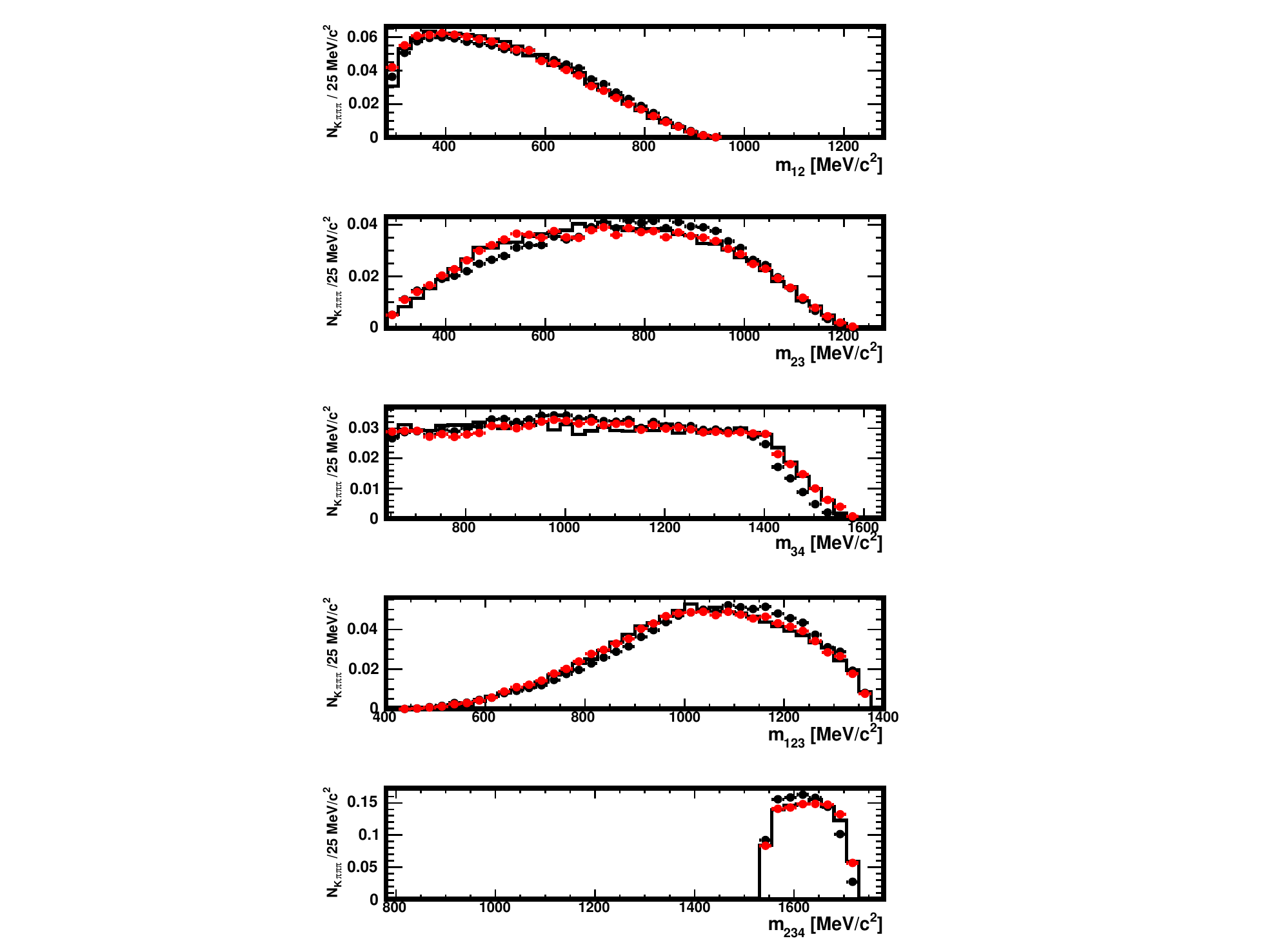}
    \vspace*{-0.5cm}
  \caption{Distributions of $m_{12}$, $m_{23}$, $m_{34}$, $m_{123}$ and $m_{234}$ in the original sample (histogram),
in the distorted one (full black circles) and in the distorted sample where the decays have been re-weighted using the $\omega_i$ weights (red), 
as explained in Sect.~\ref{subsec:ResultsK3pi}. The data used here are restricted to the region $1150 < m_{234}$ MeV$/c^2$.
The absolute normalisation is arbitrary when the correction is not applied and natural when it is applied (red).} 
  \label{fig:Distr_m234_4}
\end{figure}

\clearpage

\begin{figure}[tb]
    \hspace*{-3cm}
       \includegraphics[width=1.4\linewidth]{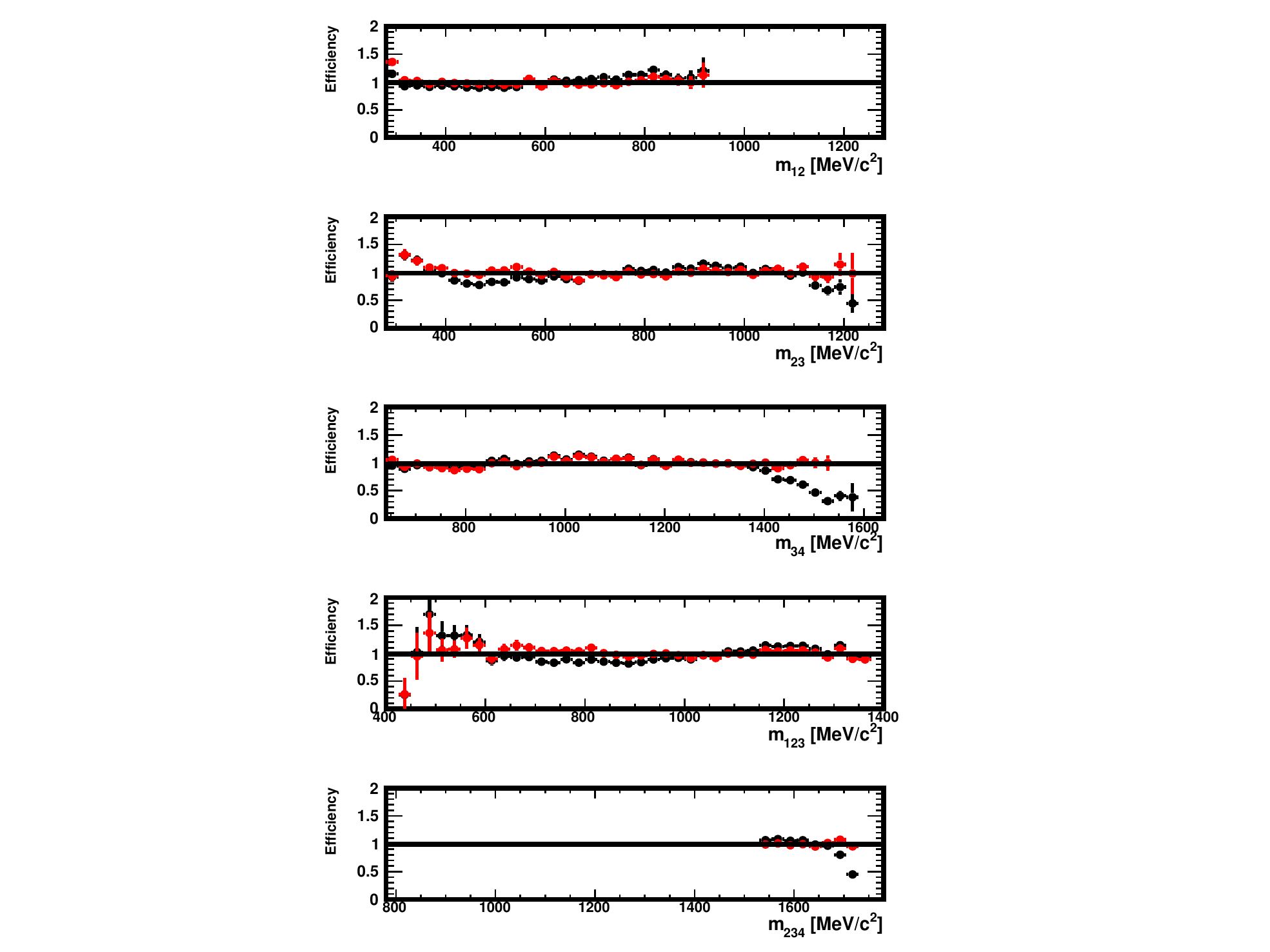}
    \vspace*{-0.5cm}
  \caption{Efficiency in $m_{12}$, $m_{23}$, $m_{34}$, $m_{123}$ and $m_{234}$ in the data generated for this study, 
in the region $1150 < m_{234}$ MeV$/c^2$. Shown are the ratios of the distributions found in the distorted and 
original samples, with no correction (black) and for decays re-weighted using $\omega_i$ weights (red)
as explained in Sect.~\ref{subsec:ResultsK3pi}. The absolute normalisation is arbitrary when the correction is not applied and natural when it is applied (red).} 
  \label{fig:Ratio_m234_4}
\end{figure}



\begin{figure}[tb]
    \hspace*{-3cm}
       \includegraphics[width=1.4\linewidth]{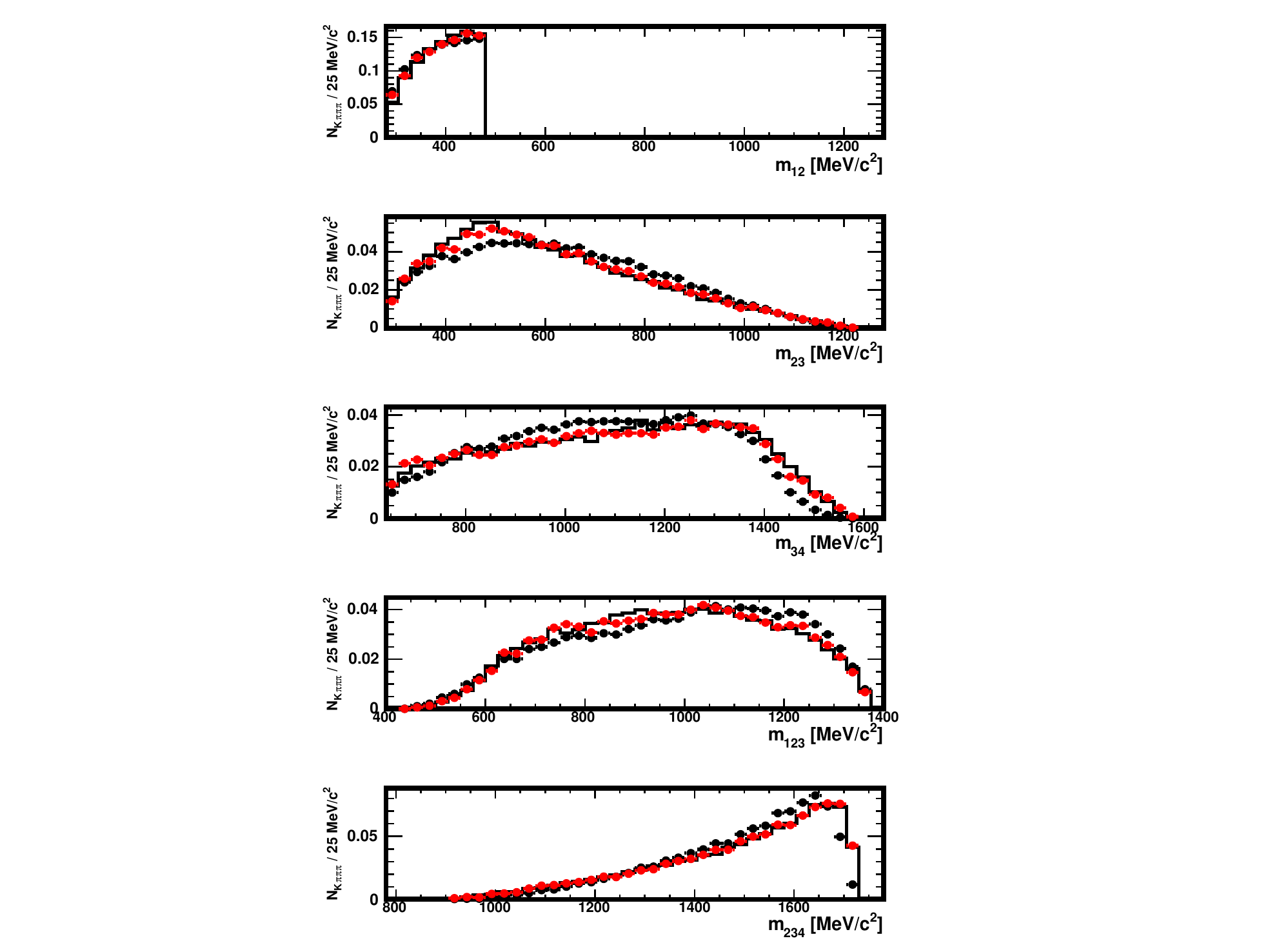}
    \vspace*{-0.5cm}
  \caption{Distributions of $m_{12}$, $m_{23}$, $m_{34}$, $m_{123}$ and $m_{234}$ in the original sample (histogram),
in the distorted one obtained with a tighter selection (full black circles) and in the distorted sample where the decays have been re-weighted using the $\omega_i$ weights (red), 
as explained in Sect.~\ref{subsec:ResultsK3pi}. The data used here are restricted to the region $ 0 < m_{12} < 480$ MeV$/c^2$.
The absolute normalisation is arbitrary when the correction is not applied and natural when it is applied (red).} 
  \label{fig:Distr_m12_1_Sel2}
\end{figure}

\clearpage

\begin{figure}[tb]
    \hspace*{-3cm}
       \includegraphics[width=1.4\linewidth]{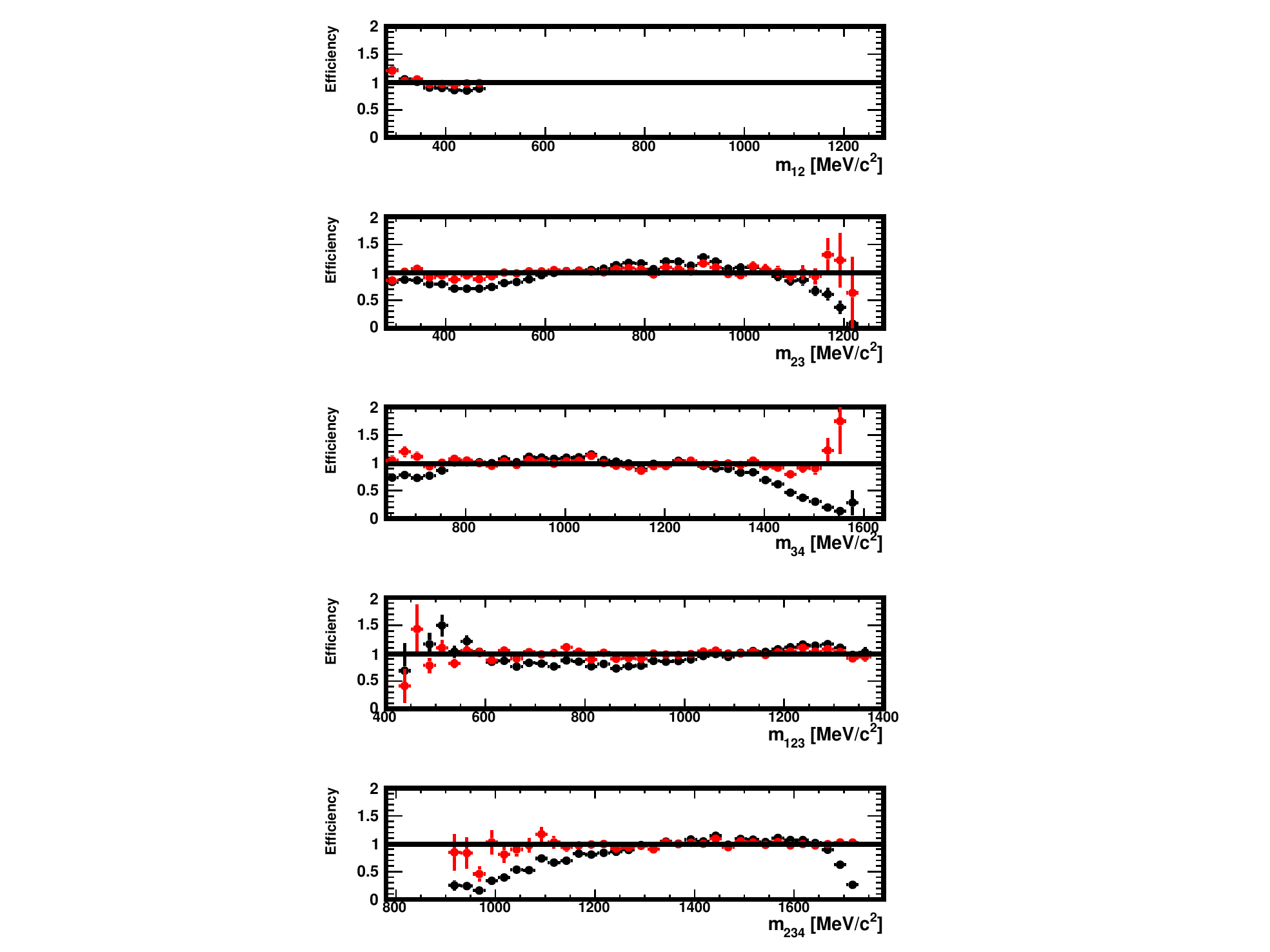}
    \vspace*{-0.5cm}
  \caption{Efficiency in $m_{12}$, $m_{23}$, $m_{34}$, $m_{123}$ and $m_{234}$ in the data generated for this study, 
in the region $0 < m_{12} < 480$ MeV$/c^2$ and with a tighter selection. Shown are the ratios of the distributions found in the distorted and 
original samples, with no correction (black) and for decays re-weighted using $\omega_i$ weights (red)
as explained in Sect.~\ref{subsec:ResultsK3pi}. The absolute normalisation is arbitrary when the correction is not applied and natural when it is applied (red).} 
  \label{fig:Ratio_m12_1_Sel2}
\end{figure}

\clearpage

\begin{figure}[tb]
    \hspace*{-3cm}
       \includegraphics[width=1.4\linewidth]{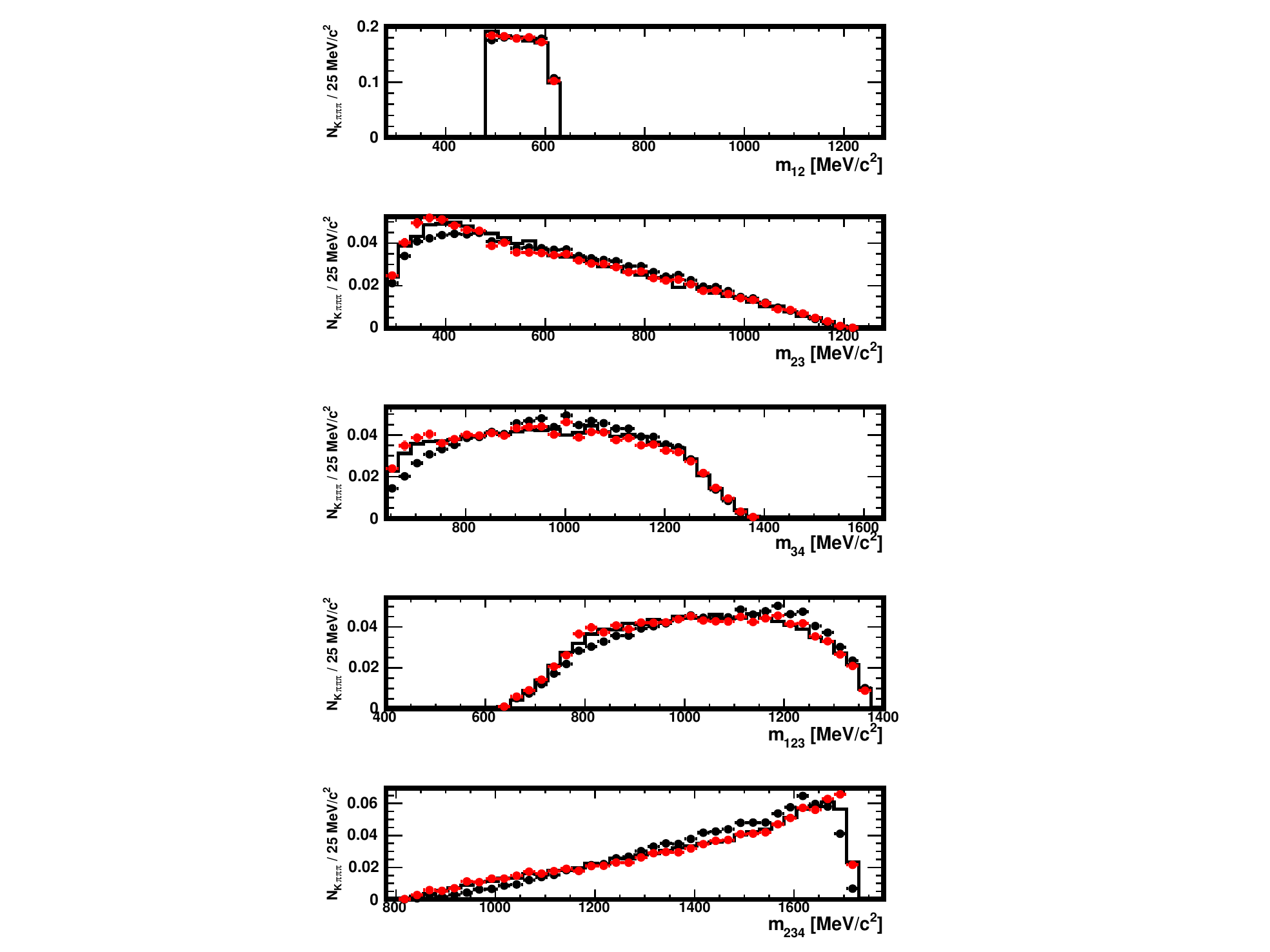}
    \vspace*{-0.5cm}
  \caption{Distributions of $m_{12}$, $m_{23}$, $m_{34}$, $m_{123}$ and $m_{234}$ in the original sample (histogram),
in the distorted one obtained with a tighter selection (full black circles) and in the distorted sample where the decays have been re-weighted using the $\omega_i$ weights (red), 
as explained in Sect.~\ref{subsec:ResultsK3pi}. The data used here are restricted to the region $480 < m_{12} < 620$ MeV$/c^2$.
The absolute normalisation is arbitrary when the correction is not applied and natural when it is applied (red).} 
  \label{fig:Distr_m12_2_Sel2}
\end{figure}

\clearpage

\begin{figure}[tb]
    \hspace*{-3cm}
       \includegraphics[width=1.4\linewidth]{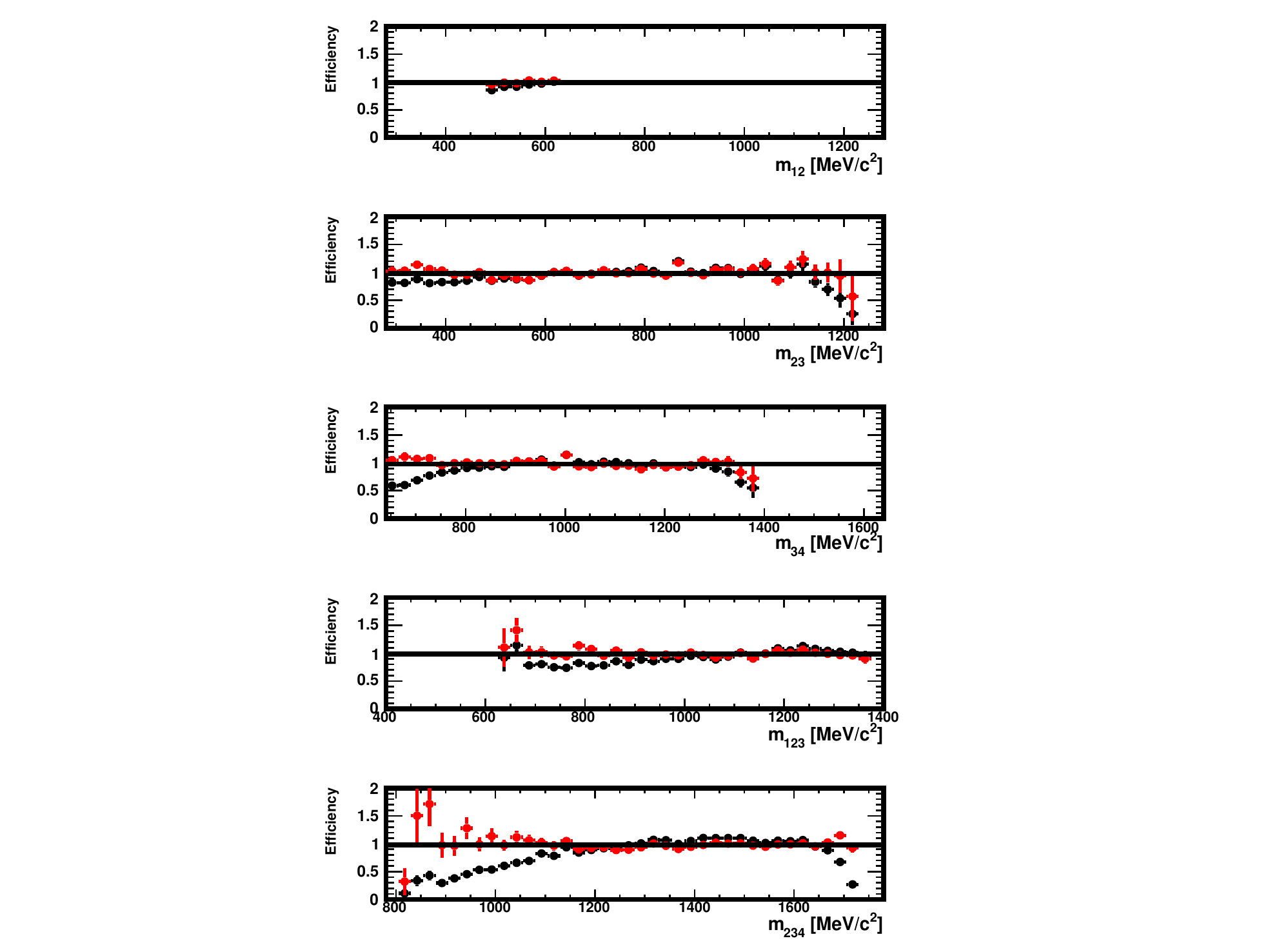}
    \vspace*{-0.5cm}
  \caption{Efficiency in $m_{12}$, $m_{23}$, $m_{34}$, $m_{123}$ and $m_{234}$ in the data generated for this study, 
in the region $ <480 m_{12} < 620$ MeV$/c^2$ and with a tighter selection. Shown are the ratios of the distributions found in the distorted and 
original samples, with no correction (black) and for decays re-weighted using $\omega_i$ weights (red)
as explained in Sect.~\ref{subsec:ResultsK3pi}. The absolute normalisation is arbitrary when the correction is not applied and natural when it is applied (red).} 
  \label{fig:Ratio_m12_2_Sel2}
\end{figure}

\clearpage

\begin{figure}[tb]
    \hspace*{-3cm}
       \includegraphics[width=1.4\linewidth]{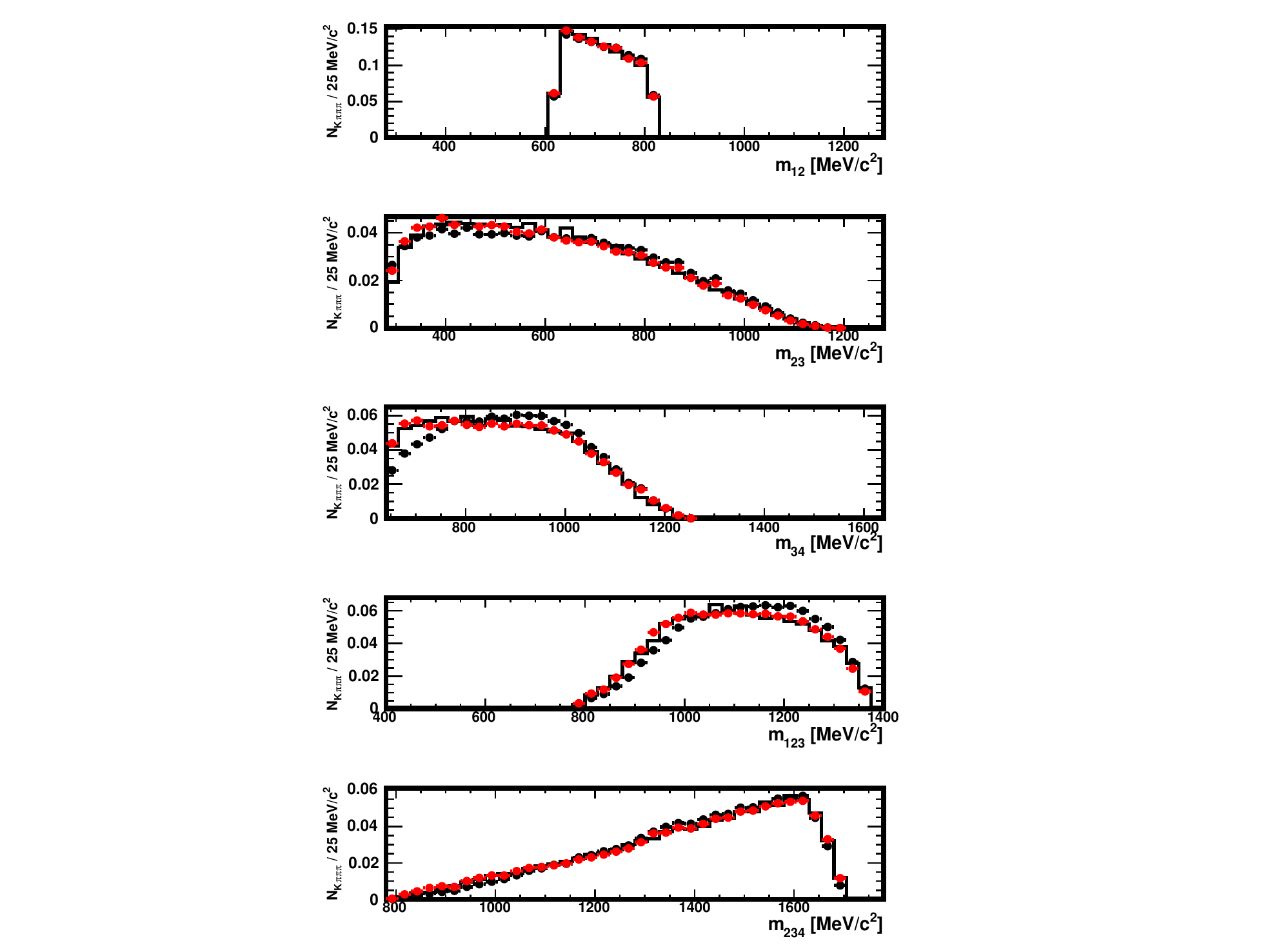}
    \vspace*{-0.5cm}
  \caption{Distributions of $m_{12}$, $m_{23}$, $m_{34}$, $m_{123}$ and $m_{234}$ in the original sample (histogram),
in the distorted one obtained with a tighter selection (full black circles) and in the distorted sample where the decays have been re-weighted using the $\omega_i$ weights (red), 
as explained in Sect.~\ref{subsec:ResultsK3pi}. The data used here are restricted to the region $620 < m_{12} < 820$ MeV$/c^2$.
The absolute normalisation is arbitrary when the correction is not applied and natural when it is applied (red).} 
  \label{fig:Distr_m12_3_Sel2}
\end{figure}

\clearpage

\begin{figure}[tb]
    \hspace*{-3cm}
       \includegraphics[width=1.4\linewidth]{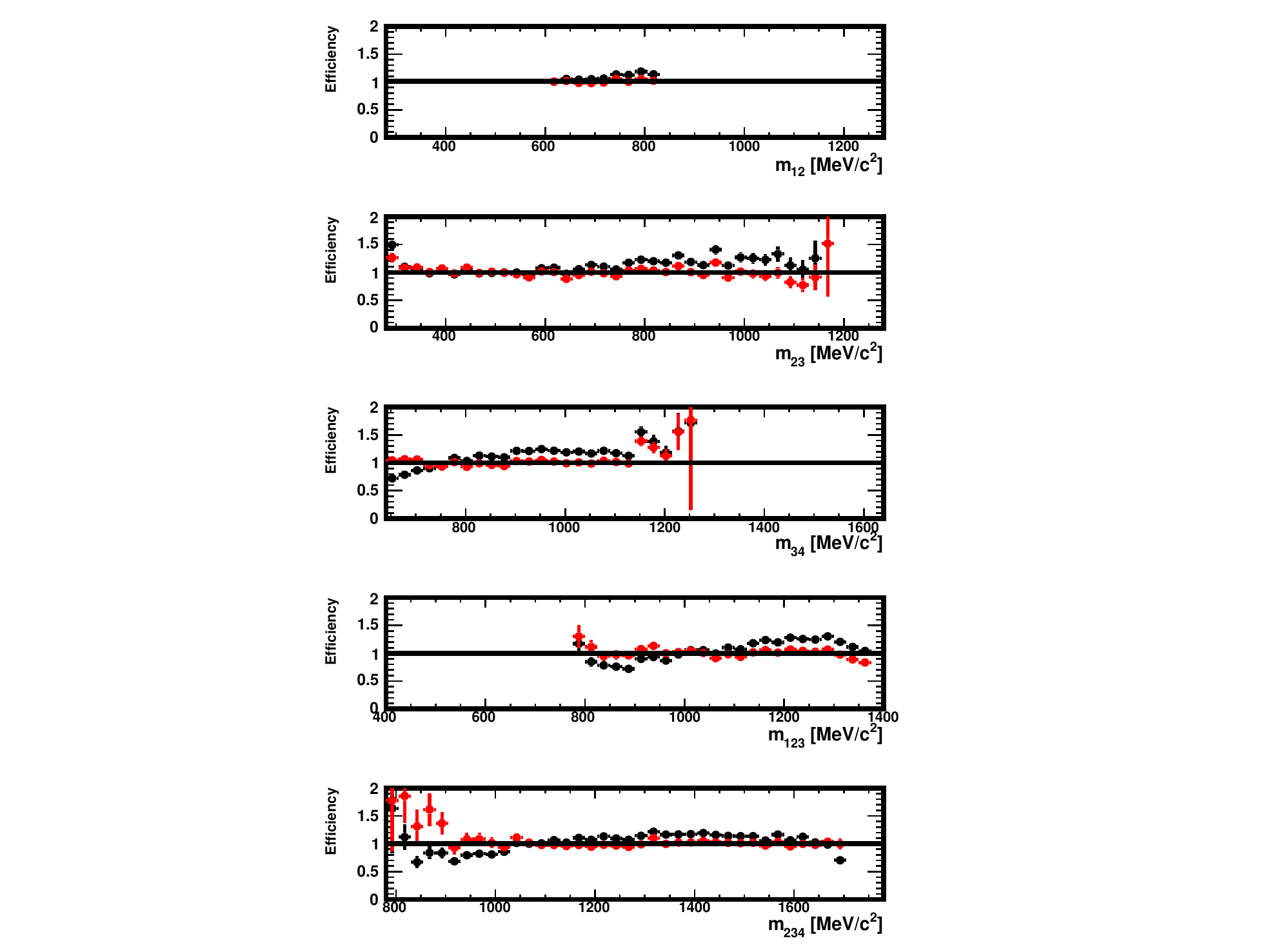}
    \vspace*{-0.5cm}
  \caption{Efficiency in $m_{12}$, $m_{23}$, $m_{34}$, $m_{123}$ and $m_{234}$ in the data generated for this study, 
in the region $620 < m_{12} < 820$ MeV$/c^2$ and with a tighter selection. Shown are the ratios of the distributions found in the distorted and 
original samples, with no correction (black) and for decays re-weighted using $\omega_i$ weights (red)
as explained in Sect.~\ref{subsec:ResultsK3pi}. The absolute normalisation is arbitrary when the correction is not applied and natural when it is applied (red).} 
  \label{fig:Ratio_m12_3_Sel2}
\end{figure}

\clearpage

\begin{figure}[tb]
    \hspace*{-3cm}
       \includegraphics[width=1.4\linewidth]{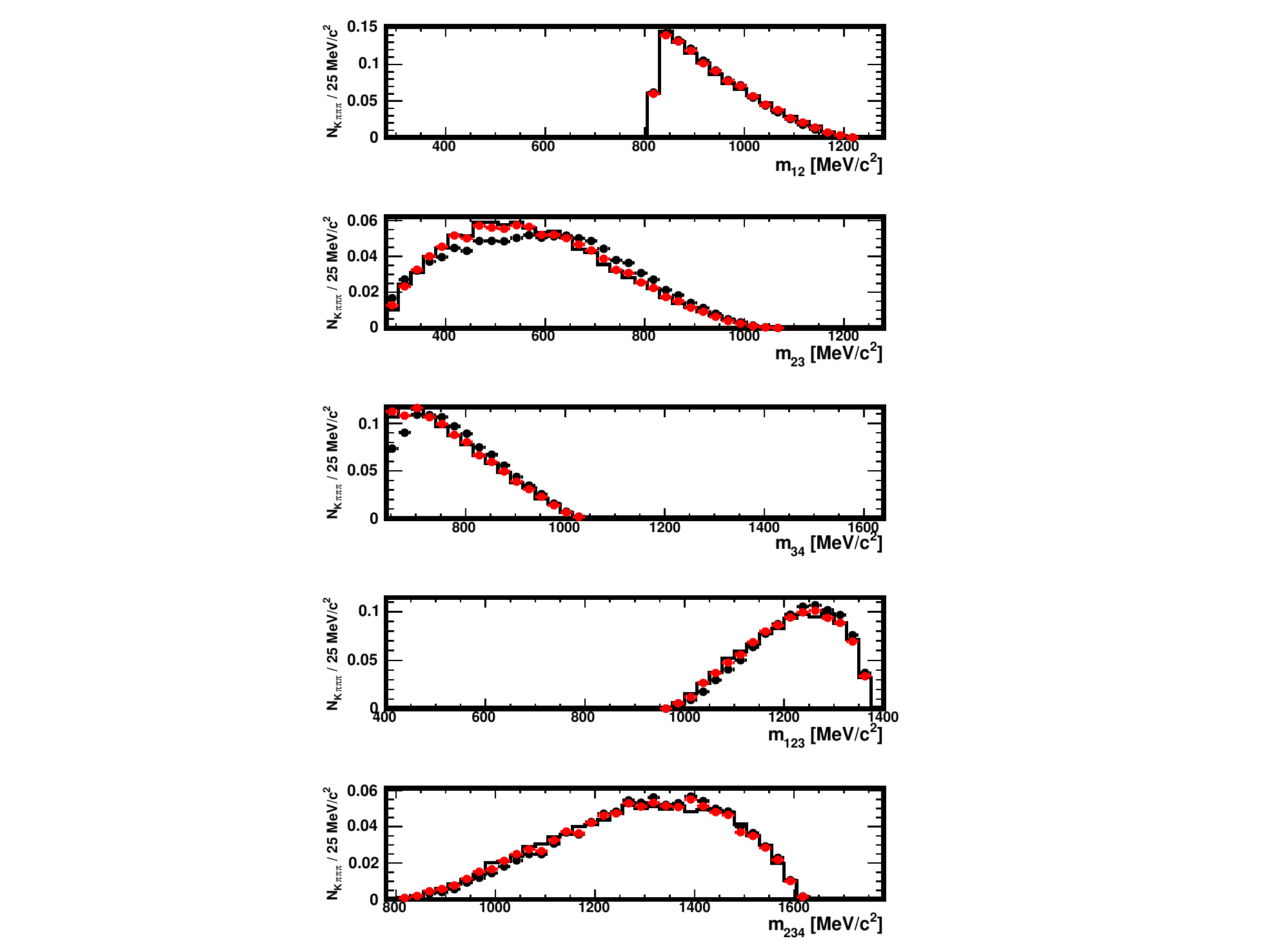}
    \vspace*{-0.5cm}
  \caption{Distributions of $m_{12}$, $m_{23}$, $m_{34}$, $m_{123}$ and $m_{234}$ in the original sample (histogram),
in the distorted one obtained with a tighter selection (full black circles) and in the distorted sample where the decays have been re-weighted using the $\omega_i$ weights (red), 
as explained in Sect.~\ref{subsec:ResultsK3pi}. The data used here are restricted to the region $820 < m_{12}$ MeV$/c^2$.
The absolute normalisation is arbitrary when the correction is not applied and natural when it is applied (red).} 
  \label{fig:Distr_m12_4_Sel2}
\end{figure}

\clearpage

\begin{figure}[tb]
    \hspace*{-3cm}
       \includegraphics[width=1.4\linewidth]{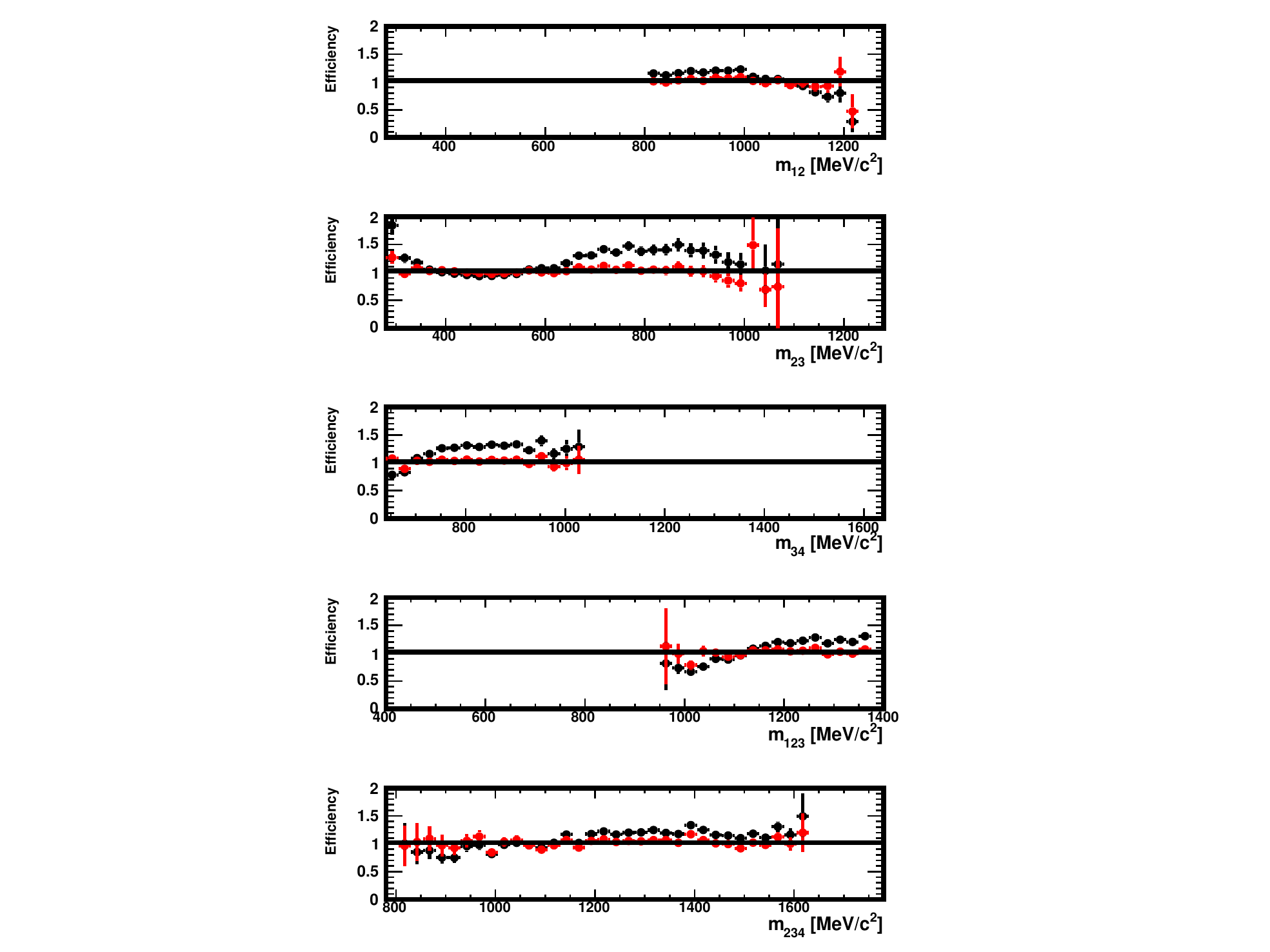}
    \vspace*{-0.5cm}
  \caption{Efficiency in $m_{12}$, $m_{23}$, $m_{34}$, $m_{123}$ and $m_{234}$ in the data generated for this study, 
in the region $820 < m_{12}$ MeV$/c^2$ and with a tighter selection. Shown are the ratios of the distributions found in the distorted and 
original samples, with no correction (black) and for decays re-weighted using $\omega_i$ weights (red)
as explained in Sect.~\ref{subsec:ResultsK3pi}. The absolute normalisation is arbitrary when the correction is not applied and natural when it is applied (red).} 
  \label{fig:Ratio_m12_4_Sel2}
\end{figure}

\clearpage

\begin{figure}[tb]
    \hspace*{-3cm}
       \includegraphics[width=1.4\linewidth]{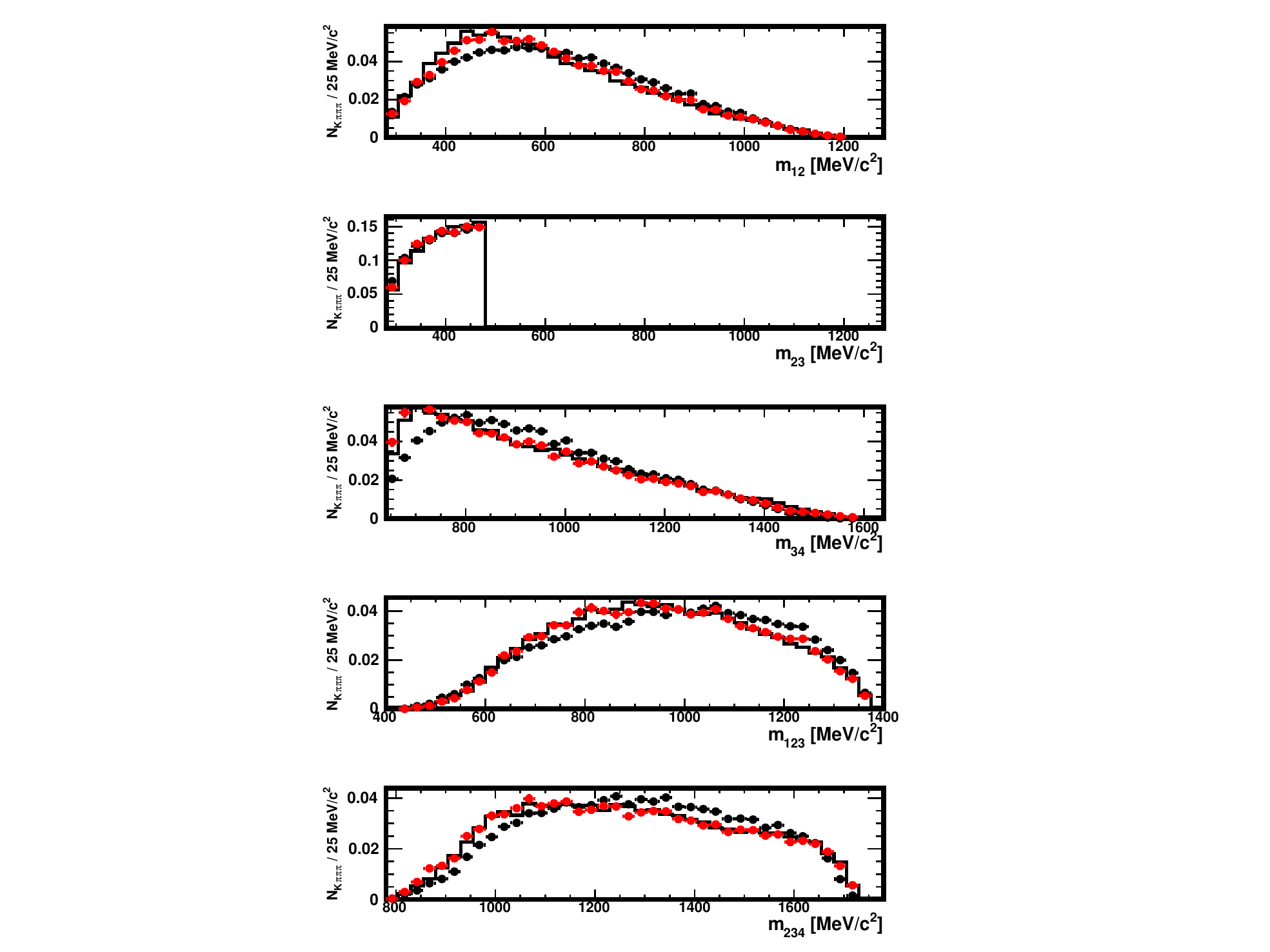}
    \vspace*{-0.5cm}
  \caption{Distributions of $m_{12}$, $m_{23}$, $m_{34}$, $m_{123}$ and $m_{234}$ in the original sample (histogram),
in the distorted one obtained with a tighter selection (full black circles) and in the distorted sample where the decays have been re-weighted using the $\omega_i$ weights (red), 
as explained in Sect.~\ref{subsec:ResultsK3pi}. The data used here are restricted to the region $ 0 < m_{23} < 480$ MeV$/c^2$.
The absolute normalisation is arbitrary when the correction is not applied and natural when it is applied (red).} 
  \label{fig:Distr_m23_1_Sel2}
\end{figure}

\clearpage

\begin{figure}[tb]
    \hspace*{-3cm}
       \includegraphics[width=1.4\linewidth]{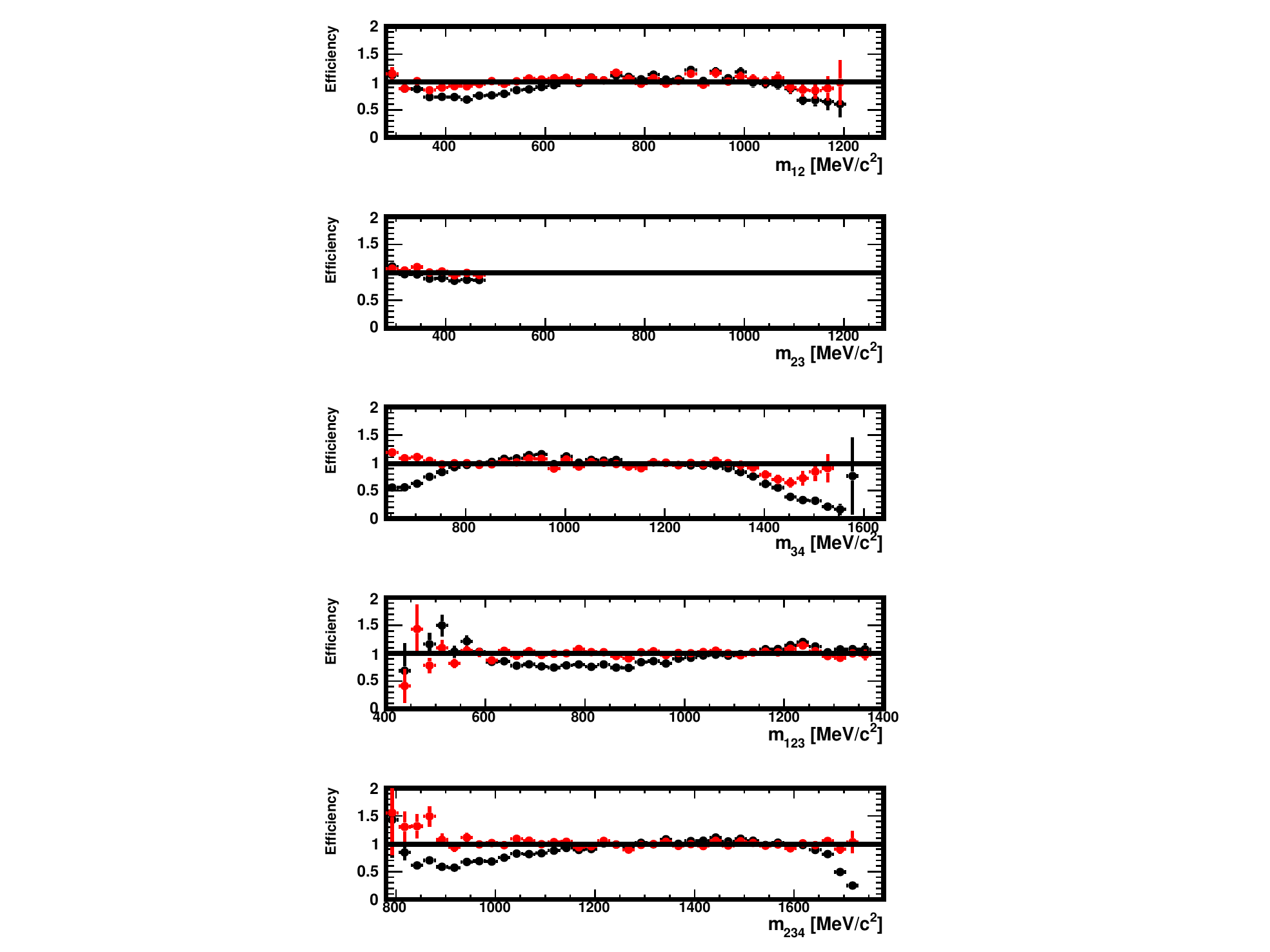}
    \vspace*{-0.5cm}
  \caption{Efficiency in $m_{12}$, $m_{23}$, $m_{34}$, $m_{123}$ and $m_{234}$ in the data generated for this study, 
in the region $0 < m_{23} < 480$ MeV$/c^2$ and with a tighter selection. Shown are the ratios of the distributions found in the distorted and 
original samples, with no correction (black) and for decays re-weighted using $\omega_i$ weights (red)
as explained in Sect.~\ref{subsec:ResultsK3pi}. The absolute normalisation is arbitrary when the correction is not applied and natural when it is applied (red).} 
  \label{fig:Ratio_m23_1_Sel2}
\end{figure}

\clearpage

\begin{figure}[tb]
    \hspace*{-3cm}
       \includegraphics[width=1.4\linewidth]{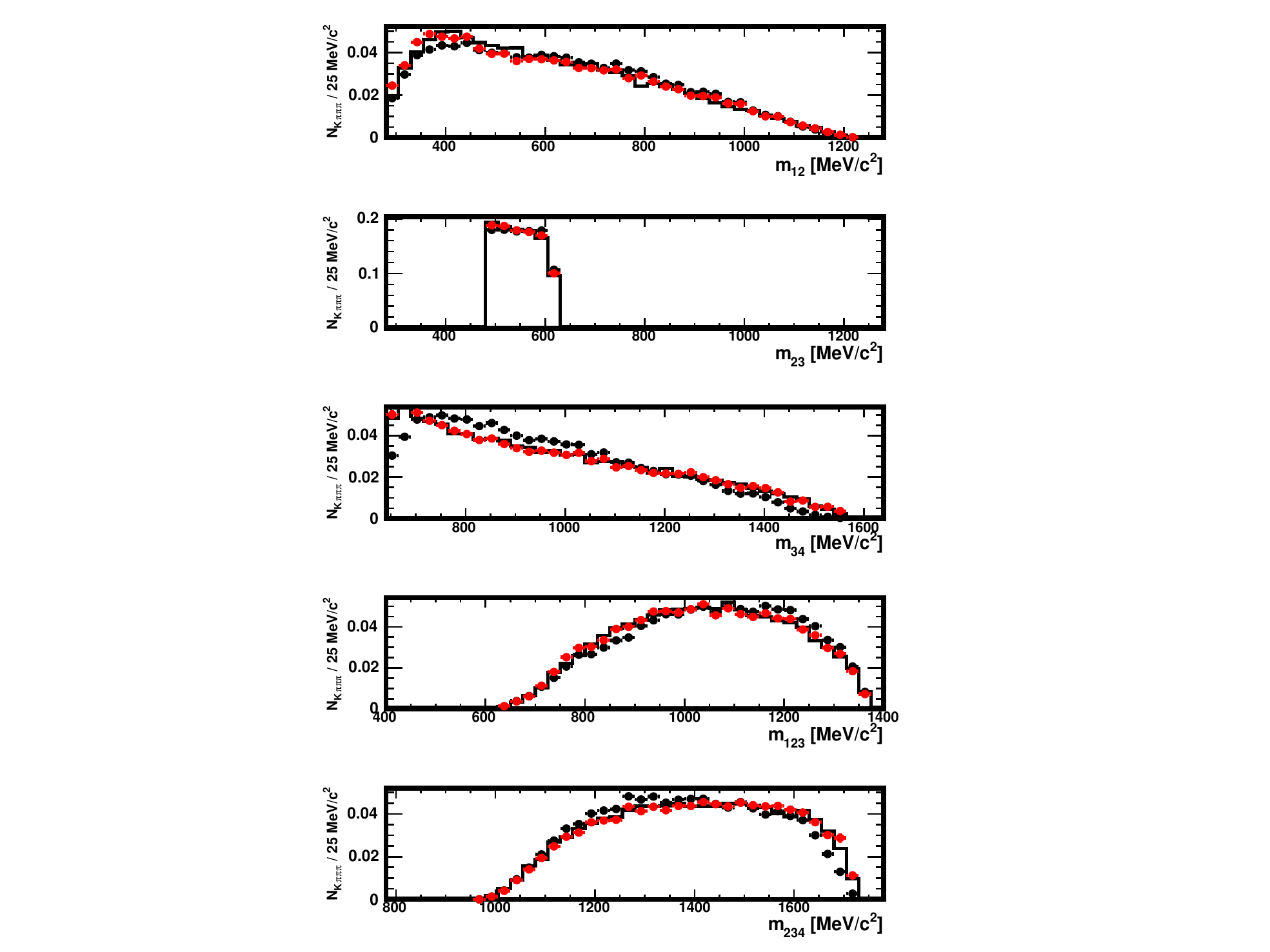}
    \vspace*{-0.5cm}
  \caption{Distributions of $m_{12}$, $m_{23}$, $m_{34}$, $m_{123}$ and $m_{234}$ in the original sample (histogram),
in the distorted one obtained with a tighter selection (full black circles) and in the distorted sample where the decays have been re-weighted using the $\omega_i$ weights (red), 
as explained in Sect.~\ref{subsec:ResultsK3pi}. The data used here are restricted to the region $480 < m_{23} < 620$ MeV$/c^2$.
The absolute normalisation is arbitrary when the correction is not applied and natural when it is applied (red).} 
  \label{fig:Distr_m23_2_Sel2}
\end{figure}

\clearpage

\begin{figure}[tb]
    \hspace*{-3cm}
       \includegraphics[width=1.4\linewidth]{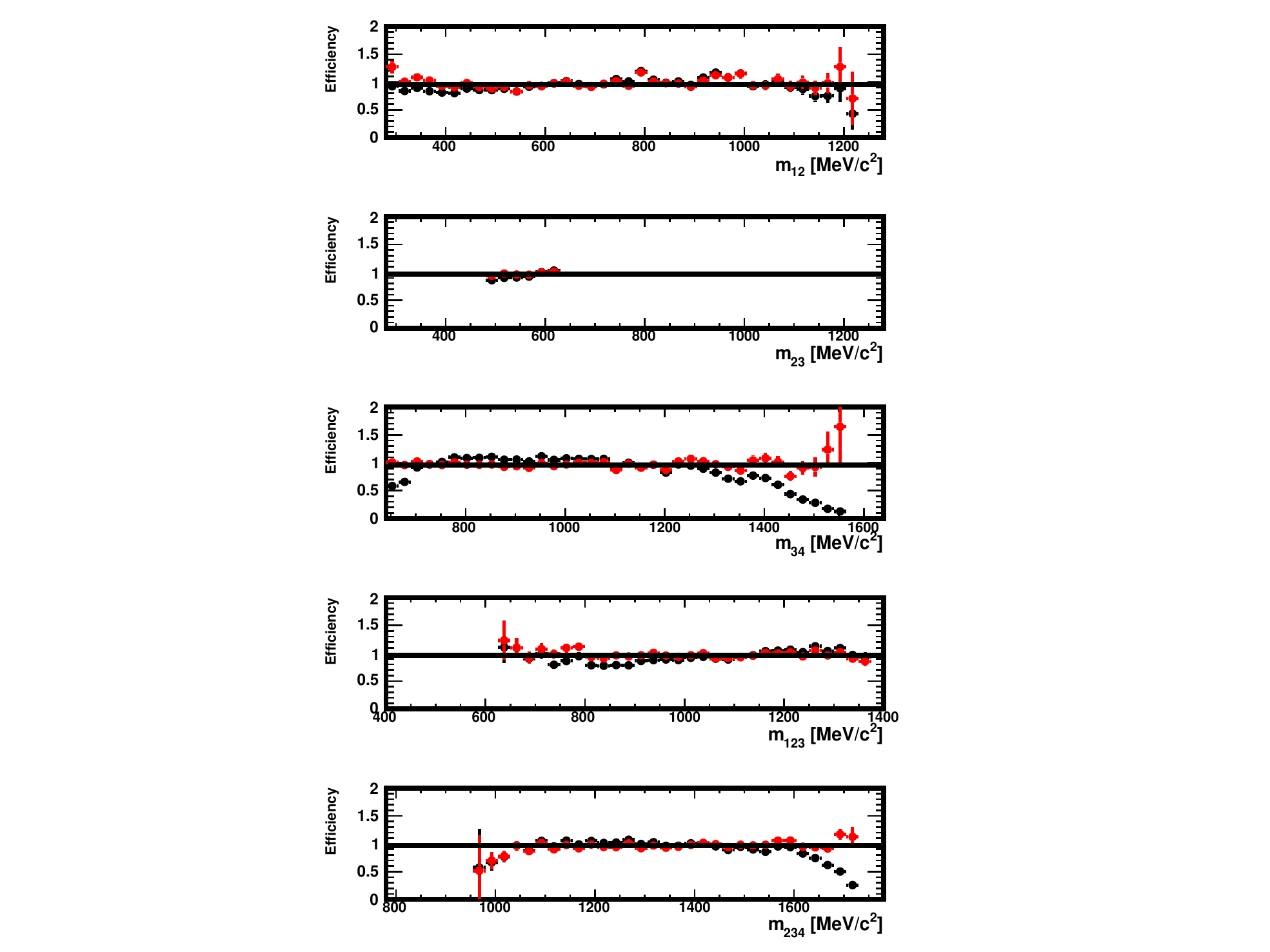}
    \vspace*{-0.5cm}
  \caption{Efficiency in $m_{12}$, $m_{23}$, $m_{34}$, $m_{123}$ and $m_{234}$ in the data generated for this study, 
in the region $ <480 m_{23} < 620$ MeV$/c^2$ and with a tighter selection. Shown are the ratios of the distributions found in the distorted and 
original samples, with no correction (black) and for decays re-weighted using $\omega_i$ weights (red)
as explained in Sect.~\ref{subsec:ResultsK3pi}. The absolute normalisation is arbitrary when the correction is not applied and natural when it is applied (red).} 
  \label{fig:Ratio_m23_2_Sel2}
\end{figure}

\clearpage

\begin{figure}[tb]
    \hspace*{-3cm}
       \includegraphics[width=1.4\linewidth]{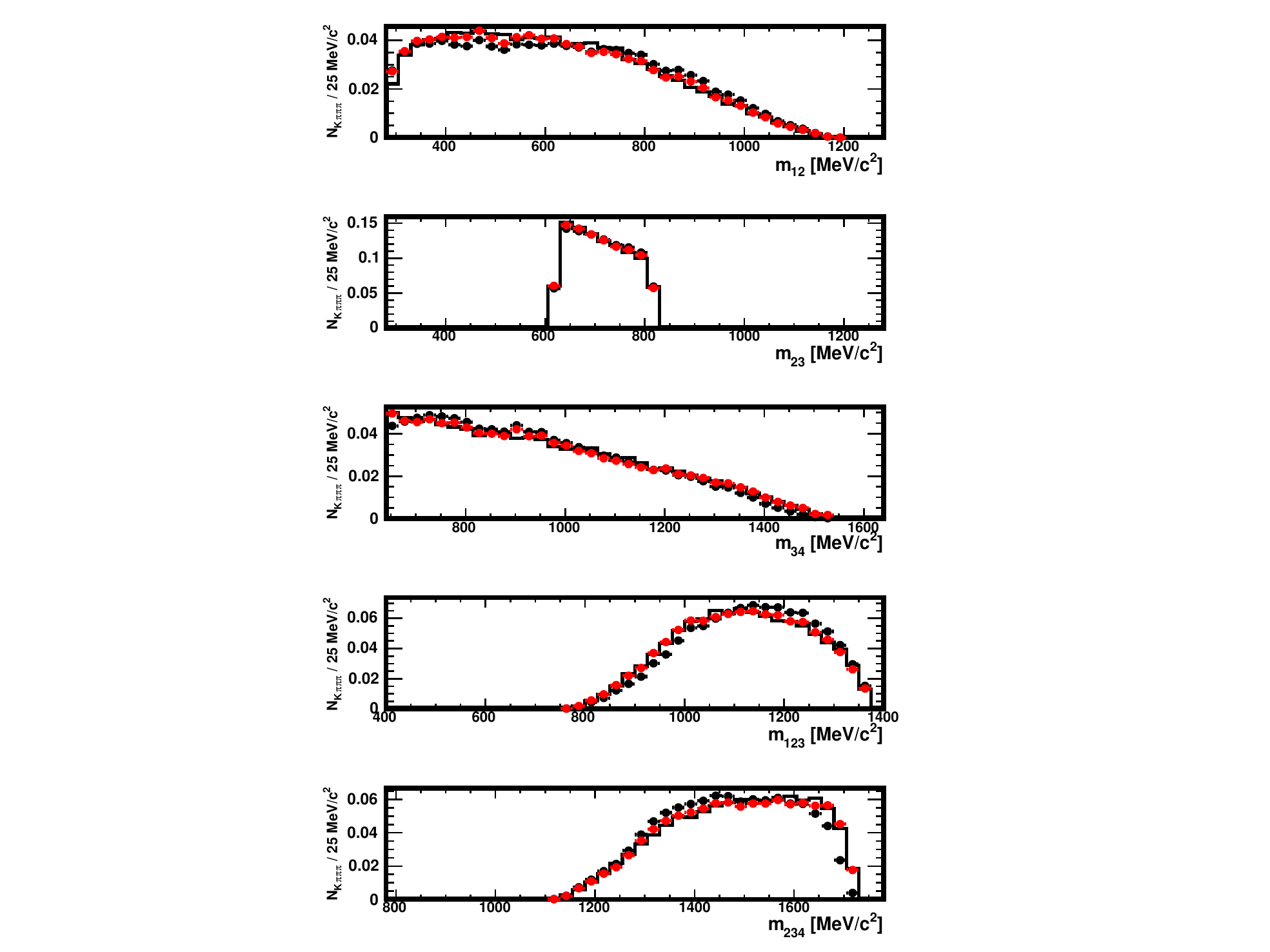}
    \vspace*{-0.5cm}
  \caption{Distributions of $m_{12}$, $m_{23}$, $m_{34}$, $m_{123}$ and $m_{234}$ in the original sample (histogram),
in the distorted one obtained with a tighter selection (full black circles) and in the distorted sample where the decays have been re-weighted using the $\omega_i$ weights (red), 
as explained in Sect.~\ref{subsec:ResultsK3pi}. The data used here are restricted to the region $620 < m_{23} < 820$ MeV$/c^2$.
The absolute normalisation is arbitrary when the correction is not applied and natural when it is applied (red).} 
  \label{fig:Distr_m23_3_Sel2}
\end{figure}

\clearpage

\begin{figure}[tb]
    \hspace*{-3cm}
       \includegraphics[width=1.4\linewidth]{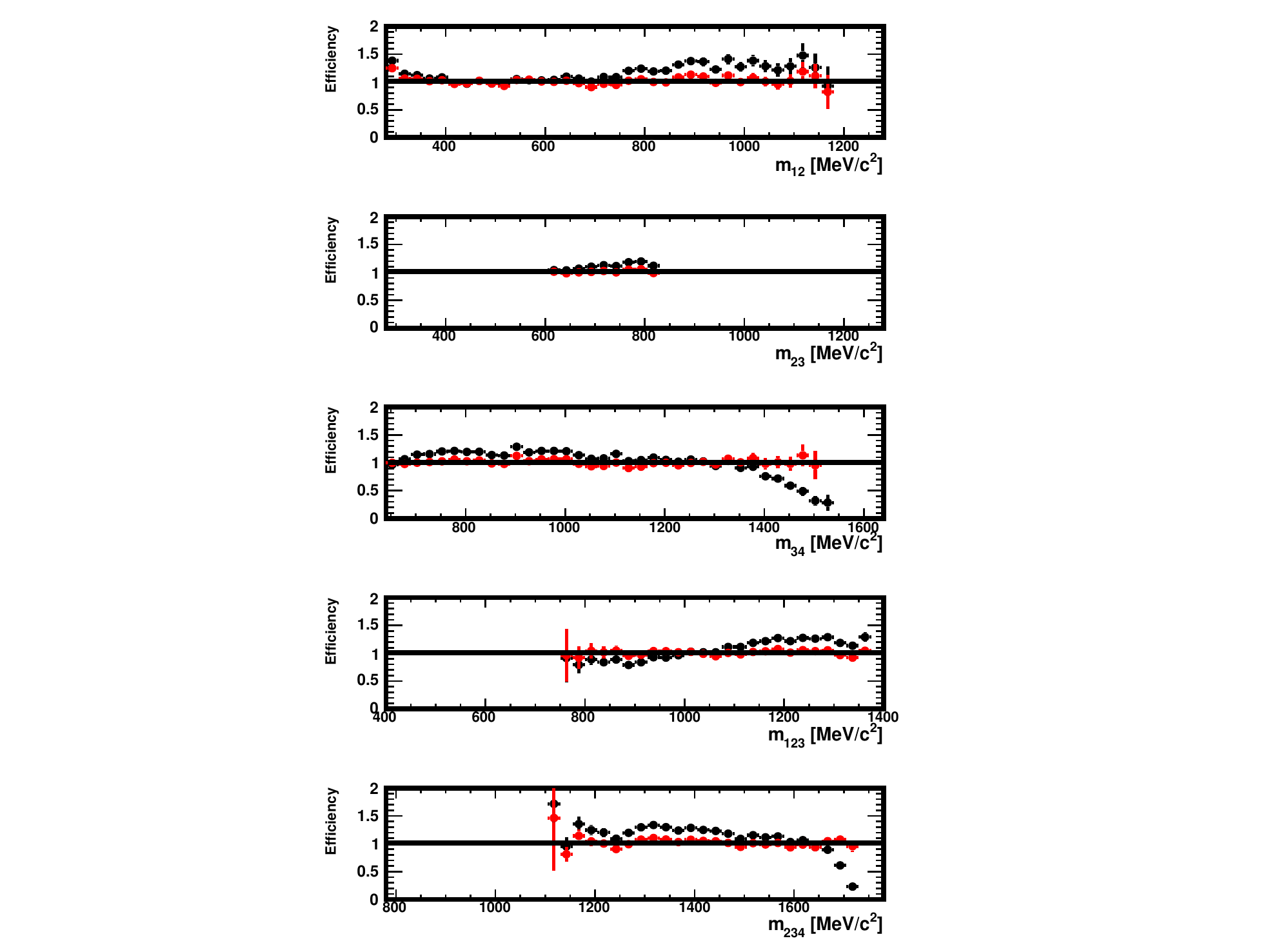}
    \vspace*{-0.5cm}
  \caption{Efficiency in $m_{12}$, $m_{23}$, $m_{34}$, $m_{123}$ and $m_{234}$ in the data generated for this study, 
in the region $620 < m_{23} < 820$ MeV$/c^2$ and with a tighter selection. Shown are the ratios of the distributions found in the distorted and 
original samples, with no correction (black) and for decays re-weighted using $\omega_i$ weights (red)
as explained in Sect.~\ref{subsec:ResultsK3pi}. The absolute normalisation is arbitrary when the correction is not applied and natural when it is applied (red).} 
  \label{fig:Ratio_m23_3_Sel2}
\end{figure}

\clearpage

\begin{figure}[tb]
    \hspace*{-3cm}
       \includegraphics[width=1.4\linewidth]{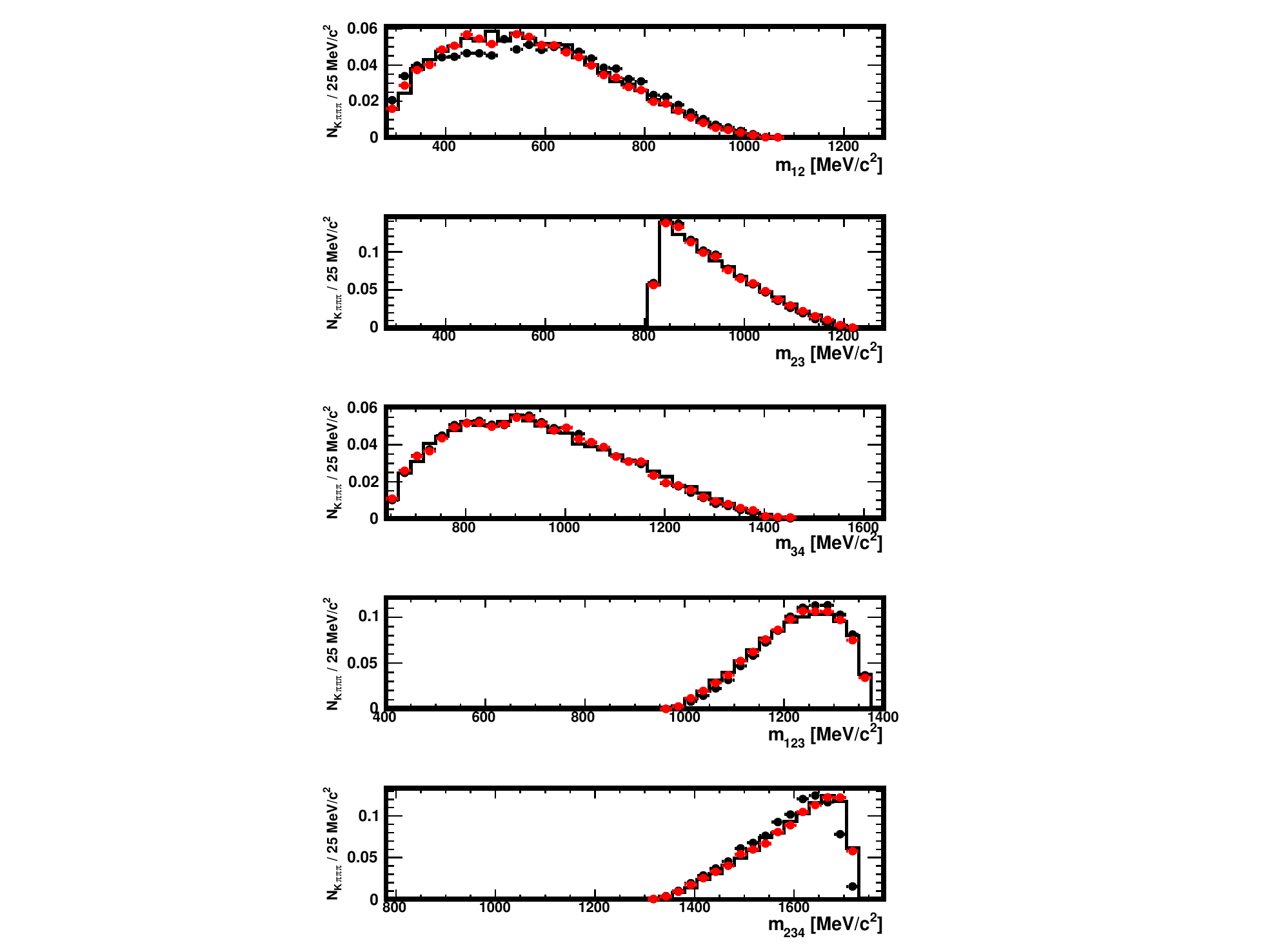}
    \vspace*{-0.5cm}
  \caption{Distributions of $m_{12}$, $m_{23}$, $m_{34}$, $m_{123}$ and $m_{234}$ in the original sample (histogram),
in the distorted one obtained with a tighter selection (full black circles) and in the distorted sample where the decays have been re-weighted using the $\omega_i$ weights (red), 
as explained in Sect.~\ref{subsec:ResultsK3pi}. The data used here are restricted to the region $820 < m_{23}$ MeV$/c^2$.
The absolute normalisation is arbitrary when the correction is not applied and natural when it is applied (red).} 
  \label{fig:Distr_m23_4_Sel2}
\end{figure}

\clearpage

\begin{figure}[tb]
    \hspace*{-3cm}
       \includegraphics[width=1.4\linewidth]{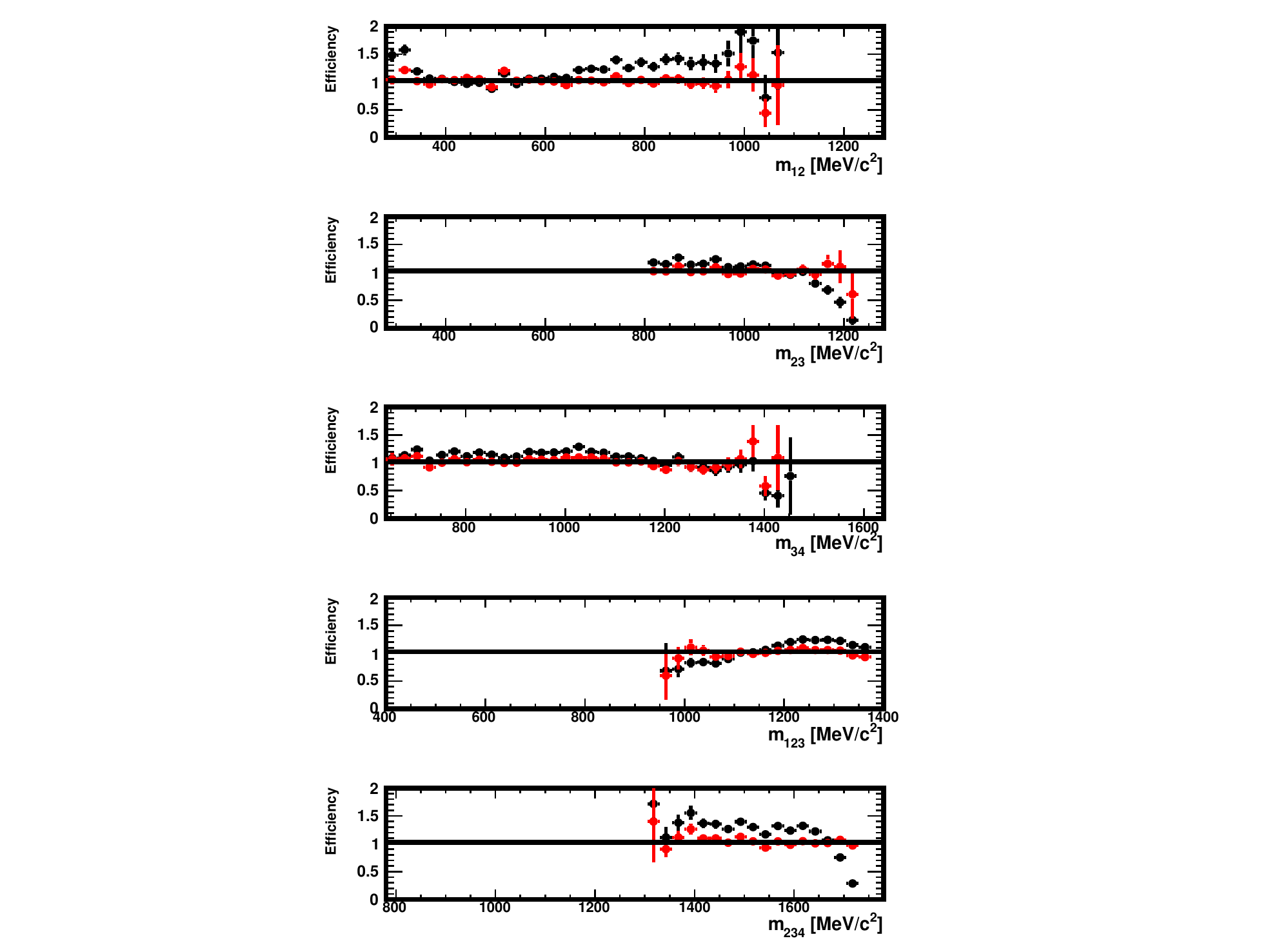}
    \vspace*{-0.5cm}
  \caption{Efficiency in $m_{12}$, $m_{23}$, $m_{34}$, $m_{123}$ and $m_{234}$ in the data generated for this study, 
in the region $820 < m_{23}$ MeV$/c^2$ and with a tighter selection. Shown are the ratios of the distributions found in the distorted and 
original samples, with no correction (black) and for decays re-weighted using $\omega_i$ weights (red)
as explained in Sect.~\ref{subsec:ResultsK3pi}. The absolute normalisation is arbitrary when the correction is not applied and natural when it is applied (red).} 
  \label{fig:Ratio_m23_4_Sel2}
\end{figure}

\clearpage

\begin{figure}[tb]
    \hspace*{-3cm}
       \includegraphics[width=1.4\linewidth]{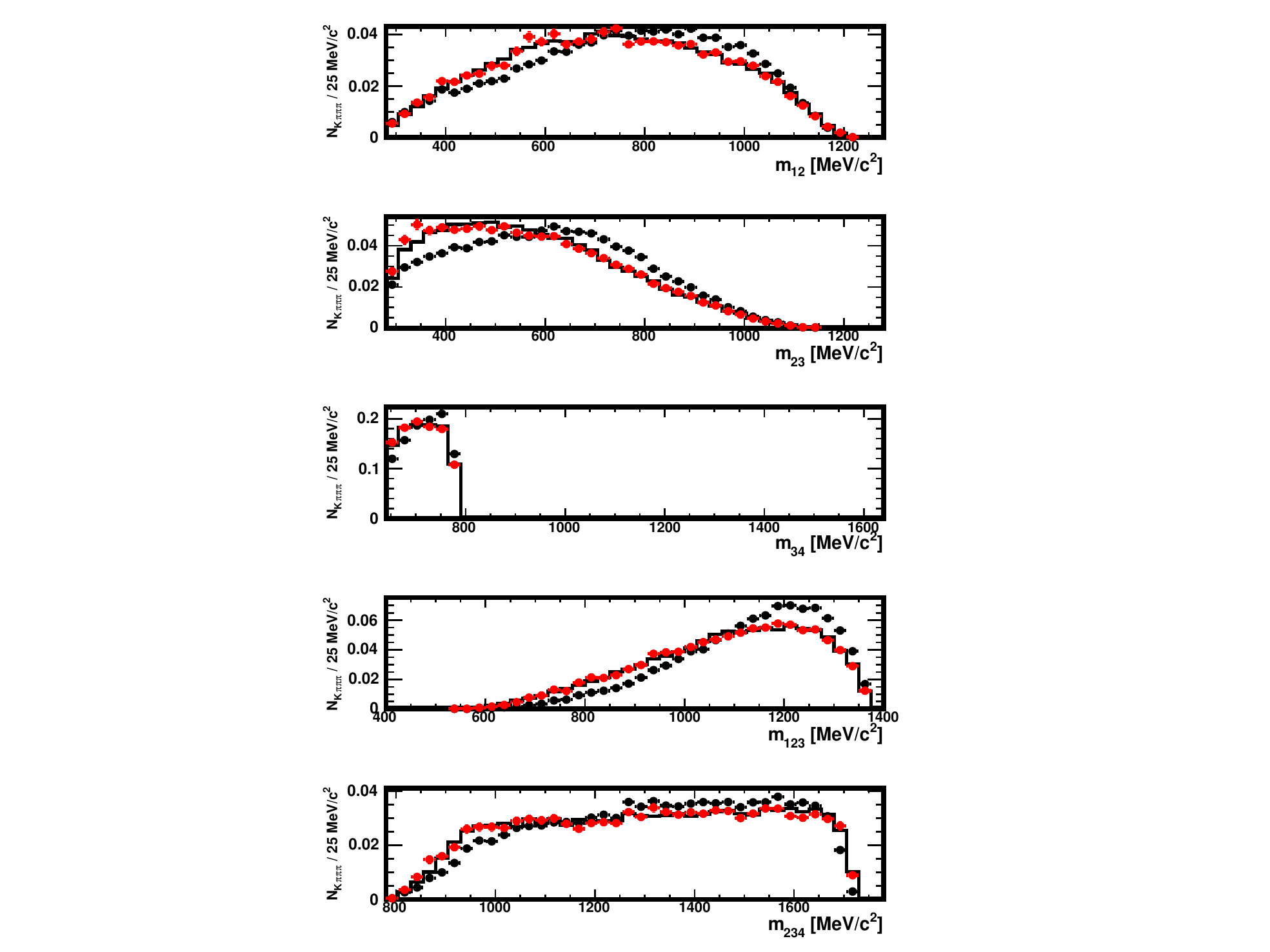}
    \vspace*{-0.5cm}
  \caption{Distributions of $m_{12}$, $m_{23}$, $m_{34}$, $m_{123}$ and $m_{234}$ in the original sample (histogram),
in the distorted one obtained with a tighter selection (full black circles) and in the distorted sample where the decays have been re-weighted using the $\omega_i$ weights (red), 
as explained in Sect.~\ref{subsec:ResultsK3pi}. The data used here are restricted to the region $ 0 < m_{34} < 780$ MeV$/c^2$.
The absolute normalisation is arbitrary when the correction is not applied and natural when it is applied (red).} 
  \label{fig:Distr_m34_1_Sel2}
\end{figure}

\clearpage

\begin{figure}[tb]
    \hspace*{-3cm}
       \includegraphics[width=1.4\linewidth]{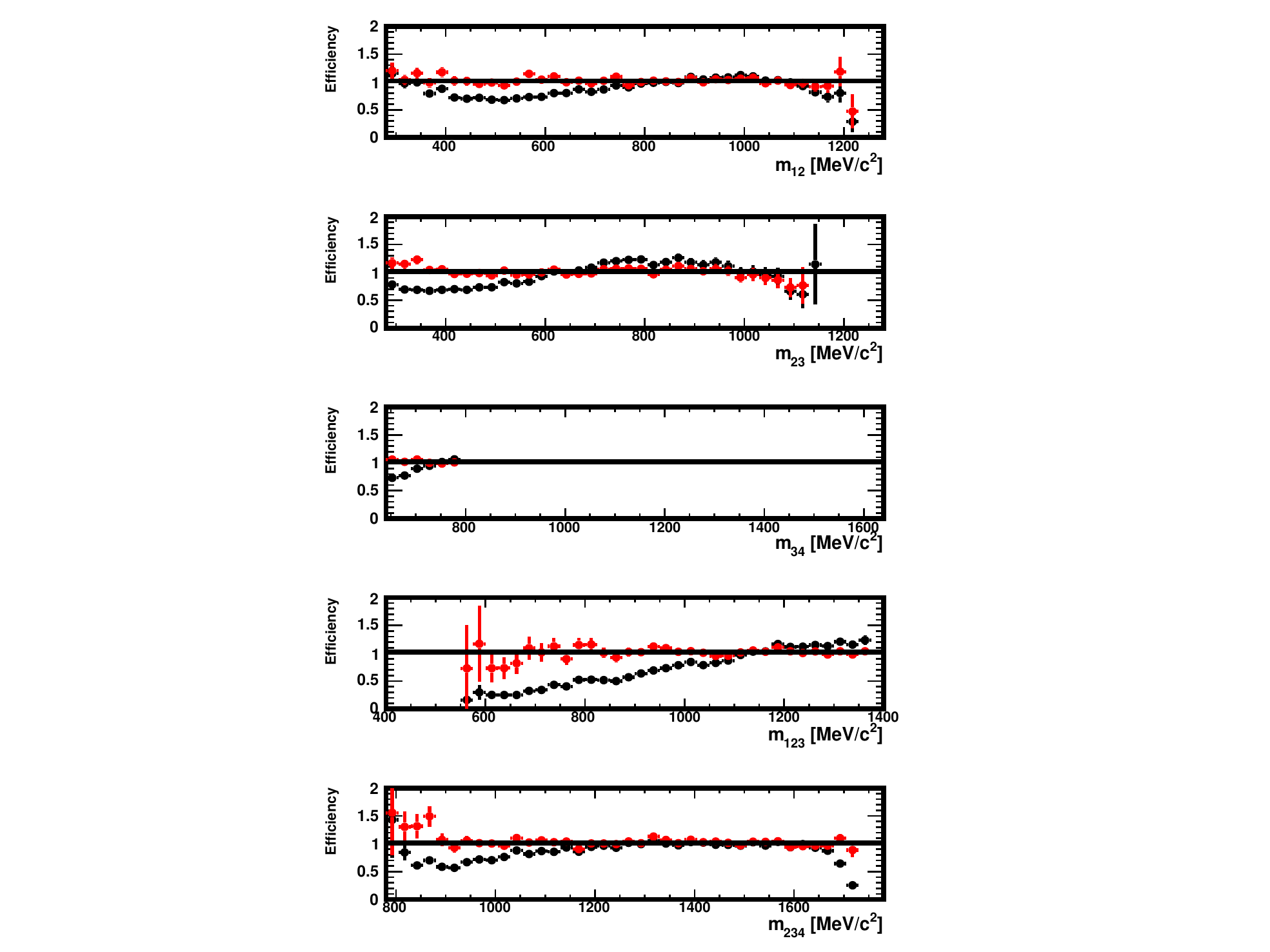}
    \vspace*{-0.5cm}
  \caption{Efficiency in $m_{12}$, $m_{23}$, $m_{34}$, $m_{123}$ and $m_{234}$ in the data generated for this study, 
in the region $0 < m_{34} < 780$ MeV$/c^2$ and with a tighter selection. Shown are the ratios of the distributions found in the distorted and 
original samples, with no correction (black) and for decays re-weighted using $\omega_i$ weights (red)
as explained in Sect.~\ref{subsec:ResultsK3pi}. The absolute normalisation is arbitrary when the correction is not applied and natural when it is applied (red).} 
  \label{fig:Ratio_m34_1_Sel2}
\end{figure}

\clearpage

\begin{figure}[tb]
    \hspace*{-3cm}
       \includegraphics[width=1.4\linewidth]{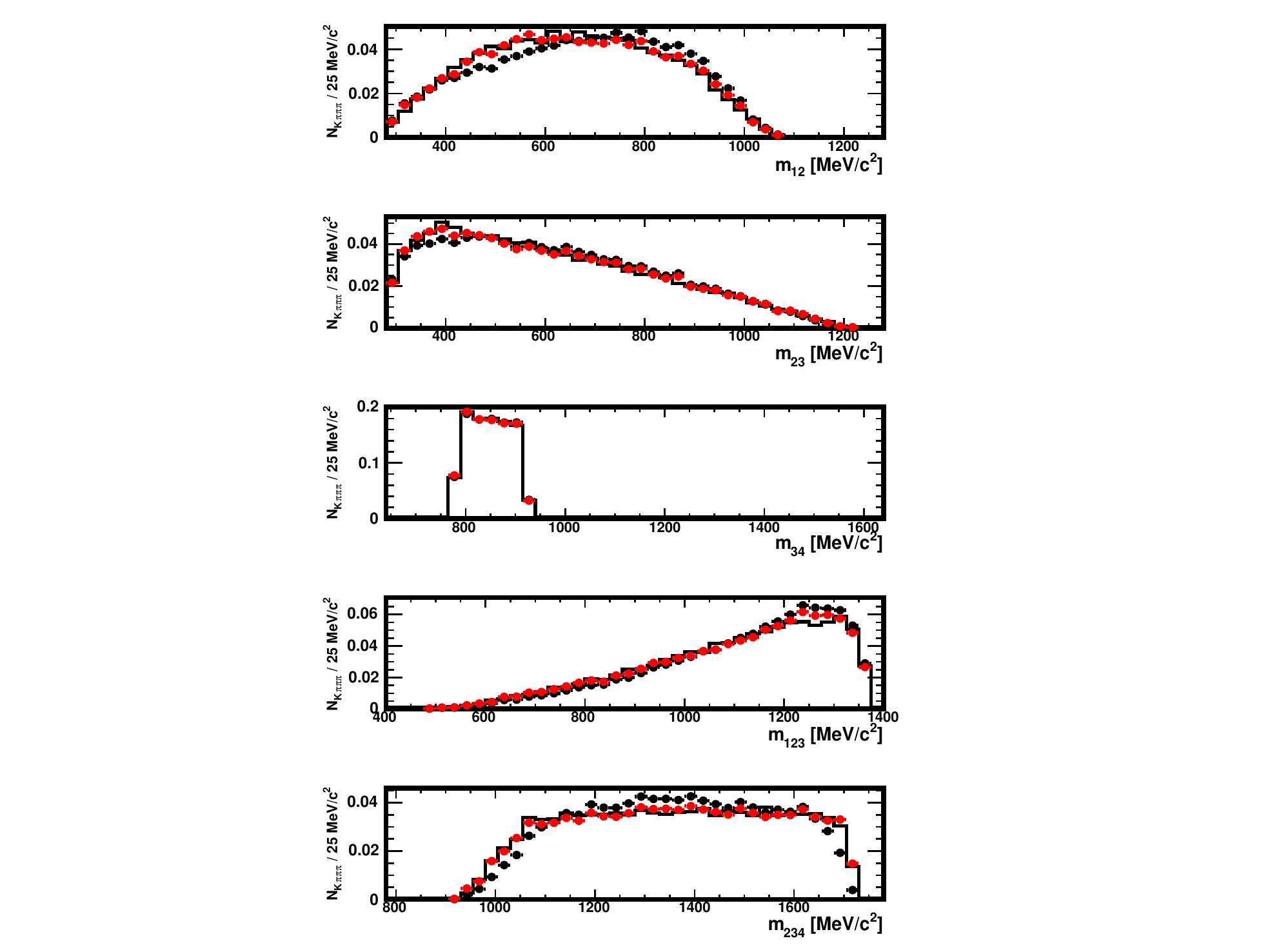}
    \vspace*{-0.5cm}
  \caption{Distributions of $m_{12}$, $m_{23}$, $m_{34}$, $m_{123}$ and $m_{234}$ in the original sample (histogram),
in the distorted one obtained with a tighter selection (full black circles) and in the distorted sample where the decays have been re-weighted using the $\omega_i$ weights (red), 
as explained in Sect.~\ref{subsec:ResultsK3pi}. The data used here are restricted to the region $780 < m_{34} < 920$ MeV$/c^2$.
The absolute normalisation is arbitrary when the correction is not applied and natural when it is applied (red).} 
  \label{fig:Distr_m34_2_Sel2}
\end{figure}

\clearpage

\begin{figure}[tb]
    \hspace*{-3cm}
       \includegraphics[width=1.4\linewidth]{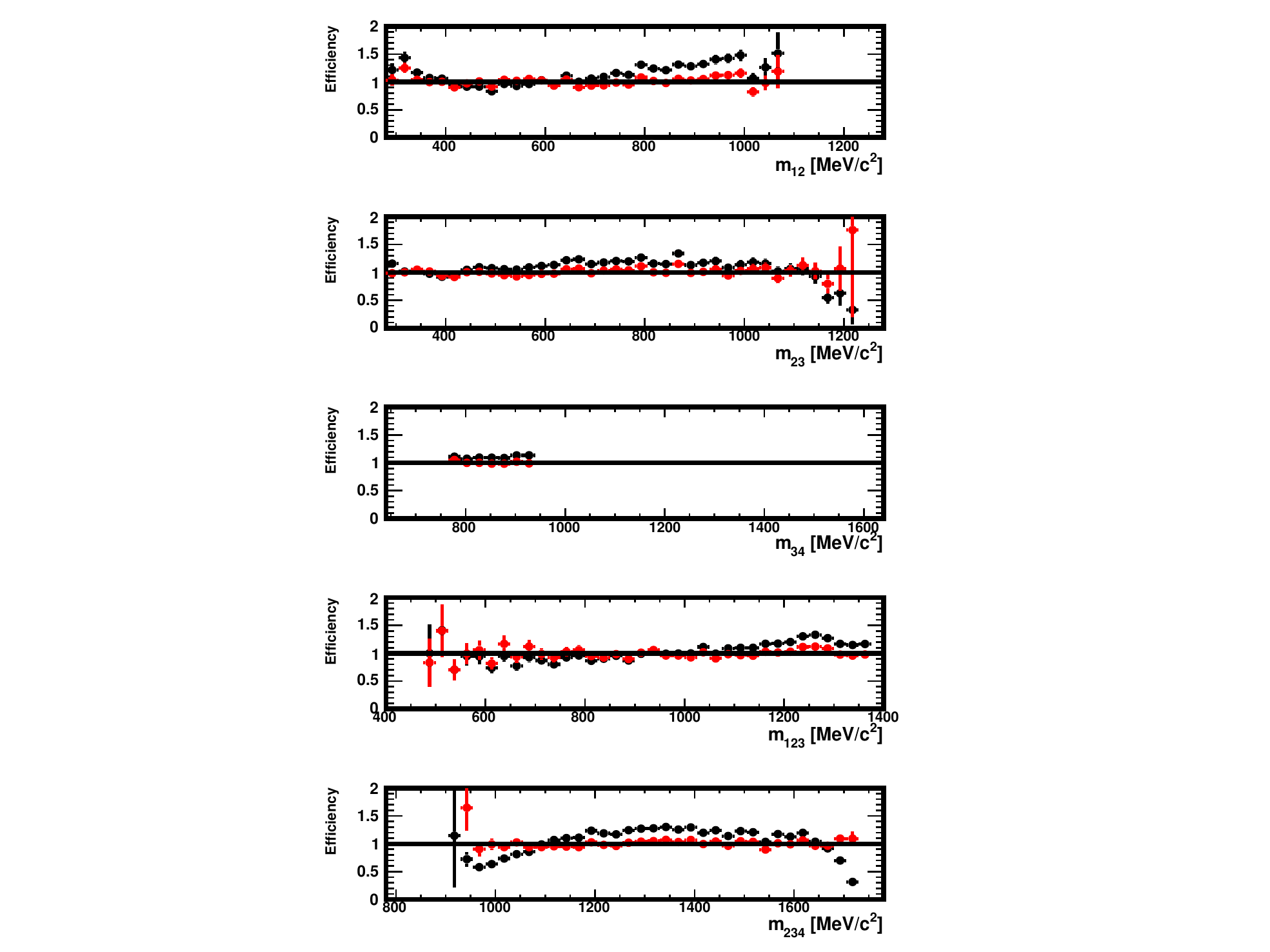}
    \vspace*{-0.5cm}
  \caption{Efficiency in $m_{12}$, $m_{23}$, $m_{34}$, $m_{123}$ and $m_{234}$ in the data generated for this study, 
in the region $ <780 m_{34} < 920$ MeV$/c^2$ and with a tighter selection. Shown are the ratios of the distributions found in the distorted and 
original samples, with no correction (black) and for decays re-weighted using $\omega_i$ weights (red)
as explained in Sect.~\ref{subsec:ResultsK3pi}. The absolute normalisation is arbitrary when the correction is not applied and natural when it is applied (red).} 
  \label{fig:Ratio_m34_2_Sel2}
\end{figure}

\clearpage

\begin{figure}[tb]
    \hspace*{-3cm}
       \includegraphics[width=1.4\linewidth]{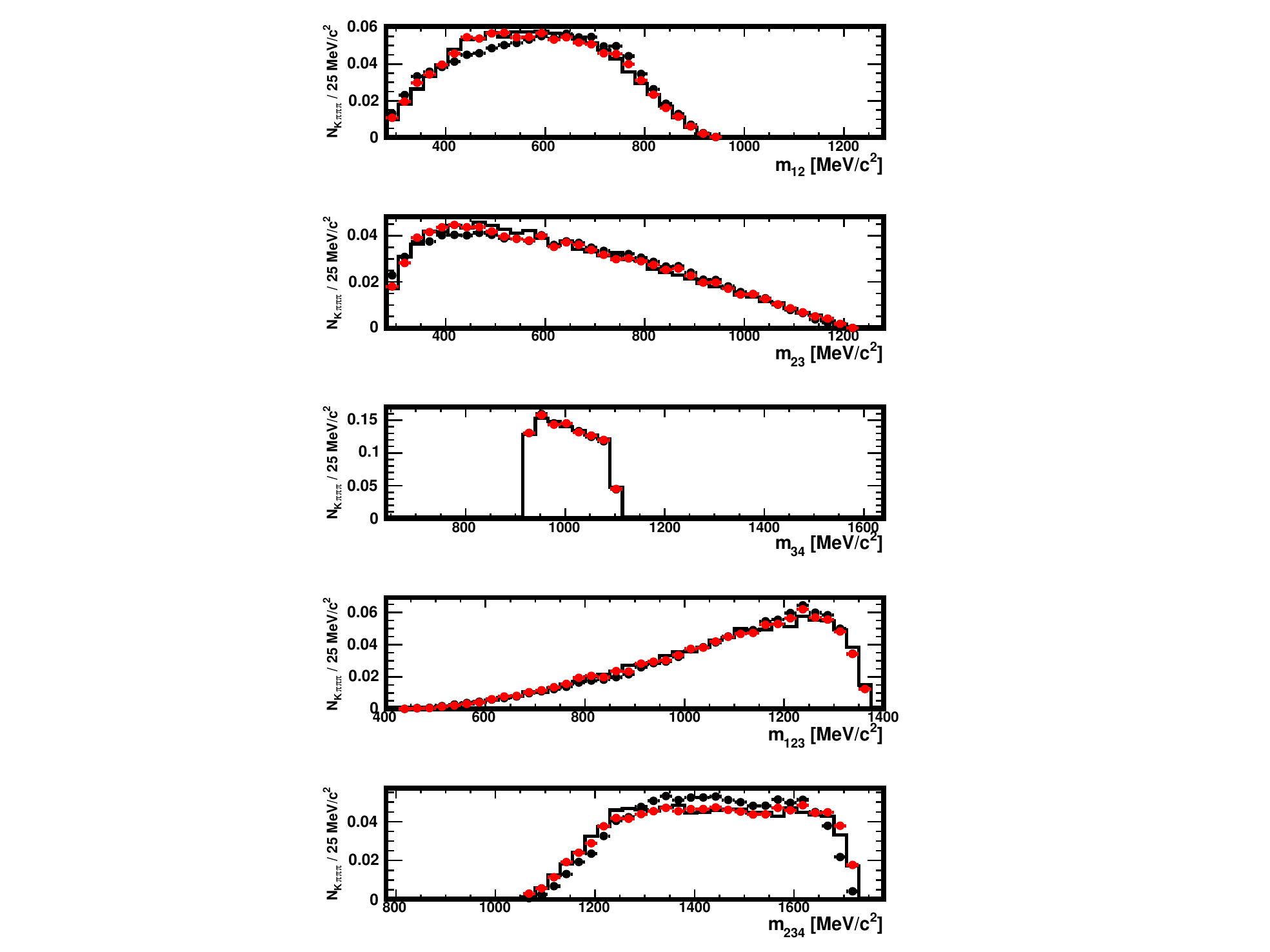}
    \vspace*{-0.5cm}
  \caption{Distributions of $m_{12}$, $m_{23}$, $m_{34}$, $m_{123}$ and $m_{234}$ in the original sample (histogram),
in the distorted one obtained with a tighter selection (full black circles) and in the distorted sample where the decays have been re-weighted using the $\omega_i$ weights (red), 
as explained in Sect.~\ref{subsec:ResultsK3pi}. The data used here are restricted to the region $920 < m_{34} < 1100$ MeV$/c^2$.
The absolute normalisation is arbitrary when the correction is not applied and natural when it is applied (red).} 
  \label{fig:Distr_m34_3_Sel2}
\end{figure}

\clearpage

\begin{figure}[tb]
    \hspace*{-3cm}
       \includegraphics[width=1.4\linewidth]{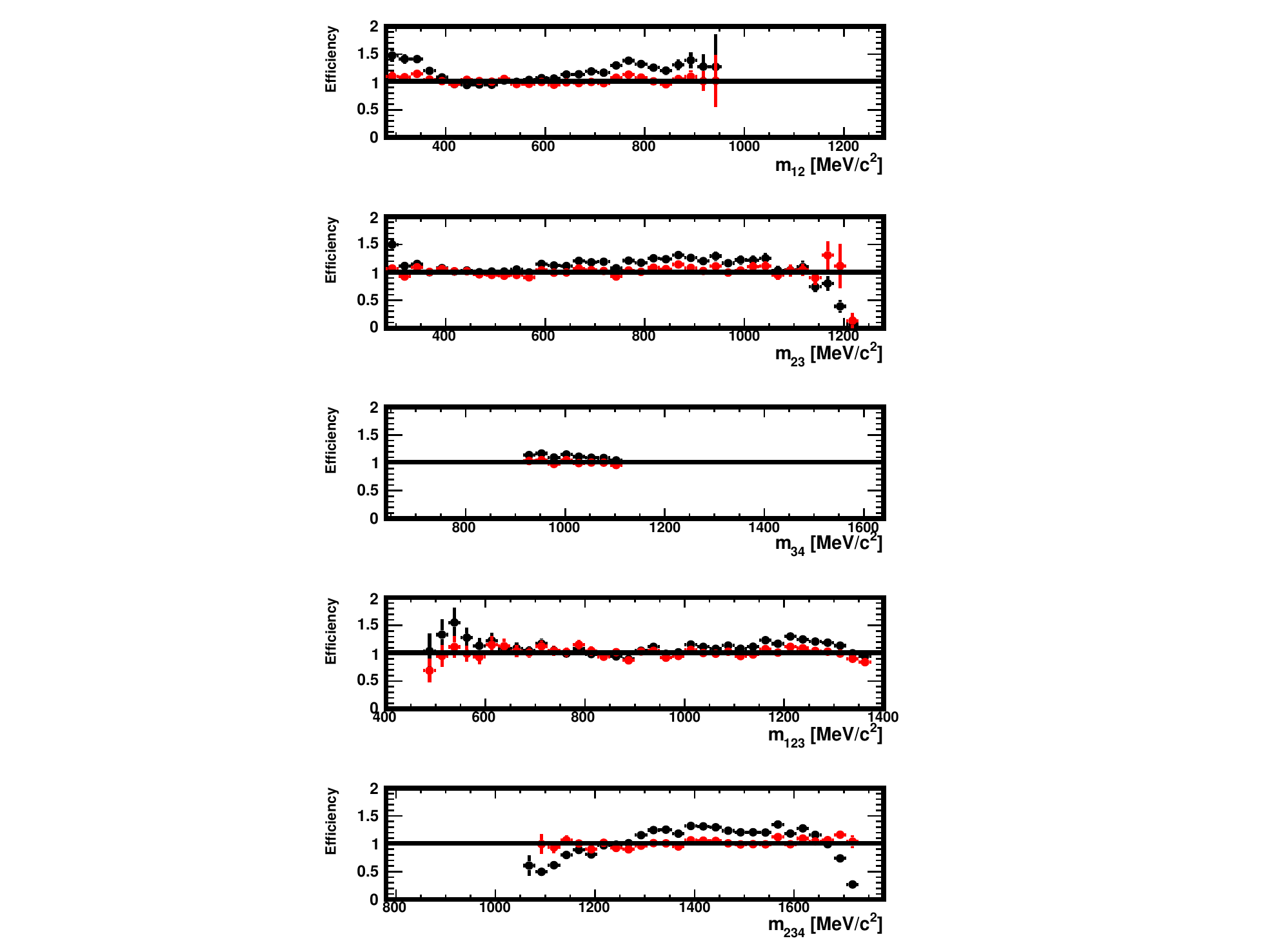}
    \vspace*{-0.5cm}
  \caption{Efficiency in $m_{12}$, $m_{23}$, $m_{34}$, $m_{123}$ and $m_{234}$ in the data generated for this study, 
in the region $920 < m_{34} < 1100$ MeV$/c^2$ and with a tighter selection. Shown are the ratios of the distributions found in the distorted and 
original samples, with no correction (black) and for decays re-weighted using $\omega_i$ weights (red)
as explained in Sect.~\ref{subsec:ResultsK3pi}. The absolute normalisation is arbitrary when the correction is not applied and natural when it is applied (red).} 
  \label{fig:Ratio_m34_3_Sel2}
\end{figure}

\clearpage

\begin{figure}[tb]
    \hspace*{-3cm}
       \includegraphics[width=1.4\linewidth]{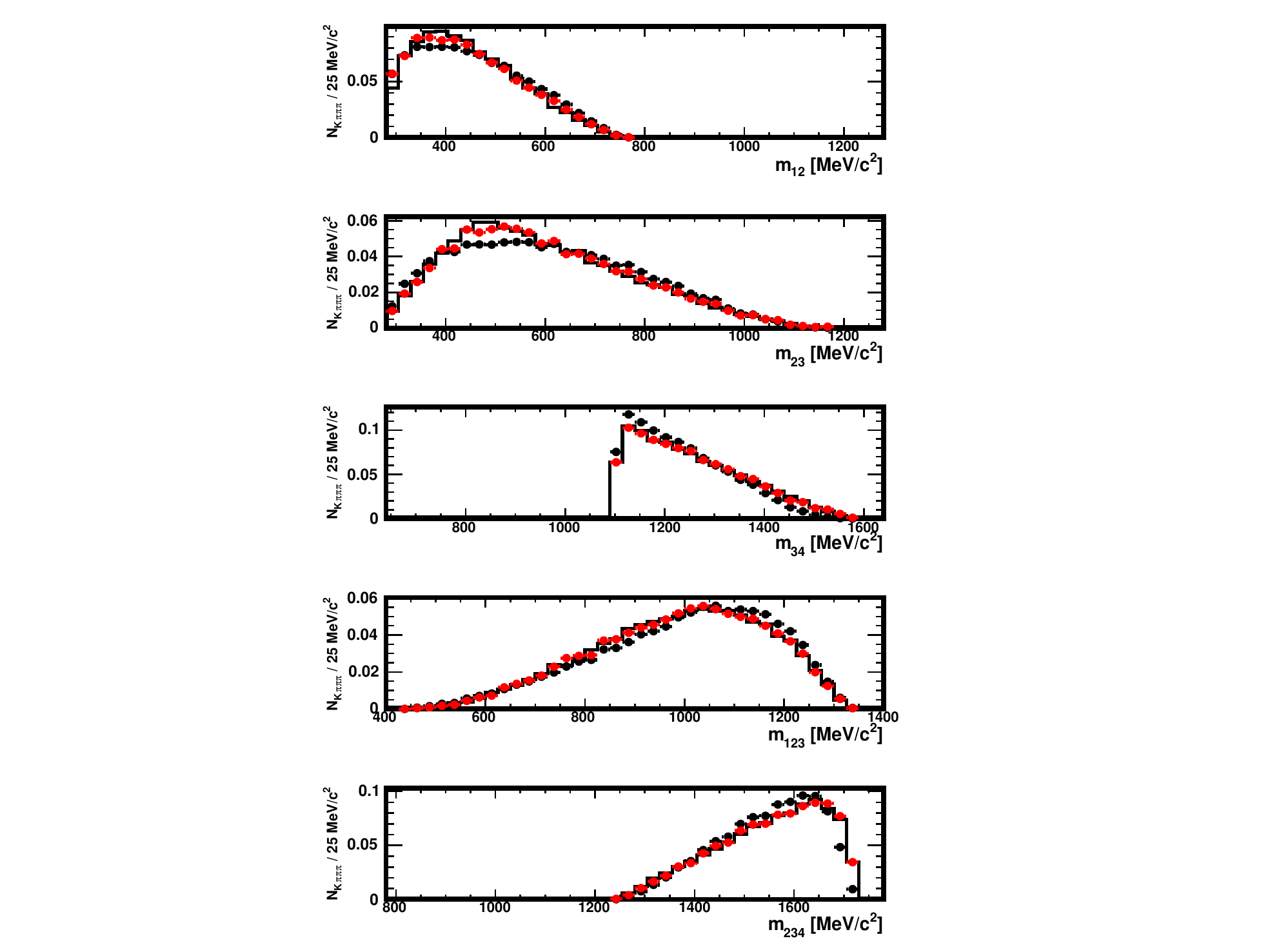}
    \vspace*{-0.5cm}
  \caption{Distributions of $m_{12}$, $m_{23}$, $m_{34}$, $m_{123}$ and $m_{234}$ in the original sample (histogram),
in the distorted one obtained with a tighter selection (full black circles) and in the distorted sample where the decays have been re-weighted using the $\omega_i$ weights (red), 
as explained in Sect.~\ref{subsec:ResultsK3pi}. The data used here are restricted to the region $1100 < m_{34}$ MeV$/c^2$.
The absolute normalisation is arbitrary when the correction is not applied and natural when it is applied (red).} 
  \label{fig:Distr_m34_4_Sel2}
\end{figure}

\clearpage

\begin{figure}[tb]
    \hspace*{-3cm}
       \includegraphics[width=1.4\linewidth]{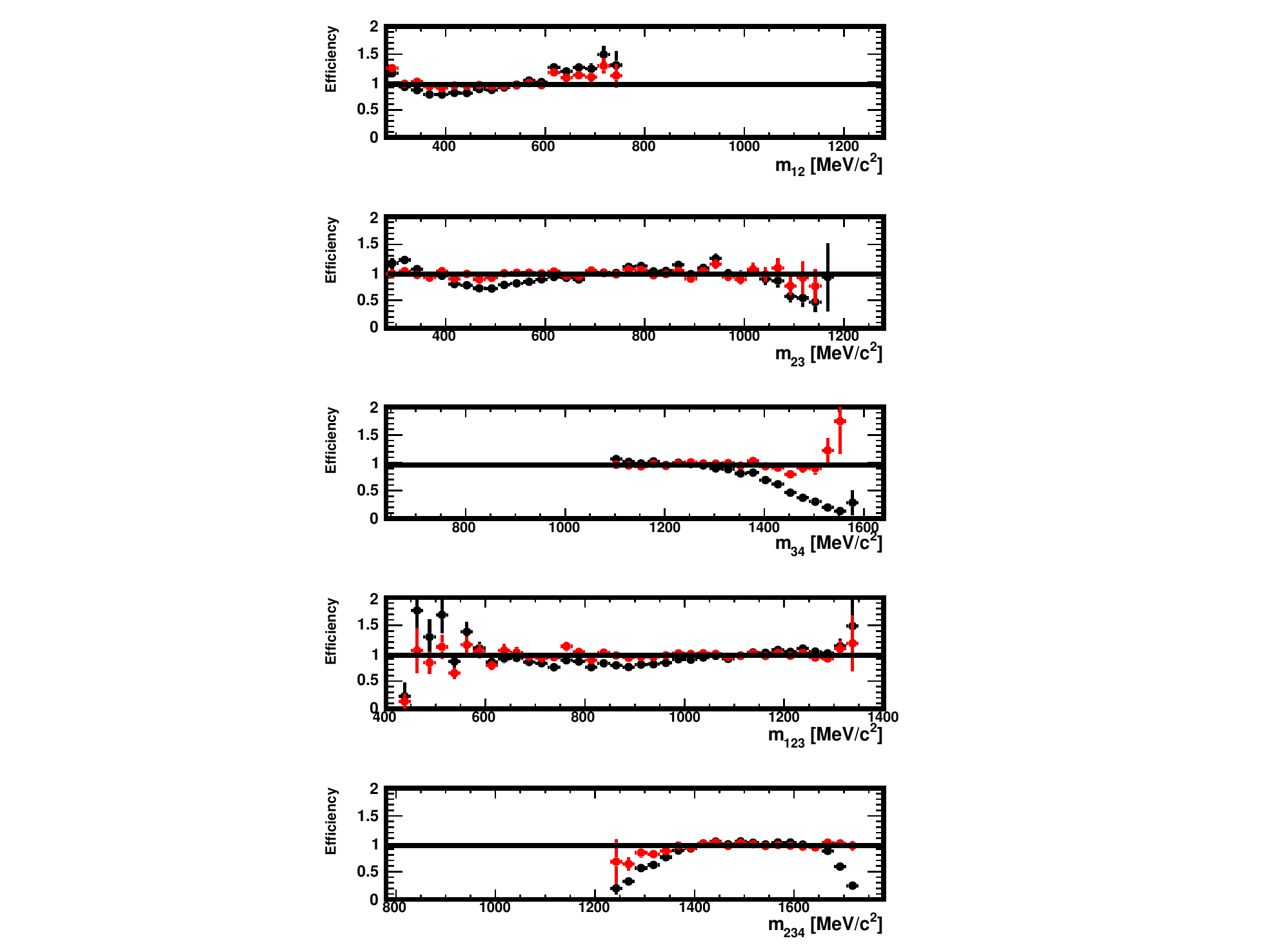}
    \vspace*{-0.5cm}
  \caption{Efficiency in $m_{12}$, $m_{23}$, $m_{34}$, $m_{123}$ and $m_{234}$ in the data generated for this study, 
in the region $1100 < m_{34}$ MeV$/c^2$ and with a tighter selection. Shown are the ratios of the distributions found in the distorted and 
original samples, with no correction (black) and for decays re-weighted using $\omega_i$ weights (red)
as explained in Sect.~\ref{subsec:ResultsK3pi}. The absolute normalisation is arbitrary when the correction is not applied and natural when it is applied (red).} 
  \label{fig:Ratio_m34_4_Sel2}
\end{figure}

\clearpage

\begin{figure}[tb]
    \hspace*{-3cm}
       \includegraphics[width=1.4\linewidth]{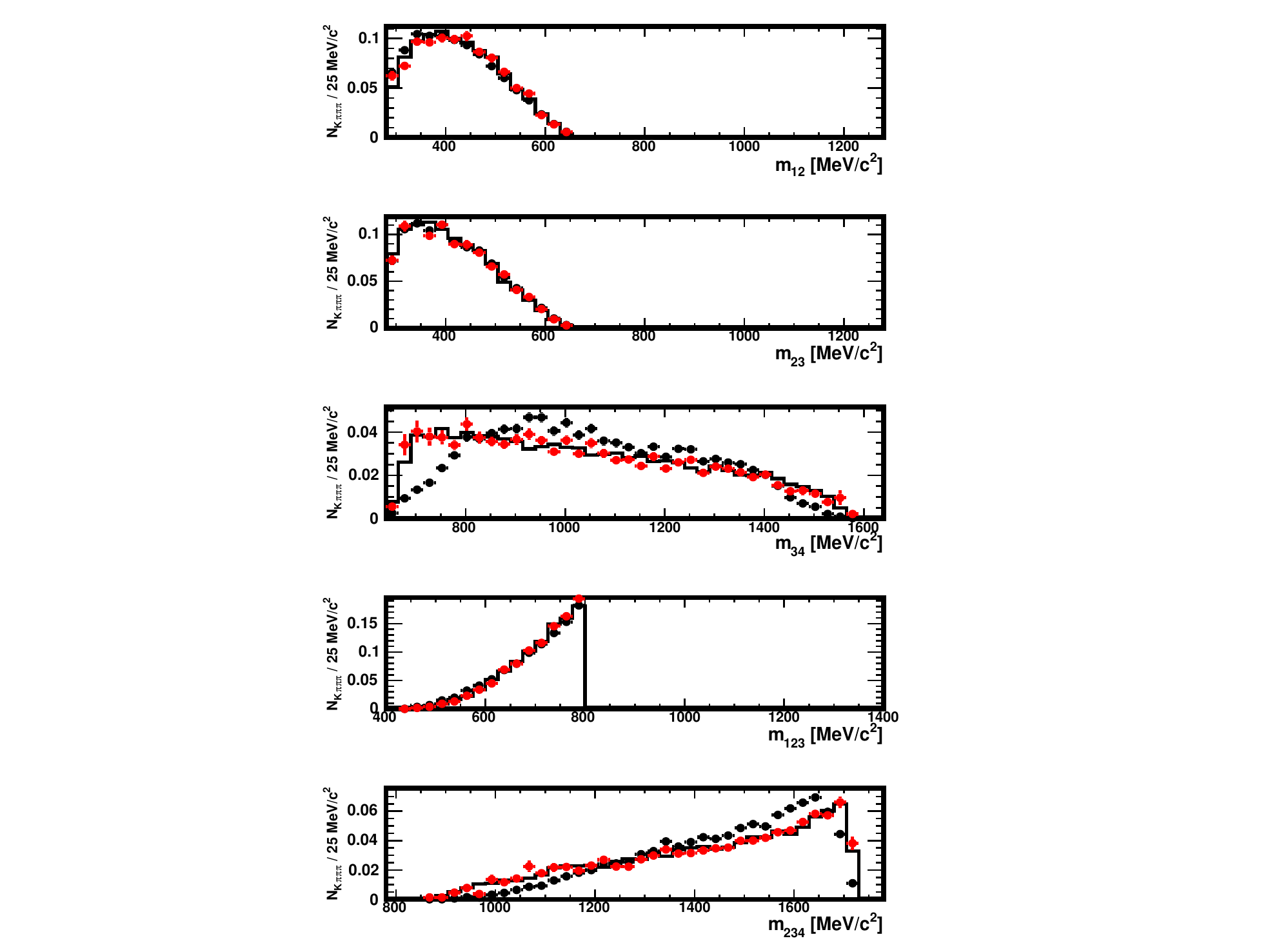}
    \vspace*{-0.5cm}
  \caption{Distributions of $m_{12}$, $m_{23}$, $m_{34}$, $m_{123}$ and $m_{234}$ in the original sample (histogram),
in the distorted one obtained with a tighter selection (full black circles) and in the distorted sample where the decays have been re-weighted using the $\omega_i$ weights (red), 
as explained in Sect.~\ref{subsec:ResultsK3pi}. The data used here are restricted to the region $ 0 < m_{123} < 800$ MeV$/c^2$.
The absolute normalisation is arbitrary when the correction is not applied and natural when it is applied (red).} 
  \label{fig:Distr_m123_1_Sel2}
\end{figure}

\clearpage

\begin{figure}[tb]
    \hspace*{-3cm}
       \includegraphics[width=1.4\linewidth]{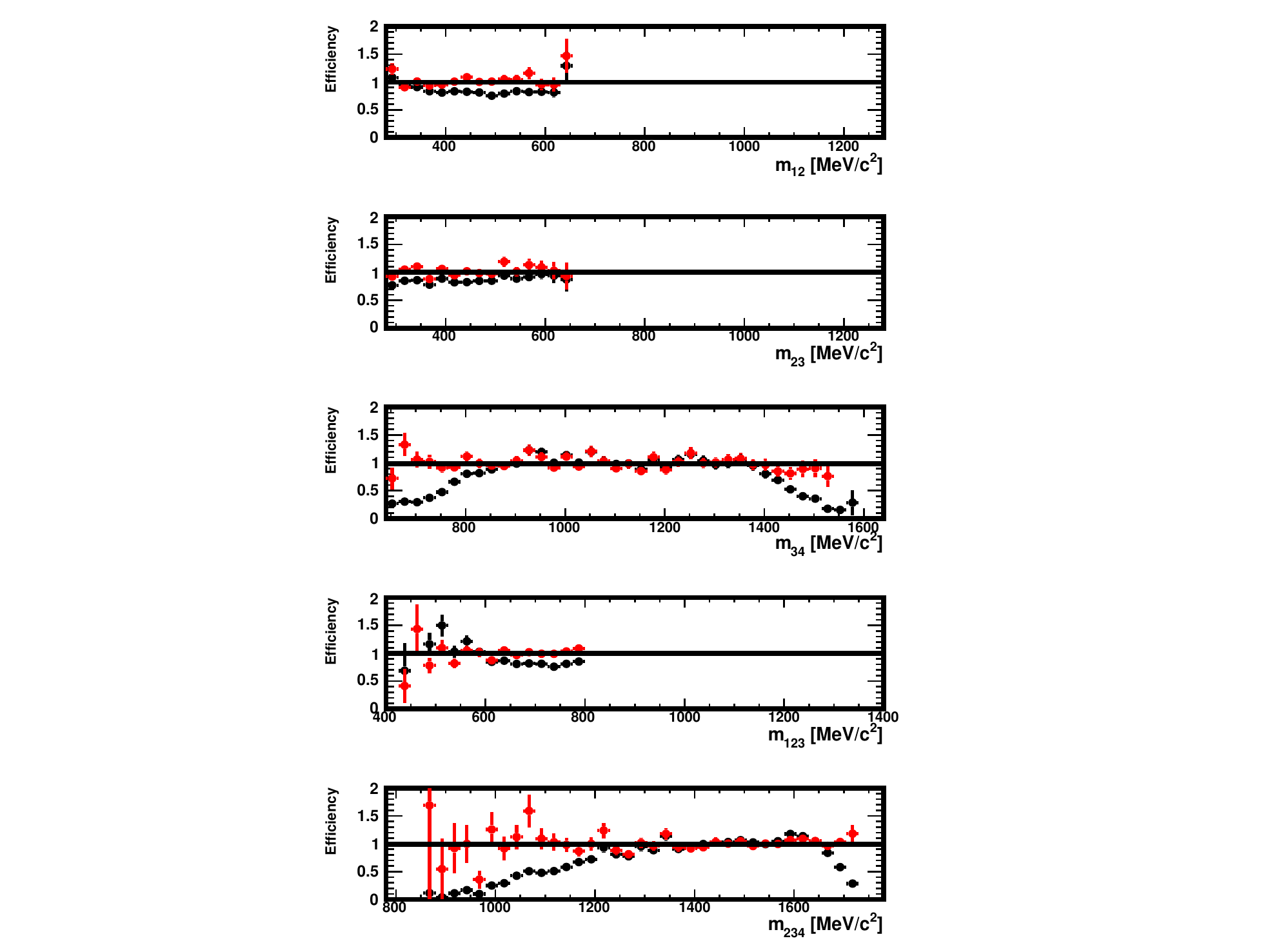}
    \vspace*{-0.5cm}
  \caption{Efficiency in $m_{12}$, $m_{23}$, $m_{34}$, $m_{123}$ and $m_{234}$ in the data generated for this study, 
in the region $0 < m_{123} < 800$ MeV$/c^2$ and with a tighter selection. Shown are the ratios of the distributions found in the distorted and 
original samples, with no correction (black) and for decays re-weighted using $\omega_i$ weights (red)
as explained in Sect.~\ref{subsec:ResultsK3pi}. The absolute normalisation is arbitrary when the correction is not applied and natural when it is applied (red).} 
  \label{fig:Ratio_m123_1_Sel2}
\end{figure}

\clearpage

\begin{figure}[tb]
    \hspace*{-3cm}
       \includegraphics[width=1.4\linewidth]{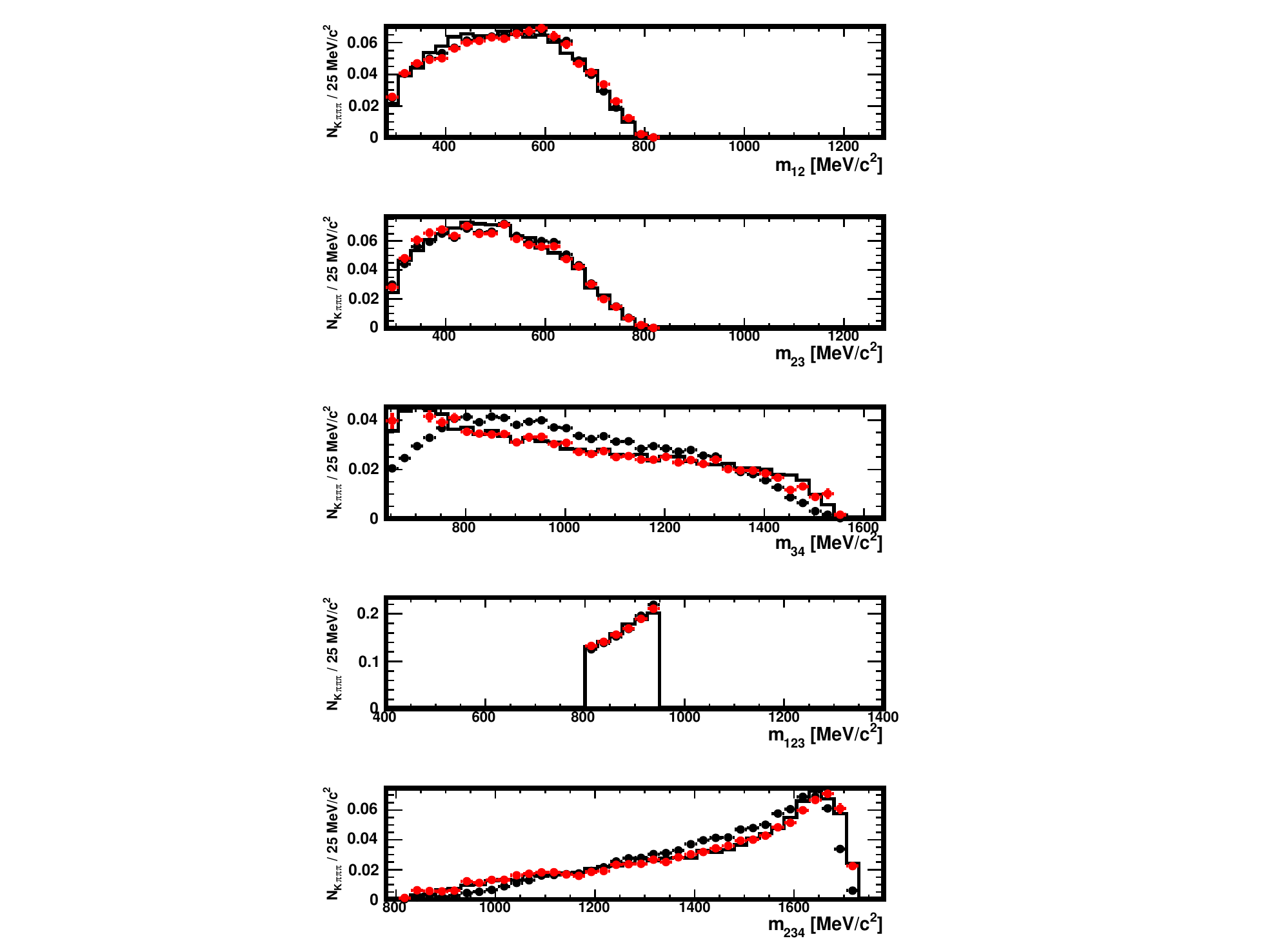}
    \vspace*{-0.5cm}
  \caption{Distributions of $m_{12}$, $m_{23}$, $m_{123}$, $m_{123}$ and $m_{234}$ in the original sample (histogram),
in the distorted one obtained with a tighter selection (full black circles) and in the distorted sample where the decays have been re-weighted using the $\omega_i$ weights (red), 
as explained in Sect.~\ref{subsec:ResultsK3pi}. The data used here are restricted to the region $800 < m_{123} < 950$ MeV$/c^2$.
The absolute normalisation is arbitrary when the correction is not applied and natural when it is applied (red).} 
  \label{fig:Distr_m123_2_Sel2}
\end{figure}

\clearpage

\begin{figure}[tb]
    \hspace*{-3cm}
       \includegraphics[width=1.4\linewidth]{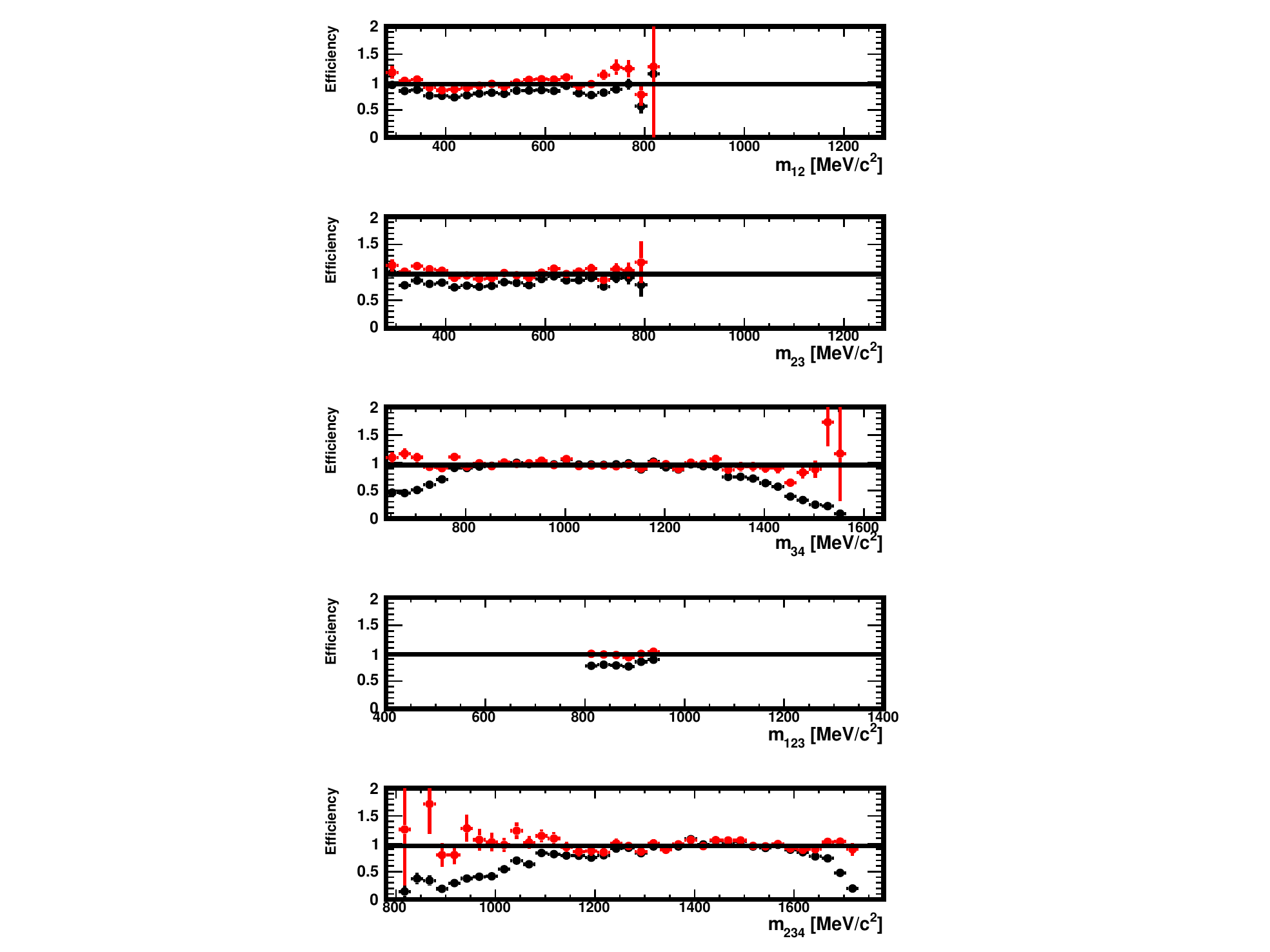}
    \vspace*{-0.5cm}
  \caption{Efficiency in $m_{12}$, $m_{23}$, $m_{34}$, $m_{123}$ and $m_{234}$ in the data generated for this study, 
in the region $ <800 m_{123} < 950$ MeV$/c^2$ and with a tighter selection. Shown are the ratios of the distributions found in the distorted and 
original samples, with no correction (black) and for decays re-weighted using $\omega_i$ weights (red)
as explained in Sect.~\ref{subsec:ResultsK3pi}. The absolute normalisation is arbitrary when the correction is not applied and natural when it is applied (red).} 
  \label{fig:Ratio_m123_2_Sel2}
\end{figure}

\clearpage

\begin{figure}[tb]
    \hspace*{-3cm}
       \includegraphics[width=1.4\linewidth]{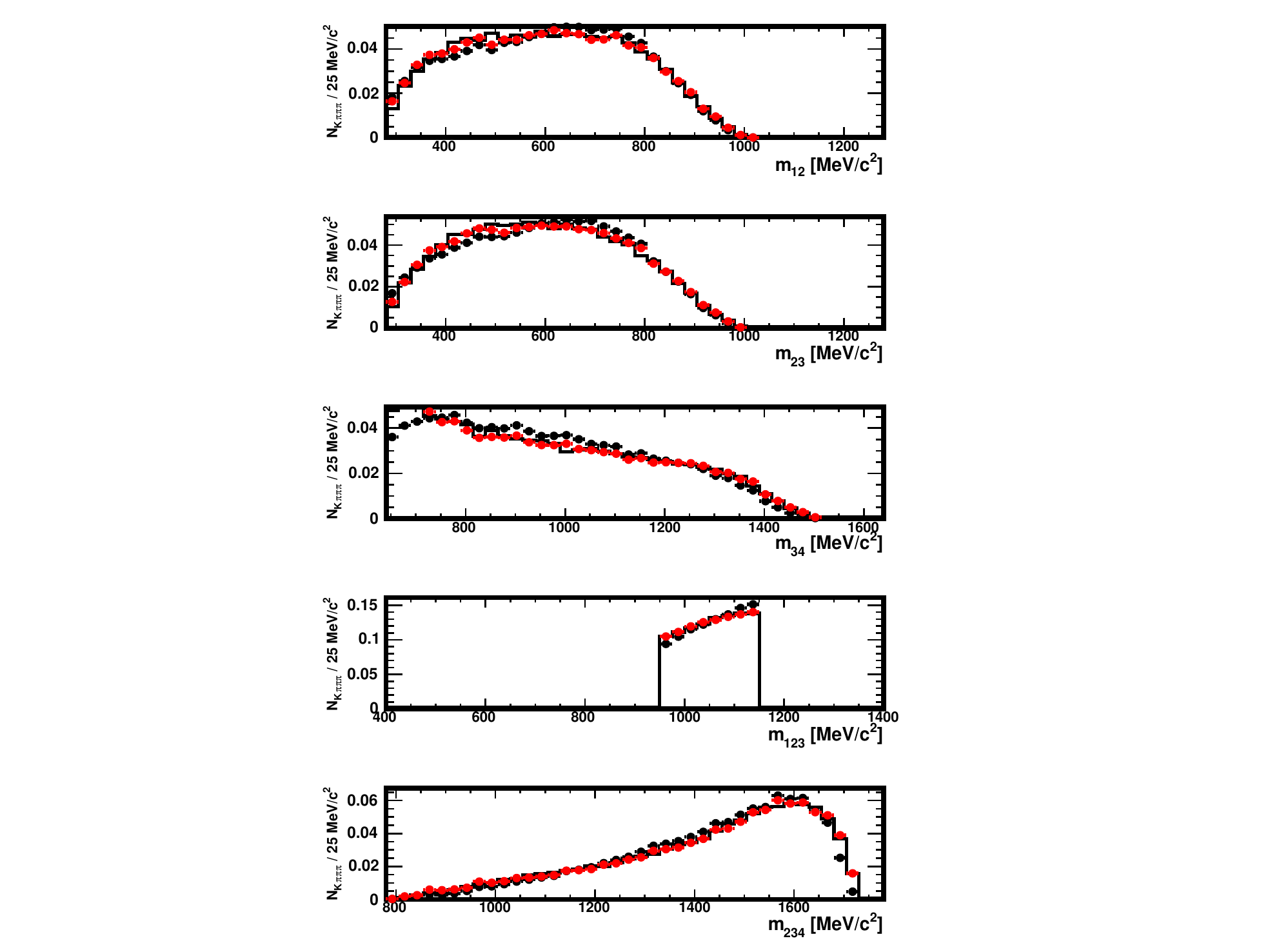}
    \vspace*{-0.5cm}
  \caption{Distributions of $m_{12}$, $m_{23}$, $m_{34}$, $m_{123}$ and $m_{234}$ in the original sample (histogram),
in the distorted one obtained with a tighter selection (full black circles) and in the distorted sample where the decays have been re-weighted using the $\omega_i$ weights (red), 
as explained in Sect.~\ref{subsec:ResultsK3pi}. The data used here are restricted to the region $950 < m_{123} < 1150$ MeV$/c^2$.
The absolute normalisation is arbitrary when the correction is not applied and natural when it is applied (red).} 
  \label{fig:Distr_m123_3_Sel2}
\end{figure}

\clearpage

\begin{figure}[tb]
    \hspace*{-3cm}
       \includegraphics[width=1.4\linewidth]{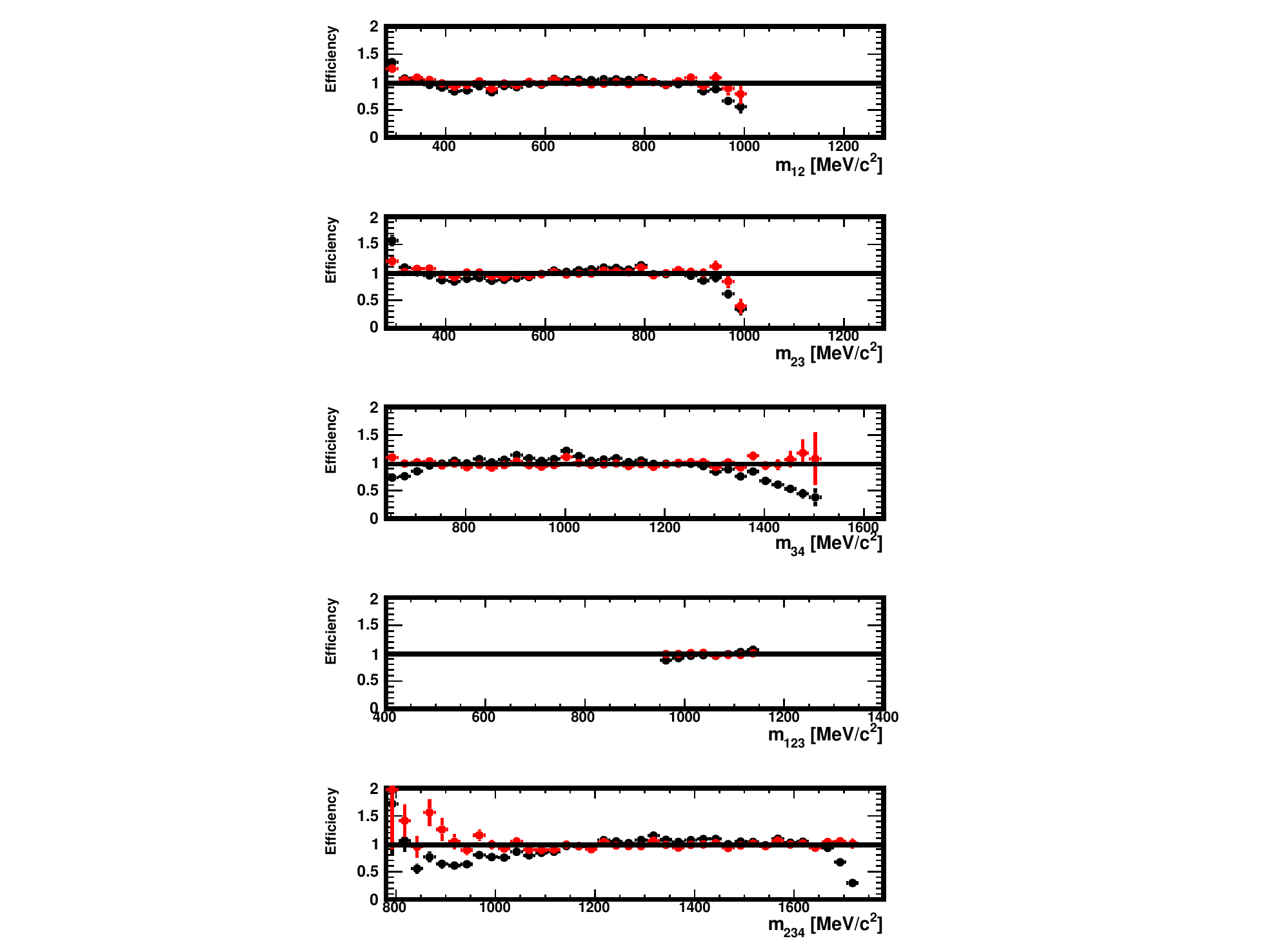}
    \vspace*{-0.5cm}
  \caption{Efficiency in $m_{12}$, $m_{23}$, $m_{34}$, $m_{123}$ and $m_{234}$ in the data generated for this study, 
in the region $950 < m_{123} < 1150$ MeV$/c^2$ and with a tighter selection. Shown are the ratios of the distributions found in the distorted and 
original samples, with no correction (black) and for decays re-weighted using $\omega_i$ weights (red)
as explained in Sect.~\ref{subsec:ResultsK3pi}. The absolute normalisation is arbitrary when the correction is not applied and natural when it is applied (red).} 
  \label{fig:Ratio_m123_3_Sel2}
\end{figure}

\clearpage

\begin{figure}[tb]
    \hspace*{-3cm}
       \includegraphics[width=1.4\linewidth]{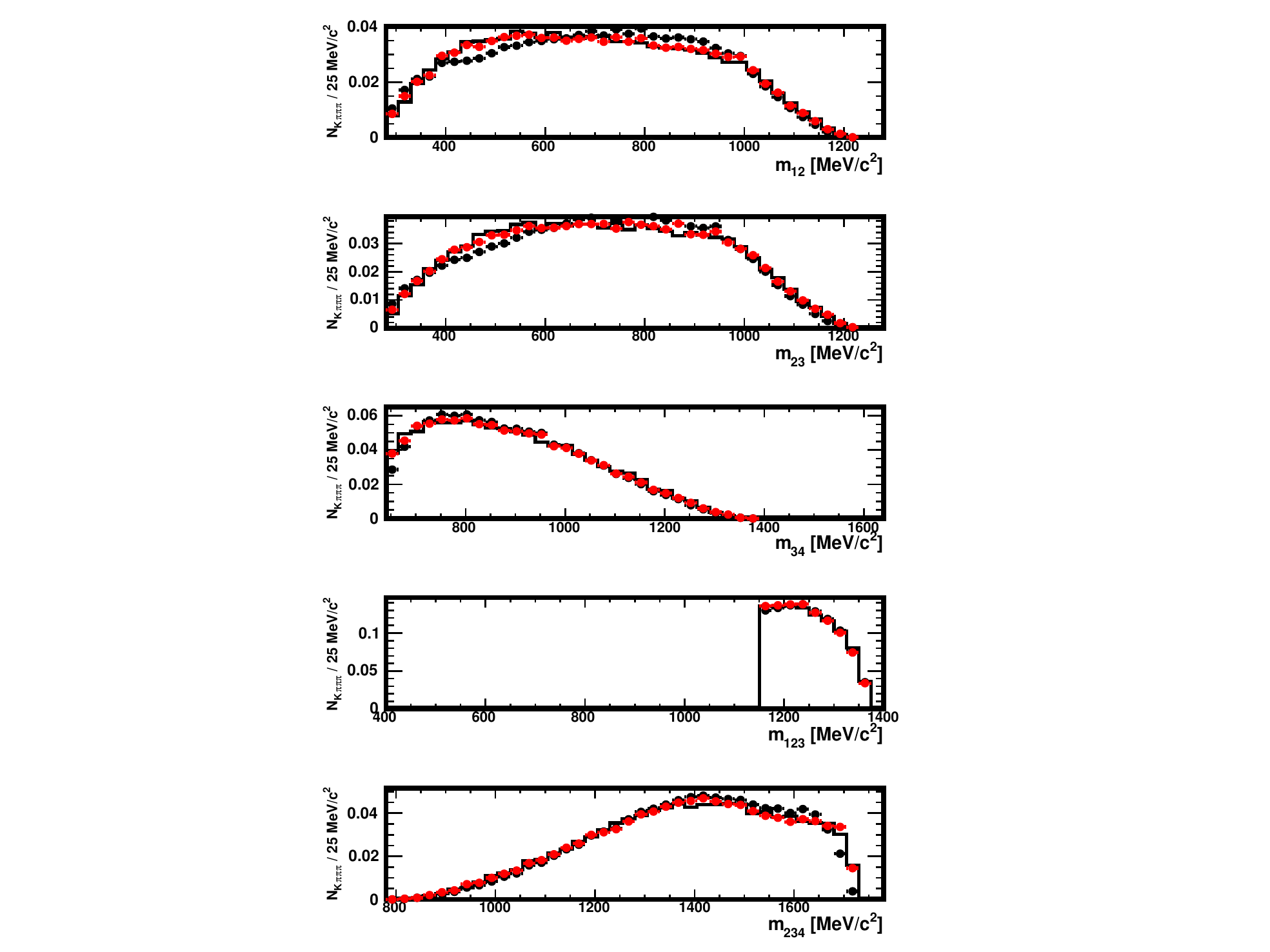}
    \vspace*{-0.5cm}
  \caption{Distributions of $m_{12}$, $m_{23}$, $m_{34}$, $m_{123}$ and $m_{234}$ in the original sample (histogram),
in the distorted one obtained with a tighter selection (full black circles) and in the distorted sample where the decays have been re-weighted using the $\omega_i$ weights (red), 
as explained in Sect.~\ref{subsec:ResultsK3pi}. The data used here are restricted to the region $1150 < m_{123}$ MeV$/c^2$.
The absolute normalisation is arbitrary when the correction is not applied and natural when it is applied (red).} 
  \label{fig:Distr_m123_4_Sel2}
\end{figure}

\clearpage

\begin{figure}[tb]
    \hspace*{-3cm}
       \includegraphics[width=1.4\linewidth]{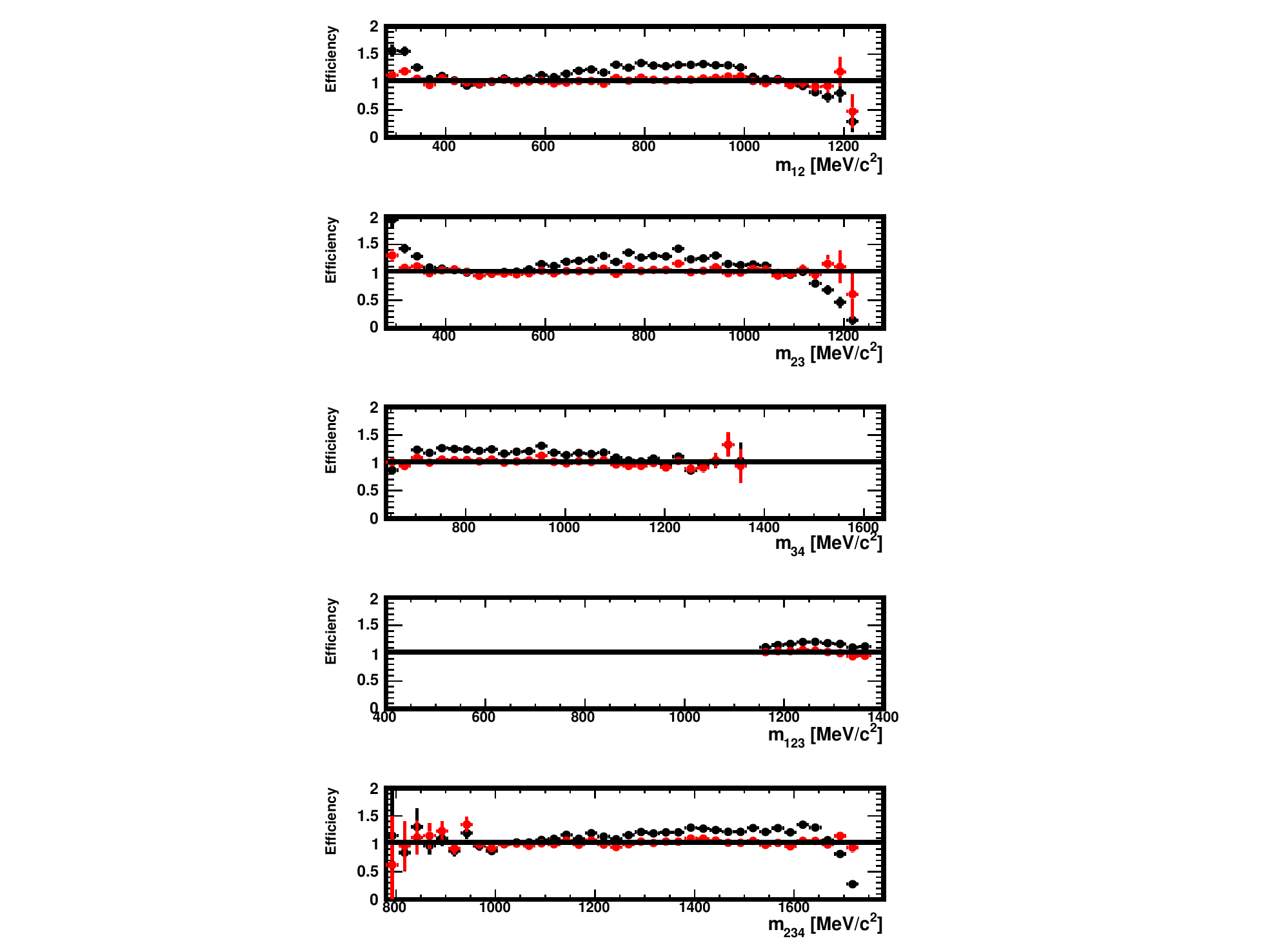}
    \vspace*{-0.5cm}
  \caption{Efficiency in $m_{12}$, $m_{23}$, $m_{34}$, $m_{123}$ and $m_{234}$ in the data generated for this study, 
in the region $1150 < m_{123}$ MeV$/c^2$ and with a tighter selection. Shown are the ratios of the distributions found in the distorted and 
original samples, with no correction (black) and for decays re-weighted using $\omega_i$ weights (red)
as explained in Sect.~\ref{subsec:ResultsK3pi}. The absolute normalisation is arbitrary when the correction is not applied and natural when it is applied (red).} 
  \label{fig:Ratio_m123_4_Sel2}
\end{figure}

\clearpage

\begin{figure}[tb]
    \hspace*{-3cm}
       \includegraphics[width=1.4\linewidth]{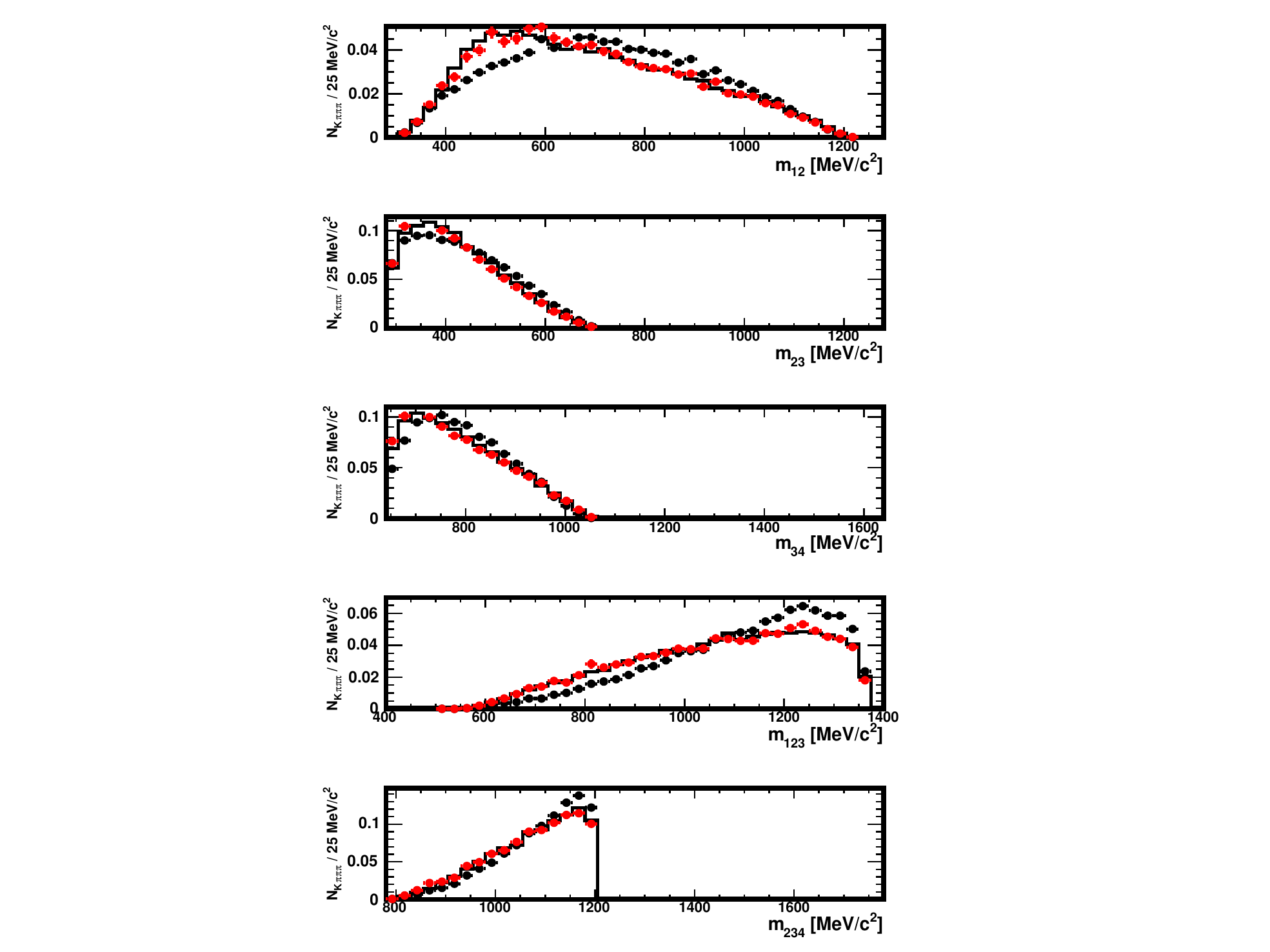}
    \vspace*{-0.5cm}
  \caption{Distributions of $m_{12}$, $m_{23}$, $m_{34}$, $m_{123}$ and $m_{234}$ in the original sample (histogram),
in the distorted one obtained with a tighter selection (full black circles) and in the distorted sample where the decays have been re-weighted using the $\omega_i$ weights (red), 
as explained in Sect.~\ref{subsec:ResultsK3pi}. The data used here are restricted to the region $ 0 < m_{234} < 800$ MeV$/c^2$.
The absolute normalisation is arbitrary when the correction is not applied and natural when it is applied (red).} 
  \label{fig:Distr_m234_1_Sel2}
\end{figure}

\clearpage

\begin{figure}[tb]
    \hspace*{-3cm}
       \includegraphics[width=1.4\linewidth]{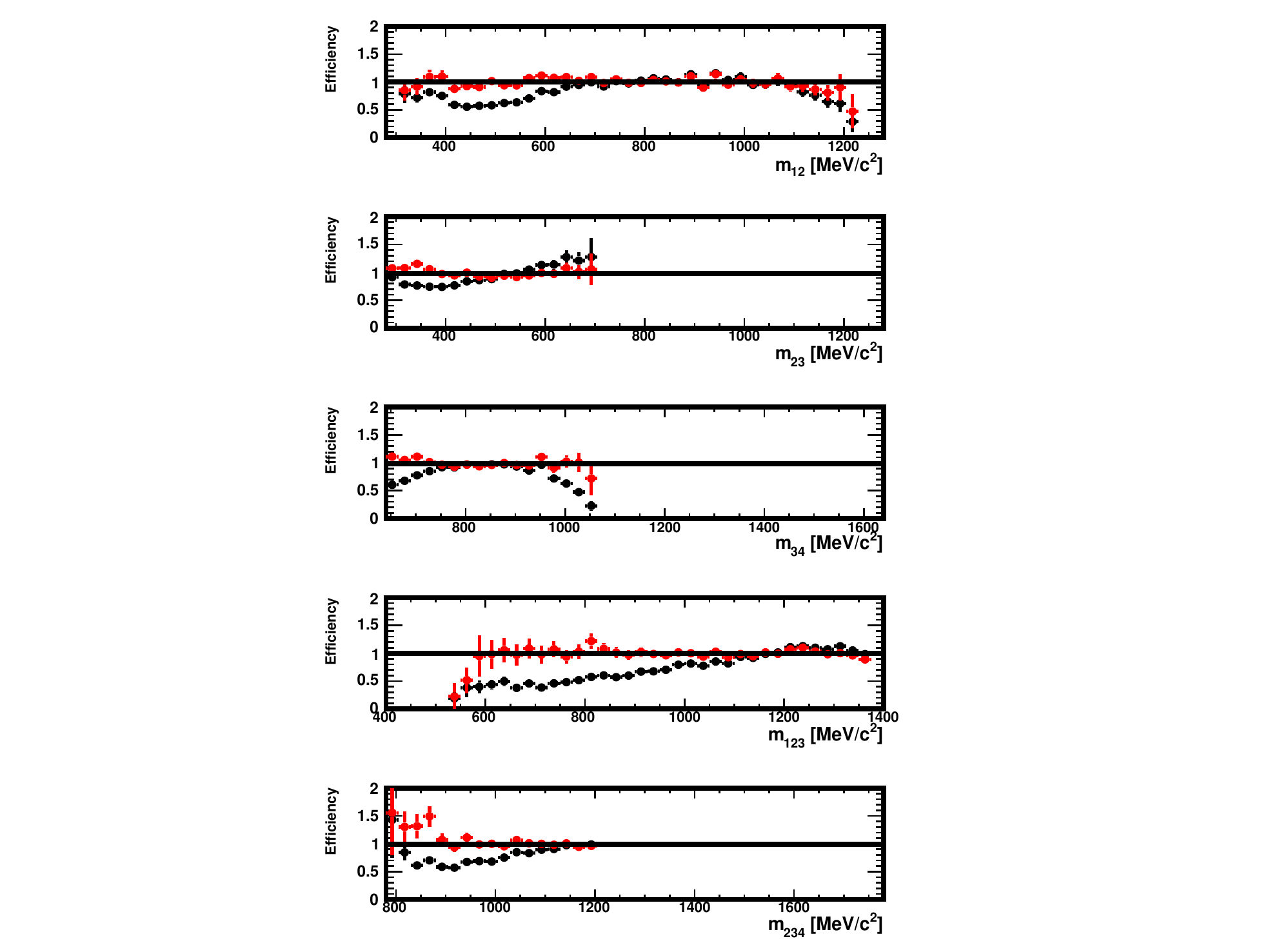}
    \vspace*{-0.5cm}
  \caption{Efficiency in $m_{12}$, $m_{23}$, $m_{34}$, $m_{123}$ and $m_{234}$ in the data generated for this study, 
in the region $0 < m_{234} < 800$ MeV$/c^2$ and with a tighter selection. Shown are the ratios of the distributions found in the distorted and 
original samples, with no correction (black) and for decays re-weighted using $\omega_i$ weights (red)
as explained in Sect.~\ref{subsec:ResultsK3pi}. The absolute normalisation is arbitrary when the correction is not applied and natural when it is applied (red).} 
  \label{fig:Ratio_m234_1_Sel2} 
\end{figure}

\clearpage

\begin{figure}[tb]
    \hspace*{-3cm}
       \includegraphics[width=1.4\linewidth]{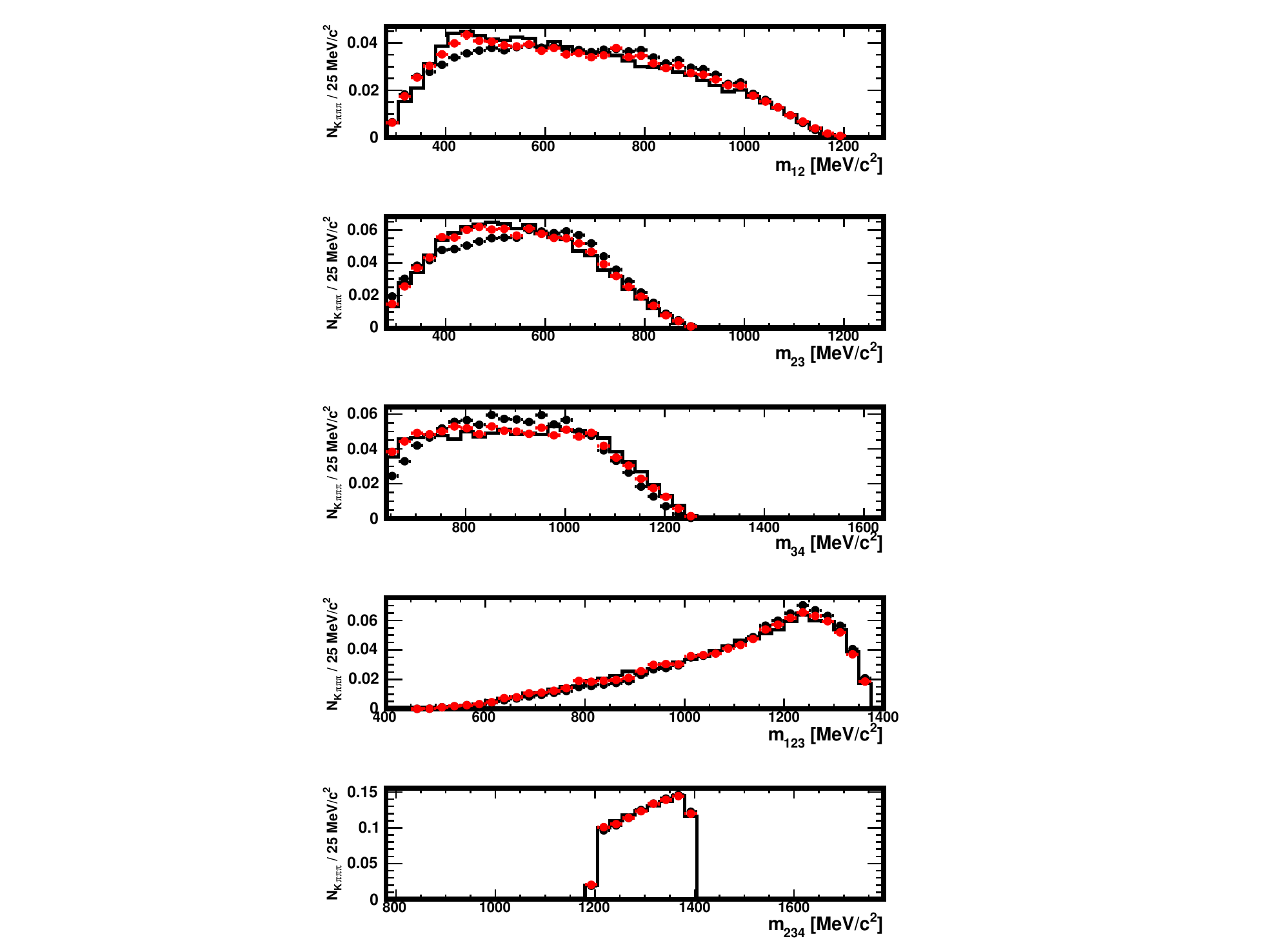}
    \vspace*{-0.5cm}
  \caption{Distributions of $m_{12}$, $m_{23}$, $m_{123}$, $m_{123}$ and $m_{234}$ in the original sample (histogram),
in the distorted one obtained with a tighter selection (full black circles) and in the distorted sample where the decays have been re-weighted using the $\omega_i$ weights (red), 
as explained in Sect.~\ref{subsec:ResultsK3pi}. The data used here are restricted to the region $800 < m_{234} < 950$ MeV$/c^2$.
The absolute normalisation is arbitrary when the correction is not applied and natural when it is applied (red).} 
  \label{fig:Distr_m234_2_Sel2}
\end{figure}

\clearpage

\begin{figure}[tb]
    \hspace*{-3cm}
       \includegraphics[width=1.4\linewidth]{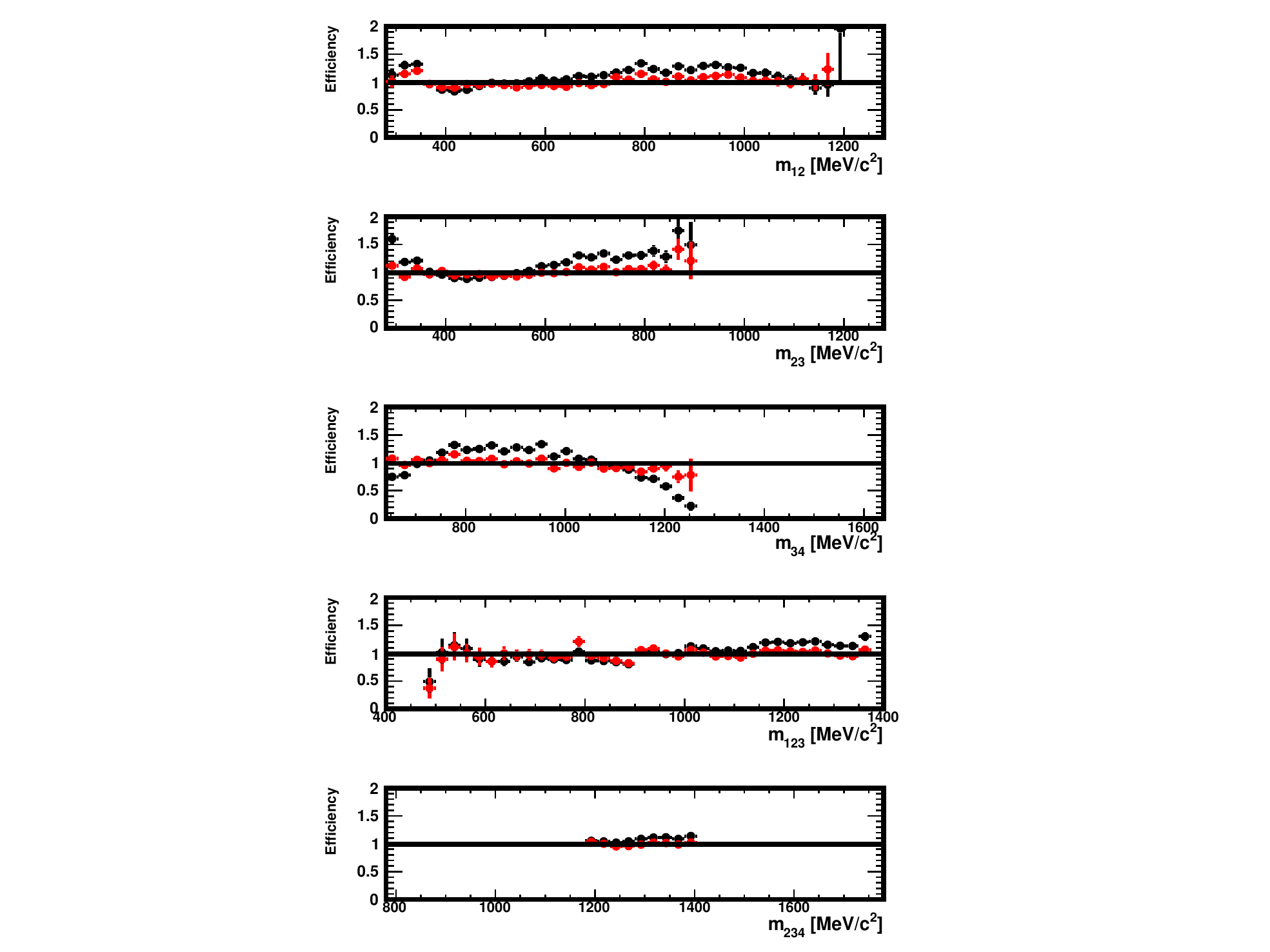}
    \vspace*{-0.5cm}
  \caption{Efficiency in $m_{12}$, $m_{23}$, $m_{34}$, $m_{123}$ and $m_{234}$ in the data generated for this study, 
in the region $ <800 m_{234} < 950$ MeV$/c^2$ and with a tighter selection. Shown are the ratios of the distributions found in the distorted and 
original samples, with no correction (black) and for decays re-weighted using $\omega_i$ weights (red)
as explained in Sect.~\ref{subsec:ResultsK3pi}. The absolute normalisation is arbitrary when the correction is not applied and natural when it is applied (red).} 
  \label{fig:Ratio_m234_2_Sel2} 
\end{figure}

\clearpage

\begin{figure}[tb]
    \hspace*{-3cm}
       \includegraphics[width=1.4\linewidth]{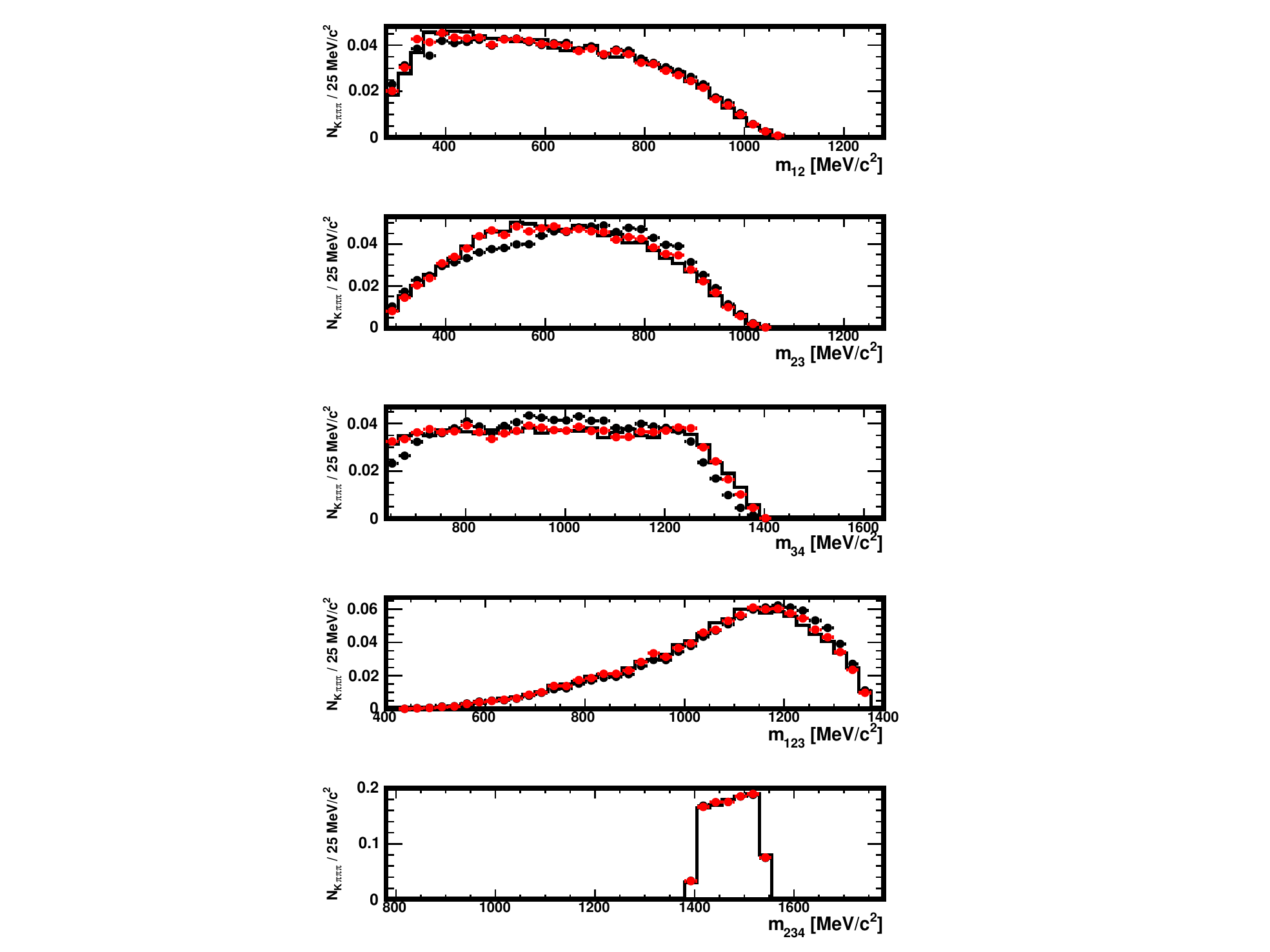}
    \vspace*{-0.5cm}
  \caption{Distributions of $m_{12}$, $m_{23}$, $m_{34}$, $m_{123}$ and $m_{234}$ in the original sample (histogram),
in the distorted one obtained with a tighter selection (full black circles) and in the distorted sample where the decays have been re-weighted using the $\omega_i$ weights (red), 
as explained in Sect.~\ref{subsec:ResultsK3pi}. The data used here are restricted to the region $950 < m_{234} < 1150$ MeV$/c^2$.
The absolute normalisation is arbitrary when the correction is not applied and natural when it is applied (red).} 
  \label{fig:Distr_m234_3_Sel2}
\end{figure}

\clearpage

\begin{figure}[tb]
    \hspace*{-3cm}
       \includegraphics[width=1.4\linewidth]{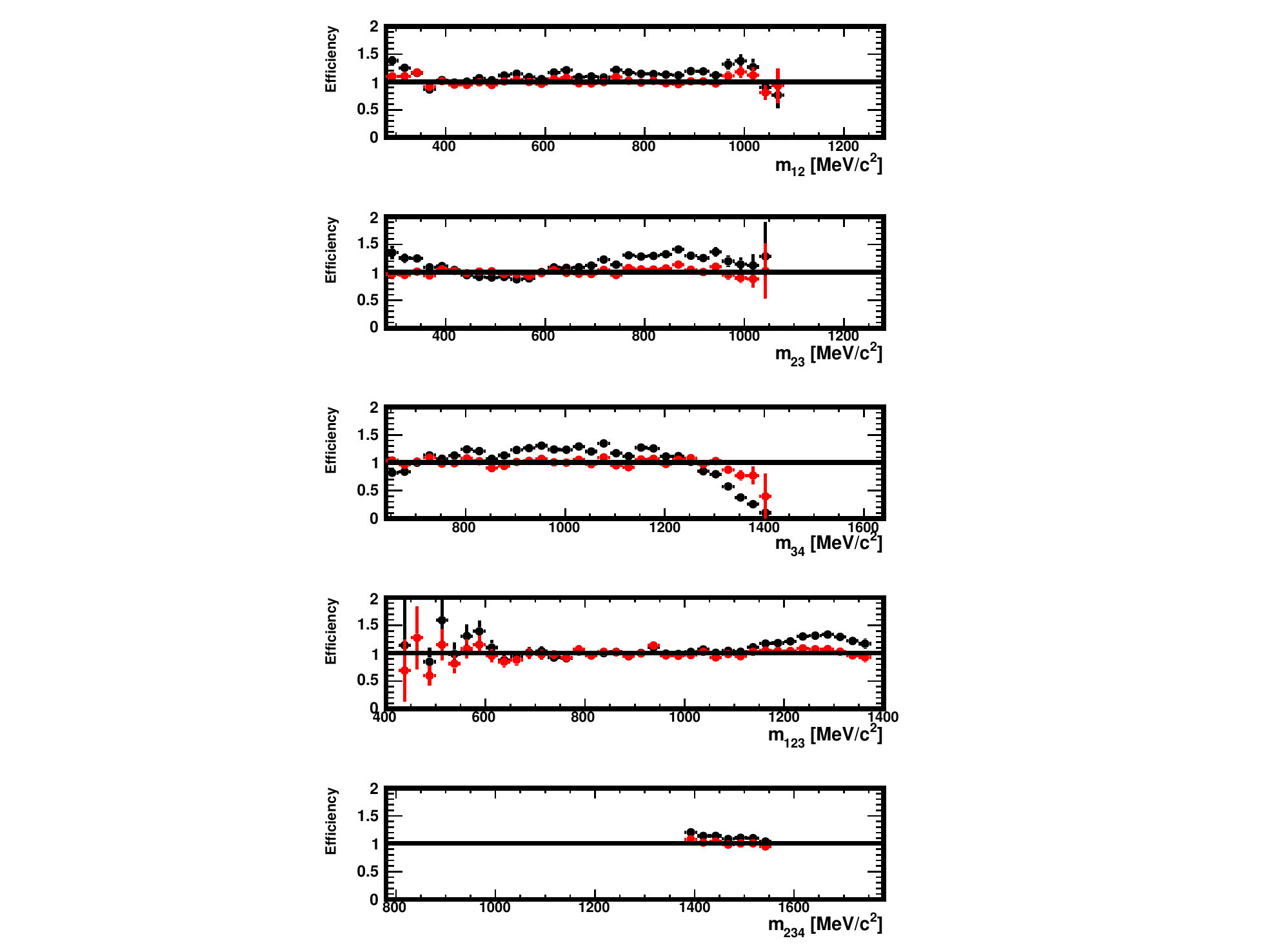}
    \vspace*{-0.5cm}
  \caption{Efficiency in $m_{12}$, $m_{23}$, $m_{34}$, $m_{123}$ and $m_{234}$ in the data generated for this study, 
in the region $950 < m_{234} < 1150$ MeV$/c^2$ and with a tighter selection. Shown are the ratios of the distributions found in the distorted and 
original samples, with no correction (black) and for decays re-weighted using $\omega_i$ weights (red)
as explained in Sect.~\ref{subsec:ResultsK3pi}. The absolute normalisation is arbitrary when the correction is not applied and natural when it is applied (red).} 
  \label{fig:Ratio_m234_3_Sel2}
\end{figure}

\clearpage

\begin{figure}[tb]
    \hspace*{-3cm}
       \includegraphics[width=1.4\linewidth]{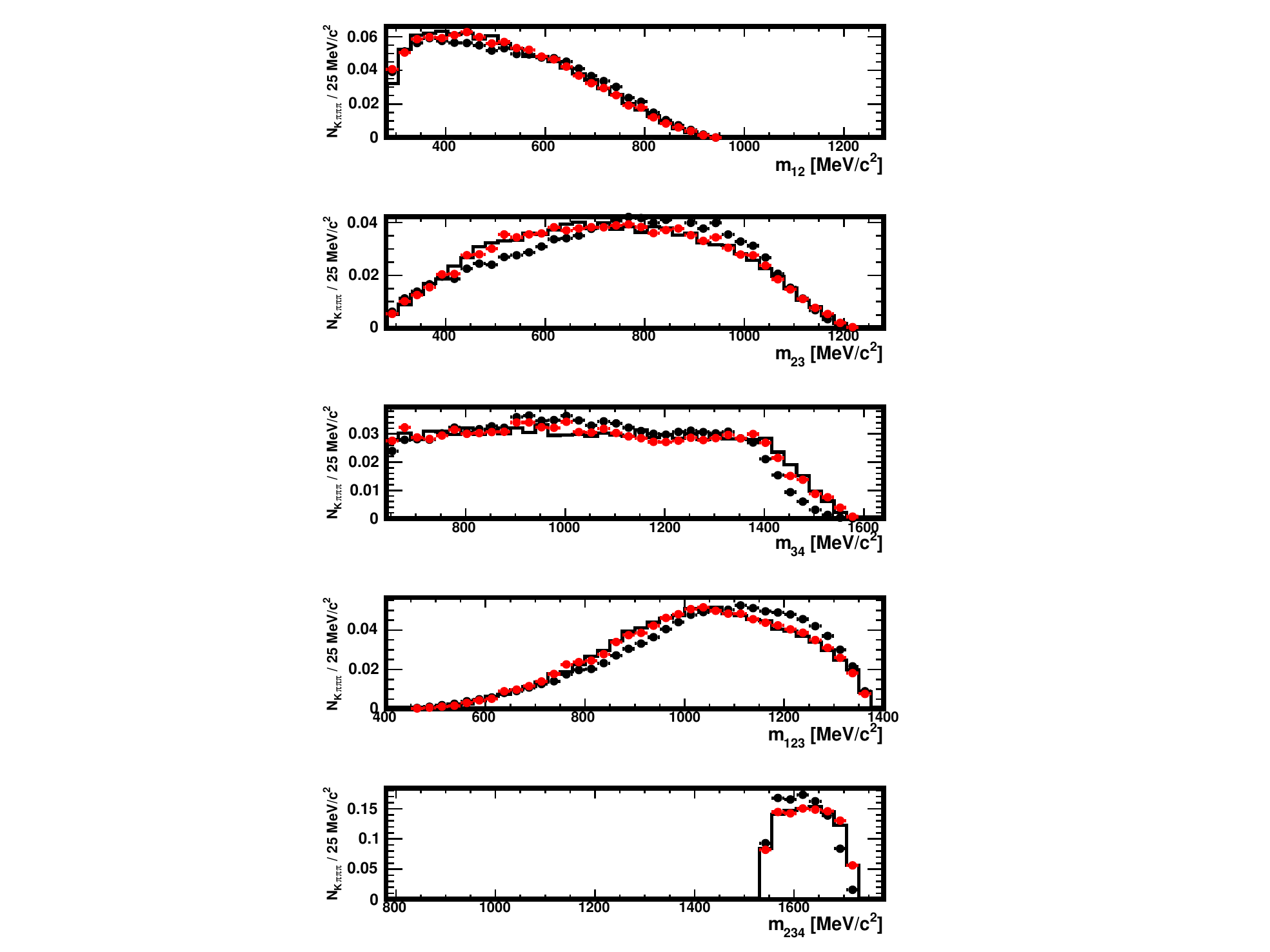}
    \vspace*{-0.5cm}
  \caption{Distributions of $m_{12}$, $m_{23}$, $m_{34}$, $m_{123}$ and $m_{234}$ in the original sample (histogram),
in the distorted one obtained with a tighter selection (full black circles) and in the distorted sample where the decays have been re-weighted using the $\omega_i$ weights (red), 
as explained in Sect.~\ref{subsec:ResultsK3pi}. The data used here are restricted to the region $1150 < m_{234}$ MeV$/c^2$.
The absolute normalisation is arbitrary when the correction is not applied and natural when it is applied (red).} 
  \label{fig:Distr_m234_4_Sel2}
\end{figure}

\clearpage

\begin{figure}[tb]
    \hspace*{-3cm}
       \includegraphics[width=1.4\linewidth]{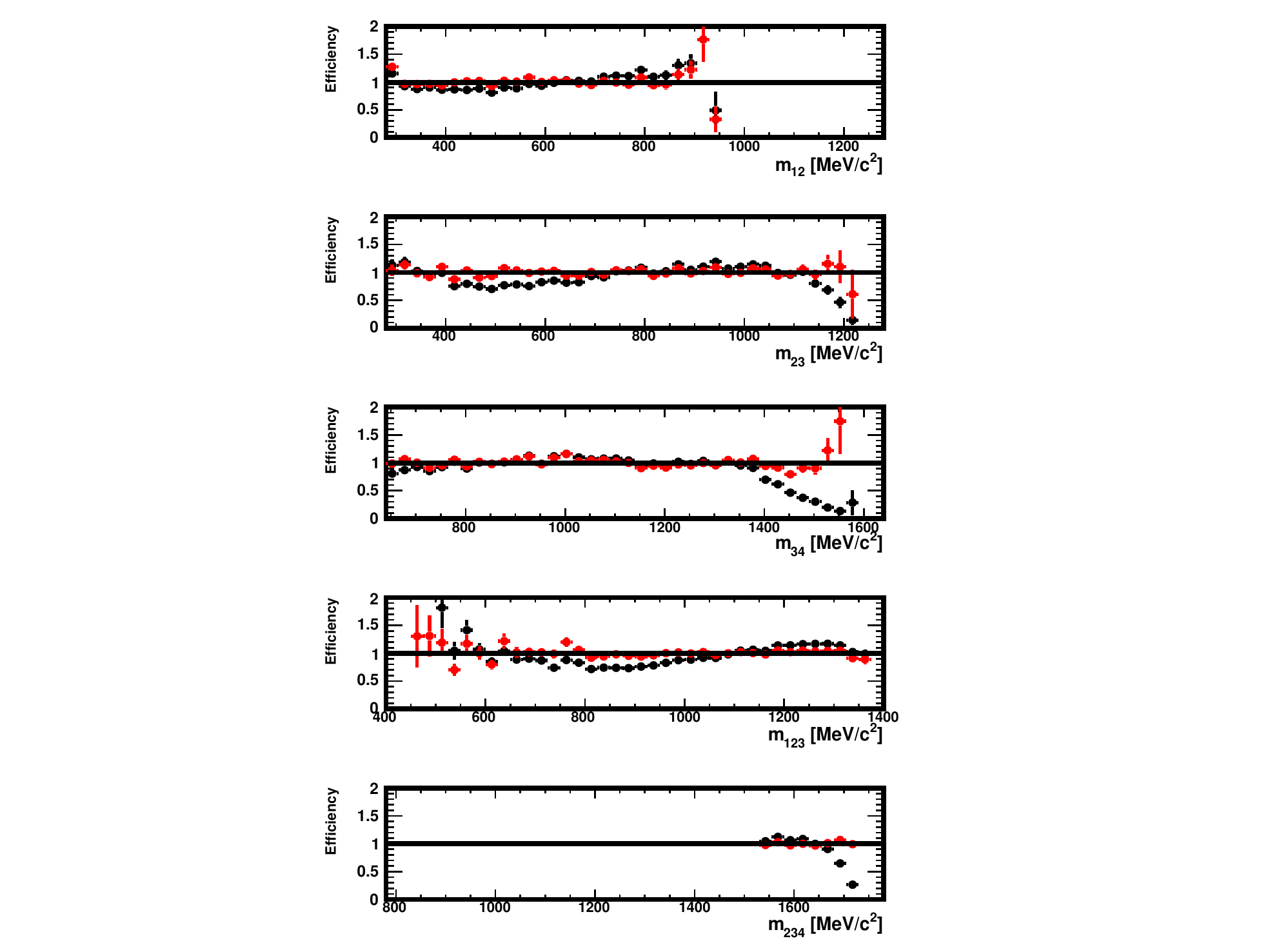}
    \vspace*{-0.5cm}
  \caption{Efficiency in $m_{12}$, $m_{23}$, $m_{34}$, $m_{123}$ and $m_{234}$ in the data generated for this study, 
in the region $1150 < m_{234}$ MeV$/c^2$ and with a tighter selection. Shown are the ratios of the distributions found in the distorted and 
original samples, with no correction (black) and for decays re-weighted using $\omega_i$ weights (red)
as explained in Sect.~\ref{subsec:ResultsK3pi}. The absolute normalisation is arbitrary when the correction is not applied and natural when it is applied (red).} 
  \label{fig:Ratio_m234_4_Sel2}
\end{figure}

\section{ Additional figures related to the \kstmumu test case}
\label{App:AppB}

\begin{figure}[tb]
    \hspace*{-2.3cm}
       \includegraphics[width=1.3\linewidth]{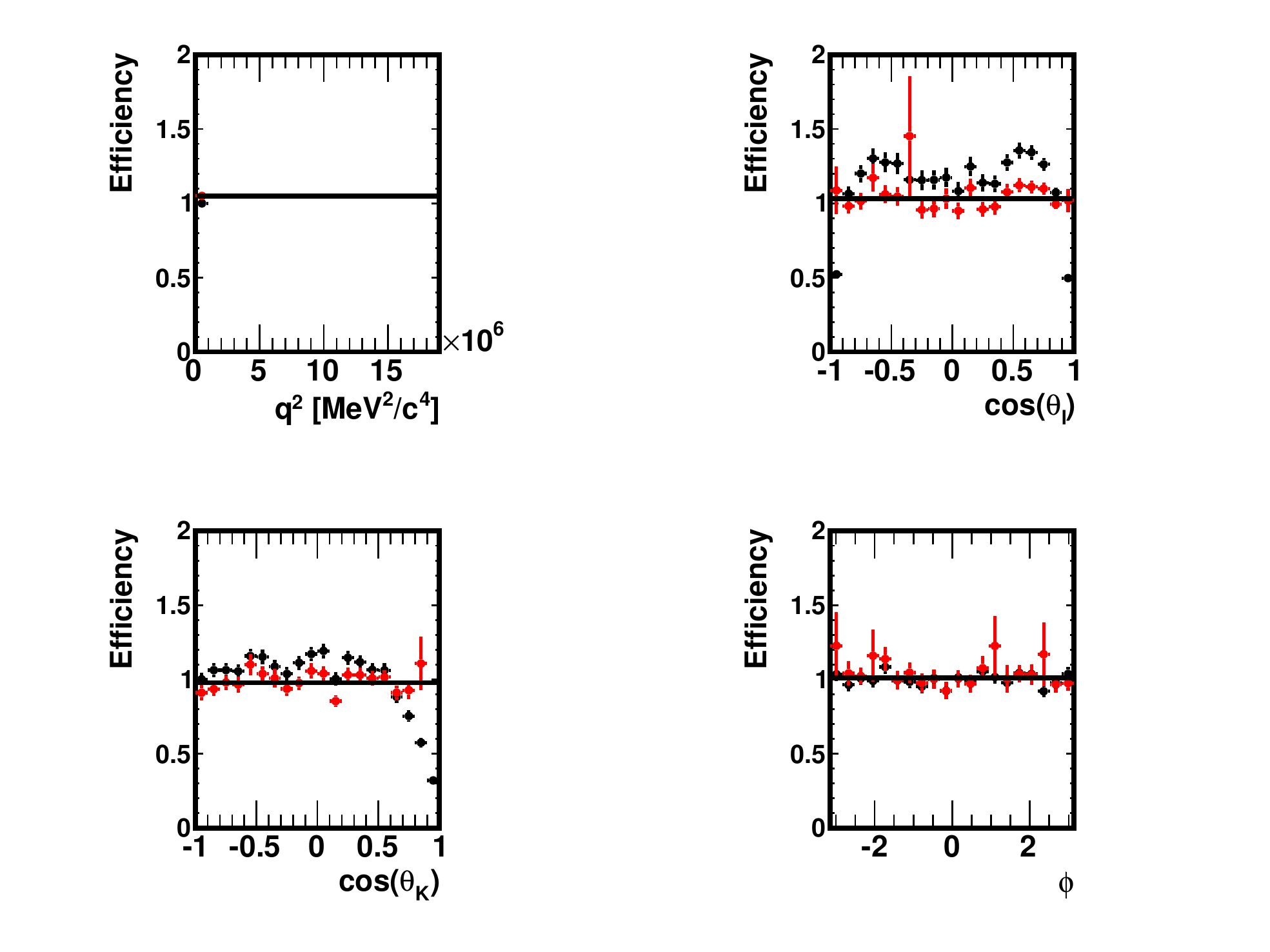}
    \vspace*{-0.5cm}
  \caption{Efficiency in $q^{2}$, cos$\theta_l$, cos$\theta_K$ and $\phi$ in the data generated for the present study, 
in the region $0.1 < q^2 < 0.98$ GeV$^{2}/c^4$. Shown are the ratios of the distributions found in the distorted and 
original samples, with no correction (black) and for decays re-weighted using $\omega_i$ weights (red)
as explained in Sect.~\ref{subsec:resultsB2KstMuMu}. The absolute normalisation is arbitrary when the correction is not applied and natural when it is applied (red).} 
  \label{fig:Ratio_q2_1}
\end{figure}

\begin{figure}[tb]
    \hspace*{-2.3cm}
       \includegraphics[width=1.3\linewidth]{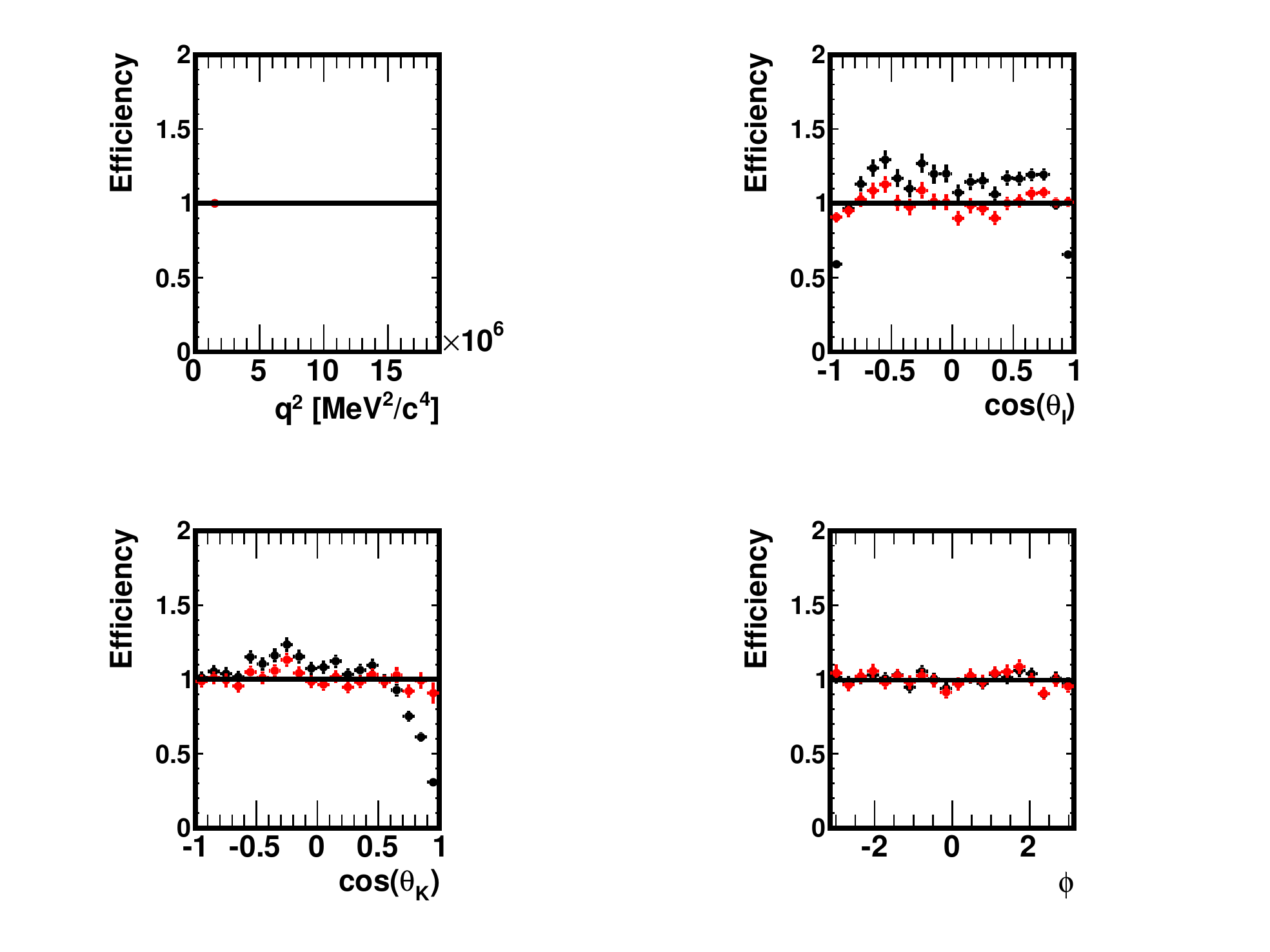}
    \vspace*{-0.5cm}
  \caption{Efficiency in $q^{2}$, cos$\theta_l$, cos$\theta_K$ and $\phi$ in the data generated for the present study, 
in the region $0.98 < q^2 < 2.0$ GeV$^{2}/c^4$. Shown are the ratios of the distributions found in the distorted and 
original samples, with no correction (black) and for decays re-weighted using $\omega_i$ weights (red)
as explained in Sect.~\ref{subsec:resultsB2KstMuMu}. The absolute normalisation is arbitrary when the correction is not applied and natural when it is applied (red).} 
  \label{fig:Ratio_q2_2}
\end{figure}

\begin{figure}[tb]
    \hspace*{-2.3cm}
       \includegraphics[width=1.3\linewidth]{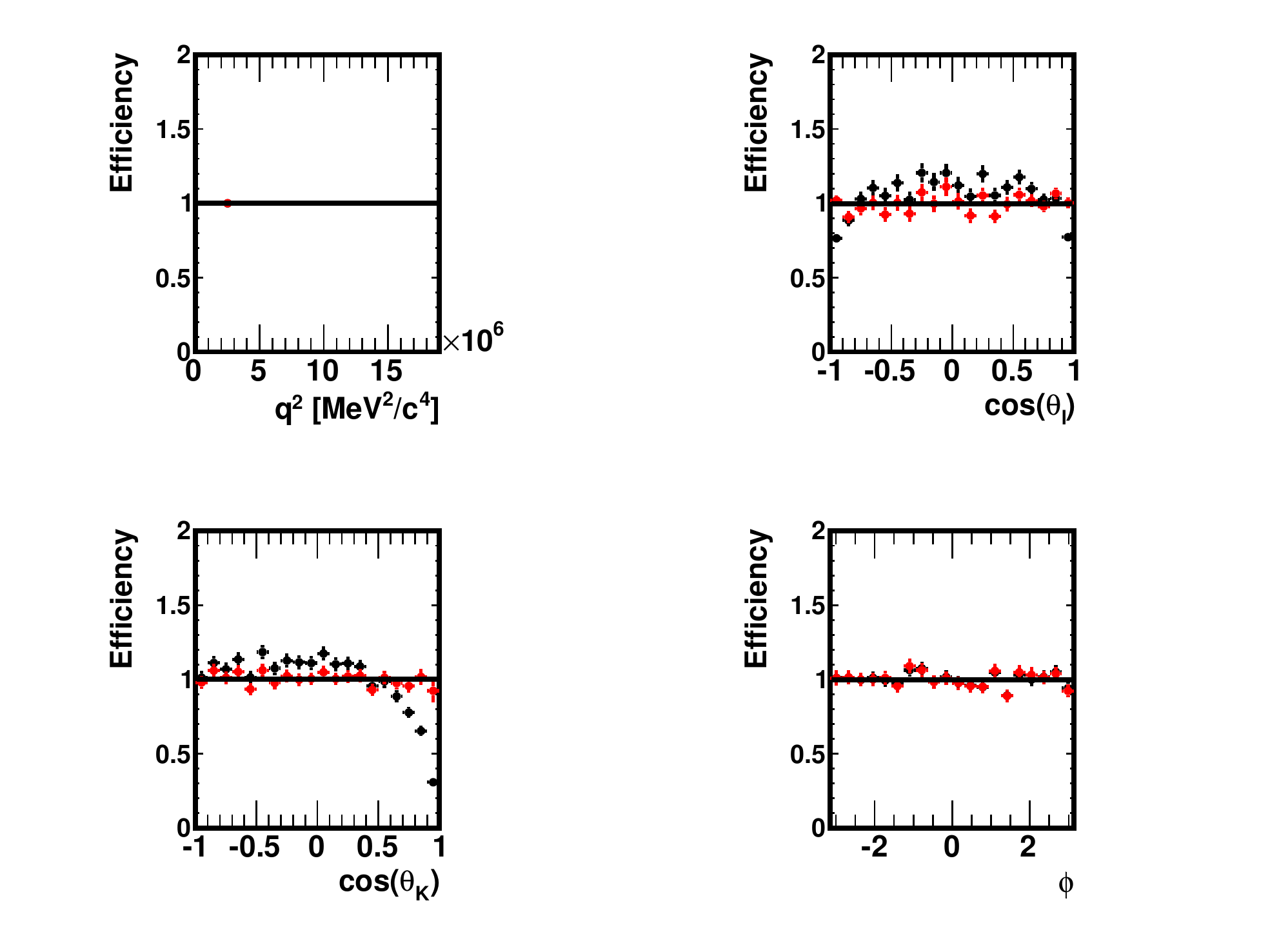}
    \vspace*{-0.5cm}
  \caption{Efficiency in $q^{2}$, cos$\theta_l$, cos$\theta_K$ and $\phi$ in the data generated for the present study, 
in the region $2.0 < q^2 < 3.0$ GeV$^{2}/c^4$. Shown are the ratios of the distributions found in the distorted and 
original samples, with no correction (black) and for decays re-weighted using $\omega_i$ weights (red)
as explained in Sect.~\ref{subsec:resultsB2KstMuMu}. The absolute normalisation is arbitrary when the correction is not applied and natural when it is applied (red).} 
  \label{fig:Ratio_q2_3}
\end{figure}

\begin{figure}[tb]
    \hspace*{-2.3cm}
       \includegraphics[width=1.3\linewidth]{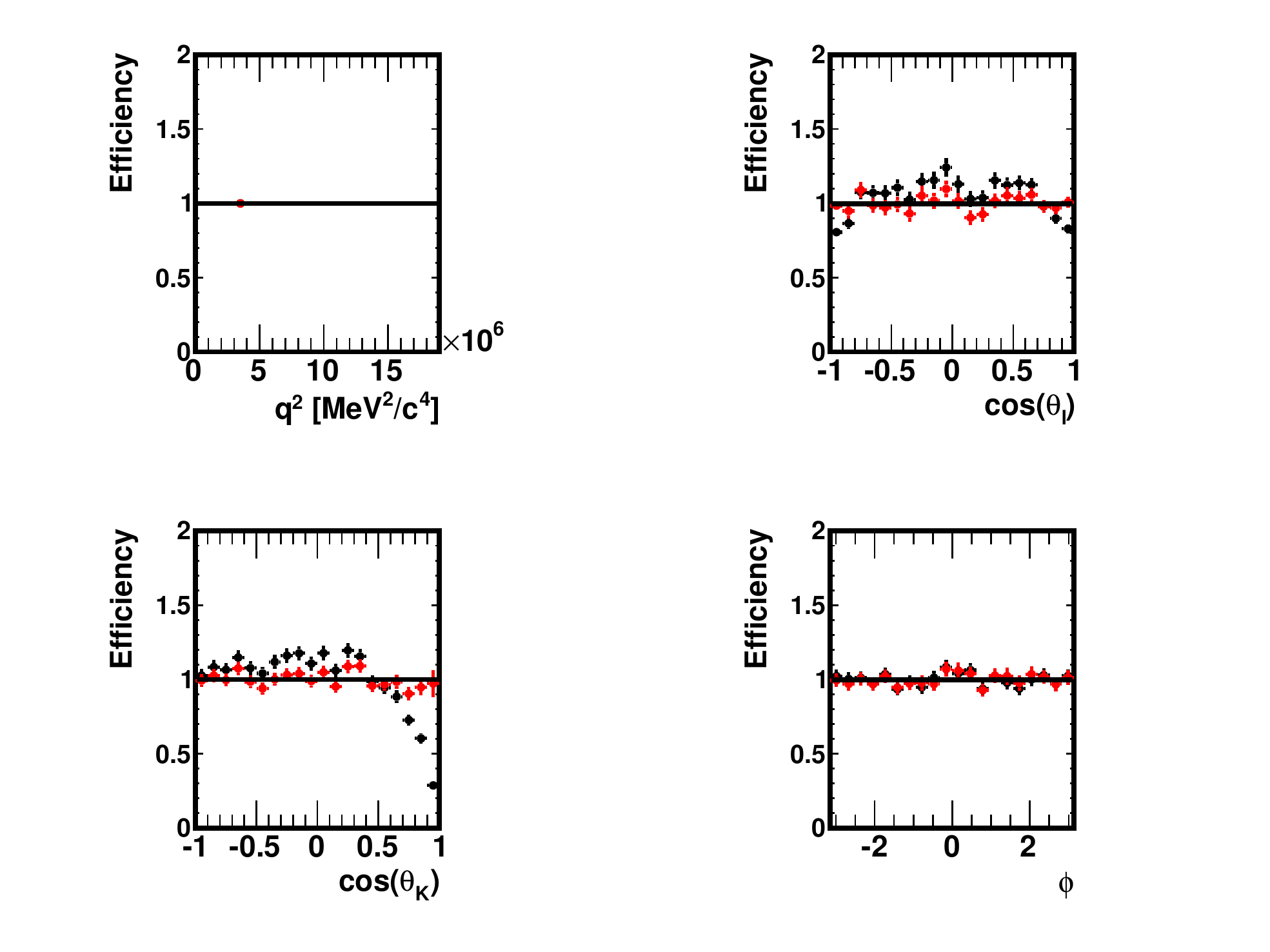}
    \vspace*{-0.5cm}
  \caption{Efficiency in $q^{2}$, cos$\theta_l$, cos$\theta_K$ and $\phi$ in the data generated for the present study, 
in the region $3.0 < q^2 < 4.0$ GeV$^{2}/c^4$. Shown are the ratios of the distributions found in the distorted and 
original samples, with no correction (black) and for decays re-weighted using $\omega_i$ weights (red)
as explained in Sect.~\ref{subsec:resultsB2KstMuMu}. The absolute normalisation is arbitrary when the correction is not applied and natural when it is applied (red).} 
  \label{fig:Ratio_q2_4}
\end{figure}

\begin{figure}[tb]
    \hspace*{-2.3cm}
       \includegraphics[width=1.3\linewidth]{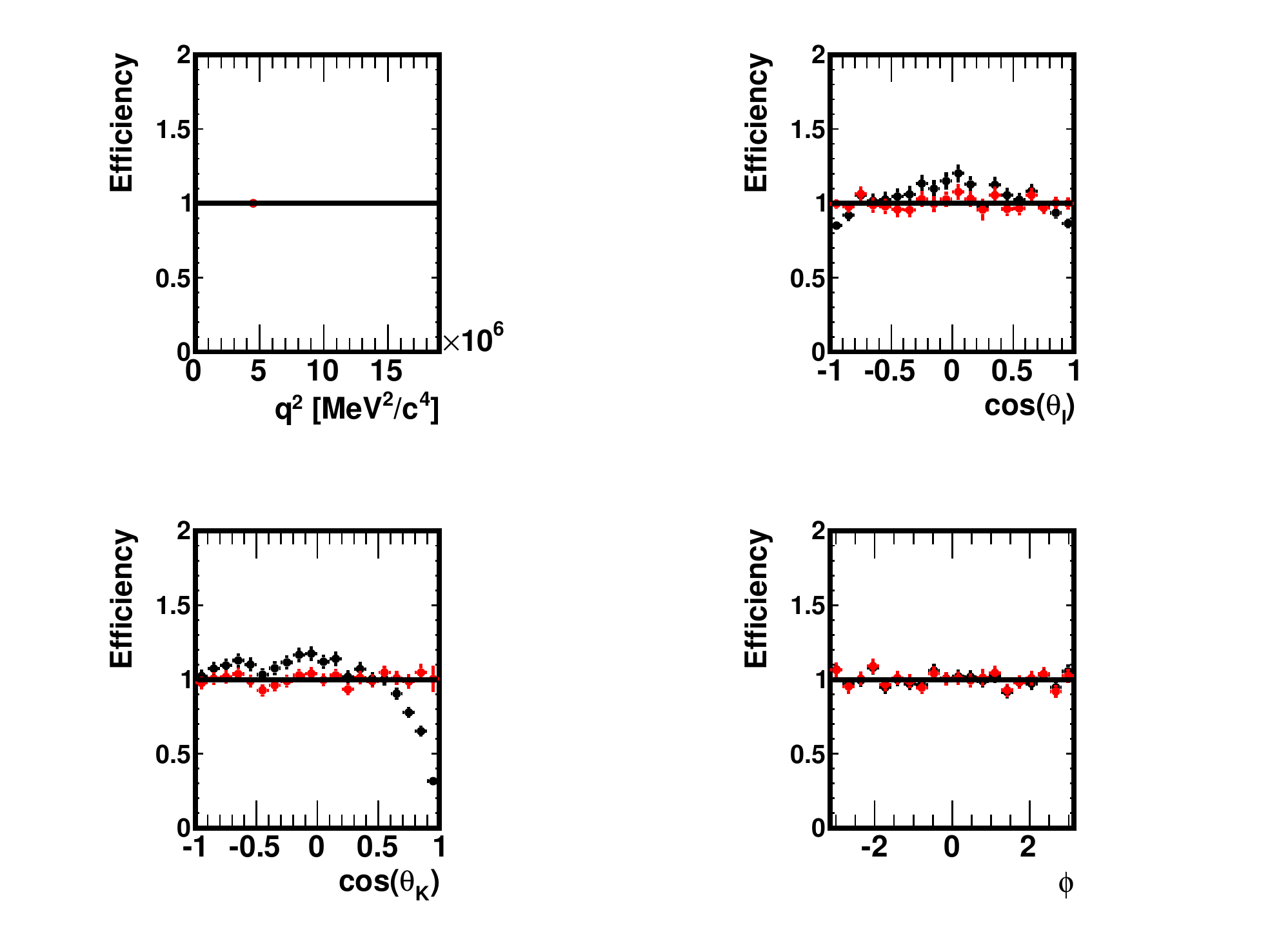}
    \vspace*{-0.5cm}
  \caption{Efficiency in $q^{2}$, cos$\theta_l$, cos$\theta_K$ and $\phi$ in the data generated for the present study, 
in the region $4.0 < q^2 < 5.0$ GeV$^{2}/c^4$. Shown are the ratios of the distributions found in the distorted and 
original samples, with no correction (black) and for decays re-weighted using $\omega_i$ weights (red)
as explained in Sect.~\ref{subsec:resultsB2KstMuMu}. The absolute normalisation is arbitrary when the correction is not applied and natural when it is applied (red).} 
  \label{fig:Ratio_q2_5}
\end{figure}

\begin{figure}[tb]
    \hspace*{-2.3cm}
       \includegraphics[width=1.3\linewidth]{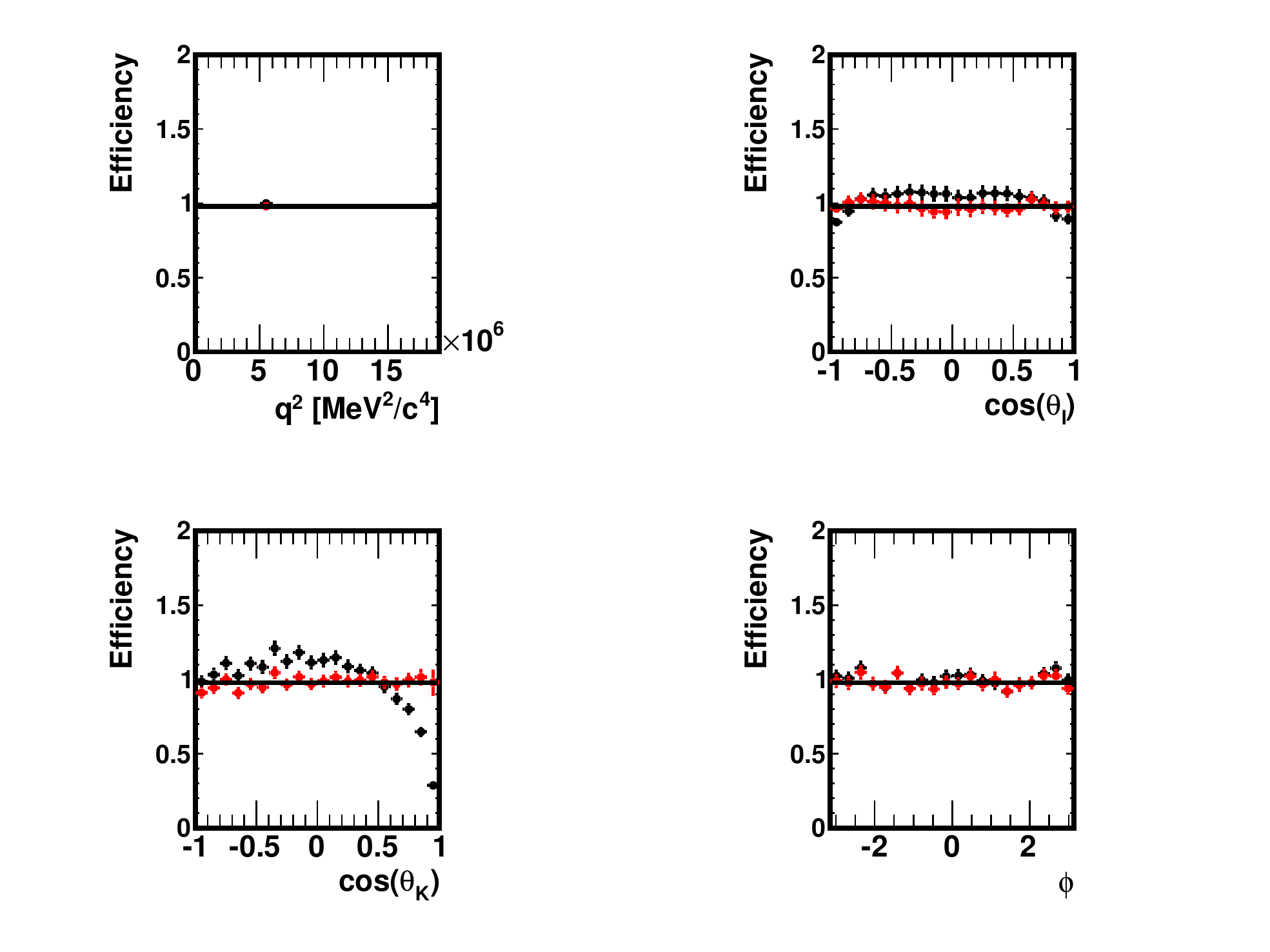}
    \vspace*{-0.5cm}
  \caption{Efficiency in $q^{2}$, cos$\theta_l$, cos$\theta_K$ and $\phi$ in the data generated for the present study, 
in the region $5.0 < q^2 < 6.0$ GeV$^{2}/c^4$. Shown are the ratios of the distributions found in the distorted and 
original samples, with no correction (black) and for decays re-weighted using $\omega_i$ weights (red)
as explained in Sect.~\ref{subsec:resultsB2KstMuMu}. The absolute normalisation is arbitrary when the correction is not applied and natural when it is applied (red).} 
  \label{fig:Ratio_q2_6}
\end{figure}

\begin{figure}[tb]
    \hspace*{-2.3cm}
       \includegraphics[width=1.3\linewidth]{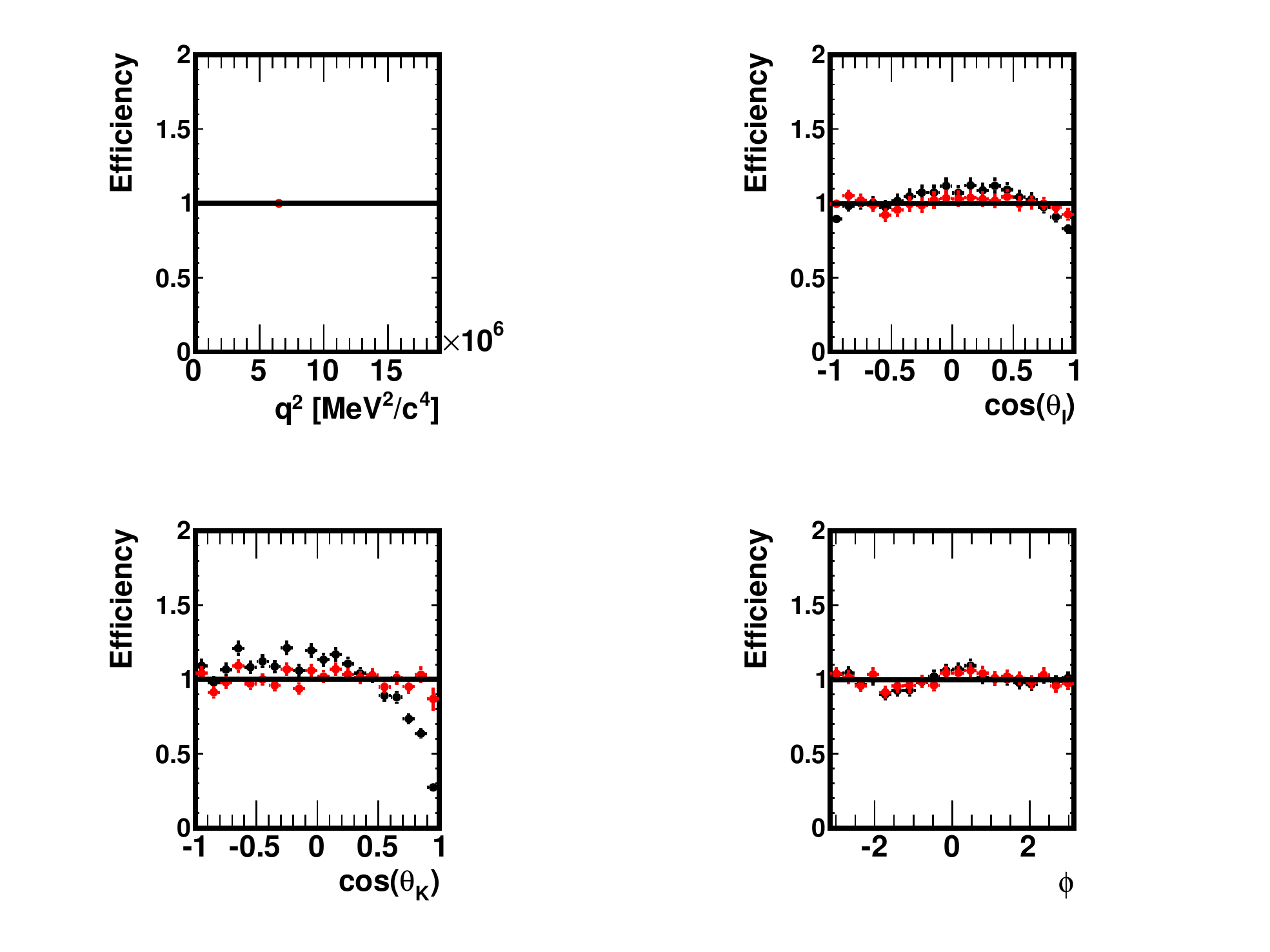}
    \vspace*{-0.5cm}
  \caption{Efficiency in $q^{2}$, cos$\theta_l$, cos$\theta_K$ and $\phi$ in the data generated for the present study, 
in the region $6.0 < q^2 < 7.0$ GeV$^{2}/c^4$. Shown are the ratios of the distributions found in the distorted and 
original samples, with no correction (black) and for decays re-weighted using $\omega_i$ weights (red)
as explained in Sect.~\ref{subsec:resultsB2KstMuMu}. The absolute normalisation is arbitrary when the correction is not applied and natural when it is applied (red).} 
  \label{fig:Ratio_q2_7}
\end{figure}

\begin{figure}[tb]
    \hspace*{-2.3cm}
       \includegraphics[width=1.3\linewidth]{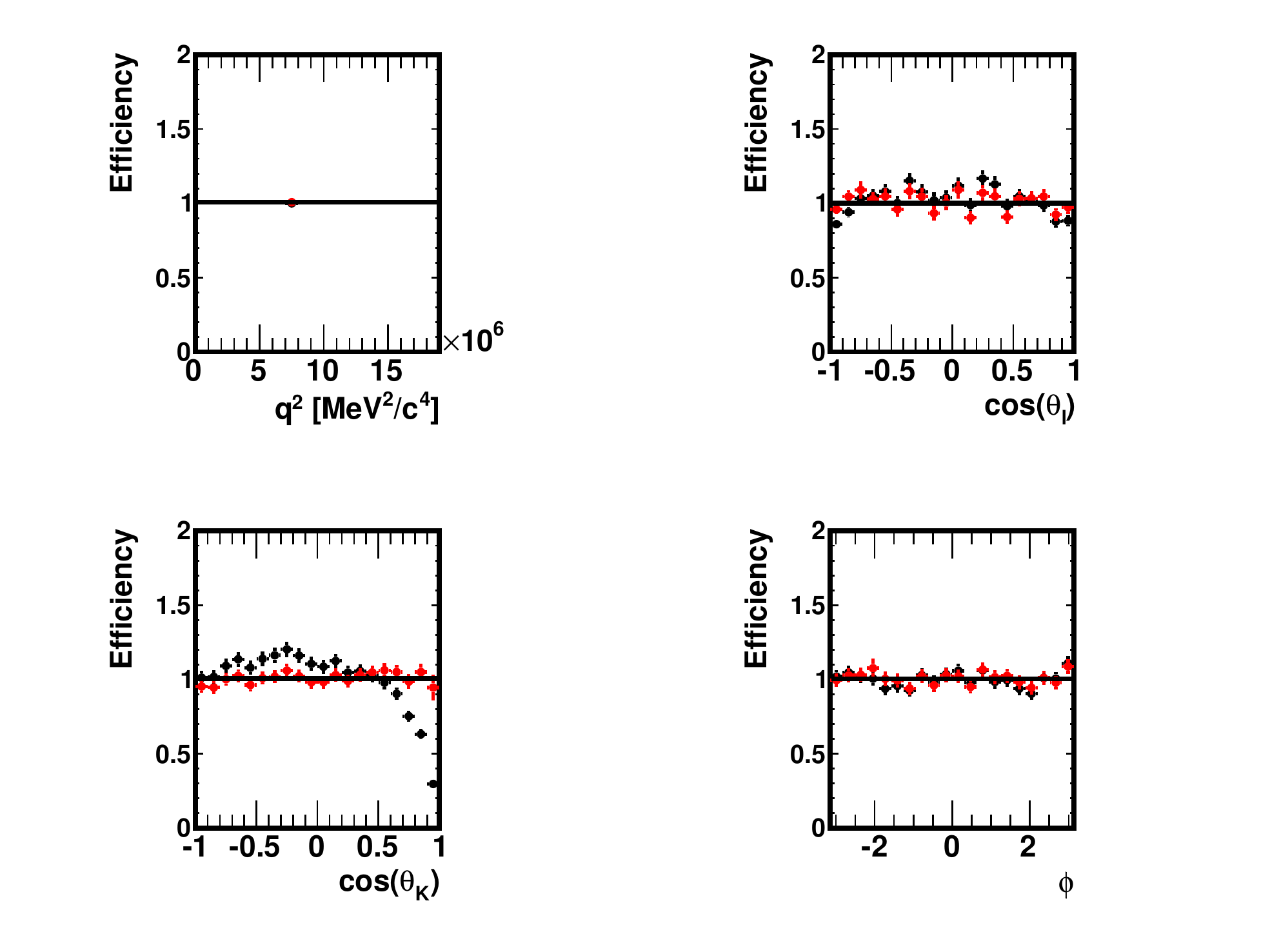}
    \vspace*{-0.5cm}
  \caption{Efficiency in $q^{2}$, cos$\theta_l$, cos$\theta_K$ and $\phi$ in the data generated for the present study, 
in the region $7.0 < q^2 <8.0 $ GeV$^{2}/c^4$. Shown are the ratios of the distributions found in the distorted and 
original samples, with no correction (black) and for decays re-weighted using $\omega_i$ weights (red)
as explained in Sect.~\ref{subsec:resultsB2KstMuMu}. The absolute normalisation is arbitrary when the correction is not applied and natural when it is applied (red).} 
  \label{fig:Ratio_q2_8}
\end{figure}

\begin{figure}[tb]
    \hspace*{-2.3cm}
       \includegraphics[width=1.3\linewidth]{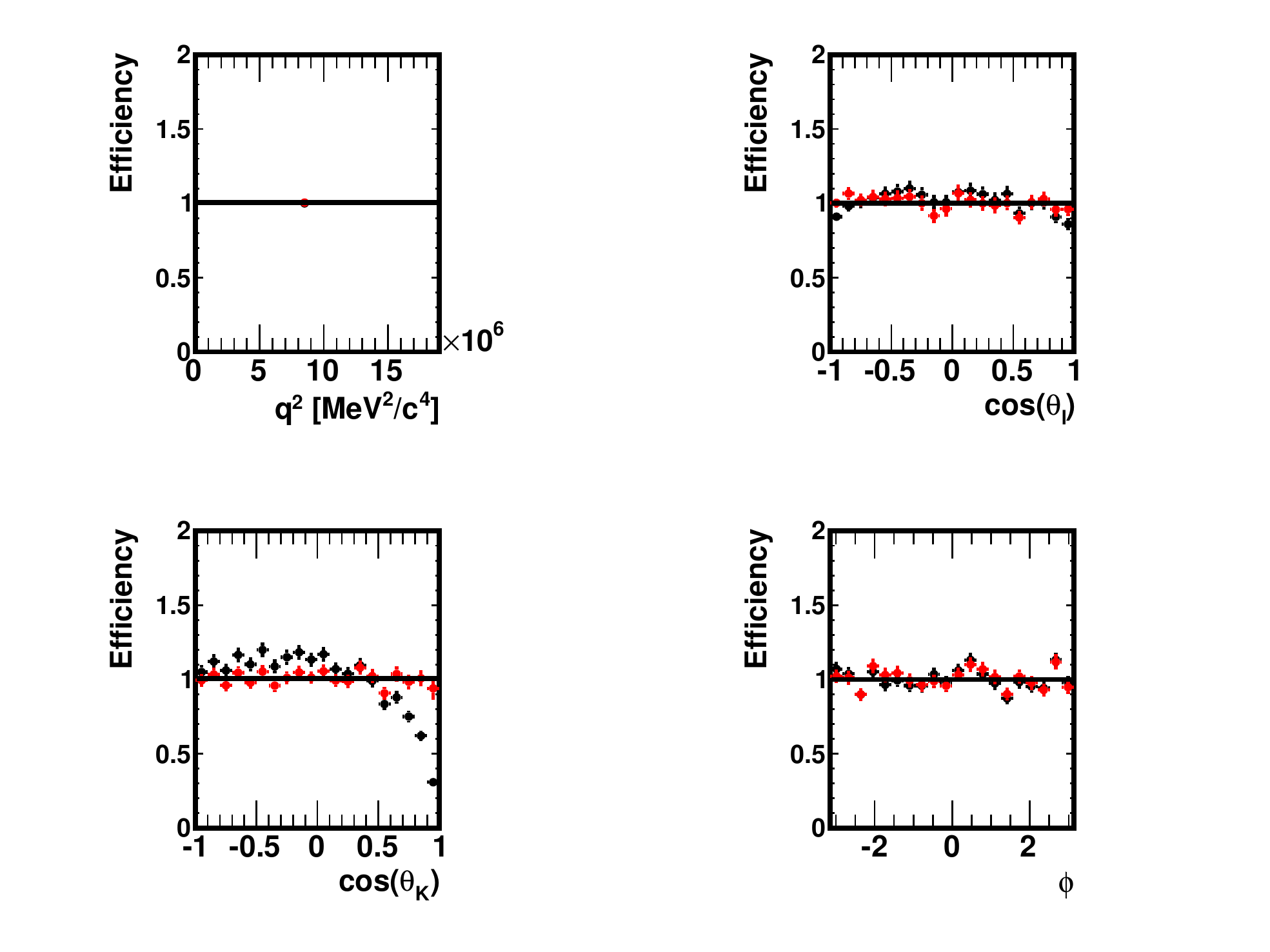}
    \vspace*{-0.5cm}
  \caption{Efficiency in $q^{2}$, cos$\theta_l$, cos$\theta_K$ and $\phi$ in the data generated for the present study, 
in the region $8.0 < q^2 < 9.0$ GeV$^{2}/c^4$. Shown are the ratios of the distributions found in the distorted and 
original samples, with no correction (black) and for decays re-weighted using $\omega_i$ weights (red)
as explained in Sect.~\ref{subsec:resultsB2KstMuMu}. The absolute normalisation is arbitrary when the correction is not applied and natural when it is applied (red).} 
  \label{fig:Ratio_q2_9}
\end{figure}

\begin{figure}[tb]
    \hspace*{-2.3cm}
       \includegraphics[width=1.3\linewidth]{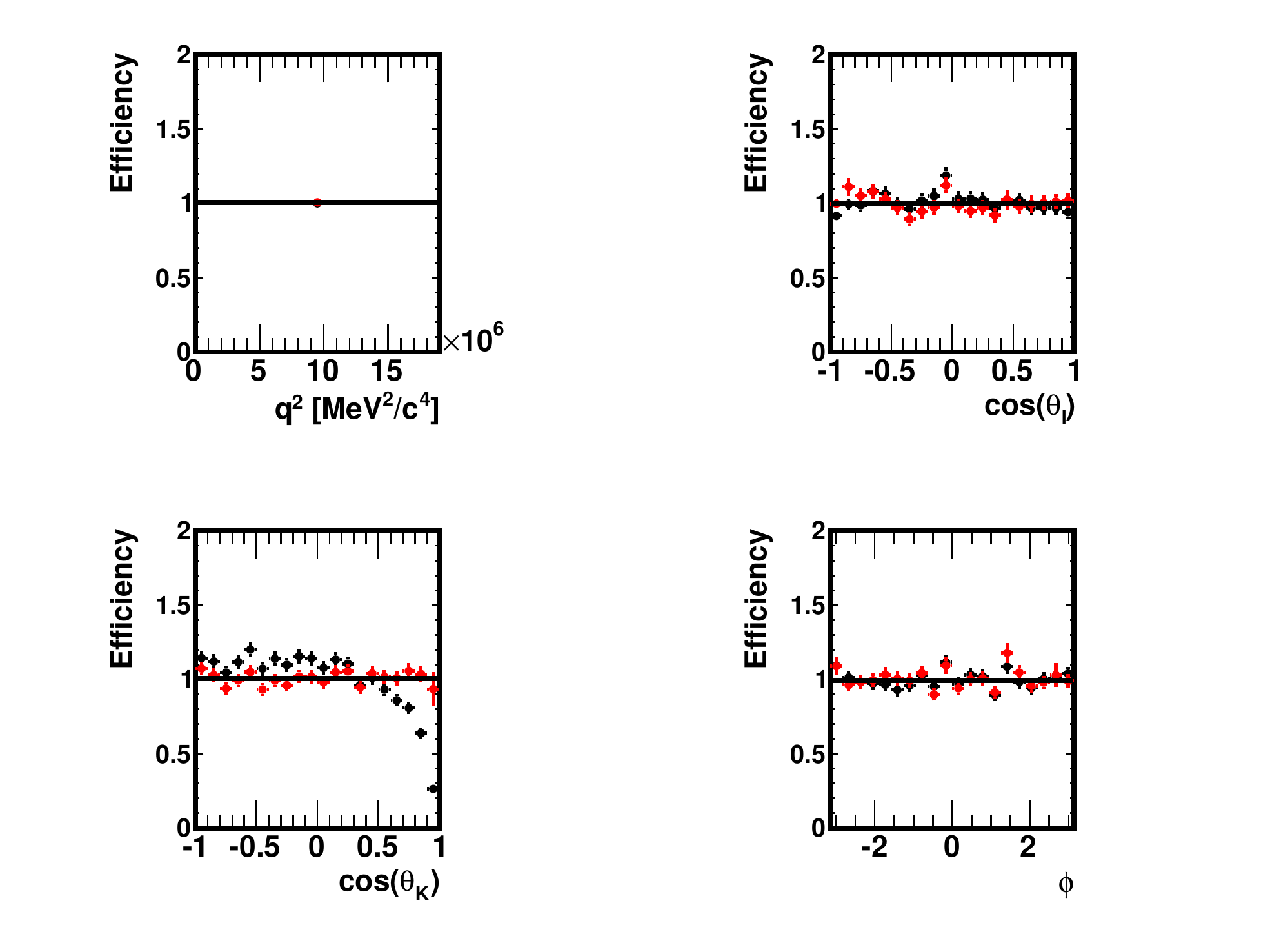}
    \vspace*{-0.5cm}
  \caption{Efficiency in $q^{2}$, cos$\theta_l$, cos$\theta_K$ and $\phi$ in the data generated for the present study, 
in the region $9.0 < q^2 < 10.0$ GeV$^{2}/c^4$. Shown are the ratios of the distributions found in the distorted and 
original samples, with no correction (black) and for decays re-weighted using $\omega_i$ weights (red)
as explained in Sect.~\ref{subsec:resultsB2KstMuMu}. The absolute normalisation is arbitrary when the correction is not applied and natural when it is applied (red).} 
  \label{fig:Ratio_q2_10}
\end{figure}

\begin{figure}[tb]
    \hspace*{-2.3cm}
       \includegraphics[width=1.3\linewidth]{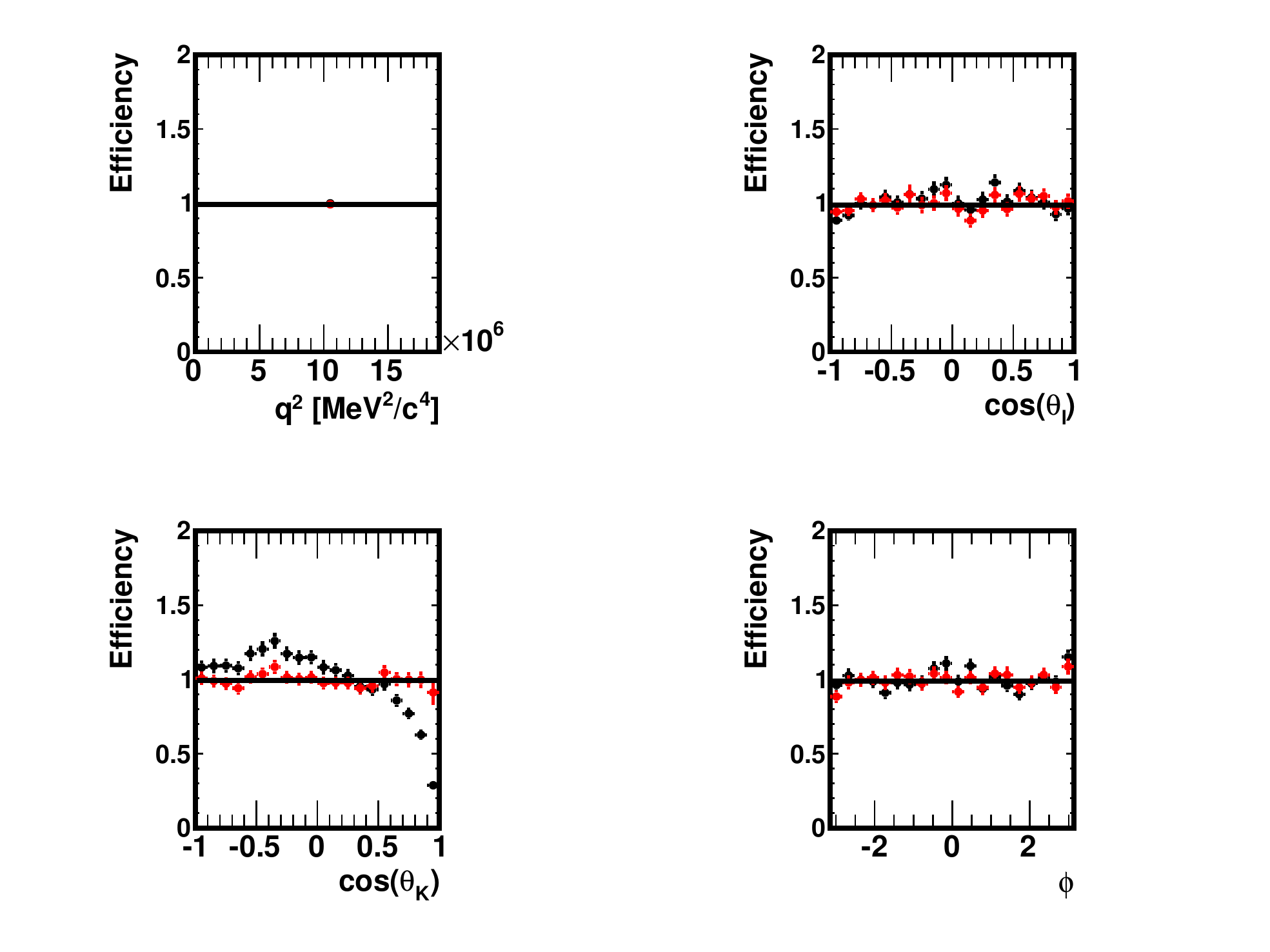}
    \vspace*{-0.5cm}
  \caption{Efficiency in $q^{2}$, cos$\theta_l$, cos$\theta_K$ and $\phi$ in the data generated for the present study, 
in the region $10.0 < q^2 < 11.0$ GeV$^{2}/c^4$. Shown are the ratios of the distributions found in the distorted and 
original samples, with no correction (black) and for decays re-weighted using $\omega_i$ weights (red)
as explained in Sect.~\ref{subsec:resultsB2KstMuMu}. The absolute normalisation is arbitrary when the correction is not applied and natural when it is applied (red).} 
  \label{fig:Ratio_q2_11}
\end{figure}

\begin{figure}[tb]
    \hspace*{-2.3cm}
       \includegraphics[width=1.3\linewidth]{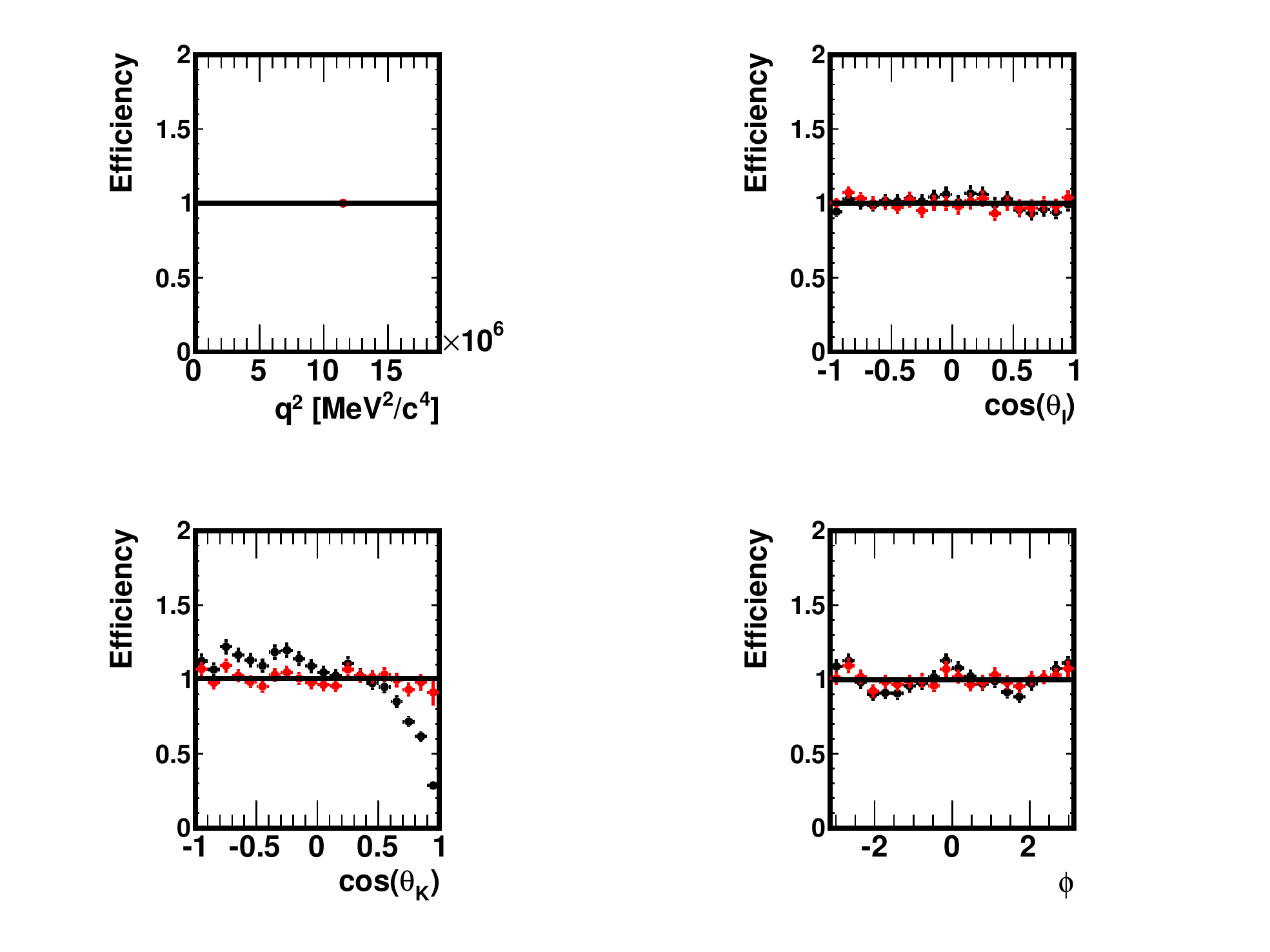}
    \vspace*{-0.5cm}
  \caption{Efficiency in $q^{2}$, cos$\theta_l$, cos$\theta_K$ and $\phi$ in the data generated for the present study, 
in the region $11.0 < q^2 < 12.0$ GeV$^{2}/c^4$. Shown are the ratios of the distributions found in the distorted and 
original samples, with no correction (black) and for decays re-weighted using $\omega_i$ weights (red)
as explained in Sect.~\ref{subsec:resultsB2KstMuMu}. The absolute normalisation is arbitrary when the correction is not applied and natural when it is applied (red).} 
  \label{fig:Ratio_q2_12}
\end{figure}

\begin{figure}[tb]
    \hspace*{-2.3cm}
       \includegraphics[width=1.3\linewidth]{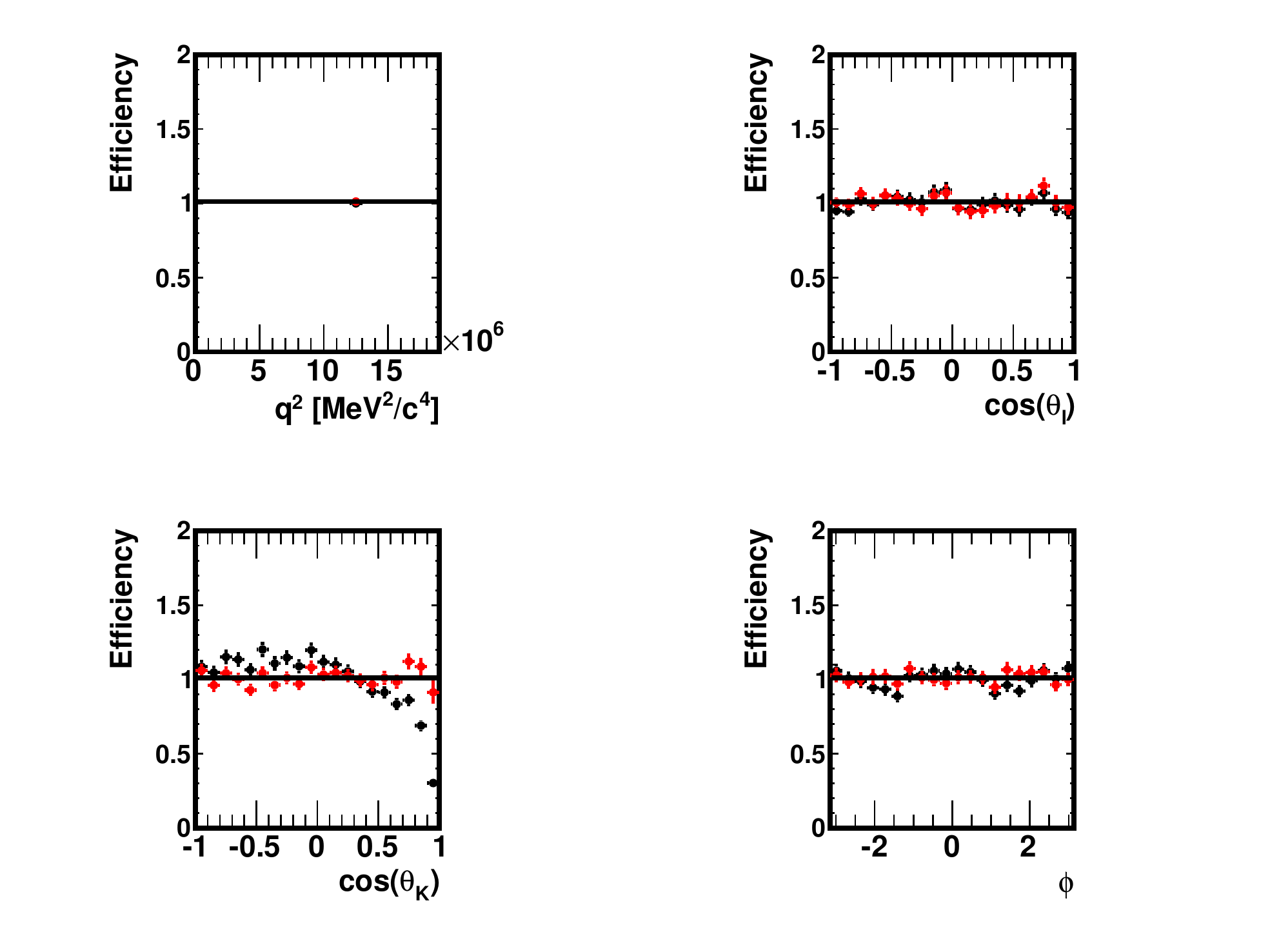}
    \vspace*{-0.5cm}
  \caption{Efficiency in $q^{2}$, cos$\theta_l$, cos$\theta_K$ and $\phi$ in the data generated for the present study, 
in the region $12.0 < q^2 <13.0 $ GeV$^{2}/c^4$. Shown are the ratios of the distributions found in the distorted and 
original samples, with no correction (black) and for decays re-weighted using $\omega_i$ weights (red)
as explained in Sect.~\ref{subsec:resultsB2KstMuMu}. The absolute normalisation is arbitrary when the correction is not applied and natural when it is applied (red).} 
  \label{fig:Ratio_q2_13}
\end{figure}

\begin{figure}[tb]
    \hspace*{-2.3cm}
       \includegraphics[width=1.3\linewidth]{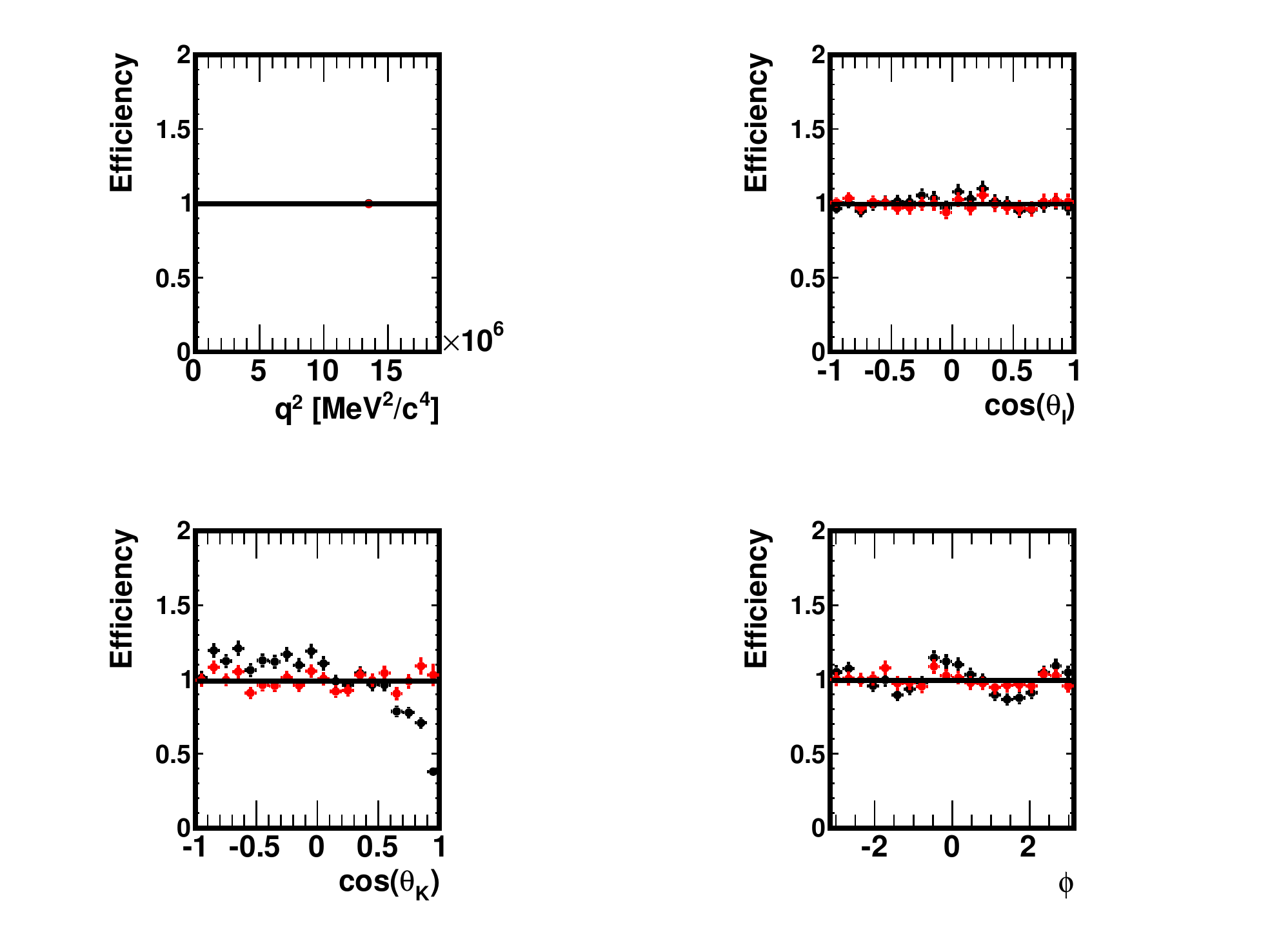}
    \vspace*{-0.5cm}
  \caption{Efficiency in $q^{2}$, cos$\theta_l$, cos$\theta_K$ and $\phi$ in the data generated for the present study, 
in the region $13.0 < q^2 < 14.0$ GeV$^{2}/c^4$. Shown are the ratios of the distributions found in the distorted and 
original samples, with no correction (black) and for decays re-weighted using $\omega_i$ weights (red)
as explained in Sect.~\ref{subsec:resultsB2KstMuMu}. The absolute normalisation is arbitrary when the correction is not applied and natural when it is applied (red).} 
  \label{fig:Ratio_q2_14}
\end{figure}

\begin{figure}[tb]
    \hspace*{-2.3cm}
       \includegraphics[width=1.3\linewidth]{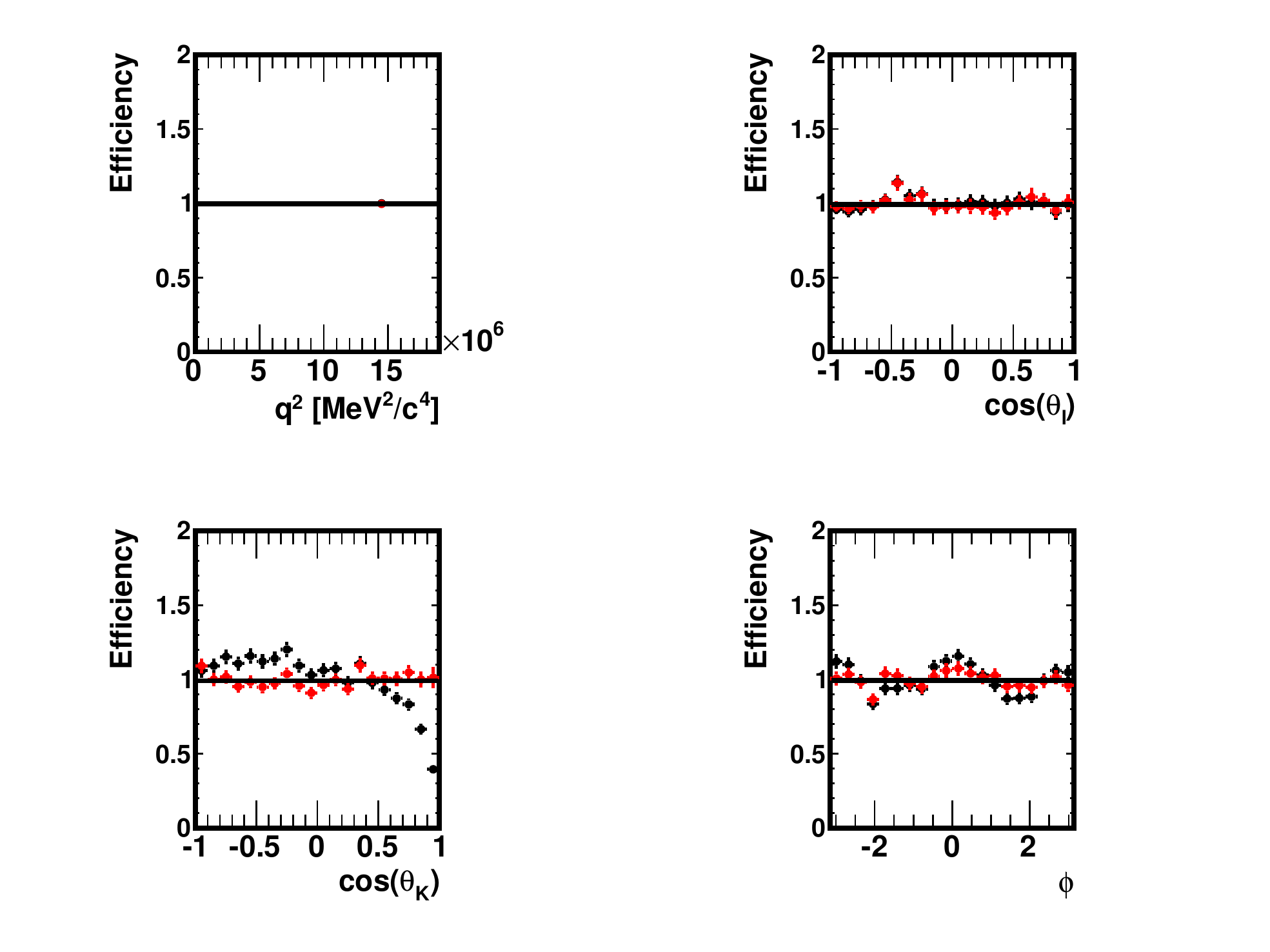}
    \vspace*{-0.5cm}
  \caption{Efficiency in $q^{2}$, cos$\theta_l$, cos$\theta_K$ and $\phi$ in the data generated for the present study, 
in the region $14.0 < q^2 < 15.0$ GeV$^{2}/c^4$. Shown are the ratios of the distributions found in the distorted and 
original samples, with no correction (black) and for decays re-weighted using $\omega_i$ weights (red)
as explained in Sect.~\ref{subsec:resultsB2KstMuMu}. The absolute normalisation is arbitrary when the correction is not applied and natural when it is applied (red).} 
  \label{fig:Ratio_q2_15}
\end{figure}

\begin{figure}[tb]
    \hspace*{-2.3cm}
       \includegraphics[width=1.3\linewidth]{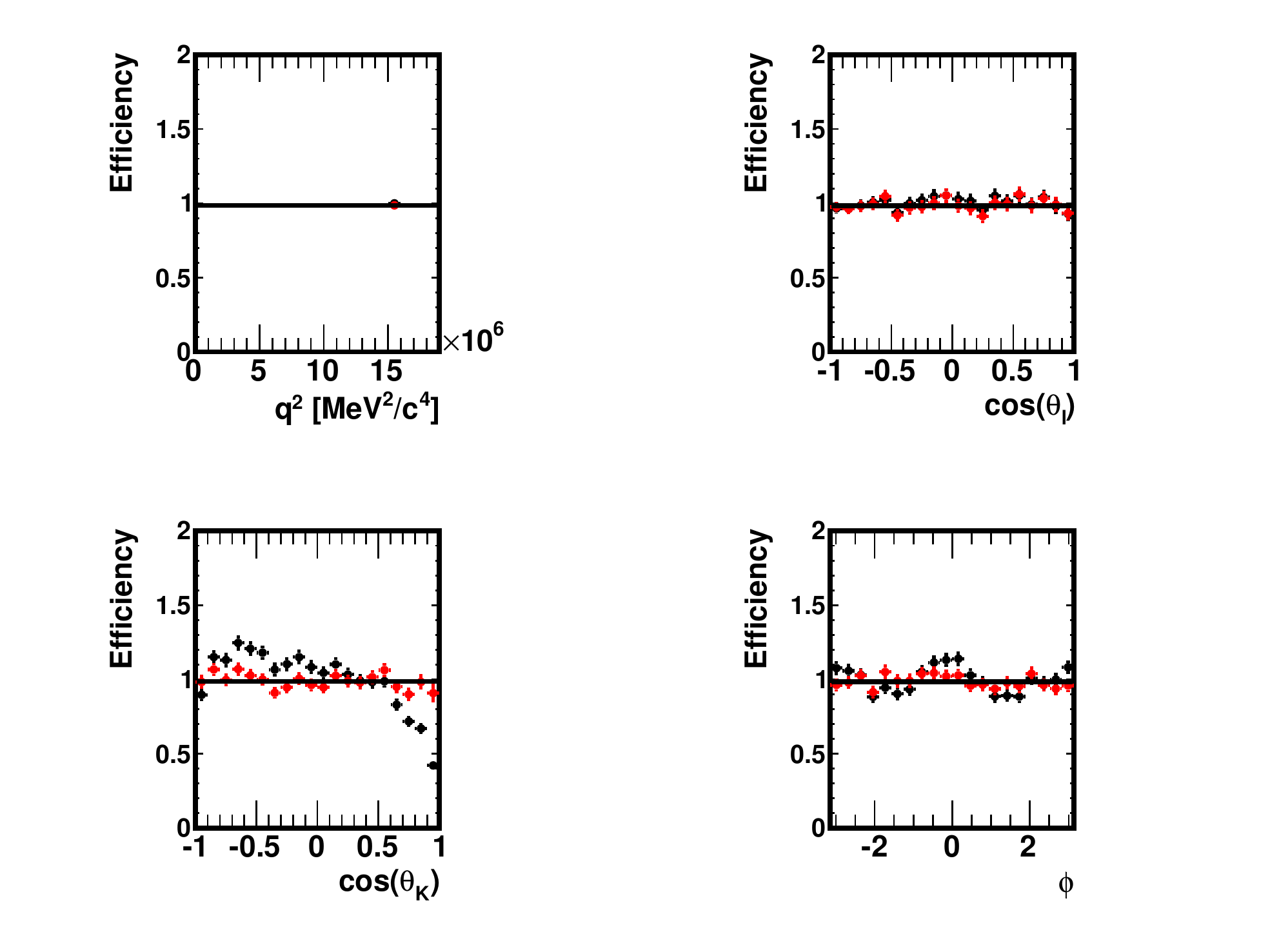}
    \vspace*{-0.5cm}
  \caption{Efficiency in $q^{2}$, cos$\theta_l$, cos$\theta_K$ and $\phi$ in the data generated for the present study, 
in the region $15.0 < q^2 < 16.0$ GeV$^{2}/c^4$. Shown are the ratios of the distributions found in the distorted and 
original samples, with no correction (black) and for decays re-weighted using $\omega_i$ weights (red)
as explained in Sect.~\ref{subsec:resultsB2KstMuMu}. The absolute normalisation is arbitrary when the correction is not applied and natural when it is applied (red).} 
  \label{fig:Ratio_q2_16}
\end{figure}

\begin{figure}[tb]
    \hspace*{-2.3cm}
       \includegraphics[width=1.3\linewidth]{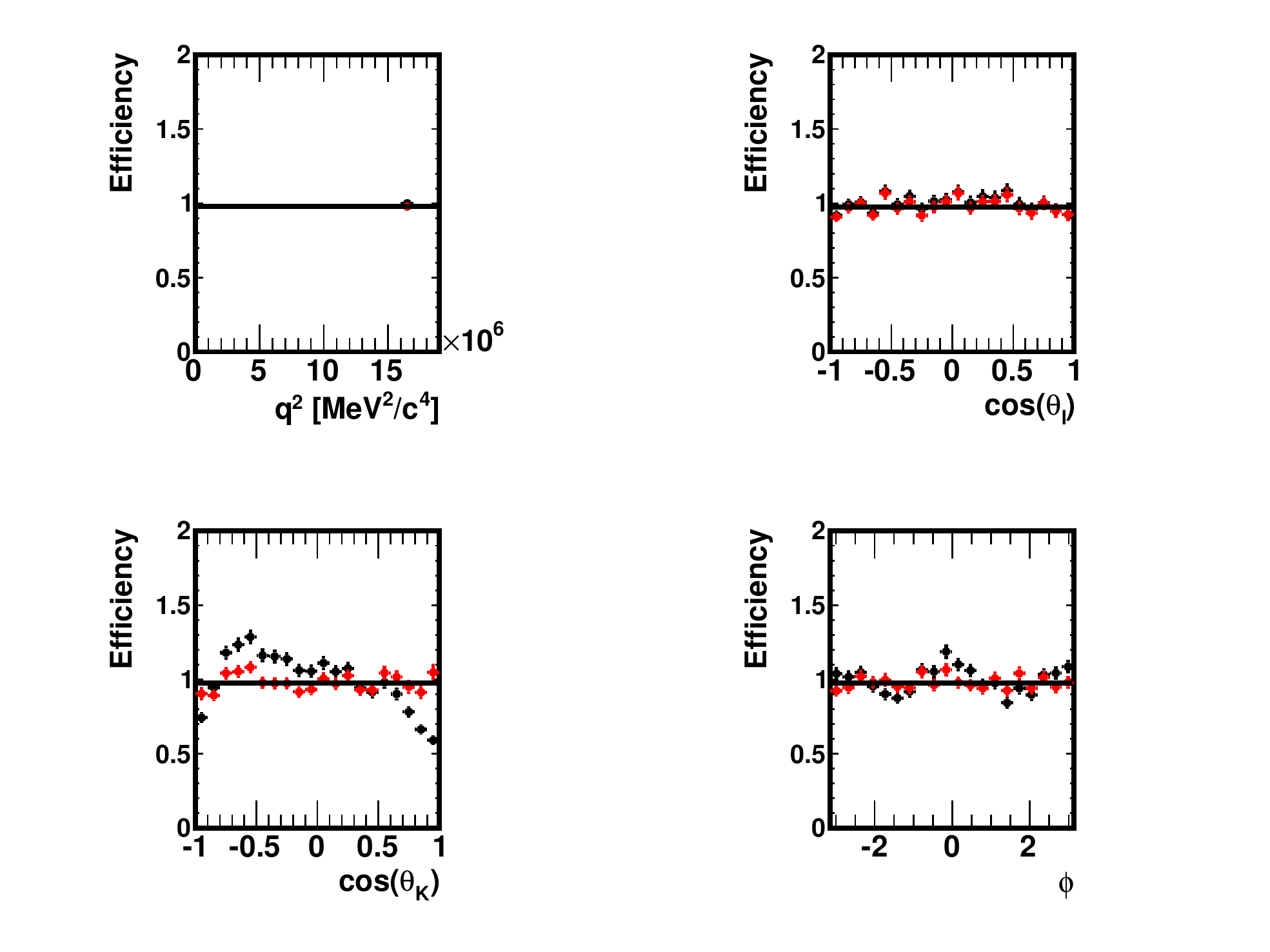}
    \vspace*{-0.5cm}
  \caption{Efficiency in $q^{2}$, cos$\theta_l$, cos$\theta_K$ and $\phi$ in the data generated for the present study, 
in the region $16.0 < q^2 < 17.0$ GeV$^{2}/c^4$. Shown are the ratios of the distributions found in the distorted and 
original samples, with no correction (black) and for decays re-weighted using $\omega_i$ weights (red)
as explained in Sect.~\ref{subsec:resultsB2KstMuMu}. The absolute normalisation is arbitrary when the correction is not applied and natural when it is applied (red).} 
  \label{fig:Ratio_q2_17}
\end{figure}

\clearpage

\begin{figure}[tb]
    \hspace*{-2.3cm}
       \includegraphics[width=1.3\linewidth]{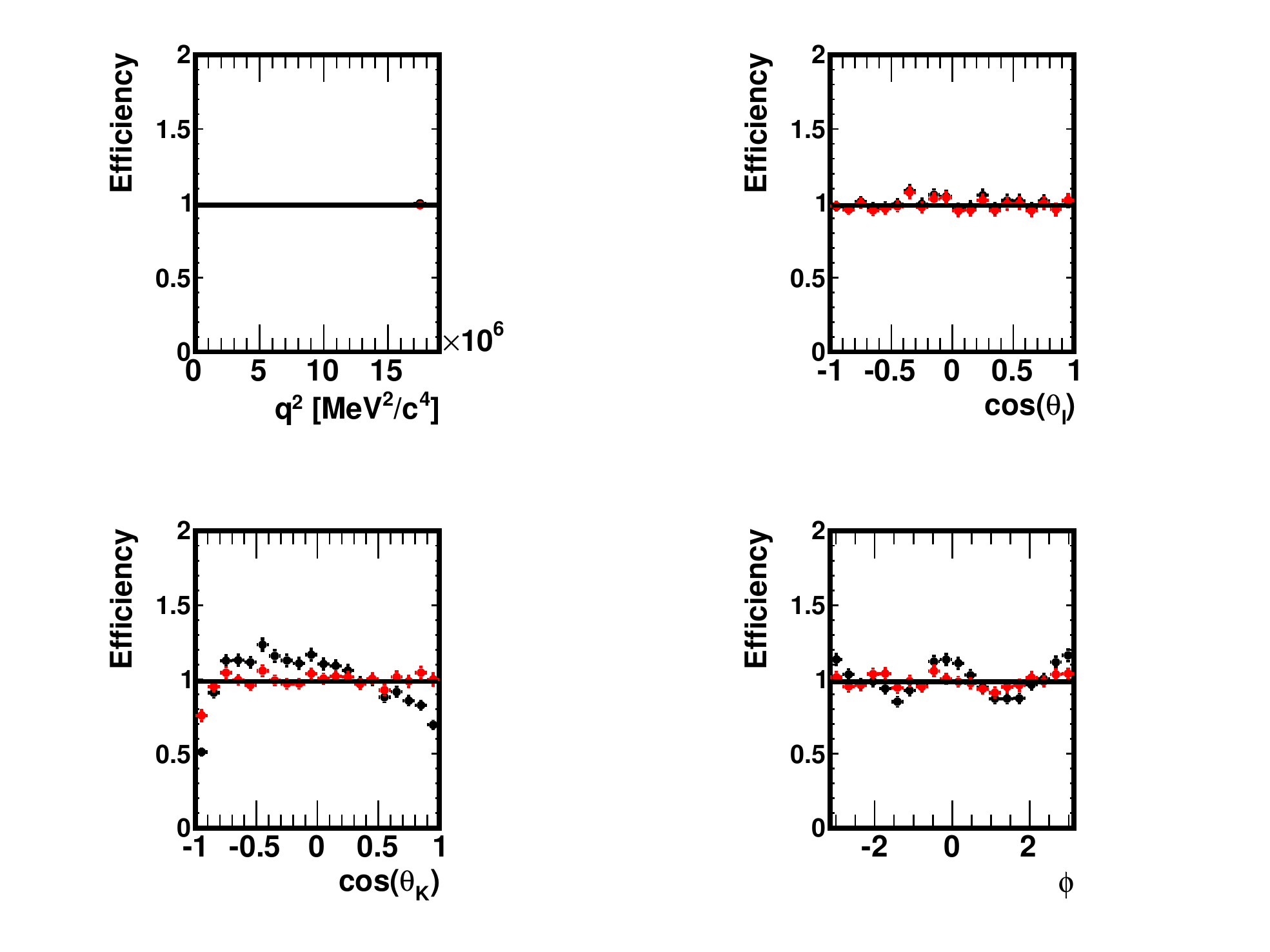}
    \vspace*{-0.5cm}
  \caption{Efficiency in $q^{2}$, cos$\theta_l$, cos$\theta_K$ and $\phi$ in the data generated for the present study, 
in the region $17.0 < q^2 < 18.0$ GeV$^{2}/c^4$. Shown are the ratios of the distributions found in the distorted and 
original samples, with no correction (black) and for decays re-weighted using $\omega_i$ weights (red)
as explained in Sect.~\ref{subsec:resultsB2KstMuMu}. The absolute normalisation is arbitrary when the correction is not applied and natural when it is applied (red).} 
  \label{fig:Ratio_q2_18}
\end{figure}

\clearpage

\begin{figure}[tb]
    \hspace*{-2.3cm}
       \includegraphics[width=1.3\linewidth]{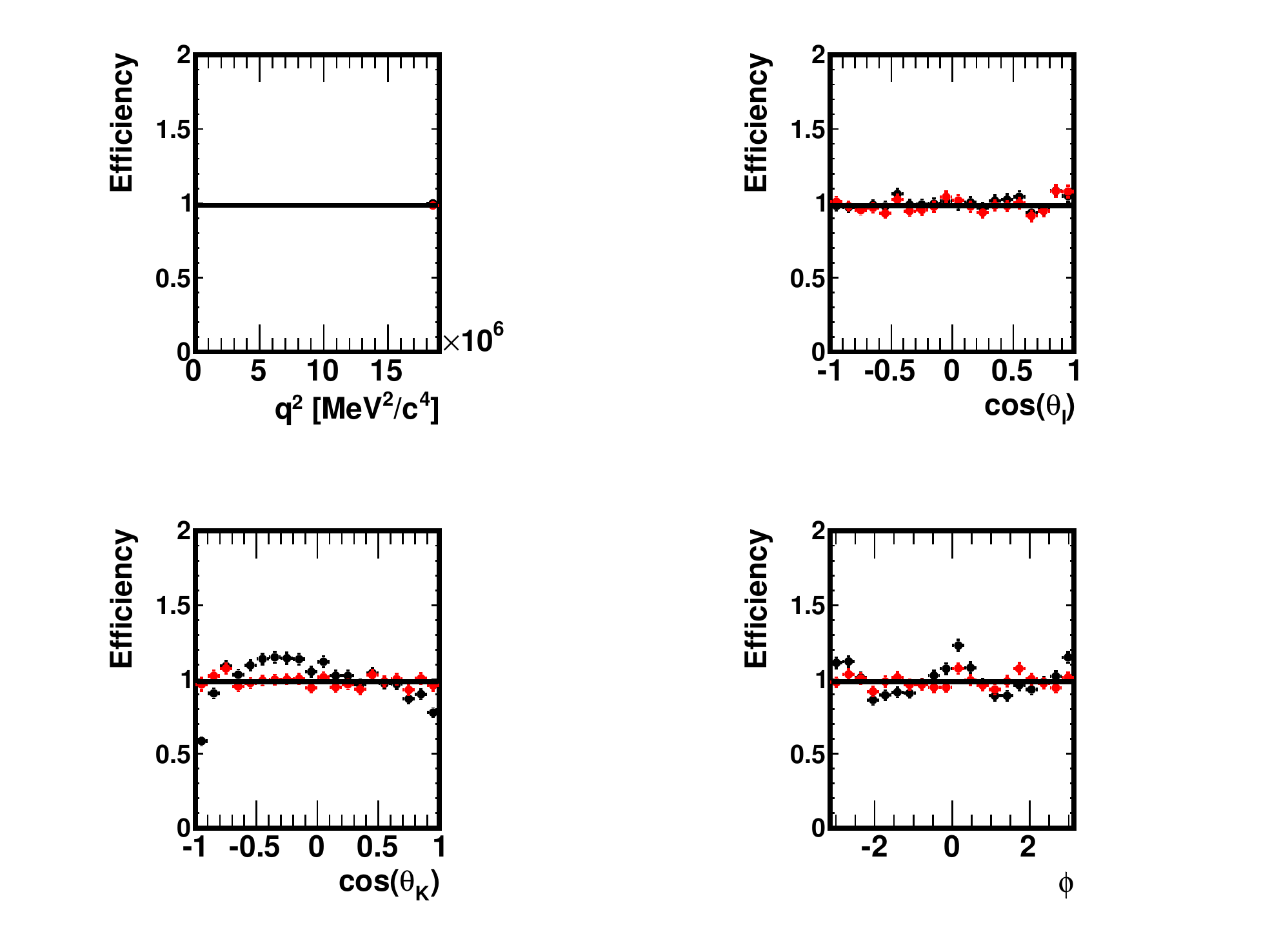}
    \vspace*{-0.5cm}
  \caption{Efficiency in $q^{2}$, cos$\theta_l$, cos$\theta_K$ and $\phi$ in the data generated for the present study, 
in the region $18.0 < q^2 < 19.0$ GeV$^{2}/c^4$. Shown are the ratios of the distributions found in the distorted and 
original samples, with no correction (black) and for decays re-weighted using $\omega_i$ weights (red)
as explained in Sect.~\ref{subsec:resultsB2KstMuMu}. The absolute normalisation is arbitrary when the correction is not applied and natural when it is applied (red).} 
  \label{fig:Ratio_q2_19}
\end{figure}

\end{appendices}


\end{document}